# ONTOLOGY-DRIVEN PROCESSING OF TRANSDISCIPLINARY DOMAIN KNOWLEDGE

## MONOGRAPH

Glushkov Institute of Cybernetics of the National Academy of Sciences of Ukraine
Taras Shevchenko National University of Kyiv

Oleksandr Palagin
Mykola Petrenko
Sergii Kryvyi
Mykola Boyko
Kyrylo Malakhov

**GLUSHKOV INSTITUTE OF CYBERNETICS**
**OF THE NATIONAL ACADEMY OF SCIENCES OF UKRAINE**

**TARAS SHEVCHENKO NATIONAL UNIVERSITY OF KYIV**

# ONTOLOGY-DRIVEN PROCESSING OF TRANSDISCIPLINARY DOMAIN KNOWLEDGE

**MONOGRAPH**
Scientific publication (issue)
Published in the author's edition

**Oleksandr Palagin**
Glushkov Institute of Cybernetics of the National Academy of Sciences of Ukraine

**Mykola Petrenko**
Glushkov Institute of Cybernetics of the National Academy of Sciences of Ukraine

**Sergii Kryvyi**
Taras Shevchenko National University of Kyiv

**Mykola Boyko**
Glushkov Institute of Cybernetics of the National Academy of Sciences of Ukraine

**Kyrylo Malakhov**
Glushkov Institute of Cybernetics of the National Academy of Sciences of Ukraine

**Team of Authors:**

***Oleksandr Palagin, Mykola Petrenko, Sergii Kryvyi, Mykola Boyko, Kyrylo Malakhov***

**Reviewers:**

***Volodymyr Shyrokov***, Academician of the National Academy of Sciences of Ukraine.

***Dmytro Lande***, Professor, Dr. of Sci, Institute for Information Recording of National Academy of Sciences of Ukraine.

The monograph discusses certain aspects of modern real-world problems facing humanity, which are much more challenging than scientific ones. Modern science is unable to solve them in a fundamental way. Vernadsky's noosphere thesis, in fact, appeals to the scientific worldview that needs to be built in a way that overcomes the interdisciplinary barriers and increases the effectiveness of interdisciplinary interaction and modern science overall. We are talking about the general transdisciplinary knowledge. In world practice, there is still no systematic methodology and a specific form of generally accepted valid scientific theory that would provide transdisciplinary knowledge. Non-linear interdisciplinary interaction is the standard of evolution of modern science. At the same time, a new transdisciplinary theory (domain of scientific research) is being de facto created and the process is repeated many times: from an individual or group of disciplines, through interdisciplinary interaction, in a direction that brings us closer to creating a holistic general scientific worldview.

The demand to create a rigorous methodology for transdisciplinary research has revealed the necessity to define the position and the role of informatics in system and technological support of transdisciplinary research and use of their results in addressing global problems of modern civilization. This conclusion is absolutely reasonable, given the system-forming role of modern informatics and the integration of information technologies into almost all Hi-Tech domains.

The global interpretation of transdisciplinarity does not exclude the utilitarian approach, i.e., the approach to effective interaction of scientific disciplines in the interests of creating a complete scientific worldview or a global system of scientific knowledge that allows to represent it in all its "diversity and complexity - spatial, temporal, informational and cognitive". This way leads to the creation of a systemology of transdisciplinary interaction as an independent domain of informatics, keeping in mind its already discussed function, as well as a set of information technologies for formulating and solving complex scientific and technical problems.

The transdisciplinary paradigm envisages the creation in the near future of a unified transdisciplinary knowledge system that provides a systematic formulation and solution of specific problems in the implementation of comprehensive projects of high complexity, social significance, conflict and competition.

# PREFACE

We are currently experiencing a significant shift from an information-based society to a knowledge-based one, driven by advanced information technologies that enable users to solve highly complex problems. These technologies are key to the rapid progress of modern civilization. However, it is critical to examine the benefits, costs, and potential risks associated with this process. How can we effectively balance long-term and short-term goals along this path? Is it possible to achieve harmony between social development and high-tech progress?

The pace of modern civilization's development exceeds even the most daring predictions of futurists and visionaries. Society and high technology are increasingly making mutual demands, which are generally justified. This interaction creates a feedback loop and the potential for reconciliation. However, achieving this harmony is not always straightforward, often due to resource constraints, lack of mutual respect, and inadequate self-esteem. Tensions persist, awaiting resolution of conflicts and problems.

A few decades ago, society had limited knowledge of computers, information technology, and virtual reality, yet it managed to function well. Today, it is difficult to imagine life without personal computers, tablets, mobile phones, or the Internet. While these technologies serve us admirably, they also consume a significant amount of our time, attention, and financial resources. People of all ages, including young children, are actively engaged in information interaction with modern technology.

The eminent Nobel Laureate Albert Schweitzer believed in the invincibility and profound purpose of human consciousness, despite the apparent hopelessness of our transient existence. He envisioned a powerful creed, " awe of life," which inspires admiration and underscores the importance of consciousness as the essence of life itself. It is consciousness that should ensure the continuity of life on Earth and put an end to the deliberate destruction and undermining of the world and the achievements of civilization. While it is easy to articulate such ideals, it is far more difficult to embody this creed and dedicate one's entire life to its pursuit, as the great A. Schweitzer did. Teaching this innate and compelling principle to humanity at large is an even greater challenge.

Yet faith in human consciousness and the fruitfulness of the credo of "reverence for life" is worth fighting for. It is for faith, not for our own lives, because we are talking about the future of our grandchildren and great-grandchildren. It is necessary to build human civilization as an automated system with controlled feedback, choosing the right criteria for the quality of the result of activity. This is not an easy challenge. Back in the Soviet era, the brilliant academician V.M. Glushkov proposed a nationwide automated system for managing the country's economy (NWAS). The NWAS was rejected, but its idea is alive and well and still fascinates with its versatility and transparency. At the same time, while the level of information technology in the 1960s was weak, today it is fully mature and ready to meet major challenges. However, in addition to economic subsystems, the NWAS should include subsystems of environmental, social, and political monitoring, i.e., everything that corresponds to today's ideas about the so-called "sustainable development" of society. How to choose the right criteria, how to direct the development of science and high technologies in the right direction?

Finally, we have reached the concept of transdisciplinarity and the role of computer science. The era of analyticism and its inherent differentiation of science is over. The real problems facing humanity today are much more complex than scientific problems.

Modern science is unable to solve them in a fundamental way. One of the reasons is the disunity of scientific disciplines and the lack of coordination of scientific teams in solving problems. There are many examples of this, but the main one is that the logic of material production and the associated utilization of natural resources have led to environmental degradation. How is it possible that everything was fine, and suddenly global problems arose? Not so well, and most importantly, not suddenly. Starting with natural philosophy, scientific differentiation over two and a half thousand years has led to the emergence of many scientific disciplines that are not related to each other, leaving a huge area of interdisciplinary space that is not covered by anything. Vernadsky's thesis about the noosphere, in fact, appeals to the scientific worldview that needs to be built in order to overcome interdisciplinary barriers and increase the effectiveness of interdisciplinary interaction and modern science in general.

We are talking about general transdisciplinary (TD) knowledge. In the world practice, there is still no systematic methodology and a specific form of generally accepted structural scientific theory that would ensure the acquisition of TD knowledge. It is about identifying new relationships between the concepts of the original scientific disciplines, establishing a new system of laws that interconnect them, and solving problems of system integration in the performance of increasingly complex tasks. At the same time, the knowledge of the original disciplines can remain unchanged, in the simplest case, included in whole or in part in the new hierarchical system of knowledge, or undergo modification through the processes of exchange of paradigmatic positions, concepts and methods of different sciences (non-linear case). Non-linear interdisciplinary interaction is the way modern science evolves. In this case, a new TD-theory (a domain of scientific research) is de facto formed and the process is repeated many times: from one or a group of disciplines, through interdisciplinary interaction, in a direction that brings us closer to creating a holistic general scientific worldview.

The need for a rigorous methodology in transdisciplinary research has underscored the importance of defining the role and place of informatics within the framework of TD research and the utilization of its outcomes to address the global challenges faced by modern civilization. This conclusion arises naturally, considering the pivotal role of modern informatics as a system-shaping force and the pervasive integration of information technology across various hi-tech domains.

The global perspective of transdisciplinarity does not dismiss the utilitarian approach, which emphasizes the effective collaboration of scientific disciplines to construct a comprehensive scientific understanding of the world or a global system of knowledge that encompasses its diverse and intricate aspects – be it spatial, temporal, informational, or cognitive. Achieving this goal necessitates the development of a transdisciplinary interaction systemology, which can be regarded as a distinct field of informatics. Such a systemology should account for its aforementioned function and encompass a suite of information technologies tailored to formulating and addressing complex scientific and technical problems.

The transdisciplinary paradigm envisions the establishment of a cohesive TD knowledge system in the foreseeable future. This system aims to offer a formalized framework for defining and addressing specific tasks within the context of comprehensive projects characterized by high complexity, social significance, conflict, and competition.

**ІНСТИТУТ КІБЕРНЕТИКИ ІМ. В.М. ГЛУШКОВА**
**НАЦІОНАЛЬНОЇ АКАДЕМІЇ НАУК УКРАЇНИ**

**КИЇВСЬКИЙ НАЦІОНАЛЬНИЙ УНИВЕРСИТЕТ**
**ІМЕНІ ТАРАСА ШЕВЧЕНКА**

# ОНТОЛОГІЧНЕ ОБРОБЛЕННЯ ТРАНСДИСЦИПЛІНАРНИХ ПРЕДМЕТНИХ ЗНАНЬ

**МОНОГРАФІЯ**
Наукове видання

**Палагін О.В.**
Інститут кібернетики імені В.М. Глушкова Національної академії наук України

**Петренко М.Г.**
Інститут кібернетики імені В.М. Глушкова Національної академії наук України

**Кривий С.Л.**
Київський Національний Університет імені Тараса Шевченка

**Бойко М.О.**
Інститут кібернетики імені В.М. Глушкова Національної академії наук України

**Малахов К.С.**
Інститут кібернетики імені В.М. Глушкова Національної академії наук України

**Колектив авторів**:

***Палагін О.В., Петренко М.Г., Кривий С.Л., Бойко М.О., Малахов К.С.***

**Рецензенти:**

***Широков В.А.***, академік НАН України.

***Ланде Д.В.***, професор, д.т.н., Інститут проблем реєстрації інформації НАН України.

У монографії розглянуті певні аспекти сучасних реальних проблем, які постають перед людством і значно складніші, ніж наукові задачі. Сучасна наука неспроможна їх кардинально вирішувати. Теза В.І. Вернадського про *ноосферу*, по суті, апелює до наукової картини світу, яку необхідно будувати для того, щоб подолати міждисциплінарні бар'єри та підвищити ефективність міждисциплінарної взаємодії та сучасної науки в цілому. Йдеться про загальне трансдисциплінарне знання. У світовій практиці поки що відсутня системна методологія та певна форма загальноприйнятої конструктивної наукової теорії, які б забезпечили отримання трансдисциплінарних знань. Нелінійна міждисциплінарна взаємодія – норма еволюції сучасної науки. При цьому де-факто формується нова трансдисциплінарна теорія (галузь наукових досліджень) і багаторазово повторюється процес: від однієї чи групи дисциплін, через міждисциплінарну взаємодію – у напрямку, що наближає до створення цілісної загальнонаукової картини світу.

Необхідність розроблення строгої методології трансдисциплінарних наукових досліджень виявила потребу у визначенні місця та ролі інформатики у системно-технологічній підтримці трансдисциплінарних досліджень та використанні їх результатів при вирішенні глобальних проблем розвитку сучасної цивілізації. Такий висновок є абсолютно закономірним, враховуючи системоутворюючу роль сучасної інформатики та інтеграцію інформаційних технологій практично у всі галузі HiTech.

Глобальне тлумачення трансдисциплінарності не заперечує утилітарного, тобто підходу до ефективної взаємодії наукових дисциплін в ім'я побудови повної наукової картини світу або глобальної системи наукових знань, що дозволяють відобразити її в усій «різноманітності та багатоскладності – просторовій, часовій, інформаційній та когнітивній». Цей шлях лежить через створення системології трансдисциплінарної взаємодії як самостійного розділу інформатики, маючи на увазі вже згадану її функцію, а також сукупність інформаційних технологій постановки та вирішення складних науково-технічних проблем.

Трансдисциплінарна парадигма передбачає побудову в найближчому майбутньому єдиної ТД-системи знань, яка забезпечує формалізовану постановку та вирішення конкретних задач при виконанні комплексних проектів високої складності, соціальної значущості, конфліктності та конкурентності.

# ЗМІСТ

## Перелік умовних скорочень

АЗ – Апаратні засоби
АЛП – Апаратний лінгвістичний процесор
АМП – Апаратний морфологічний процесор
БД – База даних
ГНД – Галузь наукових досліджень
ЕС – Елементарний сенс
ІЗ – Інженер зі знань
ІКС – Інтелектуальна комп'ютерна система
ЛКТ – Лінгвістичний корпус текстів
ЛМ – Локальна мережа
МА – Морфологічний аналіз
МД – Міждисциплінарний
МНД – Міждисциплінарні наукові дослідження
МОКС – Мовно-онтологічна картина світу
НД – Наукові дослідження
НДР – Науково-дослідна робота
НКС – Наукова картина світу
НП – Наукова публікація
НР – Науковий співробітник
ОнС БДНП – Онтологічна система оброблення баз даних наукових публікацій
ОНТ – Об'єкт нової техніки
ПБЗ – Персональна база знань
ПдГ – Предметна галузь
ПдО – Предметна область
ПК – персональний комп'ютер
ПЛІС – Програмовні логікові інтегральні схеми
ПМ – Природна мова
ПМО – Природномовний об'єкт
ПМТ – Природномовний текст
ПрП – Проблемний простір
РСДП – Системи дослідницького проєктування, що розвиваються
СА – Семантичний аналіз
САПР – Система автоматизованого проєктування
СДП – Системи дослідницького проєктування
СМ – Семантична мережа

СмС – Семантична структура
СнА – Синтаксичний аналіз
СОЗ – Сервіс-орієнтовані знання
СООУ-ІКС – Сервіс-орієнтовані онтолого-керовані інтелектуальні комп'ютерні системи
ТД – Трансдисциплінарний
ТЗ – Технічне завдання
ТОДОС – Трансдисциплінарні Онтологічні Діалоги Об'єктно-орієнтованих Систем
ТР – Технічне рішення
ТСІ – Технологія системної інтеграції
ШІ – Штучний інтелект
API – Application Programming Interface
CRF – Concept Relation Interpretation Function
IRI – Internationalized Resource Identifiers
NBIC-кластер – (N – нано, B – био, I – инфо, C – когнито)
OWL – Web Ontology Language
RDF – Resource Description Framework
RDFS – RDF Vocabulary Description Language
SPARQL – SPARQL Protocol And RDF Query Language
SW – Semantic Web
XML – EXtensible Markup Language
UML – Universal Modelling Language
URI – Uniform Resource Identifier
URL – Universal Modelling Language

# Вступ

Ми живемо в епоху переходу від інформаційного до знання-орієнтованого суспільства, основою якого є досконалі інформаційні технології, що надають користувачеві можливості вирішення задач найвищої складності. Саме такі технології забезпечують стрімкий прогрес сучасної цивілізації. Які переваги та витрати, а може, і небезпеки цього процесу? Як поєднувати на цьому шляху далекі та ближні цілі? Зрештою, чи можна знайти гармонію у розвитку суспільства та наукомістких технологій? Темпи розвитку сучасної цивілізації випереджають найсміливіші припущення футурологів та провісників. Суспільство і високі технології висувають дедалі більше взаємних, взагалі кажучи, законних претензій одне до одного, створюючи зворотний зв'язок і передумови їхнього узгодження. Не завжди це вдається, часом бракує ресурсу, взаємоповаги та належної самооцінки. Залишається напруження в очікуванні вирішення конфліктів та проблем.

Ще якихось кілька десятків років тому суспільство уявлення не мало про комп'ютери, інформаційні технології та віртуальну реальність. І, між іншим, непогано обходилося без них. Сьогодні людина не може уявити своє життя без персонального комп'ютера, планшета, мобільного телефону, Інтернету. Вони вірно служать, однак відволікаючи при цьому на себе солідну частину нашого часу, уваги та фінансових витрат. В орбіту активної інформаційної взаємодії із сучасною інформаційною технікою включилися як дорослі, і навіть першокласники і дошкільнята.

Великий нобелівець Альберт Швейцер, незважаючи на всю безвихідь примарного сьогодні, вірив у непереможність і велику місію людської свідомості, яку, зрештою, оцінить могутнє кредо, яке зачаровує своїм захопленням і очевидністю – «благоговіння перед життям». Адже свідомість – це продукт життя, його квінтесенція. Що ж, як не вона, має подбати про безперервність життя на планеті Земля, покласти край цілеспрямованій смерті, руйнації навколишнього світу та поточних досягнень цивілізації? Легко сказати, набагато важче втілити це кредо, втілити в непохитні установки, присвятивши цьому завданню все своє життя, як це зробив великий А. Швейцер. Ще важче навчити цьому природному і привабливому принципу нерозумне людство.

І все ж віра в людську свідомість і плідність кредо «благоговіння перед життям» варте того, щоб за неї боротися. Саме за віру, а не за власне життя, адже йдеться про майбутнє наших онуків та правнуків. Необхідно будувати людську цивілізацію як автоматизовану систему з керованим зворотним зв'язком, правильно вибравши критерії якості результату діяльності. Завдання не просте. Ще геніальний академік В.М. Глушков за часів радянського ладу запропонував загальнодержавну автоматизовану систему управління економікою країни (ЗДАС). ЗДАС відкинули, але її ідея жива і, як і раніше, захоплює своєю універсальністю та прозорістю. При цьому, якщо рівень інформаційних технологій у далекі 60-ті був слабкуватий, то сьогодні він цілком дозрів і готовий до здійснення великих викликів. Щоправда, крім економічних підсистем в ЗДАС мають увійти підсистеми екологічного, соціального, політичного моніторингу, тобто все те, що відповідає сьогоднішнім уявленням про так званий «стабільний розвиток» суспільства. Як коректно вибрати згадані критерії, як спрямувати розвиток науки та високих технологій у потрібне русло?

Нарешті ми підійшли до поняття трансдисциплінарності та ролі інформатики. Епоха аналітизму та властива йому диференціація науки завершена. Реальні сьогоднішні проблеми, що стоять перед людством, набагато складніші наукових проблем.

Сучасна наука неспроможна їх кардинально вирішувати. Однією з причин є роз'єднаність наукових дисциплін, нескоординованість роботи наукових колективів над комплексним вирішенням проблем. Прикладів тому – дуже багато, головний – один: логіка розвитку матеріального виробництва та пов'язана з ним утилізація природних ресурсів призвели до екологічної деградації навколишнього середовища. Як же так: все було гаразд, і раптом виникли глобальні проблеми? Не так добре, а головне – не раптом. Починаючи з натурфілософії, наукова диференціація за дві з половиною тисячі років призвела до появи множини наукових дисциплін, не пов'язаних між собою, залишивши величезний ареал міждисциплінарного простору, нічим не заповненого. Теза В.І. Вернадського про *ноосферу*, по суті, апелює до наукової картини світу, яку необхідно будувати для того, щоб подолати міждисциплінарні бар'єри та підвищити ефективність міждисциплінарної взаємодії та сучасної науки в цілому.

Йдеться про загальне трансдисциплінарне (ТД) знання. У світовій практиці поки що відсутня системна методологія та певна форма загальноприйнятої конструктивної наукової теорії, які б забезпечили отримання ТД-знань. Йдеться про виявлення нових відношень між поняттями вихідних наукових дисциплін, встановлення нової системи законів, які їх пов'язують, вирішення задач системної інтеграції при виконанні все більш складних завдань. При цьому знання вихідних дисциплін можуть залишатися незмінними, в найпростішому випадку включеними цілком або частково в нову ієрархічну систему знань, або зазнати модифікації завдяки процесам обміну парадигмальними положеннями, поняттями та методами різних наук (нелінійний випадок). Нелінійна міждисциплінарна взаємодія – норма еволюції сучасної науки.

При цьому де-факто формується нова ТД-теорія (галузь наукових досліджень) і багаторазово повторюється процес: від однієї чи групи дисциплін, через міждисциплінарну взаємодію – у напрямку, що наближає до створення цілісної загальнонаукової картини світу.

Необхідність розробки строгої методології ТД-наукових досліджень виявила потребу у визначенні місця та ролі інформатики у системно-технологічній підтримці ТД-досліджень та використанні їх результатів при вирішенні глобальних проблем розвитку сучасної цивілізації. Такий висновок є абсолютно закономірним, враховуючи системоутворюючу роль сучасної інформатики та інтеграцію інформаційних технологій практично у всі галузі Hi Tech.

Глобальне тлумачення трансдисциплінарності не заперечує утилітарного, тобто підходу до ефективної взаємодії наукових дисциплін в ім'я побудови повної наукової картини світу або глобальної системи наукових знань, що дозволяють відобразити її в усій «різноманітності та багатоскладності – просторовій, часовій, інформаційній та когнітивній». Цей шлях лежить через створення системології трансдисциплінарної взаємодії як самостійного розділу інформатики, маючи на увазі вже згадану її функцію, а також сукупність інформаційних технологій постановки та вирішення складних науково-технічних проблем.

ТД-парадигма передбачає побудову в найближчому майбутньому єдиної ТД-системи знань, яка забезпечує формалізовану постановку та вирішення конкретних задач при виконанні комплексних проектів високої складності, соціальної значущості, конфліктності та конкурентності.

# Розділ 1. Системологічні засади трансдисциплінарності

## 1.1. **Розвиток і становлення трансдисциплінарності**

Важливим фактором, що визначає місце і роль науки в житті суспільства, є те, що вона поки, на жаль, не має цілісної картини світу. У зв'язку з цим їй властиві часткові проекції буття під кутом зору приватних наукових дисциплін, які осягають ті чи інші закони природи в залежності від проблемної орієнтації досліджень.

Водночас епоха аналітизму та властива їй диференціація науки та замкнутих наукових теорій уже позаду. Стало очевидним, що реальні проблеми, які стоять перед людським суспільством, набагато складніші за наукові, і наука не в змозі їх кардинально вирішити внаслідок роз'єднаності наукових дисциплін та їх спеціалізації, слабкої координації наукових колективів та їх тематики, відсутністю системного моніторингу та загальної формалізованої мови подання знань.

Однією з головних завдань трансдисциплінарних досліджень (ТД-досліджень) є забезпечення ефективної ТД-взаємодії на всіх етапах життєвого циклу вирішення фундаментальних і прикладних наукових проблем. Окрім завдань інфраструктурної підтримки ТД-досліджень на перший план виходять завдання їх методологічного супроводу та забезпечення процесів інтеграції, конвергенції та уніфікованого формалізованого представлення ТД-знань та операцій над ними. Не останню роль відіграє системологічна підготовка навичок та розширення діапазону світогляду ТД-дослідників. Тут доречно говорити про *етос* трансдисциплінарності як сутнісному феномену і понятійній метафорі, як загальній основі взаєморозуміння представників різних наукових дисциплін. З точки зору ТД-проблематики чітко помітні вузькість та неефективність існуючого підходу до підготовки кадрів вищої кваліфікації, формування структури та переліку спеціальностей у вузах, управлінню знаннями загалом.

Трансдисциплінарність як поняття, апелює до загальної наукової картини світу, яка відображає реальний світ у всьому його різноманітті і багатозв'язності – просторової, часової, інформаційної та когнітивної [1, 2]. Виходячи з принципу невичерпності матерії, відобразити у всій «вичерпній» повноті (з боку спостерігача) таке різноманіття неможливо навіть теоретично. Але природознавство і, зокрема, її найбільш просунута дисципліна – фізика давно перехворіли на механічний «лапласівський» детермінізм з його причинно-наслідковим методологічним принципом, а точніше претензією на загальну універсальність [3]. При вирішенні наукових та практичних проблем аналізу та синтезу складних систем так чи інакше доводиться долати зазначені протиріччя за рахунок втрати інформації шляхом спрощення постановки зазначених проблем, використання ймовірнісних підходів, встановлення принципів міждисциплінарної взаємодії у вигляді узгодження понять, методів досліджень та інтерпретації їх результатів. Таким чином, шлях до трансдисциплінарності лежить через створення *системології міждисциплінарної взаємодії* (у світлі еволюції наукових теорій) як самостійної галузі знань чи окремого розділу інформатики, маючи на увазі її системоутворюючу функцію. Тим більше, що інформатика володіє, крім чіткого математичного базису, також і технологіями постановки та вирішення складних науково-технічних проблем.

Таким чином, сутність трансдисциплінарного підходу до дослідження комплексних науково-технічних проблем полягає в ефективному забезпеченні двоєдності концепцій поглиблення конкретних знань у предметній області (ПдО), з одного боку, та розширення охоплення проблеми, виходячи з реальності єдності світу, та прагнення відтворити цілісну наукову картину світу (НКС) – з іншого. Його реалізація полягає у з'ясуванні нових закономірностей за результатами системної інтеграції вихідних наукових теорій шляхом обміну поняттями та методами різних наук, формуванні нових понять, категорій, нових наукових теорій, що узагальнюють вихідні та розширюють діапазон трансдисциплінарності у напрямку побудови глобальної інтегрованої системи знань, яка не просто фіксує наукову картину світу, але й є активним середовищем, що забезпечує вирішення конкретних науково-технічних задач (шляхом занурення в неї формалізованих задач) та розвиток самої системи знань.

### 1.1.1 *Визначення і сутність понять, пов'язаних з дисциплінарними дослідженнями*

З точки зору класифікації наукових підходів корисно вибрати такий критерій як ступінь повноти пізнання навколишнього світу. Тоді всі підходи можна звести до чотирьох основних видів: дисциплінарний, міждисциплінарний (МД), мультидисциплінарний і трансдисциплінарний.

Під міждисциплінарністю розуміється інтеграція наукових областей через взаємне проникнення загальних понять. Термін "мультидисциплінарність" означає таке дослідження, коли об'єкт, процес чи явище вивчається одночасно у різних аспектах одразу кількома дисциплінами. "Трансдисциплінарні дослідження характеризуються перенесенням когнітивних схем з однієї дисциплінарної області в іншу" [4–6].

Нами зазначені терміни розуміються залежно від "розподілу" понять та наукових дисциплін за онтологічними рівнями ієрархії, що передбачає різні схеми їх взаємодії. Такий розподіл є суттєвим при розгляді системології ТД-взаємодії, системної інтеграції предметних дисциплін та формуванні "кластерів конвергенції" при виконанні трансдисциплінарних проєктів та їх інформаційно-технологічній підтримці.

Такі поняття як Ноосфера, Об'єкт, Процес, Система, Інформація, Природа, Суспільство, Людина, Предметна галузь, Наука, Наукова діяльність, Наукова картина світу, Техніка, Технологія тощо відносяться до рівня категорій. Такі поняття як Філософія, Фізика, Математика, Біологія, Хімія, Медицина, Гуманітарні та Соціальні науки, Інформатика, Нанотехнології тощо відносяться до рівня доменів наукових дисциплін. Множина наукових дисциплін, які є напрямками, розділами та підрозділами доменів, відноситься до рівня наукових дисциплін. До цього рівня також належать і так звані "подвійні" дисципліни, наприклад біоінформатика, фізична хімія та складніші їх комбінації.

Сказане пояснюється схемою, показаною на рис. 1.

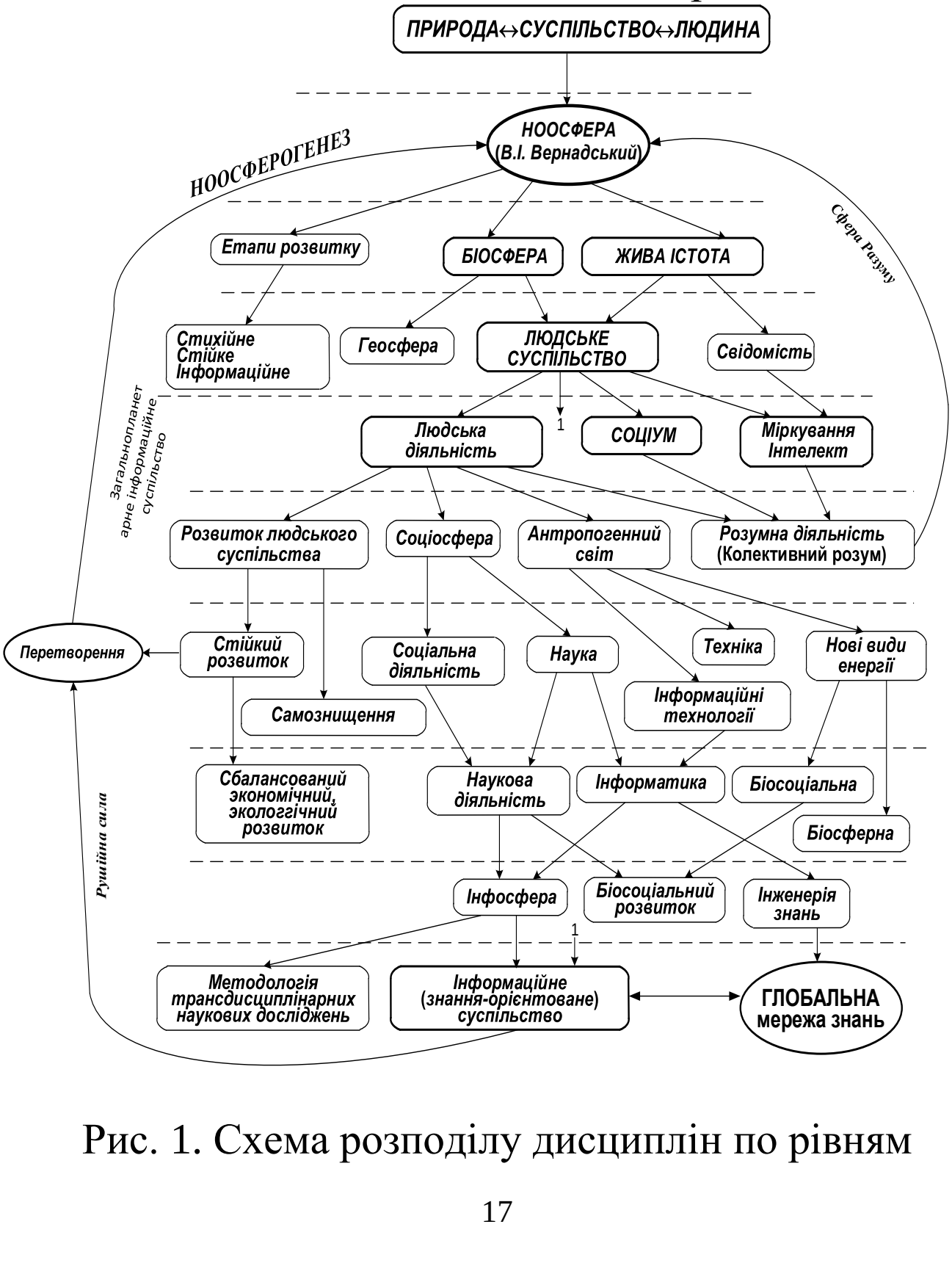

Рис. 1. Схема розподілу дисциплін по рівням

Дисциплінарність дозволяє науці поступово розвиватися у межах предметних напрямів, а дисциплінарний підхід ділить навколишній світ на окремі предметні області. Якщо розв'язання задачі чи проблеми виходить за межі можливостей дисциплінарних підходів, то прийнято вважати, що воно знаходиться "на стику наукових дисциплін". В процесі поступального розвитку дисциплінарного підходу виникає протилежність, що з одного боку, зумовлює накопичення дисциплінарних знань, а з іншого – встановлення природного обмеження повноти пізнання навколишнього світу. Вихід із становища позначений такою тезою: "якщо не можливо вийти межі дисциплінарного напрямку, то можливо розширити область застосування дисциплінарної методології" [7]. У свою чергу розширення сфери застосування дисциплінарної методології призвело до появи міждисциплінарних та мультидисциплінарних наукових підходів, які склали такі рівні класифікації наукових підходів. Процес розвитку цих підходів призвів до того, що метафора "стик дисциплін" поступово набула вигляду міждисциплінарних і мультидисциплінарних напрямів, кожен з яких має свої особливості вирішення завдань, що стоять перед ним.

Міждисциплінарність передбачає інтеграцію кількох наукових дисциплін. Одна з них відіграє роль провідної та результати МД-дослідження завжди трактуються цією провідною дисципліною. Особливість МД-підходу полягає в тому, що він допускає пряме перенесення методів дослідження з однієї наукової дисципліни в іншу. Перенесення методів у цьому випадку обумовлений виявленням подібностей досліджуваних предметних областей. МД-підхід призначений, перш за все, для вирішення конкретних дисциплінарних проблем, у вирішенні яких будь-яка конкретна дисципліна має концептуальні та методологічні труднощі.

Синергетична парадигма як розділ МД-підходу в ієрархії знань займає особливе місце. З одного боку, вона апелює до цілісності та інтегрального представлення, системно визначаючи ефекти взаємодії об'єктів, процесів та суб'єктів, а з іншого, – акцентує увагу на нелінійності, нестійкості та появу атракторів, що змінюють у результаті багаторівневу організацію та поведінку системи. В обох випадках вона виражається сукупністю формальних моделей самоорганізації та спрямована на відтворення наукової картини світу, що особливо важливо під час переходу до трансдисциплінарного підходу у наукових дослідженнях та втілення парадигми глобального еволюціонізму. Наукова картина світу при цьому може бути як ТД-онтологія, що увібрала у собі як онтології окремих дисциплін, так і методи останніх, включаючи варіанти їх перехресного впливу. Трансдисциплінарність дозволяє побудувати єдину ТД-методологію аналізу і синтезу, включивши її до загальнонаукової картини світу. Докладніше проблеми синергетики розглянуті в [8].

Сказане відповідно до схеми, представленої на рис. 1, можна інтерпретувати в такий спосіб. МД-підхід стосується рівня наукових дисциплін та зв'язків рівня доменів наукових дисциплін, наприклад "Штучний інтелект → Обчислювальна техніка → Програмування → Інформатика → Медицина → Медична інформатика".

Мультидисциплінарність не передбачає перенесення методів дослідження з однієї дисципліни в іншу. Усі дисципліни зберігають свою предметну спрямованість.

При мультидисциплінарному підході дослідники формують узагальнену картину предмета дослідження, стосовно якої її дисциплінарні картини постають як її рівноправні частини. Накопичення результатів мультидисциплінарних досліджень у подібних областях дисциплінарних знань призводить до появи нових мультидисциплінарних дисциплін, наприклад, таких як фізико-хімічна біологія, екологія тощо. Своє практичне застосування мультидисциплінарний підхід знайшов, зокрема, у роботі експертних груп.

ТД системний підхід використовує знання, сформовані та накопичені дисциплінарними, міждисциплінарними та мультидисциплінарними підходами. Трансдисциплінарність має забезпечувати координацію та інтеграцію дисциплінарних знань на основі єдиного аксіоматичного підходу (генеральних систем трансдисциплінарності). Так спочатку уявляли собі трансдисциплінарність Ж. Піаже та Е. Янч [9–12].

Трансдисциплінарність – це дослідницька стратегія, яка перетинає дисциплінарні кордони і розвиває холістичне (пріоритетний розгляд цілого по відношенню до його частин) бачення. Трансдисциплінарність у вузькому значенні означає інтеграцію різних форм і методів дослідження, включаючи спеціальні прийоми наукового пізнання для вирішення наукових проблем. Трансдисциплінарність у широкому сенсі означає єдність знання поза конкретних дисциплін.

Кожна наукова дисципліна має у своїй основі блок базових понять, що відповідає принципу їх необхідності та достатності. В [10] для трансдисциплінарності пропонуються такі базові поняття: простір, інформація, час, система, функція, енергія, розвиток і мета. Запропоновані ТД базові поняття є ядром категоріального рівня понять – основи онтологій верхнього рівня, що використовуються при постановці проблем та формуванні складних ТД-проєктів.

Якісні характеристики розглянутих вище наукових підходів представлені у табл. 1.1.

Зазначимо зміни у структурі науки, зумовлені трансформацією дисциплінарно організованої науки у ТД-дослідження: виділення наступних ознак постнекласичного етапу, зміна характеру наукової діяльності, обумовлена революцією у засобах отримання та зберігання знань (комп’ютеризація науки, зрощення науки з промисловим виробництвом) тощо); підвищення значення економічних та соціально-політичних факторів та цілей; зміна самого об’єкта дослідження – відкриті системи, що саморозвиваються (“людинорозмірні” об’єкти, прикладом яких є об’єкти біотехнологій, екологічні системи, біосфера тощо) [13–15].

Табл. 1.1 Якісні характеристики наукових підходів

| **Найменування підходу** | **Проблема** | **Якісні показники** | | | | |
|---|---|---|---|---|---|---|
| | | Використання методик і методів | Інтеграція методик і методів | Феномен | Влас. мова | Області дослідження |
| Дисциплінарний | Внутрішньо-дисциплінарна | Свої | – | Всередині конкретної ПдО | + | Конкретна ПдО |
| Міждисциплінарний | Декількох ПдО | Запозичені з інших ПдО | + | Всередині науковий | – | Дві і більше ПдО |
| Мультидисциплінарний | Декількох ПдО | Об’єднання між дисциплінарних | + | Поза конкретних ПдО | – | Об’єднання міждисциплінарних ПдО |
| Трансдисциплінарний | Багатофакторна Природи і Суспільства | Функціональний синтез методологій різних ПдО | + (найбільш сильна) | Погранична сфера науки і буття | – | Дисципліни природної і соціальної сфер, Людина. Кооперація пізнавальної і інноваційної діяльності |

ТД-дослідження, захоплюючи зони прикордонних (демаркаційних) ареалів наукових дисциплін, інтегрують сутнісні основи останніх, утворюючи кластери конвергенції, в яких відбувається потужна синергетична взаємодія за рахунок взаємопроникнення парадигм і конкретних поточних результатів кожної з дисциплін у той чи інший кластер. Зазначена взаємодія відбиває цілісність реального світу, будучи стимулом і водночас гарантією успішності ТД-досліджень, пов'язаних із нею практичних проєктів, нетривіальності і значимості їх результатів [16].

Важливим кроком у напрямку трансдисциплінарності, як згадувалося вище, є формування перспективних самодостатніх кластерів трансдисциплінарних досліджень, які забезпечують врахування наслідків (ризиків) та взаємний вплив основних факторів та зворотних зв'язків у процесі теоретичного аналізу, цілеспрямованих фізичних експериментів та реалізації глобальних системних проєктів сталого розвитку людського суспільства, збереження довкілля, розвитку науки загалом тощо. [17].

Яскравим прикладом технонауки може слугувати кластер NBIC-конвергенції (N – нано, B – біо, I – інфо, C – когно) та МІІЗ-кластер, що перетинається з ним (М – мова, І – інформація, І – інтелект, З – знання). Інформатика привносить у ці кластери як системоутворюючу, так і комп'ютерно-технологічну компоненти. Головні проривні напрямки у зазначених кластерах, це: стирання граней між живими та неживими системами, наноробототехніка з її численними додатками, глобальні суперкомп'ютерні агломерації з високим рівнем штучного інтелекту. До них слід додати єдину розподілену ТД-систему знань як глобально-комунікаційний варіант загальнонаукової картини світу та наступний етап розвитку існуючих Інтернет та Semantic Web.

Передбачається, що ТД-дослідження мають забезпечити: ефективну ТД-взаємодію на всіх етапах життєвого циклу вирішення фундаментальних та прикладних наукових проблем; методологічний супровід та забезпечення процесів інтеграції, конвергенції та уніфікованого формалізованого подання ТД-знань та операцій над ними; загальну основу взаєморозуміння представників різних наукових дисциплін; включення людини у внутрішньонауковий контекст спочатку через врахування параметрів спостереження (фізика мікросвіту), потім через включення соціальних та гуманітарних факторів, що призведе до зближення природничо-гуманітарного знання; проходження технологічними проєктами соціальної експертизи [16-21].

Таким чином, ТД-дослідження – це якісно новий етап інтегрованості науки та суспільства. Для його завершення науковому співтовариству належить ще багато чого розробити, зокрема:

– загальнонаукову картину світу, включаючи предметні дисципліни, та відповідну їй глобальну мережу ТД-знань;

– метатеорію та метамову трансдисциплінарності, одним із претендентів на таку мову є метамова нормальних форм знань, описана в [22, 23];

– системологію ТД-взаємодії, образно-понятійний апарат та моделі, можливості яких дозволяли б, з одного боку, охопити всі фактори, що формують складну проблему та впливають на неї, а з іншого боку, – виявити та врахувати механізми, за допомогою яких здійснюється цей вплив;

– метод (або сукупність методів) системного дослідження, що забезпечує доступ до всієї дисциплінарної інформації та її аналізу, зрозумілий та доступний фахівцям будь-якої наукової дисципліни;

– перспективні та самодостатні кластери конвергенції, що становлять ядро шостого технологічного укладу [21];

– методи проведення експериментів, що дозволяють вивчати багатофакторний вплив об’єктів пізнання та оцінки їх результатів;

– способи постановки та вирішення складних багатофакторних проблем у науці, техніці та технології.

Перелічені задачі є узагальненими і, своєю чергою, включають ряд підзадач.

### 1.1.2 ***Рух знань***

Процес розвитку знань у будь-якій предметній галузі та загалом у науці можна охарактеризувати параметрами, похідними від простору та часу, і показаний на рис. 2.

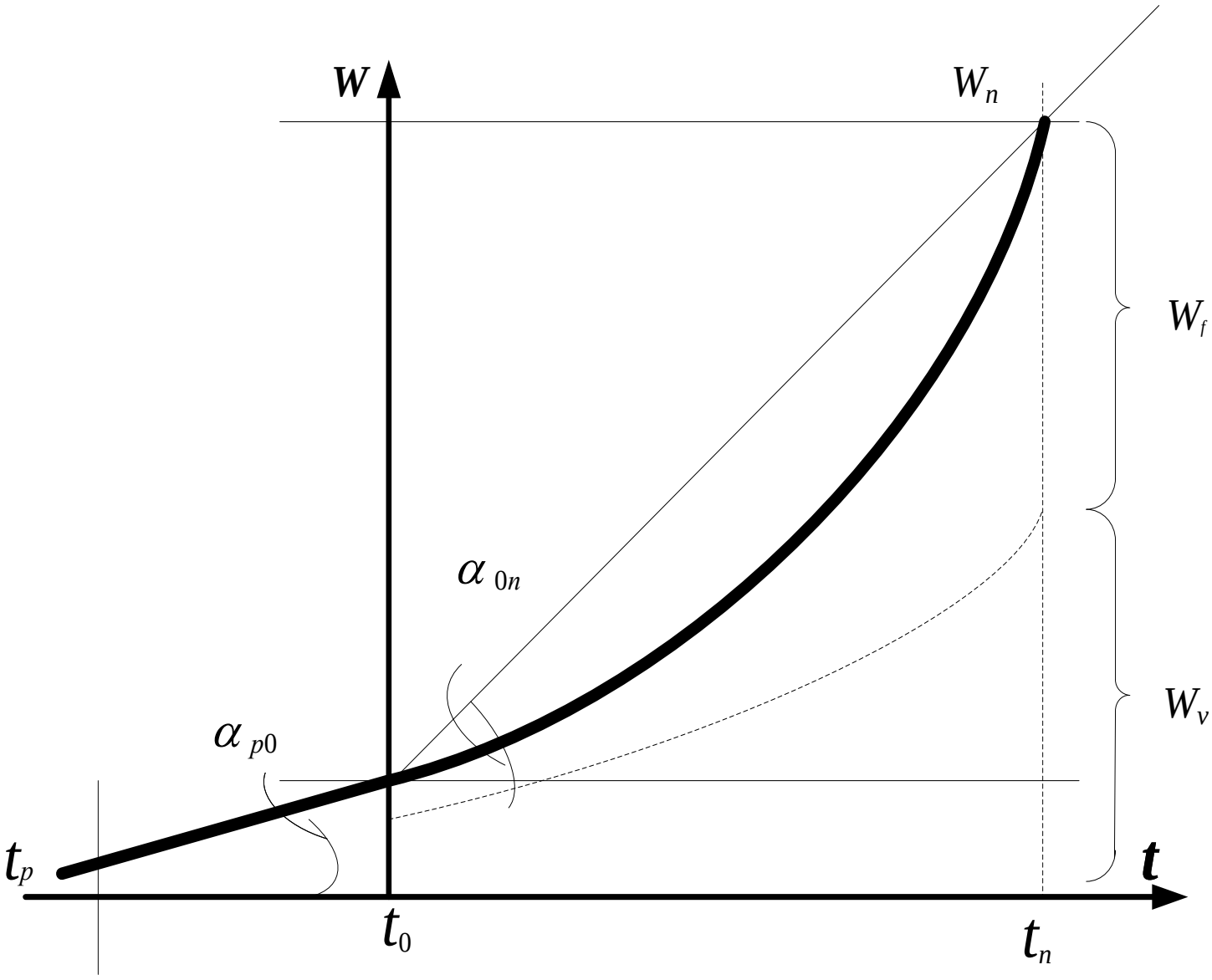

Рис. 2. Залежність $W = W(t)$

де $W$ – об'єм знань, $W(t)$ – швидкість зростання об'єму знань.

Очевидним фактом є те, що швидкість зростання об'єму знань у ранньому періоді $(t_p > t < t_0)$ розвитку науки загалом чи будь-якої її предметної області значно відставала від сучасної, $V_p < V_n$. Тут $t < t_p$ – період зародження, $t_0$ – поточний час, $t_0 > t < t_n$ – прогнозний період.

З точністю до похибки лінеаризації співвідношення між середніми швидкостями зростання об'єму знань у періоди $t_p \div t_0$ та $t_0 \div t_n$ кількісно визначається нерівністю $\tan\alpha_{0n} > \tan\alpha_{p0}$.

Об'єм знань та об'єм інформації, необхідної для їх подання $I(W)$, пропорційні між собою, але не рівні. Співвідношення між ними в будь-який момент часу значною мірою визначається співвідношенням між вербальною $(W_v)$ і формалізованою $(W_f)$ компонентами: $W = W_v + W_f$.

Під формалізованим поданням знань розуміється аналітична, таблична та графічна форми. Очевидно, що формалізоване подання компактніше за вербальне і більш придатне для оперативного оброблення. У часі відбуваються три процеси:

– збільшення сумарного об'єму знань (і інформації, необхідної для їх подання),
– збільшення формалізованої компоненти знань,
– заміна старих знань новими знаннями.

Співвідношення $K_f = \dfrac{W_f}{W}$ визначає ступінь формалізації знань і залежить від предметної області та моменту часу $t$, що розглядається. Сумарний об'єм інформації $I(W)$, який представляє знання в даній предметній області, на певному відрізку часу може стабілізуватися за рахунок процесу перекладу вербальної компоненти знань у компактне формалізоване подання зі збереженням змістовності знань як таких. Цей процес частково компенсує поточний приріст об'єму знань та інформації.

Як зазначалося вище, істотною відмінністю у розвитку загальних знань є наявність двох протилежних тенденцій: диференціації дисциплін наукового знання та його інтеграції. Якщо ранній період $(t_p \div t_0)$ мала місце лише перша з них (йшов процес утворення нових дисциплін), то сучасний період $\left(t_0 \div t_n\right)$ характеризується присутністю обох тенденцій з явною інтенсифікацією процесів інтеграції наукових дисциплін. Вони пов'язані зі створенням багатопрофільних дослідницьких організацій та народженням згаданих трансдисциплінарних кластерів конвергенції як у розвитку відповідних теорій, так і в реалізації складних науково-технічних проєктів.

Уніфікація та конструктивізація знань та їх представлень – один із найбільш складних та важливих етапів формулювання трансдисциплінарних проблем та проведення ТД-досліджень [24, 25]. Це зумовлено не тільки складністю, великою розмірністю задач та даних, надзвичайним обсягом системно-аналітичної складової на тлі міждисциплінарних відмінностей (диференціації та замкнутості сучасних наукових теорій), а й відсутністю методології трансдисциплінарної взаємодії. Головне – у нетривіальності самого переходу від прямих методів наукових досліджень до досліджень на основі професійного управління знаннями (knowledge management) та, звичайно ж, засобів їхньої методологічної та інформаційно-технологічної підтримки.

У цьому плані під час постановки ТД-досліджень велику роль відіграють метазнання. Так подання будь-якого об'єкта $(x)$ може бути полегшена тим, що будь-який інший об'єкт $(y)$, який часто подається, і що відповідно до [26] можливе визначення умовної складності $K_S(x \mid y) = \min l(s)$, де $s$ – спосіб подання, $l(s)$ – довжина опису об'єкта. Якщо умовна складність $K_S(x \mid y)$ значно менша, ніж безумовна складність $K_S(x)$, то природно припустити, що в об'єкті $(y)$ міститься деяка інформація (знання) про об'єкт $(x)$.

Різниця:

$$I_S(x \mid y) = K_S(x) - K_S(x \mid y) \tag{1}$$

приймається за міру інформації (знань) про $(x)$, що міститься в $(y)$.

Колмогорівські оцінки безпосередньо пов'язані з головною процедурою отримання знань з даних, характерною для наукових досліджень незалежно від їх області. У ТД-дослідженнях вони мають принципово глибинний характер. Мова йде про механізм узагальнення на основі метасистемних переходів [27] у багаторівневій системі знань $(S)$, умовно вираженій у вигляді:

$$S = \sum_i S_i,\ i = 1, 2, \ldots, m; \tag{2}$$

$$S_i = C_i + \sum_j S_{ij},\ j = 1,\ 2, ...,\ n\,,$$

де:

$n$ – число узагальнюючих понять-об'єктів на кожному *i*-му рівні, $n \in \mathbb{N}$;

$m$ – число рівнів, $m \in \mathbb{N}$;

$S_i$ – концептуальна система *i*-го рівня, $i \in \mathbb{N}$;

$C_i$ – механізм керування процедурою узагальнення на *i*-му рівні, $i \in \mathbb{N}$.

Очевидно, конструювання механізму (2), що лежить в основі методології наукових досліджень, безпосередньо пов'язане зі створенням концептуально-понятійного каркасу відповідних наукових теорій, в якості якого може слугувати сукупність формальних комп'ютерних онтологій конкретних предметних областей досліджень.

### 1.1.3 *Трансдисциплінарність, ноосферна концепція і наукова картина світу*

Як зазначалося вище, сучасний етап розвитку науки та її застосунків носить явно ТД-характер. "Цей факт зумовив необхідність розробки строгої методології ТД наукових досліджень, розширення мережі ТД міжнародних центрів та шкіл, нарешті, визначення місця та ролі інформатики у системно-технологічній підтримці ТД-досліджень та використання їх результатів під час вирішення глобальних проблем розвитку сучасної цивілізації. Наголосимо ще раз, що ТД-парадигма передбачає побудову в найближчому майбутньому загальнонаукової картини світу або, що те саме, – єдиної ТД-системи знань, яка забезпечує формалізовану постановку та вирішення конкретних завдань при виконанні комплексних проєктів високої складності, соціальної значущості, конфліктності та конкурентності" [16]. Іншими словами, йдеться про метанауку, здатну пояснювати процеси і явища навколишнього світу, що постійно еволюціонує, а також вирішувати все більш складні прикладні проблеми природи і суспільства.

Саме при переході до суспільства знань та ТД знання-орієнтованим технологіям по-справжньому проявляється системоутворююча роль інформатики. Шлях до трансдисциплінарності лежить через створення системології МД-взаємодії (у світлі еволюції наукових теорій) як самостійної галузі знань.

Терміни: ноосферогенез, трансдисциплінарність, онтолого-керовані системи, віртуальна парадигма, інформаційно-когнітивна підтримка наукових досліджень, персоніфіковані бази знань, смарт-проєкти, Інтернет-

речі – чітко окреслюють, насамперед, предметну галузь інформатики та інформаційних технологій XXI століття, орієнтовану безпосередньо на етап розвитку людського суспільства, заснований на економіці знань. Насправді ж процес ноосферогенезу, за В.І. Вернадським, зачіпає глибші аспекти взаємодії у системі «Людина-Природа». Він апелює до наукової думки, а, отже, до когнітивних ресурсів людського розуму та НКМ, побудова якої неможлива без ТД-підходу до науки та людської цивілізації загалом [16]. На рис. 3 наведено варіант категоріального рівня ноосфери.

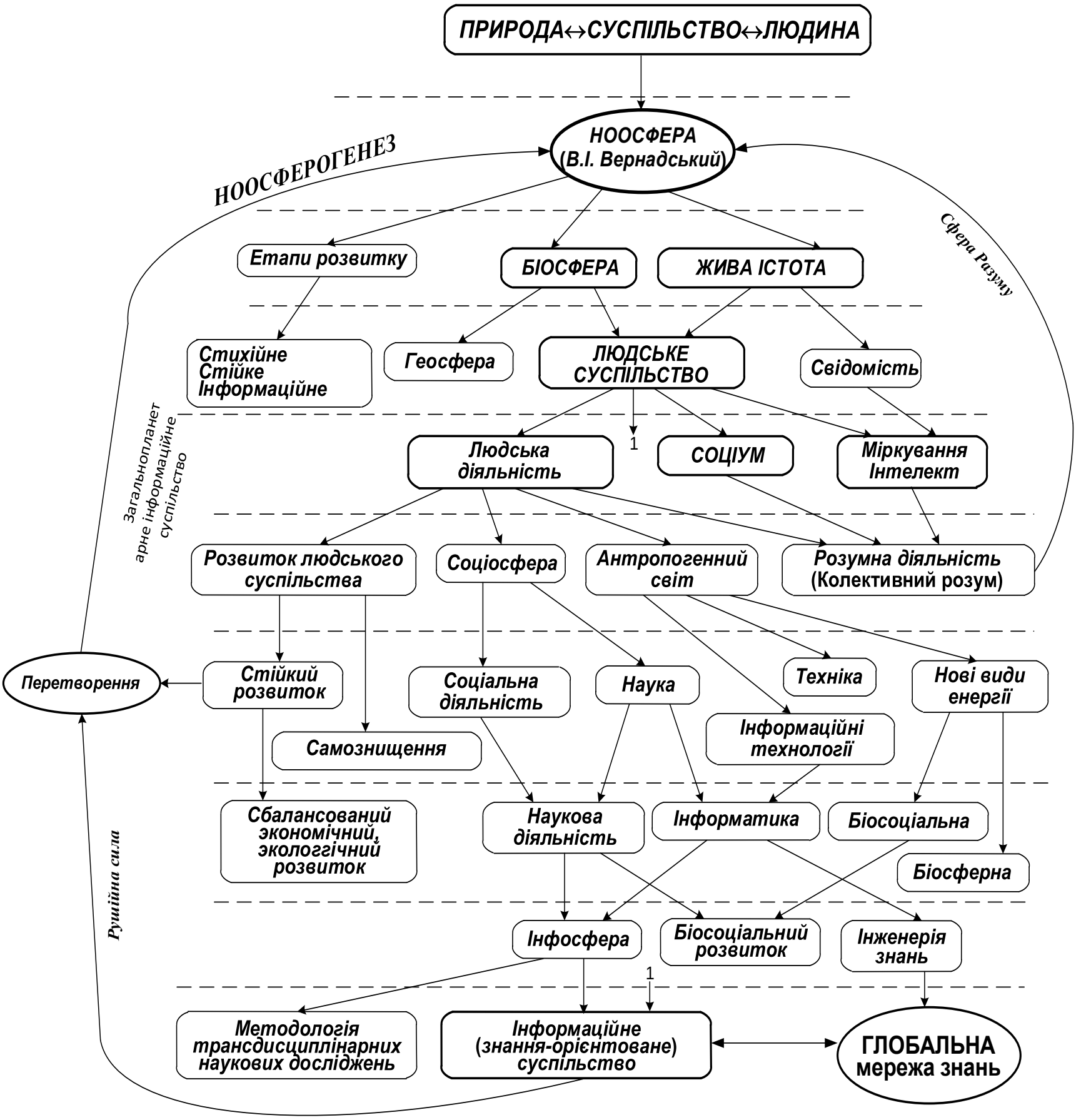

Рис. 3. Категоріальний рівень ноосфери

На рисунку представлені основні концепти ноосферної парадигми, серед них головними є тріада “знання-орієнтована концепція → колективний розум → сталий розвиток суспільства”.

ТД-дослідження дозволяють усвідомити складність вирішуваних проблем, взяти до уваги різноманітність уявлень про життєвий світ і поставлені проблеми, пов’язати абстрактне та конкретне знання з використанням засобів “онтологічного інжинірингу” та глобальної мережі ТД-знань.

ТД-дослідження часто носять виражено проєктний характер, коли для вирішення деякої життєво важливої проблеми формується тимчасова робоча група, яка включає в себе представників різних галузей – вчених, техніків, політиків, представників громадськості тощо [19].

Методологічно ТД-дослідження включають ряд основних стадій:

1) аналіз проблеми, її ідентифікація та структурування;
2) розробка недостатнього теоретичного матеріалу, необхідного для вирішення проблеми;
3) побудова кластерів конвергенції предметних областей;
4) вибір (або розроблення нової) мови для формалізованого подання знань предметних областей та відповідних комп’ютерних онтологій;
5) практична реалізація проєкту.

Зазначені стадії організовані в лінійній послідовності, але мають зворотні зв’язки.

В [28] виділено чотири загальні принципи та характеристики ТД-досліджень:

– принцип розбиття складної проблеми на простіші задачі з урахуванням виду знання та учасників проєкту;

– принцип розвитку дослідження через рекурсивність: ТД-дослідження організуються циклічно, у формі багаторазових рекурсивних циклів, що дозволяє коригувати проміжні результати дослідження;

– досягнення ефективності отриманого результату за допомогою конкретизації та орієнтації на застосування до конкретних соціальних практик, цільових груп, співвіднесення з подібними дослідженнями в інших проєктах тощо;

– досягнення результатів через відкриту змагальність та плюралізм.

### 1.1.4 *Реалізація концепції та мережа ТД-знань*

Сучасні інструментальні інформаційні технології з текст-процесингом, семантичним аналізом та узагальненням змістового контенту дозволяють значною мірою автоматизувати процес опису знань предметних галузей. Кожен такий опис є ланцюжком “онтологія – формальне викладення наукової теорії – прикладна система" [16]. Тоді архітектуру єдиної трансдисциплінарної мережі (МТЗ) знань можна представити у вигляді рис. 4. На ньому:

$O_1 \div O_n$ – онтології предметних областей,

$T_1 \div T_n$ – формальне представлення наукових теорій,

$ПС_1 \div ПС_n$ – відповідні прикладні системи,

ОКР – онтологія категоріального рівня,

ГІІ – глобальна інформаційна інфраструктура (next generation network).

Роль ОКР полягає у забезпеченні міждисциплінарної взаємодії на рівні загальної мови категорій. Роль онтологій предметних знань полягає у реалізації онтологічного управління лише на рівні архітектури інтелектуальної комп'ютерної системи (ІКС) чи ГІІ.

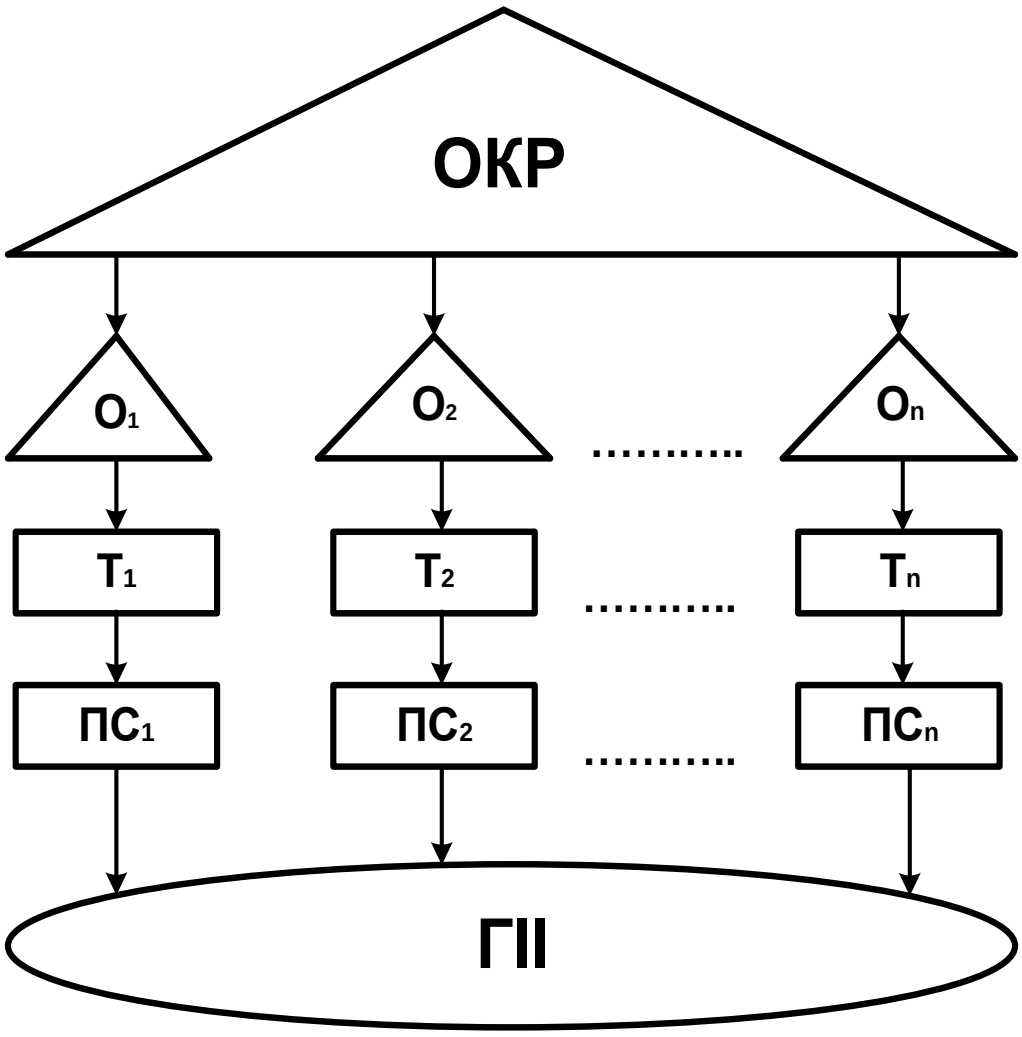

Рис. 4. Архітектура мережі трансдисциплінарних знань

Мережа трансдисциплінарних знань поки що відсутня, є, по суті, надбудова над існуючою Інтернет-мережею, яка, у свою чергу, еволюціонує в напрямку Semantic Web. З іншого боку, в останні роки отримали інтенсивний розвиток сенсорні вимірювальні мережі, забезпечуючи Інтернет первинною інформацією про навколишній світ. В [29] йдеться про єдину сенсорну мережу CeNCE-Central Nervons System for Earth, яка повинна будуватися на основі стандартів відкритих універсальних протоколів та інтерфейсів і володіти властивостями самоорганізації, самовідновлення та динамічної адаптації структури залежно від змін зовнішнього середовища. Розвитку глобальної сенсорної мережі (мережі первинної інформації) сприяє сучасний стан та амбітні плани промисловості та технологій мікро- та нано-електромеханічних систем (сенсорів), тенденції їх інтелектуалізації, оснащення засобами радіозв'язку [29, 30].

В результаті сучасний Інтернет перетворюється на багаторівневу глобальну інформаційну мережу, яка об'єднує воєдино засоби та технології отримання первинної інформації про сучасний навколишній світ і data mining, а також системи отримання та оброблення формалізованих знань, надання на їх основі сервісів у вигляді рішень конкретних задач користувача. Така глобальна багаторівнева інформаційна мережа має вигляд, як показано на рис. 5.

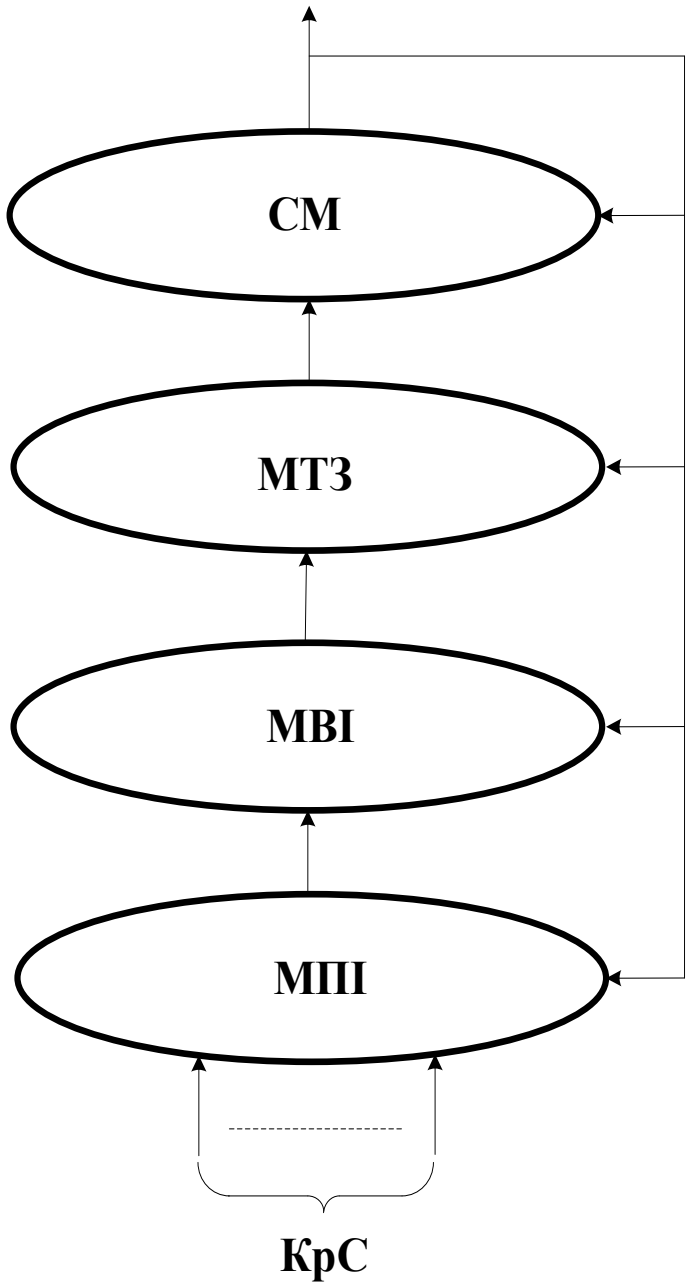

Рис. 5. Глобальна багаторівнева інформаційна мережа

На рисунку: МПІ – мережа первинної інформації;

МВІ – мережа вторинної інформації (Semantic Web);

МТЗ – мережа трансдисциплінарних знань;

СМ – сервісна мережа;

КрС – користувацькі системи.

Петля зворотного зв’язку відзначає можливість використання певних сервісів, що надаються КрС, для управління мережами нижніх рівнів, включаючи спеціальні центри оброблення даних сенсорів та сенсорних вимірювальних мереж (входять до складу МПІ).

Втілення в життя трансдисциплінарної концепції передбачає комплекс наукових досліджень та дослідницьких проєктів (крім задекларованих вище), спрямованих на розробку:

– систем управління процесами ТД-досліджень (підсистем моніторингу, управління знаннями, управління науково-технічними програмами);

– засобів та систем формалізованого подання знань, методів їх оброблення, накопичення, інтеграції та сервісного супроводу;

– проблемно-орієнтованих комплексів, включаючи автоматизовані робочі місця наукових дослідників усіх рангів;

– систем економізації знань та підтримки інтелектуальної власності;

– прикладних систем різного призначення (управління наукою, економікою, галузями народного господарства, створення науково-інноваційних центрів та віртуальних організацій, медико-економічний моніторинг, електронні курси, персональні бази знань тощо).

## 1.2. **Проблеми організації трансдисциплінарних наукових досліджень**

Головний акцент у підготовці та проведенні наукових досліджень, насамперед трансдисциплінарних, з урахуванням досягнень сучасної інформатики, робиться на розвиток можливостей інтелектуальних інформаційних технологій. Йдеться як про ефективні методи отримання, подання та оброблення знань як результату науково-дослідної роботи (НДР), так і про ефективну організацію процесу НДР на всіх етапах її життєвого циклу [31]:

1) постановка наукової проблеми;
2) проведення власне науково-дослідної роботи щодо вирішення цієї проблеми;
3) узагальнення, оформлення та презентація результатів НДР;
4) використання результатів.

Головне завдання першого етапу – формування багаторівневого проблемно-орієнтованого простору, де кожен із рівнів орієнтований на відповідний аспект вирішення проблеми. Незалежно від орієнтації основні процедури та обсяг роботи визначаються обсягом та якістю пошуку, підбору та оброблення релевантного матеріалу в широкому інформаційному просторі. Аналіз та узагальнення цього матеріалу дозволяє визначити вихідний стан наукової проблеми та оцінити підходи у вирішенні аналогічних наукових проблем.

Дуже важливим і продуктивним, особливо в прикладних НДР, є вибір правильної інноваційної стратегії. Останні відрізняються різними моделями інноваційного процесу, які безпосередньо впливають на планування та організацію НДР. В даний час є велике розмаїття методів, засобів і процедур, що забезпечують виконання цього етапу НДР. В [31], зокрема, розглянуто модель формулювання інноваційної стратегії, засновану на процесі зменшення невизначеності у системі вимог "ринок – технічні засоби – терміновість виконання проєкту". З урахуванням її положень Project management, як складова частина Knowledge management, при трансдисциплінарному підході в масштабах наукової галузі або науково-дослідної (проєктної) організації в кінцевому рахунку зводиться до коректного переходу від сукупності цілей до сукупності проєктів (множина $P$), представлених на множині дисциплін $D$, де кожному $p_i \in P, i = \overline{1,n}$ відповідає підмножина дисциплін $D_i \subset D$, а його елементи $d_{ij} \in D_i$, $i = \overline{1,n},\ j = \overline{1,m}$ відзначають дисципліни, знання яких використовуються при виконанні проєкту $p_i, i = \overline{1,n}$. Підмножина $D_i$ називають кластером конвергенції відповідних дисциплін щодо проєкту $P_i$, а елементи $d_{ij} \in D_i$ приймають значення $\{0,1\}$, залежно від факту використання знань тієї чи іншої дисципліни у проєкті $P_i$. Проєкти, що мають загальні кластери конвергенції, за інших рівних умов є претендентами на першочергове виконання через ресурсні переваги.

При цьому очевидно, що з одного боку, джерелом будь-якої наукової проблеми є комплекс протиріч, що лежать в основі існуючої наукової теорії, виявлення якого є результатом глибокого аналізу предметної області. З іншого боку, коректне формулювання складної наукової проблеми є саме по собі досить складною науковою проблемою, яка спирається на методологію системного підходу та виявлення цільової проблемної ситуації.

Саме на цьому етапі повинні зіграти свою вирішальну роль методи та засоби онтологічного аналізу, що включають ефективні процедури онтологічного пошуку первинної інформації та їх когнітивного оброблення, аж до автоматизації побудови онтологічного опису предметної області (областей).

Задача другого етапу НДР розбивається, як правило, на такі підетапи:

а) уточнення моделі проблемної ситуації та наукової проблеми;

б) інтегроване подання вихідних знань та формування робочих гіпотез;

в) нарощування нових знань як результат перевірки та відбору гіпотез.

Для нього характерно використання методів продуктивного морфологічного класифікування, динамічного моделювання та аналітичного конструювання.

Враховуючи трансдисциплінарний характер сучасних наукових програм і проєктів, актуальними видаються методи аналізу сутнісних процесів конвергенції предметних областей та пов'язаних з нею процесів когнітивної взаємодії, на рівні понятійних структур. Йдеться про формування згаданих вище онтологічних кластерів, а також про механізм понятійного колайдера (аналог зіштовхування елементарних частинок у прискорювачі), який дозволяє сформувати часто несподівані, але продуктивні множини понятійних комбінацій та структур. Аналіз та фільтрація останніх мають стати завданнями систем когнітивного конструювання, яким притаманні засоби та процедури нейролінгвістичного, лексикографічного та знання-орієнтованого аналізу та синтезу.

Дуже важливим розділом другого етапу, особливо для задач дослідницького проєктування, є експериментальний розділ. Сучасне інтегроване середовище проєктувальника-експериментатора включає різні процедури логіко-евристичного аналізу технічних систем які проєктуються, випробувальні комплекси, графічні системи, засоби програмування тощо.

Третій етап безпосередньо пов'язаний з ефективним узагальненням та поданням результатів НДР. Його головним цільовим завданням є побудова формалізованої системи знань, яка, з одного боку, формує структуру проблемної наукової теорії, а з іншого – забезпечує ефективне впровадження результатів НДР, тобто реалізацію обраної на першому етапі інноваційної стратегії. Таке об'ємне інтегральне подання результатів НДР повинне наблизити їх до потенційних користувачів. І тому доцільно виробити єдиний стандарт електронного подання знань, отриманих в результаті виконання НДР. Варіант такого стандарту можна подати у вигляді, зображеному на рис. 6, де: *O* – онтологічний опис предметної області (онтограф, тезаурус термінів, система логікового виведення); *V* – образна компонента онтологічного опису (*3D* графіка, мультимедійне подання матеріальних об'єктів, *VR* – об'єкти віртуальної реальності); *T* – подання предметної області лише на рівні формальної теорії; *L* – повний лінгвістичний корпус, що представляє предметну область, *S* – підсистема сервіс-орієнтованої архітектури, множина сервісів, що надаються користувачеві (з урахуванням рангу користувачів); *W* – корпоративний Web-портал із певною кількістю сервісів, *U* – користувач, 1 та 2 – підсистеми електронного представлення.

Електронне подання результатів НДР, виходячи із рис. 6, є відкритою інформаційною системою в сенсі IEE POSIX 1003.0 з відкритими специфікаціями на інтерфейси та функціями розширюваності, масштабованості, інтероперабельності, переносимості застосунків тощо. Розроблення зазначеного стандарту на подання результатів НДР є суттєвим кроком в напрямку реалізації парадигми *E*-science, де знання подані в явній, конструктивній уніфікованій формі, готовій для вирішення конкретних прикладних задач, а процес наукових досліджень вимагає об’єднання зусиль низки наукових колективів з розподілом ресурсів між ними та інтенсифікацією процесів обміну результатами досліджень, коли традиційні підходи не в змозі ефективно підтримувати ці процеси та відповідні обсяги інформації. Таке об’єднання всіх ресурсів та ефективної підтримки на рівні загальної інформаційної інфраструктури в масштабах галузі науки (зокрема, академічної) передбачає створення робочих місць дослідників, оснащених усім необхідним як на рівні комунікаційного середовища (локальні та глобальні мережі), так і на рівні їх системного та інформаційно-технологічного забезпечення.

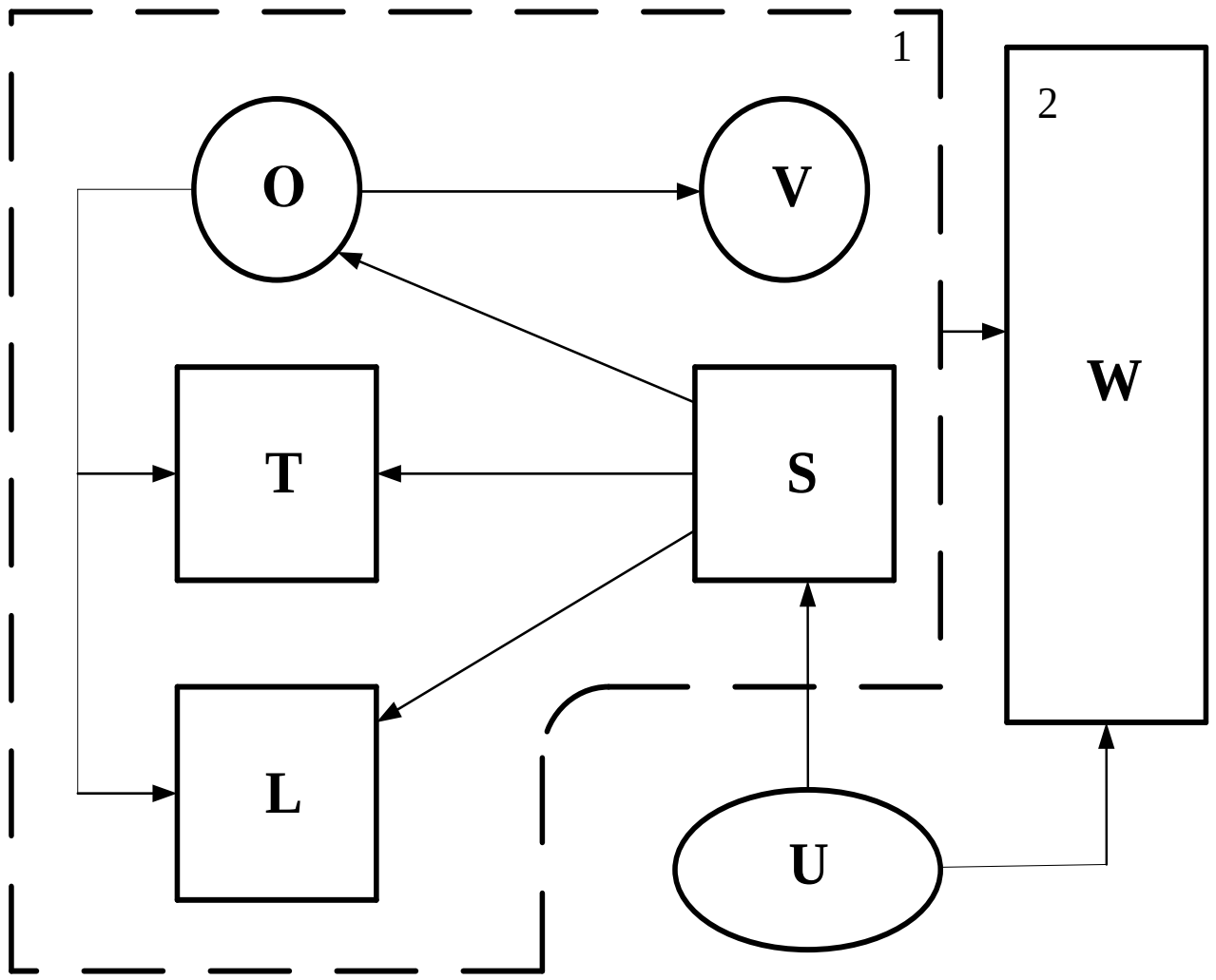

Рис. 6. Стандарт електронного представлення результатів НДР

Сучасні системи підтримки наукових досліджень крім індивідуальних платформ використовують, як правило, два основні взаємодоповнюючі типи інформаційних систем колективного користування [32]: Grid-технології та хмарні обчислення.

Grid-комп’ютинг – це сучасна технологія розподіленого оброблення інформації на основі розвиненої мережевої інфраструктури та інструментарію прозорого управління гетерогенними ресурсами, моніторингу даних, підтримки грід-сервісів та безпеки, систем віртуальної взаємодії колективних користувачів.

Хмарні обчислення – це технологія розподіленого оброблення інформації, в якого комп'ютерні ресурси надаються користувачеві в режимі Інтернет-сервісу у вигляді сервісних центрів, віртуальних платформ (як правило пов'язаних у мережу) на правах оренди у компаній-власників ресурсів (наприклад, сервіси Google Apps/Docs, Amazon ECR, Microsoft Office Web). Потенціал хмарних обчислень дуже високий та відповідає загальній тенденції глобалізації. Особливу привабливість «хмарі» забезпечує її доступність (будь-яка точка Інтернету).

Дані типи систем колективного користування можуть стати довгостроковою основою ресурсної підтримки трансдисциплінарних наукових досліджень, незалежно від спрямованості останніх [33]. Обов'язковою компонентою сервісних центрів має бути формалізована система знань (і даних), що представляють онтолого-керовану наукову картину світу, яка забезпечує несуперечливу постановку та вирішення задач НДР з урахуванням проблем екології, сталого розвитку, загальної траєкторії цивілізації тощо.

Таким чином, проблеми ефективної підтримки трансдисциплінарних наукових досліджень призводять до формування та системного аналізу сервіс-орієнтованої парадигми ноосферогенезу, що задається ланцюжком: ноосферогенез – трансдисциплінарність – інформатика – онтологічна концепція – наукова картина світу – перспективні інформаційні технології. Сутнісна функція, місце та послідовність концептів у цьому ланцюжку чітко визначені і, по суті, становлять методологічну основу сучасних наукових досліджень як основи розвитку цивілізації. Ноосферологія, як зазначалося вище, – це цілісна сукупність знань і гармонійна взаємодія у системі «Людина – Природа» під керівництвом свідомості людини та волі людини. Вона органічно пов'язана з науковою та технологічною компонентами розвитку цивілізації. Розвиток науки перейшов від стадії диференціації до стадії інтеграції, забезпечивши можливість реалізації трансдисциплінарної концепції розвитку науки, яка апелює до наукової картини світу під час постановки та проведення наукових досліджень та виконання складних дослідницьких проєктів. Без неї немислимий цілеспрямований позитивний процес ноосферогенезу. Тут виконує свою місію інформатика як системоутворююча галузь знань. В її надрах зародилася онтологічна концепція, сутність якої полягає у формальному онтологічному описі предметних областей та наукової картини світу загалом (завдання поки що не виконано, але процес розпочато). Нарешті, сучасні інформаційні технології вже стали сьогодні основою практично всіх Hi Tech та побудови знання-орієнтованого суспільства, здатного вирішити всі суттєві протиріччя розвитку сучасної (технологічної) цивілізації. Процес інтелектуалізації інформаційних технологій орієнтований насамперед на складні фундаментальні та прикладні наукові дослідження, ось чому системи knowledge engineering та knowledge management займають особливе місце в розділі інформатики, який зветься «Штучний інтелект» [34, 35].

**Висновки за розділом**

1. Епоха аналітизму та властива їй диференціація науки та замкнутих наукових теорій уже позаду. Стало очевидним, що реальні проблеми, які стоять перед людським суспільством, набагато складніші за наукові, і наука не в змозі їх кардинально вирішити внаслідок роз'єднаності наукових дисциплін та їх спеціалізації, слабкої координації наукових колективів та їх тематики, відсутністю системного моніторингу та загальної формалізованої мови подання знань. Однією з головних завдань трансдисциплінарних досліджень є забезпечення ефективної ТД-взаємодії на всіх етапах життєвого циклу вирішення фундаментальних і прикладних наукових проблем. Окрім завдань інфраструктурної підтримки ТД-досліджень на перший план виходять завдання їх методологічного супроводу та забезпечення процесів інтеграції, конвергенції та уніфікованого формалізованого представлення ТД-знань та операцій над ними.

2. Сутність трансдисциплінарного підходу до дослідження комплексних науково-технічних проблем полягає в ефективному забезпеченні двоєдності концепцій поглиблення конкретних знань у предметній області, з одного боку, та розширення охоплення проблеми, виходячи з реальності єдності світу, та прагнення відтворити цілісну наукову картину світу – з іншого.

3. Дисциплінарність дозволяє науці поступово розвиватися у межах предметних напрямів, а дисциплінарний підхід ділить навколишній світ окремі предметні області. Якщо розв'язання задачі чи проблеми виходить за межі можливостей дисциплінарних підходів, то прийнято вважати, що воно знаходиться «на стику наукових дисциплін».

4. Міждисциплінарність передбачає інтеграцію кількох наукових дисциплін. Одна з них відіграє роль провідної та результати МД-дослідження завжди трактуються цією провідною дисципліною.

5. Трансдисциплінарність – це дослідницька стратегія, яка перетинає дисциплінарні кордони і розвиває холістичне (пріоритетний розгляд цілого по відношенню до його частин) бачення. Трансдисциплінарність у вузькому значенні означає інтеграцію різних форм і методів дослідження, включаючи спеціальні прийоми наукового пізнання для вирішення наукових проблем.

Трансдисциплінарність у широкому сенсі означає єдність знання поза конкретних дисциплін.

6. Розвиток NBIC-кластера конвергенції відкриває широкі, поки що повністю не оцінені можливості глобального знання-орієнтованого Інтернету, а з ним і всієї сучасної цивілізації. Очевидно, цей розвиток йтиме шляхом створення спочатку прикладних розподілених систем у конкретних предметних областях (Інтернет-речей, smart-системи в телемедичній, екологічному моніторингу, інформаційному супроводі товарів і послуг, енергетичних системах, комунальних службах тощо). Центральне місце в них займуть Grid-, Blockchain-технології та Cloud-computing, а також віртуальні організації, структури та сервіси.

7. Вчення В.І. Вернадського про ноосферу за своєю суттю апелює до НКМ, яку необхідно будувати для того, щоб подолати МД бар'єри та підвищити ефективність міждисциплінарної взаємодії та сучасної науки загалом. Йдеться створення загального ТД-знания.

8. Сучасний Інтернет перетворюється на багаторівневу глобальну інформаційну мережу, яка об'єднує воєдино засоби та технології отримання первинної інформації про сучасний навколишній світ і data mining, а також системи отримання та оброблення формалізованих знань, надання на їх основі сервісів у вигляді рішень конкретних завдань користувача.

9. Проблеми ефективної підтримки трансдисциплінарних наукових досліджень призводять до формування та системного аналізу сервіс-орієнтованої парадигми ноосферогенезу, що задається ланцюжком: *ноосферогенез – трансдисциплінарність – інформатика – онтологічна концепція – наукова картина світу – перспективні інформаційні технології*. Сутнісна функція, місце та послідовність концептів у цьому ланцюжку чітко визначені і, по суті, становлять методологічну основу сучасних наукових досліджень як основи розвитку цивілізації. Ноосферологія, як зазначалося вище, – це цілісна сукупність знань і гармонійна взаємодія у системі «Людина – Природа» під керівництвом свідомості людини та волі людини.

**Перелік посилань до розділу 1**

1. Киященко Л.П. Феномен трансдисциплинарности: опыт философского анализа. Santalka, Filosofia, 2006. Т. 14, № 1. С. 17–37.
2. Мокий В.С. Методологии трансдисциплинарности–4. Нальчик: Институт трансдисциплинарных технологий, 2011. 59 с.
3. Лебедев С.А., Кудрявцев И.К. Детерминизм и недетерминизм в развитии естествознания. *Вестник Московского университета*. Серия 7. Философия. 2005. № 6. С. 3–20.
4. Князева Е.Н. Трансдисциплинарные стратегии исследования. *Вестник ТГПУ (TSPU Bulletin).* 2011. № 10 (112). С.193–201.
5. Киященко Л.П., Гребенщикова Е.Г. Современная философия науки: трансдисциплинарные аспекты. М.: МГМСУ, 2011. 172 с.
6. Кургаєв О.П. Еволюція структури об'єкта науки. Кибернетика и системный анализ, 2016. Том 52, № 2. С. 11–21. DOI: https://doi.org/10.1007/s10559-016-9813-6.
7. URL: http://www.anoitt.ru/cabdir/about_td.php. (дата звернення: 23.06.2021).
8. Palagin A.V., Petrenko N.G. Development and formation of transdisciplinary and interdisciplinary research and the role of computer science. Дніпро: ДВНЗ УДХТУ, Computer modeling: Analisis, Control, Optimization. 2018. – №1 (3). – С.46-70. URL: http://kmauo.org/wp-content/uploads/2018/1/Palagin.pdf. (дата звернення: 23.06.2021).
9. Jean Piaget. L'épistémologie des relations interdisciplinaires, in Léo Apostel et al, 1972.
10. Мокий В.С. Методология трансдисциплинарности-4. Нальчик: АНОИТТ, 2017. 112 с.
11. Erich Jantsch. Vers l'interdisciplinarité et la transdisciplinarité dans l'enseignement et l'innovation, in Léo Apostel et al, 1972.
12. Erich Jantsch. Technological Planning and Social Futures, Cassell/Associated Bussiness Programmes, London, 1972.
13. Гуреев П.М. Современная наука и методология трансдисциплинарности. КиберЛенинка – научная электронная библиотека. URL: https://cyberleninka.ru/search?q= современная +наука+и+методология+ +трансдисциплинарности. (дата звернення: 26.02.2018).
14. Колесникова И. А. Трансдисциплинарная стратегия исследования непрерывного образования. Непрерывное образование: XXI век. 2014. Вып. 4(8) URL:http://lll21.petrsu.ru/journal/article.php?id=2261 (дата

звернення: 07.01.18).

15. Степин В.С. Теоретическое знание, 1989. URL: http://philosophy.ru/library/stepin/index.html (дата звернення: 08.01.2018).
16. Palagin A. V. Transdisciplinarity Problems and the Role of Informatics. Cybernetics and Systems Analysis volume 49, pages 643–651 (2013). DOI: https://doi.org/10.1007/s10559-013-9551-y.
17. Prayd V., Medvedev D.A. *Filosofskiye nauki (Philosophical sciences)*. 2008. (1.): 97–117.
18. Черникова И.В. Трансдисциплинарные методологии и технологии современной науки. Вопросы философии, 2015. №4. С. 26–35.
19. Киященко Л.П. Когнитивная инновация в фокусе философии трансдисциплинарности. Знание. Понимание. Умение, 2012. № 2. С. 34–49.
20. Gibbons M., Nowotny H., Limoges C., Schwartzman S., Scott P., Trow M. The new production of knowledge: The dynamics of science and research in contemporary societies. L., 1994.
21. Киященко Л.П., Моисеев В.И. Философия трансдисциплинарности. Рос. акад. наук, Ин-т философии. М.: ИФРАН, 2009. 205 с.
22. Kurgaev, A.F., Grigoriev, S.N. Metalanguage of Normal Forms of Knowledge. Cybern Syst Anal 52, 839–848 (2016). DOI: https://doi.org/10.1007/s10559-016-9885-3.
23. Кургаев А.Ф. Нормальные формы знаний / А.Ф. Кургаев, С.Н. Григорьев: Доповіді НАН України, 2015. № 11. С. 36–43. URL: http://dspace.nbuv.gov.ua/handle/123456789/97945. (дата звернення: 23.03.2022).
24. Palagin A.V., Kurgayev A.F. Interdisciplinary scientific research: optimization of system-information support. Visn. Nac. Akad. Nauk Ukr.2009. (3): 14-15. [in Ukrainian]. URL: http://dspace.nbuv.gov.ua/bitstream/handle/123456789/3466/ statta.pdf. (дата звернення: 23.03.2022).
25. Palagin A., Kurgaev A. The Problem of Scientific Research Effectiveness. International Journal "Information Theories and Applications" (ITHEA). Vol. 17. Number 1, 2010. pp. 88–99.
26. Колмогоров Н.А. Теория информации и теория алгоритмов. Изд-во "Наука", 1987. 304 с.
27. Турчин В.Ф. Феномен науки: Кибернетический подход к эволюции. Изд. 2-е. М.: ЭТС, 2000.
28. Pohl C., Hadorn G. Principles for Designing Transdisciplinary Research.

München, 2007. URL: http://www.transdisciplinarity.ch. (дата звернення: 23.03.2022).

29. Майская В. Амбициозные планы промышленности МЭМС. Электроника: НТБ, 2012. № 8. С.100–105.
30. Новиков В.Н., Федулева М.В. Распределенные измерительные системы на основе сетевых технологий: Датчики и системы, 2012. № 9. С. 38–41.
31. Палагин А.В. Современные информационные технологии в научных исследованиях: Искусственный интеллект. 1999. №2. С. 20–33.
32. Згуровський М.З., Петренко А.И. GRID-технології для E-науки і освіти: Наукові вісті НТУУ "КПІ" 2009-2. С.10–17 URL: https://ela.kpi.ua/bitstream/123456789/35951/1/2009-2-2.pdf. (дата звернення: 23.03.2022).
33. A.F. Kurgaev, A.V. Palagin. Concerning the information support for research. Visn. Nac. Akad. Nauk Ukr. 2015. (8): 33–48. DOI: https://doi.org/10.15407/visn2015.08.033.
34. Palagin O.V.. Transdisciplinarity, informatics and development of modern civilization. *Visn. Nac. Akad. Nauk Ukr*. 2014. (7): 25-33. DOI: http://doi.org/10.15407/visn2014.07.025/ .
35. Palagin, A.V. An Ontological Conception of Informatization of Scientific Investigations. Cybern Syst Anal 52, 1–7 (2016). DOI: https://doi.org/10.1007/s10559-016-9793-6 .

# Розділ 2. Онтологічний підхід та комп'ютерні онтології предметних знань

## 2.1. Онтологічний підхід та віртуальна парадигма

Онтологічний підхід до оброблення інформації та подання знань з'явився як засіб створення єдиного стандарту формалізації знань в різнорідних предметних областях [1].

Розвиток онтологічних методів дозволив створити ефективні засоби побудови знання-орієнторованих систем і, що дуже важливо, технологічну базу системології трансдисциплінарної взаємодії й онтологічного інжинірингу як розділу сучасного штучного інтелекту.

Онтологічний інструментарій дозволяє побудувати мовно-онтологічну картину світу (МОКС) (різновидність лексикографічної системи), розглядаючи її як складову наукової картини світу, яка є основою трансдисциплінарної концепції наукових досліджень. У цьому випадку МОКС виконує функції категоріальної надбудови баз знань в конкретних предметних областях й інтегрованих сховищах знань. Очевидно, що системна інтеграція знань виконується з врахуванням специфічних формально-методологічних вимог і критеріїв при формуванні достовірних тверджень і виведень, а категоріальний рівень з відповідною системою базових відношень представляє верхній рівень ієрархії МОКС. На всякій онтології фіксується набір типів операцій над поняттями.

Онтологічний підхід надає користувачу цілісний системний погляд на предметну область або кластер предметних областей, що складають складний дослідницький проєкт. Онтологічні моделі знань дозволяють будувати класи, об'єкти, функціональні процедури і, на сам кінець, формальні теорії, а онтологічні технології забезпечують побудову науково-дослідницьких і корпоративних інформаційно-аналітичних систем від багатофакторного аналізу вихідних інформаційних ресурсів до систем колективного прийняття рішень і управління знаннями.

Особливість сучасного етапу розвитку науки полягає в тому, що процес побудови наукової картини світу і мережі трансдисциплінарних знань відстає від потреб складних системних трансдисциплінарних проєктів як в плані управління процесом ТД-досліджень (підтримка всіх етапів життєвого циклу науково-дослідницької роботи), так і в плані управління знаннями (процедури формалізації, узагальнення, актуалізації й оцінки знань). Відсутні ефективні методи інтеграції знань різних предметних областей. Не дивлячись на це, стихійно почався та продовжується процес кластеризації (формування кластерів конвергенції) наукових дисциплін і технологій, об'єднаних спільними цілями розвитку, факторами впливу й зворотними зв'язками (вище згаданий NBIC-кластер [1, 2]).

Цей процес супроводжується утворенням нових наукових теорій і дисциплін і апелює до *канонічної форми* визначення понять, що дозволяє в результаті логікових операцій над поняттями (а паралельно і над їх визначеннями) створювати нові поняття. Головними з них є операції узагальнення й обмеження. Вказування головної частини змісту поняття має вид підведення такого, що визначається, під ближнє родове поняття на основі видоутворюючих (суттєвих і таких, що визначається) ознак: $X_{ij} = A_j X_i$, де $X_i$ і $X_{ij}$ – відповідно родове (таке, що визначає) і видове (таке, що визначається) поняття, $A_j$ – множина видових ознак. Родо-видова дефініція є найбільш представницькою, але не єдиною. Існують й інші види дефініцій: генетичні, операціональні, аксіоматичні, контекстуальні, індуктивні та ін. Відмітимо лише, що строгість визначення понять напряму визначає якість знань, а отже повноту опису предметних областей і наукових теорій. Особливу роль в ТД-системах знань відіграє формування *ієрархії базових категорій* (категоріальна стратифікація), так як вона є системотворчою.

## 2.2. **Образно-понятійні моделі предметних знань**

Наукова картина світу припускає багатовимірне його представлення, а тому наряду з понятійними компонентами онтограф повинен мати і їх образні еквіваленти. Тут доречно відмітити, що онтологічна парадигма почала і розвивалась практично одночасно з віртуальною. Сьогодні в повсякденну практику увійшли такі поняття, як віртуальний світ, віртуальна організація, віртуальна лабораторія, віртуальна система, віртуальна адресація тощо. Парадигма віртуальності точно зв'язує в єдиній моделі не тільки об'єкти інформаційних техніки і технологій, але і соціоекономічні і соціокультурні феномени, становлячись найбільш впливовою останні два десятиліття. У зв'язку з цим з'явилася мова моделювання віртуальної реальності VRML – Virtual Reality Model Language, такий стандарт представлення трьохвимірної (в тому числі реального часу) графіки, яка дозволяє описувати об'єкти і сцени. У подальшому вона була витіснена форматом X3D (Extensible 3D), хоча і продовжує використовуватися в деяких застосунках (наприклад, САПР – системах автоматизованого проєктування). В результаті традиційна понятійна онтологія еволюціонує в напрямку образно-понятійної (ОП), а кожна вершина онтографа представлена іменем не тільки поняття, а й відповідного образа (рис. 1).

Безперечною перевагою ОП-онтологій є наглядне представлення понять матеріальних об’єктів навколишнього світу та можливість їх використання у задачах розпізнавання образів як апріорну інформацію, а точніше специфікації елементів складних об’єктів (конструкцій). Не менш перспективне їх використання для представлення нематеріальних об’єктів, зокрема в сучасних системах когнітивної графіки (наприклад, для використання в інтересах таких нових науково-прикладних напрямів, як «Електронна свідомість», «Автоматизований синтез сценаріїв» тощо.

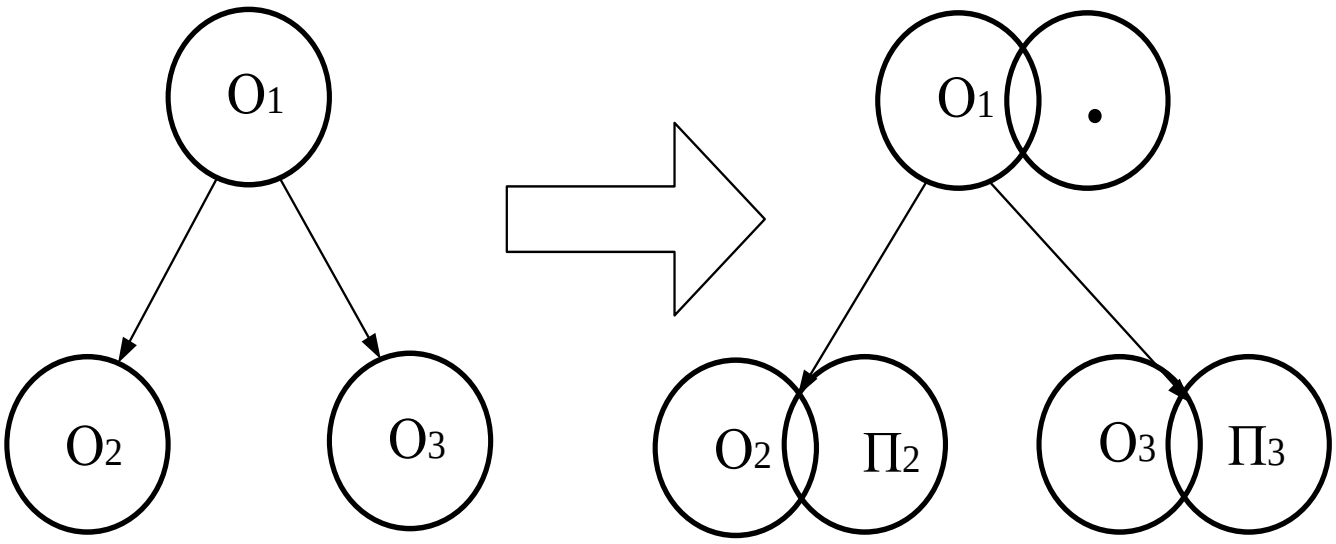

Рис. 1. Фрагмент ОП-онтографа

2.3. **Архітектура інтелектуальних комп’ютерних систем**

Теорія і практика створення та використання систем, заснованих на знаннях, – найбільш актуальний і такий, що інтенсивно розвивається, напрямок Computer Science, який дозволяє підвищити ефективність створення та застосування комп’ютерних технологій, прикладних систем й інструментальних засобів. Складність зазначеної проблеми визначається, зокрема, складністю побудови, організації та використання великих баз формалізованих знань, а також залученням цілої низки наукових теорій (логіки, комп’ютерної лінгвістики, нейрокібернетики, теорії семантичних мереж тощо), які, очевидно, повинні сприяти вирішенню проблеми добування, формального подання, оброблення та системної інтеграції знань, склавши концептуально-методологічну основу *теорії трансдисциплінарних наукових досліджень* [3–5].

Створення онтолого-керованих ІКС для таких досліджень тісно пов’язане з розробкою теоретичних засад та методології проєктування, що включають фундаментальні принципи побудови узагальненої архітектури та структури системи, формальну модель та методологію проєктування онтології предметної області та подання знань загалом, узагальнені алгоритми процедур оброблення знань тощо. У свою чергу, кожна з переличених складових загальної методології проєктування пов'язана з вирішенням складних науково-технічних проблем. Наприклад, розробка онтології ПдО визначає концептуалізацію онтологічних категорій, розробку та вдосконалення ієрархічних структур сутностей на всіх рівнях, побудову формальної системи аксіом та обмежень. Комплексне вирішення зазначених задач проєктування має підвищити роль онтологічних (концептуальних) знань під час рішення конкретних задач у прикладних областях [3].

Виконання переличених вище функцій передбачає звернення інтелектуальної комп'ютерної системи до вбудованого онтологічного графа, вершини та ребра якого навантажені цими функціями, як показано на рис. 2.

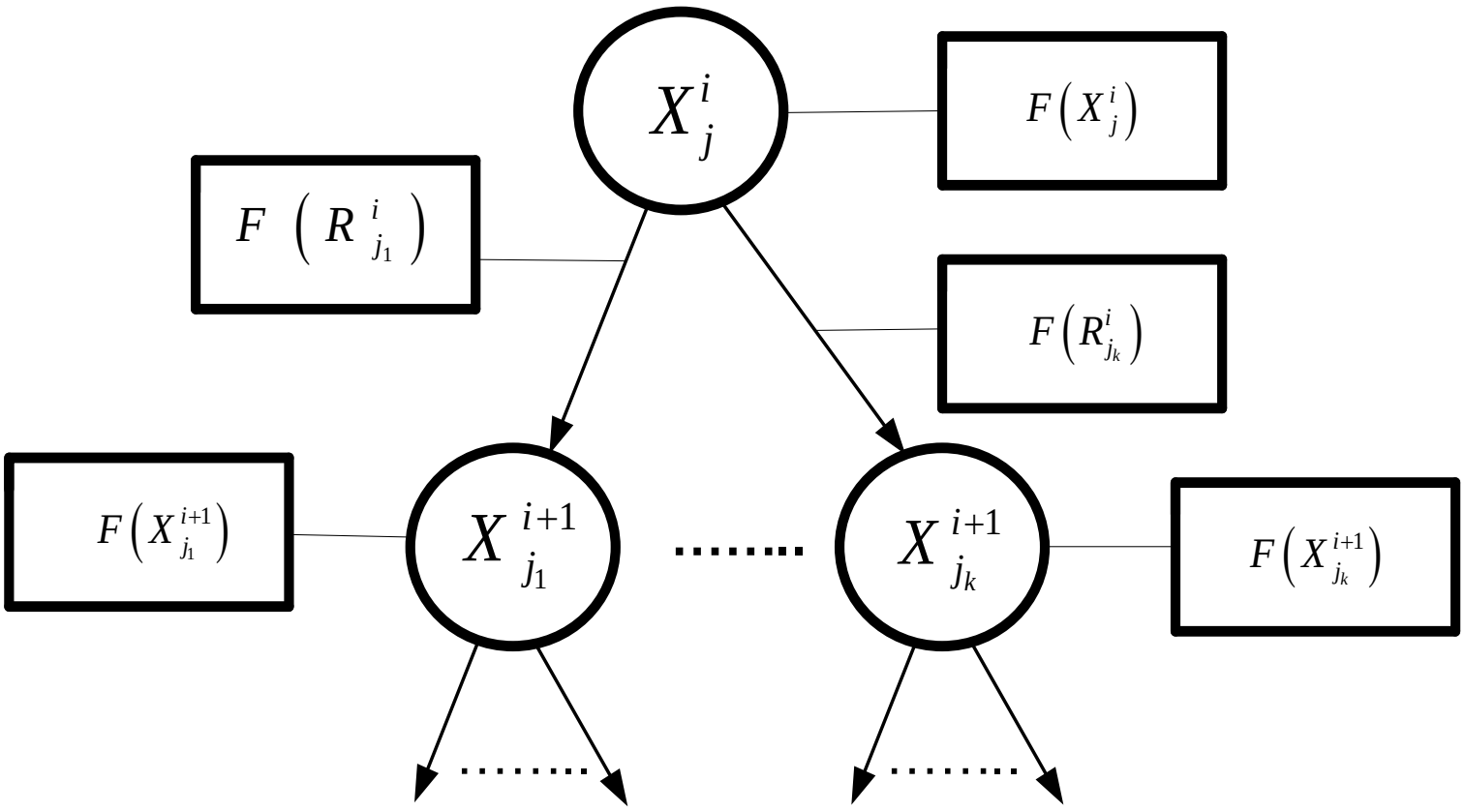

Рис. 2. Фрагмент навантаженого онтографа

Тут: $\{F_j^i\}\, i, j = 1, 2, ...$ множина функцій (сервісів), які приписані вершинам графа із множини $\{X_j^i\}\, i, j = 1, 2, ...$, де $i, j$ – поточні номери рівнів і вершин на кожному рівні відповідно. Аналогічно приписуються функції, що належать ребрам онтографа $\{F(R_{j_k}^i)\}$. Активація функцій відбувається у момент перебування онтографа у станах $X_j^i$ – для функцій $\{F(X_j^i)\}$ і в момент переходу $X_j^i \to X_{j_k}^{i+1}$ – для функцій $\{F(R_{j_k}^i)\}$ відповідно.

Паралельний розвиток концепції систем із сервіс-орієнтованою (СО) та онтолого-керованою архітектурами привів, зрештою, до таких понять та об'єктів, як сервіс-орієнтовані онтології (СОО) та знання (СОЗ), а також сервіс-орієнтовані онтолого-керовані інтелектуальні комп'ютерні системи (СООУ-ІКС). З'явившись як інструмент розробки програмного забезпечення, СО-архітектури (СОА) передбачають використання відкритих *XML* стандартів, Web-сервісів та їх застосунків, включаючи взаємодію з користувачем. На відміну від традиційних СОА системи класу СООУ–ІКС розробляються для операцій із конструктивно поданими знаннями (сервіси знань). Стандарти на уніфіковане представлення сервісів знань та їх інтерфейси поки що відсутні. Очевидно, при їх розробці сьогодні необхідно орієнтуватися на особливості СОО й існуючі протоколи Web-сервісів.

При створенні онтолого-керованих ІКС слід виділити три аспекти досліджень – онтологічний, системологічний та методологічний.

Розвиток архітектури інтелектуальних комп'ютерних систем доцільно розглядати з позиції двоєдності зовнішньої (орієнтованої на користувача) та внутрішньої (інтелектуалізація та багаторівневий розподіл функцій) архітектур [6]. Їх гармонійний взаємозв'язок забезпечує сумарну ефективність ІКС.

Побудова ефективної архітектури знання-орієнтованих інформаційних систем бачиться на шляху конструктивного використання таких розділів сучасної інформатики, як:

– knowledge-processing,

– прагматична модель мовної свідомості,

– віртуальна парадигма та її застосунки.

Узагальнена схема функціонування інтелектуальної інформаційної комп'ютерної системи для наукових досліджень (як електронного еквівалента свідомості), може бути виражена продукційним ланцюжком: "вхідний сигнал → система знань → реакція".

ІКС має попередньо сформульовані цілі (далекі та ближні) та установки (формуються на основі пріоритетів та критеріїв, вироблених у режимі зворотного зв'язку в процесі взаємодії із зовнішнім інформаційним середовищем). Основою предметної діяльності ІКС є система знань, яку можна представити у вигляді підсистеми загальних знань, що взаємодіє з множиною підсистем знань у предметних областях.

На рис. 3 представлена архітектура ІКС, в яку закладено механізм саморозвитку бази знань (БЗ) у предметній області.

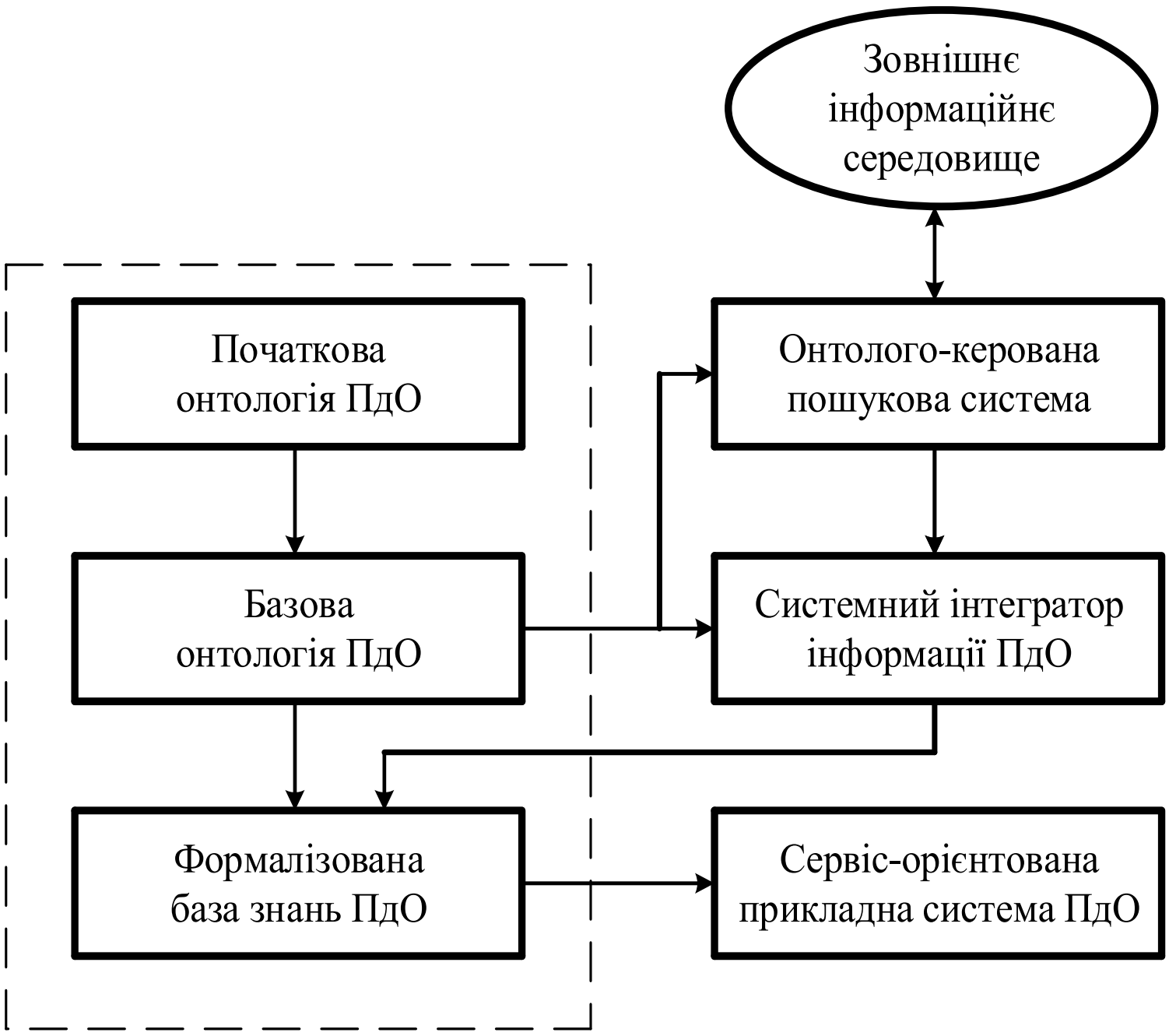

Рис. 3. Архітектура ІКС, яка розвивається

Цей механізм заснований на онтологічному управлінні процесами пошуку релевантної інформації у зовнішньому інформаційному просторі та побудові формалізованої бази знань (ФБЗ). При цьому розвиток ФБЗ здійснюється трьома шляхами [6]:

– за рахунок добування нових фактів і знань із зовнішнього інформаційного середовища, (Інтернет);

– за рахунок генерації нових знань на основі існуючих з використанням механізму виведення;

– у ході цілеспрямованого творчого процесу користувача.

Функціонування ІКС відбувається у двох режимах:

а) відпрацювання цільових завдань (зовнішніх і внутрішніх): активізація процесу, актуалізація інформації, релевантній одній чи декільком предметним областям, та розміщення їх у пам’яті, вирішення проблемної ситуації (рішення задачі), вироблення, систематизація та видача результуючих продукцій (у разі знання-орієнтованої предметної діяльності – нарощування знань);

б) вдосконалення ІКС як інформаційної системи у відповідності із загальною стратегією розвитку: інвентаризація та систематизація знань

(розширення метазнань), формалізація та когнітивізація представлень, онтологізація та інтерпретаційне розширення системи знань, об'єму реакцій та асоціативних зв'язків.

Можна виділити два основні класи задач (як загального, так і науково-технічного плану), які постають перед ІКС:

а) задана множина $X$ вихідних об'єктів. Відомий алгоритм $F$ (спосіб, функціонал, відображення) рішення задачі. Необхідно визначити шукану множину $Y$ об'єктів, таких, що $Y = F(X)$.

б) задані вихідна $S_i$ і цільова $S_0$ ситуації. Необхідно визначити спосіб (шлях) переходу $R: S_i \rightarrow S_0$. Зокрема у випадку: $S_i \approx X$ ; $S_0 \approx Y$ ; $R \approx F$ .

Серед завдань класу (б) існує підклас так званих проблемних задач, де визначення шуканого переходу пов'язане з вирішенням складної (науково-технічної) проблеми. На практиці існує множина задач, які носять змішаний характер. Наприклад, $F$ не повністю визначений у випадку (а), або є деяка апріорна інформація про $R$, зате не повністю визначена ситуація $S_0$ у випадку (б) тощо.

В основі проєктування ІКС лежить *технологія системної інтеграції* (ТСІ), яка ґрунтується на сукупності методів і засобів, що забезпечують виконання всіх етапів життєвого циклу створення об'єктів нової техніки та технологій на основі типових проєктних рішень [6]. ТСІ – це основний інструмент проєктування, що застосовується на всіх етапах життєвого циклу і представляє набір методів, інструментальних засобів та формалізованих процедур:

– побудова компонентів більш високого рівня (і системи в цілому) із компонентів більш нижчого рівня;

– забезпечення взаємодії цих компонентів на всіх етапах створення системи, починаючи з етапів проєктування та закінчуючи етапами її виробництва й експлуатації;

– систематизації типових проєктних рішень, створюючи бібліотеки стандартизованих функціональних вузлів і блоків.

Методологічною основою сучасної ТСІ є метод формалізованих специфікацій, технологічною – міжнародні стандарти на системи, програмно-апаратні компоненти, характеристики інтерфейсів, технологічні операції й експлуатаційні норми.

## 2.4. **Інтелектуальні інформаційні технології онтолого-орієнтованої підтримки трансдисциплінарних досліджень**

Методологія наукових досліджень і конструювання механізму МД-взаємодії при вирішенні складних науково-технічних проблем пов'язані зі створенням концептуально-понятійного каркасу наукових теорій. Таким каркасом може бути сукупність формальних комп'ютерних онтологій конкретних предметних областей [3].

Онтології – це суть понятійні системи, а понятійне мислення – найдосконаліша форма функціонування свідомості та інтелекту [2]. Прототипом такої системи можуть слугувати знання-орієнтовані інформаційні системи з онтолого-керованою архітектурою, що активно розвиваються.

Загальне завдання онтології – компенсувати відсутність стандартів на подання знань при взаємодії користувача з інформаційними системами та останніх між собою, а також інтеграція знань предметних областей як головне завдання ТД наукових досліджень.

Основні онтолого-керовані функції реалізуються в спеціальному класі знання-орієнтованих комп'ютерних систем, а саме в класі онтолого-керованих інтелектуальних комп'ютерних систем [7]. Основою предметної діяльності таких систем є система знань, яку можна представити у вигляді підсистеми загальних знань, що взаємодіє з безліччю підсистем знань у предметних областях.

Реалізація онтологічної концепції передбачає розробку:

– повний опис предметних знань, для чого необхідно: 1) сформувати та обробити (виконати семантичний аналіз) інтегрований лінгвістичний корпус текстів (ЛКТ) із заданих предметних областей; 2) побудувати онтографи (множини $X$, $R$) для кожної ПдО (див. 2.5); 3) формалізувати описи знань у вигляді наукової теорії; 4) виконати процедури оброблення та інтеграції предметних знань із застосуванням систем семантичного аналізу вихідних текстових матеріалів та аналітичної обробки і подання. Прикладом може слугувати інструментарій ТОДОС (Трансдисциплінарні Онтологічні Діалоги Об'єктно-орієнтованих Систем), який подається як комплекс програмно-інформаційних та методичних засобів управління знаннями з використанням підходів онтологічного управління корпоративними інформаційними ресурсами [8]. Функціональність ТОДОС може розширюватися, забезпечуючи реалізацію низки важливих та корисних функцій, таких як:

– зручний інтерфейс, що забезпечує спільну роботу групи експертів із різних галузей знань;

– побудова інтегрованого лінгвістичного корпусу текстів, що описує знання різних ПдО, включаючи: наповнення онтологічної бази знань публікацій наукового дослідника; індексацію та розмітку оригіналів публікацій для повнотекстового пошуку та збереження знань та даних в системі управління базами даних; лематизацію словоформ; екстракцію тексту з оригіналів публікацій, поданих у вигляді документів у форматах pdf, doc та docx; автоматичне формування представницького корпусу вихідних текстів для нейромережевого аналізу з автоматичним визначенням назви, авторів та кількості сторінок публікацій; повнотекстового пошуку за текстовими файлами;

– засоби побудови та формалізації інтегрованих онтологій предметних областей, подання знань у вигляді наукової теорії, перевірка їх коректності;

– побудова категоріального рівня понять та забезпечення його взаємодії з рівнем предметних знань;

– підтримка процесу побудови глобальної мережі ТД-знань і процедур її освоєння та використання;

– побудова самодостатніх кластерів конвергенції, необхідних для виконання ТД-проєкту;

– підтримка процесу виконання ТД-проєкту на всіх етапах життєвого циклу;

– оцінка соціальної значущості проєкту.

Після такого доопрацювання комплексу ТОДОС стає можливим вирішити задачу повного опису знань ПдО, життєвий цикл якого можна представити ланцюжком *“Семантичний аналіз інтегрованого ЛКТ → Побудова онтографів для кожної ПдО → Виділення континууму елементарних сенсів на певній мові Semantic Web → Узагальнення елементарних сенсів на основі науково-онтологічної та мовно-онтологічної картин світу → Формалізація предметних знань → Формування та вирішення прикладних задач з наданням користувачеві сукупності відповідних сервісів*”.

Онтологія реальної онтолого-керованої інформаційної системи містить у загальному випадку три ієрархічно пов’язані компоненти: метаонтологію, яка оперує концептами загального характеру (онтологію категоріального рівня), предметну онтологію та онтологію застосунків. Більш змістовно онтологія предметних дисциплін розглянута у [3, 9].

Більш детально категоріальний та доменний рівні знань у науково-онтологічній картині світу представлені на рис. 4. Зрозуміло, що це аспекти (навіть основні) зазначених рівнів знань однією схемою відобразити неможливо. Тому тут наголос зроблено на категорії “*Природа → Суспільство → Людина → Знання про світ → ТД наукові дослідження*”. На ньому представлено три узагальнені рівні.

Власне рівень категорій, структурований на шість підрівнів (0÷5) відповідно з категоріальними відношеннями.

Рівень доменів предметних та наукових дисциплін. Він поділений на три підрівні (1÷3), в якому відображені, переважно, домени галузі науки. Починаючи з цього рівня семантичні відношення між поняттями вже піддаються деякому науковому осмисленню. У своїй більшості вони представлені відношеннями типу “бути цілим”, “бути частиною”, “бути родом” і “бути видом”.

Рівень дисциплін представлений лише кореневими вершинами предметних дисциплін і технологій шостого технологічного устрою. На цьому рівні переважають відношення типу “ціле-частина”. Більш детально онтології певних предметних дисциплін та категоріальний рівень семантичних відношень між поняттями розглянуті у [9].

Достатньо повний аналіз онтологій верхнього рівня наведено, наприклад, в [3].

Коротко опишемо пераховані вище рівні.

*Категоріальний рівень*. Непорушне правило побудови будь-якого онтографа – вказати кореневу вершину-поняття, яке включає у свій об’єм всі вершини-поняття, що лежать нижче. У даному випадку такою вершиною є категорія “Всесвіт” (В) (рівень 0). Формально, В є тип найвищого (нульового) рівня категоризації, не має ніякої диференціації [3].

На рівні 1 розташовані категорії “Космос”, “Ноосфера”, “Об’єкт” та “Процес”. Категорія “Космос” на представленому онтографі (див. рис. 4) докладно не розглядається (на даному етапі). Категорія “Ноосфера” представлена онтографом (див. рис. 4). Категорії “Об’єкт” та “Процес” включені для поділу понять на статичні та динамічні типи. На онтографі відсутні категорії “Матеріальне” та “Абстрактне”, тому що тематична спрямованість онтографа передбачає включення до нього, в основному, абстрактних понять.

Категорії підрівнів 2÷4 відображають сутнісну основу ТД наукових досліджень, а категорії підрівня 5 їх деталізують.

Рівень доменів відіграє важливу роль при формуванні “кластерів конвергенції” – органічно пов’язаної сукупності наукових теорій, сучасних технологій і досягнень технічної галузі. Зазначена пов’язаність яскраво проявляється у згаданому вище NBIC-кластері технологій.

Рівень дисциплін конкретизує предметні дисципліни, наукові теорії та технології.

Далі спробуємо описати (хоча б у загальному вигляді) принципи взаємодії понять, розташованих на всіх рівнях.

Схема категоріального рівня понять (див. рис. 4) структурована у відповідності з онтологічними принципами, логіковим законом зворотного відношення між об'ємом і змістом понять і вимогами-твердженнями, що носять рекомендаційний характер [3].

Визначення категорій в ОКУ виконано за функціональним принципом, тобто вони відрізняються специфічною функцією в теорії пізнання. Як випливає з вищесказаного, будь-яке поняття в онтології пов'язане з будь-яким іншим поняттям у тій же онтології:

– за об'ємом – за допомогою семантичних відношень $R$ (зв'язків) до ближнього загального поняття (або категорії);

– за змістом (для суттєвих ознак) – за допомогою функцій інтерпретації $F$ та аксіом $A$, властивих тому ж ближньому загальному поняттю.

Узагальнення у понятті (у формі поняття) є операція створення поняття, узагальнено представляє предмети, що спостерігаються, чи можливі того чи іншого класу. А операція узагальнення поняття визначається як перехід від одного поняття до іншого, ширшого за об'ємом за рахунок виключення зі змісту вихідного поняття якої-небудь з ознак, що становлять видову відмінність узагальнених у цьому понятті предметів [10].

Зі сказаного вище випливає, що перехід в онтографі з рівня n на $(n-1)$ рівень у деякому вигляді є окремим випадком узагальнення поняття, а перехід в онтографі з рівня $(n-1)$ на $n$ рівень – обмеження поняття.

Приклади інтелектуальних технологій та комп'ютерних систем, таких як “Системи дослідницького проєктування”, “Багатоагентні технології та нейрокомп'ютинг”, “«Grid-комп'ютинг та хмарні обчислення” розглянуті в [11].

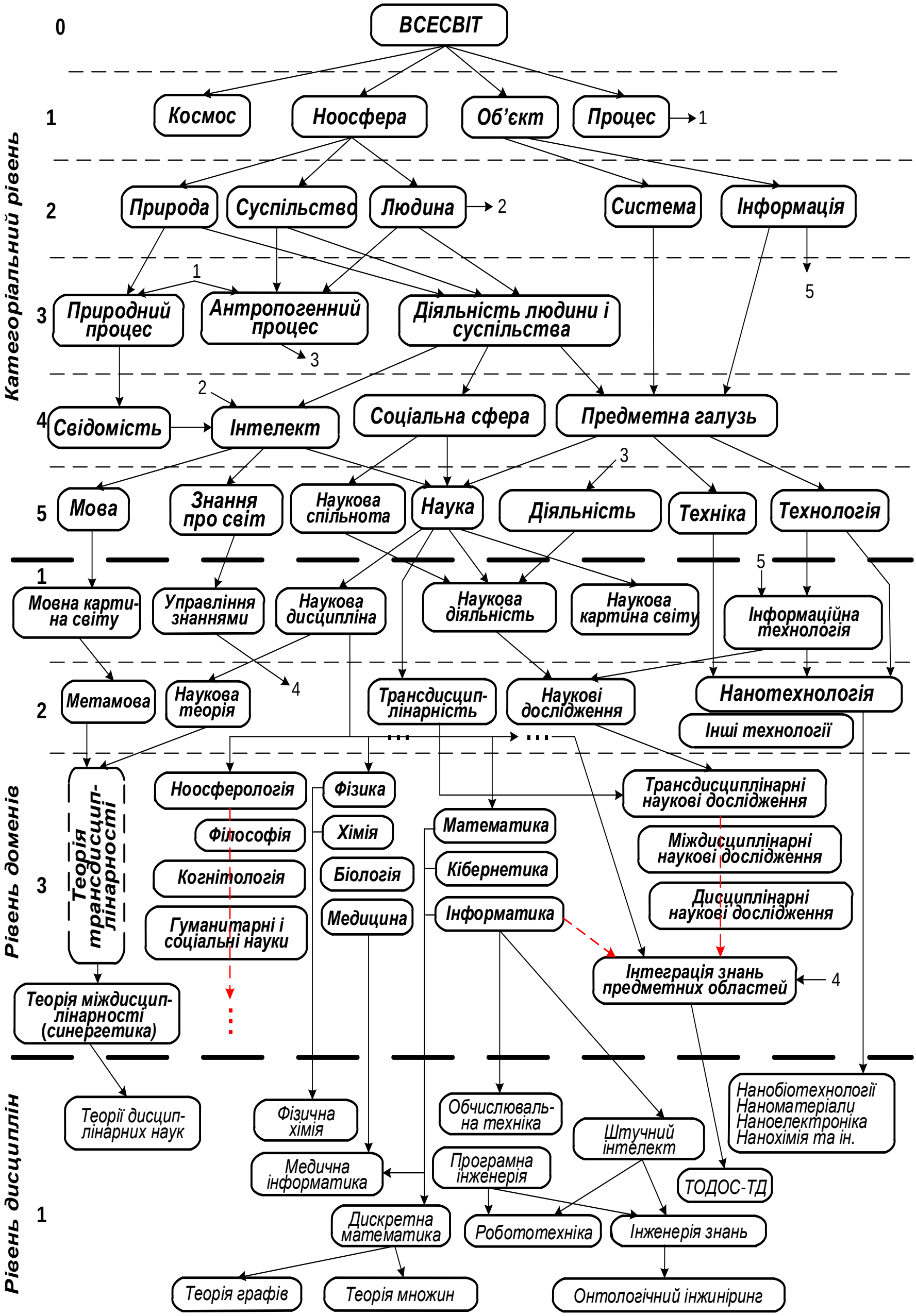

Рис. 4. Онтограф категоріального рівня представлення наукових теорій

2.4.1 ***Інтелектуальні технології підтримки ТД-досліджень***

На теперішній час розвиток науки характеризується посиленням тенденції інтеграції у вивченні об'єктів. Це тому, що сучасна наука досліджує системи, які складноорганізовані і саморозвиваються, і потребують кооперативної взаємодії різних наукових дисциплін. Так, екологія, загальна теорія систем, кібернетика, інформатика, соціобіологія є прикладами комплексу природничих, технічних та гуманітарних досліджень. Зокрема відомий підхід до опису складної реальності, який пов'язаний з ідеями побудови штучного інтелекту, зокрема з його розділами: нейрокомп'ютинг, розпізнавання образів, багатоагентні системи, прийняття рішень та експертні системи, інтелектуальні інформаційні системи, що розвиваються [11].

На рис. 5 у загальному вигляді представлено схему взаємодії пари «Трансдисциплінарність ↔ Інформатика».

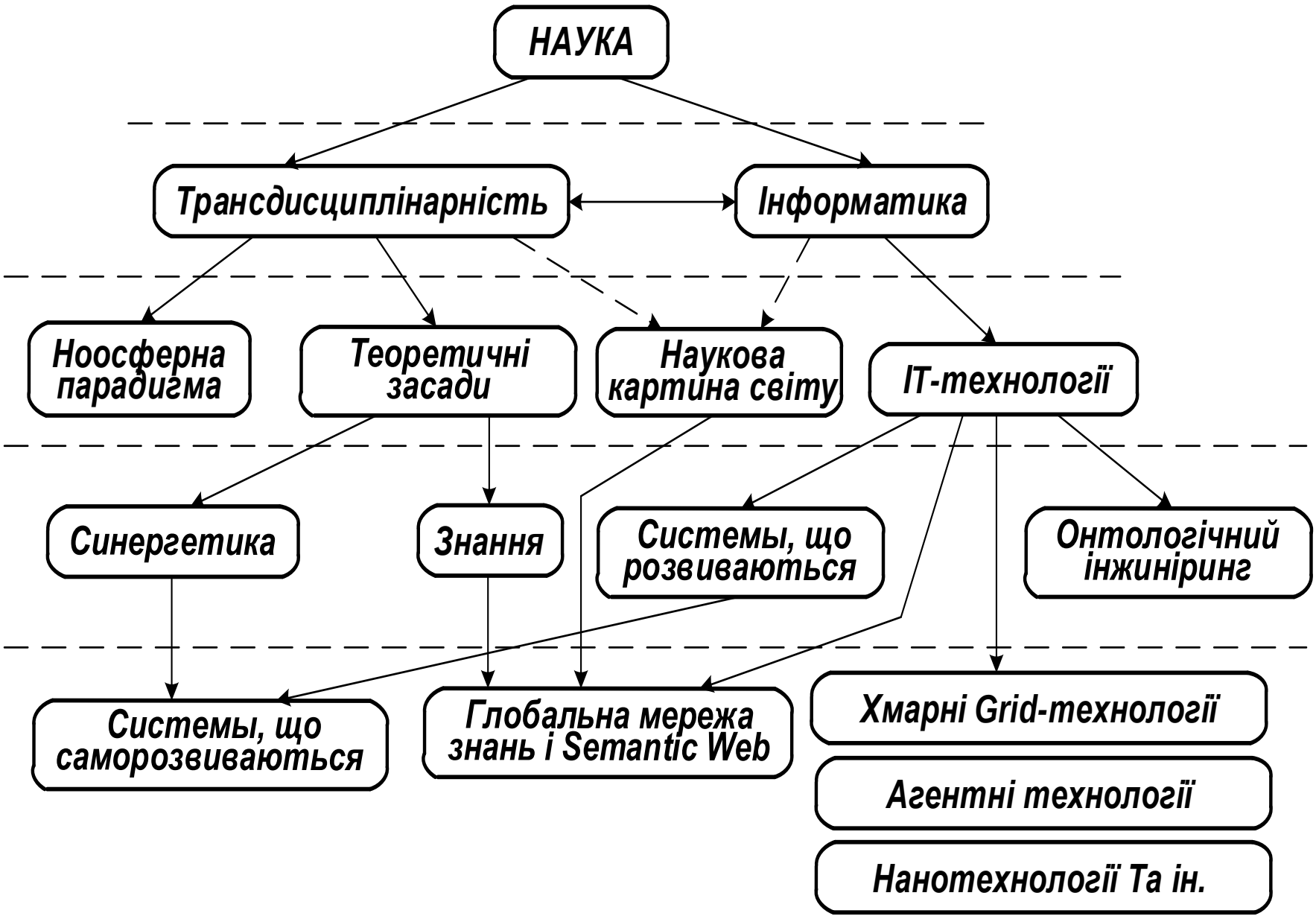

Рис. 5. Схема взаємодії пари «Трансдисциплінарність ↔ Інформатика»

Як зазначалося вище, сфери застосування ТД-досліджень постійно розширюються, що, у свою чергу, потребує постійного вдосконалення інформаційних (у тому числі і супер-комп'ютерних) технологій їхньої підтримки. При цьому до вимог додається соціальна складова, для якої переважають параметри надійності, ефективності та безпеки.

При розгляді процесу природного розвитку науки та підвищення вимог до неї з боку суспільства основою управління ТД-дослідженнями має стати інтегрована інформаційно-технологічна система, яка забезпечить організаційні процеси, моніторинг наукових досліджень, регламентує всі етапи їх життєвого циклу та електронного документообігу, аналіз та оцінювання результатів досліджень, прийняття рішень та визначення актуальних напрямів тощо. В результаті створюється загальний інтегрований простір ТД-знань, де синергетично можуть взаємодіяти численні колективи професіоналів різних предметних областей, які зосередять увагу на вирішенні найбільш важливих ТД науково-практичних проблем [12].

Ядро інтегрованих інформаційних технологій для ТД наукових досліджень складають системно інтегровані бази конструктивно представлених знань, розподілені знання-орієнтовані послуги, що забезпечують високоорганізований доступ до інформаційних та обчислювальних ресурсів, виконання таких функцій: виявлення закономірностей, підтримка прийняття рішень, віртуалізація кооперативної взаємодії, аутсорсинг, застосування сучасних методів обробки мультимедійних інформаційних ресурсів у віртуальному гіперпросторі.

Перехід від недетермінованого режиму продукування та використання знань суб’єктами наукового процесу до режиму ефективного управління знаннями, представленими в уніфікованій формі на всіх етапах їх життєвого циклу, забезпечить зростання результативності та якості наукових досліджень. При цьому усталене знання стане інтелектуальним капіталом, а суб’єкти науки – безпосередніми учасниками економічної діяльності суспільства, що створить сприятливі умови для стимулювання розвитку, як самої науки, так і креативного суспільства [12]

Трансдисциплінарність у свою чергу висуває вимогу інтеграції наукових дисциплін на основі формалізму, єдиного для всіх предметних областей. В якості такого виступає, як вказувалось вище, формальна комп’ютерна онтологія. Звідси процес ноосферогенезу спирається на парадигматичний кортеж “ноосфера → НКМ → трансдисциплінарність → онтологічна концепція → прикладні інтелектуальні системи та технології”.

Отже, сформулюємо відкриту (доповнювану і таку, що розвивається) *систему вимог* до інформаційних технологій підтримки трансдисциплінарних досліджень.

1. Нові комп’ютерні технології мають бути побудовані на основі знань, адекватних процесам вирішення проблем у науці, природі та суспільстві.
2. Безперервне вдосконалення, як самих технологій, так і методів маніпулювання ними.
3. Інформація та знання мають властивість ідемпотентності, а, отже, інформаційні технології повинні підтримувати цю властивість (information sharing and knowledge sharing).
4. Повинні підтримуватися перспективні сучасні методи оброблення даних, інформації та знань, у тому числі:
   – тісна взаємодія з Grid-, Хмарними та Суперкомп’ютерними технологіями;
   – здатність до оброблення великих об’ємів даних (Big Data);
   – багатофакторність аутентифікації;
   – орієнтація на Green computing (екотехнології).

## 2.5. **Системно-онтологічний аналіз ПдО. Моделі комп’ютерних онтологій**

Під системним аналізом зазвичай розуміється вид дослідження, при якому реальний чи уявний об’єкт розчленовується на складові частини (елементи) і потім досліджуються ці елементи і зв’язки між ними. Аналіз предметної області становить особливий вид наукової діяльності, в результаті якої будується інтерпретаційна модель предметних знань (в широкому сенсі) [13–15]. В процесі аналізу останні поділяються на інваріантні та прагматичні знання, концептуальні складові яких представляють онтологічні знання ПдО.

Системний аналіз є важливим методологічним інструментом, який багатократно використовується на різних етапах системного проєктування. Роль системного аналізу у системному проєктуванні, як правило, обмежена рішенням задач дослідницького характеру, які зводяться до побудови моделей, оцінкам параметрів системи, що проєктується, і процесів, пошуку варіантів оптимальних рішень системного проєкту та порівняльному їх аналізу. Необхідність в системному аналізі виникає завжди, коли ставиться велика народногосподарська проблема, в тому числі і проблема науково-технічного характеру [13–15].

Основним напрямком дослідження в рамках проєкту системного аналізу є визначення законів функціонування системи, формування варіантів структури системи (декількох альтернативних алгоритмів, що реалізують заданий закон функціонування) і вибір найкращого варіанта як системи в цілому, так і кожної її компоненти. Пошук рішень здійснюється шляхом рішення задач декомпозиції, аналізу досліджуваної системи і синтезу системи, яка вирішує поставлену проблему. При цьому використовується ряд загальних принципів системного аналізу: кінцевої мети, вимірювання, єдності, зв'язності, модульної побудови, ієрархії, функціональності, розвитку, децентралізації і врахування невизначеності [13, 14].

Загальний підхід до вирішення науково-технічної проблеми може бути представлений як ітераційний цикл із послідовності етапів (рис. 6), направлених на створення знання-орієнтованої інформаційної системи (ЗОІС) як засобу вирішення певної проблеми.

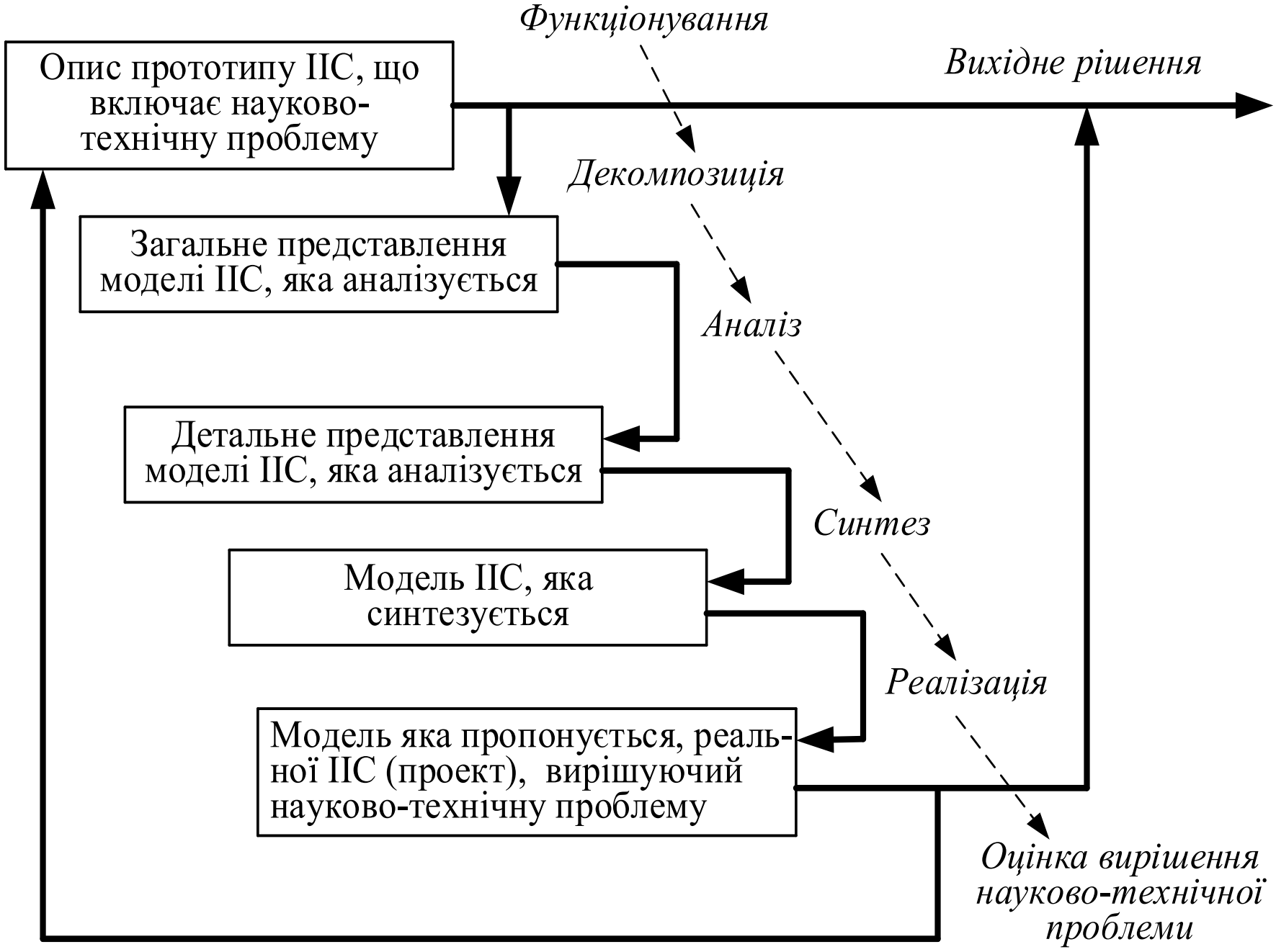

Рис. 6. Загальна схема рішення науково-технічної проблеми

Деякі ідеї з розробки методології проєктування онтології ПдО беруть свій початок в літературі з об'єктно-орієнтованого підходу (ООП), що виник як технологія програмування великих програмних продуктів [16]. Проте розробка онтологій, як ієрархічної структури понять (концептів) відрізняється від проєктування об'єктів, як класів і відношень в об'єктно-орієнтованому програмуванні. Останній зосереджується головним чином на методах опису класів – програміст приймає проєктні рішення, засновані на *операторних* властивостях класу, тоді як розробник онтології приймає ці рішення, ґрунтуючись на *структурних* властивостях класу. В результаті структура понять і відношення між поняттями в онтології відрізняються від структури класів об'єктів подібної ПдО в об'єктно-орієнтованій програмі [17]. Крім того, при розробці онтології внутрішній вміст поняття есплікується завжди, в той час як в об'єктно-орієнтованому програмуванні найчастіше застосовується метод інкапсуляції як спосіб обмеження доступу до внутрішнього вмісту об'єкта.

Системний підхід до пізнання орієнтує аналітика на розгляд будь-якої ПдО з позицій закономірностей системного цілого та взаємодії складових його частин. Системність знань виходить з багаторівневої ієрархічної організації будь-якої сутності, тобто всі об'єкти, процеси та явища можна розглядати як множину дрібніших підмножин (ознак, деталей) і, навпаки, будь-які об'єкти можна (і треба) розглядати як елементи більш високих класів узагальнень.

Розглянемо комп'ютерну (формальну) онтологію предметної області, онтологію верхнього рівня і онтологію домену предметних областей. Остання (в тому числі) має важливе значення для об'єднання (інтеграції) концептуальних знань близьких предметних областей або реалізації технології системної інтеграції міждисциплінарних наукових знань. Ми також виділяємо *початкову онтологію ПдО*, яка є домінантою, що ініціює реалізацію технології автоматизованої побудови онтології ПдО.

Тематика і практична спрямованість наших досліджень (онтологія як засіб побудови баз знань трансдисциплінарних наукових досліджень) зумовлює таке визначення комп'ютерної онтології ПдО [18].

*Визначення* 1. *Комп'ютерна онтологія ПдО* – це:

1) ієрархічна структура кінцевої множини понять, що описують задану предметну область;
2) структура є онтограф, вершинами якого є поняття, а дугами – семантичні відношення між ними;
3) поняття і відношення інтерпретуються відповідно до загальнозначущих функцій інтерпретації, взятих з електронних джерел знань заданої ПдО;
4) визначення понять і відношень виконується на основі аксіом і обмежень області дії;
5) формально онтограф описується на одній з мов опису онтологій;

6) функції інтерпретації та аксіоми описані в деякій підходящої формальної теорії.

В загальному випадку онтологію деякої ПдО формально представляють впорядкованою трійкою [9, 18–20]:

$$O = \langle X, R, F \rangle, \quad (1)$$

де $X, R, F$ – скінченні множини відповідно: $X$ – концептів (понять, термінів) предметної області, $R$ – відношень між ними, $F$ – функцій інтерпретації $X$ та/або $R$.

Розгляд граничних випадків множин (1): $R = \text{O}; R \neq \text{O}; F = \text{O}; F \neq \text{O}$ у всіх чотирьох комбінаціях значень $R$ і $F$ дає різні варіанти онтологічних конструкцій, починаючи від простого словника і таксономії до формальної структури концептуальної бази знань для високоінтелектуальних знання-орієнтованих систем. Детальний розгляд різних зазначених комбінацій з їх змістовною інтерпретацією виконано в [21].

За своєю функціональною повнотою і ступеню формальності розрізняють три види онтологій: проста, повна (чи строга) і множина проміжних або неповних онтологій.

Проста онтологія – це така онтологія, в якій $R = \text{O}; F \neq \text{O}$. Вона слугує (в основному) для однозначного сприйняття науковою спільнотою понять у відповідній прикладній області.

*Строга* або *повна онтологія* $R \neq \text{O}; F \neq \text{O}$ – це така онтологія, в якій множини концептів та концептуальних відношень максимально повні, а до функцій інтерпретації додаються аксіоми, визначення та обмеження. При цьому опис всіх компонент представлений на деякій формальній мові, яка доступна для їх інтерпретації комп'ютером. Схема формальної моделі повної онтології описується четвіркою: [18]:

$$O = \langle X, R, F, A(D, R_S) \rangle \quad (2)$$

де:

$X$ – множина концептів;

$R$ – множина концептуальних відношень між ними;

$F : X \times R$ –скінченна множина функцій інтерпретації, заданих на концептах і/або відношеннях;

$A$ – скінченна множина аксіом, що використовуються для запису завжди істинних висловлювань (визначень і обмежень);

$D$ – множина додаткових визначень понять;

$R_S$ – множина обмежень, що визначають область дії понятійних структур.

Повна (комп'ютерна) онтологія є (формальним) вираження концептуальних знань про предметну область і за своєю значимістю порівнюється з базою знань ЗОІС.

Одним з поширених варіантів неповної онтології є структура виду $O = \langle X, R \rangle$, де множина $F$ в явному вигляді відсутня $(F = \varnothing)$, в припущенні, що концепти $x_j \in X$ загальновідомі (визначені за умовчуванням) або (і) досить повно інтерпретовані відношеннями $R$.

Відомо, що засоби інформатики, які проєктуються, відповідно до їх проблемної орієнтації базуються на певній сукупності фундаментальних принципів, методик та алгоритмів. Парадигма комп'ютерних онтології, що розвивається у взаємодії з методами і засобами системного аналізу, поклала початок розвитку нової гілки методів системного аналізу ПдО – системно-онтологічного аналізу (підходу) [18].

Центральною ідеєю системно-онтологічного підходу є розробка онтологічних засобів підтримки рішення прикладних задач – поліфункціональної онтологічної системи. Така система (точніше, її концептуальна частина) описується двійкою (3), що включає онтологію ПдО (складається з онтології об'єктів і онтології процесів) і онтологію задач [3, 22, 23].

$$OнC = \langle O^{ПдО}, \left(O^{O}, O^{П}\right), O^{З} \rangle \qquad (3)$$

На рис. 7 представлена схема онтологій-компонентів предметної області і проблемного простору (ПрП). ПрП – це модель всіх таких аспектів або компонент ПдО, з якими пов'язані (опосередковано або безпосередньо) знання, необхідні для вирішення різних задач у цій ПдО. Будь-який ПрП складається з двох блоків: інваріантної (відносно незмінної) частини і множини змінних частин, відповідних окремим задачам. У складі інваріантної частини, наприклад в методології SMEE (Structured Methodology for Elicitation of Expertise), виділяють сім типів компонент: об'єкти, інструменти, оператори, операції, кінцеві продукти, побічні продукти і обмеження [24]. Ці типи компонент – суть поняття, які добре групуються в онтології об'єктів і процесів, представлених на рис. 7:

– $O^{O}$ – онтологія множини об'єктів (понять, концептів) ПдО, яка розглядається як ієрархічна структура класів, підкласів та елементів класів;

– $O^{П}$ – онтологія множини процесів ПдО, яка розглядається як ієрархічна структура процесів, підпроцесів, дій та операцій;

– $O^{З}$ – онтологія сукупності задач (типових наборів), які можуть бути поставлені і вирішені в ПдО. Розглядається як ієрархічна структура задач, підзадач, процедур і операторів.

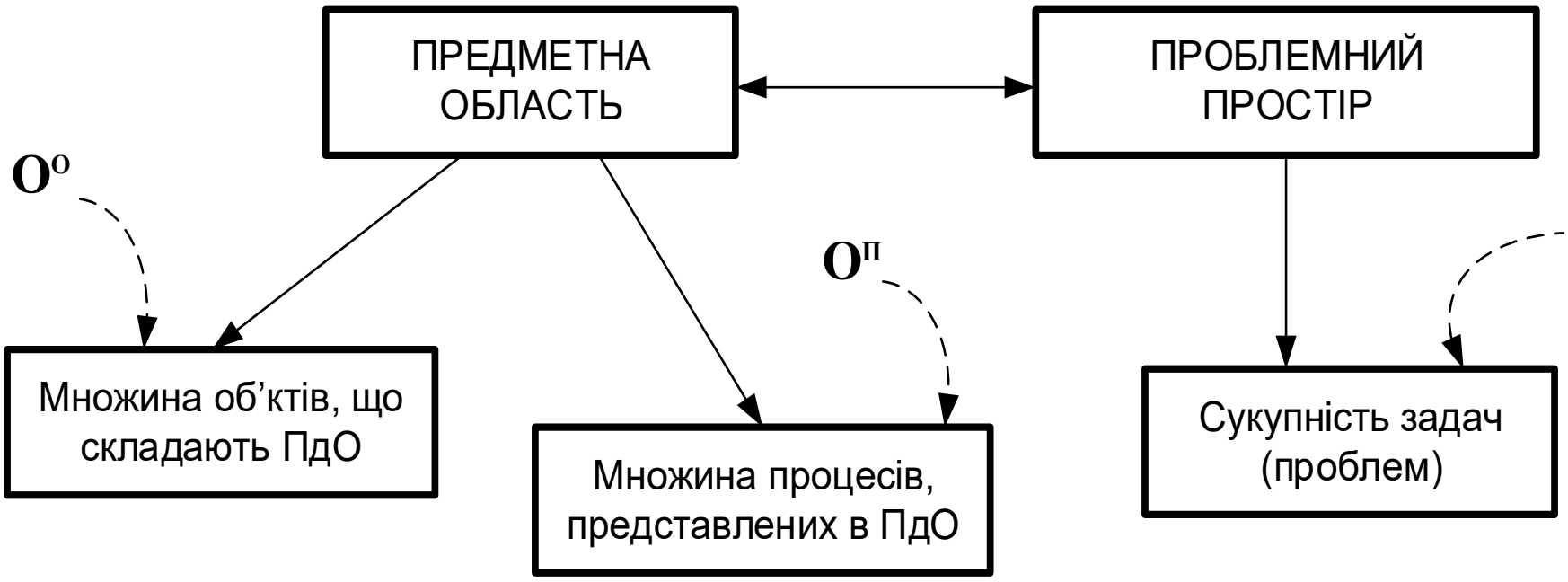

Рис. 7. Схема онтологій-компонентів предметної області

## 2.6. **Методологія розробки комп'ютерної онтології ПдО**

### 2.6.1 ***Загальний підхід до проєктування онтографа ПдО***

Для рішення задачі синтезу онтографа ПдО зазвичай задається сама предметна область і множина природномовних об'єктів, які складають лінгвістичний корпус текстів заданої ПдО. При цьому ЛКТ є максимально повним і включає, зокрема, тлумачні словники, енциклопедії, тезауруси та інші довідникові матеріали, що описують задану ПдО.

### 2.6.2 ***Онтологія об'єктів ПдО***

*Попередній аналіз предметної області*

У всі методології включений етап попереднього аналізу ПдО або складання змістовного технічного завдання на проєктування [18, 20, 25–31]. Цей етап (як і процес проєктування бази знань ПдО в цілому) має складний аналітичний характер і полягає в багаторазовому абстрагуванні, в результаті якого з усього різноманіття сторін і властивостей сутностей предметної області виділяються найбільш істотні, релевантні конкретним задачам [32]. Онтологічні системи покликані зробити знання колективним надбанням широкого кола осіб, дати потужний інструмент для фіксації, придбання та оброблення знань, перевірки їх на несуперечність, повноту тощо. Крім того, складається систематизоване подання знань про ПдО, виявляються джерела формування елементів множин і процедур, задач, що виконуються в аналізованій ПДО. Складається і документується словник термінів ПдО.

Етап попереднього аналізу ПдО включає:

– обґрунтований вибір точного (і достатнього) фрагмента ПрП, щодо якого будуть ставитися і вирішуватися задачі користувача;

– вибір методів і процедур системно-онтологічного аналізу, якими, зокрема, можуть бути абстрагування і конкретизація, композиція і декомпозиція, структурування, кластеризація і класифікація, тестування і верифікація;

– складання детального словника термінів і його розбиття на підмножини термінів-об'єктів, термінів-процесів і термінів, які іменують задачі та методи.

Якщо предметна область (і проблемний простір) представляє складну систему, то слід розглянути питання про попередній етап проєктування на основі методик IDEF, які доповнюють описані вище кроки проєктування.

Як правило, методика зводиться до алгоритму, який носить ітеративний характер. Для процесу розробки необхідно передбачити ряд “контрольних точок” для перевірки отриманих результатів на відповідність обраним критеріям. Зазначені критерії повинні мати співвідношення із заданими критеріями на проєктування бази знань ПдО, оскільки створення останньої є метою для розробників.

*Визначення* 2. Під *онтологією об'єктів* предметної області розуміється четвірка [18]:

$$O^O = \left\langle X^O, R^O, F^O, A^O\left(D, R_S\right)\right\rangle \tag{4}$$

де $X = \left\{ x_1, x_2, ..., x_i, ..., x_n \right\}$, $i = \overline{1, n}, n = Card\ X$ – скінченна множина концептів (понять-об'єктів) заданої ПдО;

$$R = \left\{ R_1, R_2, ..., R_k, ..., R_m \right\}, R \subseteq X_1 \times X_2 \times ... \times X_n,\ k = \overline{1, m},\ m = Card\ R,$$

– скінченна множина семантично значущих відношень між поняттями-об'єктами ПдО. Вони визначають тип взаємозв'язку між поняттями. У загальному випадку, відношення поділяють на загальнозначущі (з яких виділяють, як правило, відношення часткового порядку) і конкретні відношення заданої ПдО;

$F : X \times R$ – скінченна множина функцій інтерпретації, заданих на поняттях-об'єктах і/або відношеннях;

$A$ – скінченна множина аксіом, яка складається з множини визначень $D_i^l$ і множини обмежень $R_{s_i^t}$ для поняття $X_i$. Визначення записуються у вигляді тотожно істинних висловлювань, які можуть бути взяті, зокрема, з тлумачних словників ПдО. В них можуть бути зазначені додаткові взаємозв'язки понять $X_i$ з поняттями $X_j$. В множині обмежень $R_{s_i}$ можуть бути задані обмеження на інтерпретацію відповідних понять $X_i$.

Онтологія визначає загальновживані, семантично значимі "понятійні одиниці знань", якими оперують дослідники і розробники знання-орієнтованих інформаційних систем. Вона відокремлює "статичні" і "динамічні" компоненти знань ПдО від операціональних знань. На відміну від знань, закодованих в алгоритмах, онтологія забезпечує їх уніфіковане і багаторазове використання різними групами дослідників, на різних комп'ютерних платформах при вирішенні різних задач.

*Побудова компонент онтології*

Онтологія ПдО – це концептуальна модель реального світу, і її поняття повинні відображати цю реальність.

*Побудова множини* $X$ вважається найбільш важливим моментом при розробці онтології ПдО. Вона повинна бути обов'язково не пустою. Співвідношення між *Card X, Card R* і *Card F* характеризують онтологію за функціональною ознакою.

Для добре опрацьованих предметних областей за основу множини елементів $\{ x_i \}$ може бути взято вміст відповідних тлумачних словників. В іншому випадку слід скласти повний список термінів, в якому вказати (причому перетин об'ємів і сенсів понять в такому попередньому списку не суттєве):

– чим є кожен термін – поняттям-класом предметів або конкретним поняттям;

– вказати для кожного терміну можливі суттєві відношення з іншими термінами зі списку;

– описати можливі істотні властивості понять.

Відомо, що в будь-якій предметній області існують терміни-синоніми. Для них в онтології відводиться тільки одне поняття, в аксіомах якого може бути вказаний синонімічний ряд термінів. Іншими словами, синоніми одного і того ж поняття не представляють різні класи.

Далі слід уточнити і визначити остаточний список класів-понять, імена яких будуть входити в онтологію, що розроблюється, і є вершинами онтографа. Також слід прийняти єдині *правила присвоювання* імен поняттям і їх властивостям.

Потім, можливо, слід повторити деякі фрагменти процесу аналізу ПдО (з прив'язкою до складеного списку понять), виконані на попередньому етапі. Відзначимо, що в число зазначених вище "контрольних точках" (точок входу ітерації) повинно бути включено завершення розробки будь-якого компонента онтології.

В результаті має бути отриманий повний список істотних для заданої ПдО (і передбачуваних застосунків) понять і їх машинно-інтерпретовані формулювання [3].

*Побудова множини* $R$ також заснована на результатах попереднього етапу аналізу ПдО. По суті, потрібно встановити між елементами $x_i \in X$ семантичні *k*-арні відношення $R_k \subseteq X^k$. Іншими словами, необхідно побудувати множину ребер, що зв'язують вузли *орієнтованого графа*. В якості вузлів онтографа виступає множина понять ПдО. Вершиною (або вершинами) онтографа є родове поняття, яке не має надкласу, а найнижчий рівень представляють конкретні поняття (примітиви), тобто не мають видових понять в заданій ПдО.

На практиці множину $R$ на початковому етапі подають деяким узагальненим відношенням "*вище-нижче*". Відомо кілька підходів для розробки ієрархії класів: процес *низхідної* розробки, процес *зростаючої* розробки і *комбінований* процес розробки. Останній найчастіше використовується розробниками, так як він є більш природнім, спочатку оперує поняттями середнього рівня, до яких найчастіше звертаються розробники. Потім ці поняття узагальнюються і обмежуються.

Наприкінці даного підетапу слід співвіднести розроблені класи та їх ієрархії з результатами попереднього аналізу ПдО. Зокрема, уточнюються залежності для конкретних пар $(x_i, x_j)$. В процесі співвіднесення (і побудови ієрархії) слід враховувати, що [33]:

– прямі підкласи в ієрархії повинні розташовуватися на одному рівні узагальнення;

– клас може бути підкласом декількох класів, і тоді він може наслідувати властивості від всіх цих класів;

– якщо клас має тільки один прямий підклас, то, можливо, при

моделюванні допущена помилка або онтологія неповна;

– якщо у даного класу є більше дюжини (іноді говорять про число 7) підкласів, то, можливо, необхідні додаткові проміжні класи;

– в онтології число класів співвідноситься з числом передбачуваних застосунків.

І слід пам'ятати, що не існує єдиної правильної ієрархії класів.

Описана побудова онтографа є спеціальним видом класифікації понять ПдО – онтологічною класифікацією.

_Побудова множин_ $F$ і $A$. Залежно від функціональної орієнтації онтології, що проєктується, множини $F$ і $A$ можуть інтерпретуватися по-різному [17, 22, 27–29]:

1) $A \equiv F$ – множина аксіом тотожна множині функцій інтерпретації. В цьому випадку встановлюються істотні зв'язки між компонентами онтології і варіантами її використання. Основним призначенням такої онтології є однозначна інтерпретація понять, що входять в онтологію, спільнотою користувачів;

2) множина аксіом $A \subseteq F$ не тотожна множині функцій інтерпретації. В аксіомах задаються а) базові функції (підмножина $F$ ) або б) додаткові відношення (не є елементами множини $R$) між поняттями, обмеження та умови, які аналізуються в машині виведення ОнС і використовуються в процесі вирішення задач;

3) множина аксіом $A \supseteq F$ не тотожна множині $F$. Функції інтерпретації розглядаються як спеціальний вид відношень на множину понять $F: x_1 \times x_2 \times ... \times x_{n-1} \Rightarrow x_n$. В цьому випадку встановлюються істотні зв'язки між вже розробленими компонентами онтології і сукупністю задач передбачуваного застосунку (застосунків). Онтології з таким поданням використовуються в діалогових системах, в яких результатом є одне зі значень двоелементної множини або ім'я предиката.

В кінцевому рахунку, незалежно від того, яке з цих формулювань буде прийняте, ефективність розробленої онтології буде визначатися кінцевими результатами застосунків.

Крім того, з повного списку відібраних в онтологію термінів не всі представляють поняття. Існують терміни (наприклад, рольові), які відповідають властивостям певних класів-понять. Такі властивості слід прив'язати до опису самого загального класу, до якого вони належать. А підкласи цього класу будуть наслідувати вказану властивість (звичайно, якщо між ними встановлено певне відношення часткового порядку).

*Властивості понять* мають певні значення, такі як *тип* значень, *потужність* значень, *дозволені* значення (для даного класу) та інші. Наприклад, значення бувають з одиничною потужністю, потужністю без обмежень і потужністю з деяким допустимим інтервалом.

На основі побудованих множин кортежу можна синтезувати концептуальну модель ПдО, наприклад, за допомогою відомого інструментального засобу Protégé і отримати формальний опис розробленої онтології на одній з мов опису онтологій, а також графічне представлення онтографа.

### 2.6.3 *Онтологія процесів ПдО*

Синонімами онтологій об’єктів і процесів є відповідно статична і динамічна онтології ПдО. У науково-технічній літературі, коли говорять про онтології ПдО, то мається на увазі її статична складова. Саме компоненти останньої найбільш розроблені, як в літературі з філософії, так і в конкретних описах ряду предметних областей. Поведінковий опис сутностей-процесів найчастіше виконується у вигляді графічних діаграм і природномовних описів. Розробка ж бази знань не є прямою метою зазначених методик. Тому методики розробки онтології процесів практично невідомі, хоча в деяких відомих онтологіях верхнього рівня [30–33] сутність поняття “Процес” розглянута досить детально.

На рис. 8 представлений синтезований онтограф, який представляє схему початкової ділянки поняття "Процес", а саме тієї її частини, яка відповідає процесам в науково-технічних предметних областях (гілки онтографа “Соціальний процес”, “Матеріальний процес” і їм подібні не розглядаються).

Категорія *Процес* розглядається як *Дійсність* і *Подійність*, на відміну від категорії *Об’єкт*, що характеризується як *Дійсність* і *Тривалість* [34]. В першу чергу *Процес* розглядається як залежна від часу категорія і потім поділяється за видами змін, наявністю початкових і кінцевих точок тощо. Далі *Процес* поділяється на безперервний і дискретний. Перший з них характеризується наявністю експліцитних початкової і кінцевої точок або без явної вказівки цих точок. Другий вид процесу вказує, що зміни відбуваються дискретними кроками, названими подіями, які чергуються з періодами спокою, названими станами.

Наведена схема початкового розвитку онтології процесів не відображає всіх характеристик (підстав розгалужень в онтографі) категорії “Процес”, навіть для тієї її частини, що представлена на рис. 8.

*Визначення* 3. Під *онтологією процесів* предметної області розуміється трійка [3]:

$$O^{\Pi} = < \Pi, R^{\Pi}, F^{\Pi} > \qquad (5)$$

де:

$\Pi = \{nn_1, nn_2, ..., nn_l, ..., nn_L\}, \quad \forall l = (1 \div L)$ – скінченна множина концептів (понять-процесів) заданої ПдО;

$R^{\Pi} = \{r_1^n, r_2^n, ..., r_k^n, ..., r_K^n\}, \quad \forall k = (1 \div K)$ – скінченна множина семантично значущих відношень між поняттями-процесами ПдО. Вони визначають тип взаємозв’язку між процесами;

$F^{\Pi} = \Pi \times R^n = \{f_g^n\} = \{nn_l\} \times \{r_k^n\}, \quad \forall g = (1 \div G)$ – скінченна множина функцій інтерпретації, заданих на поняттях-процесах і/або відношеннях.

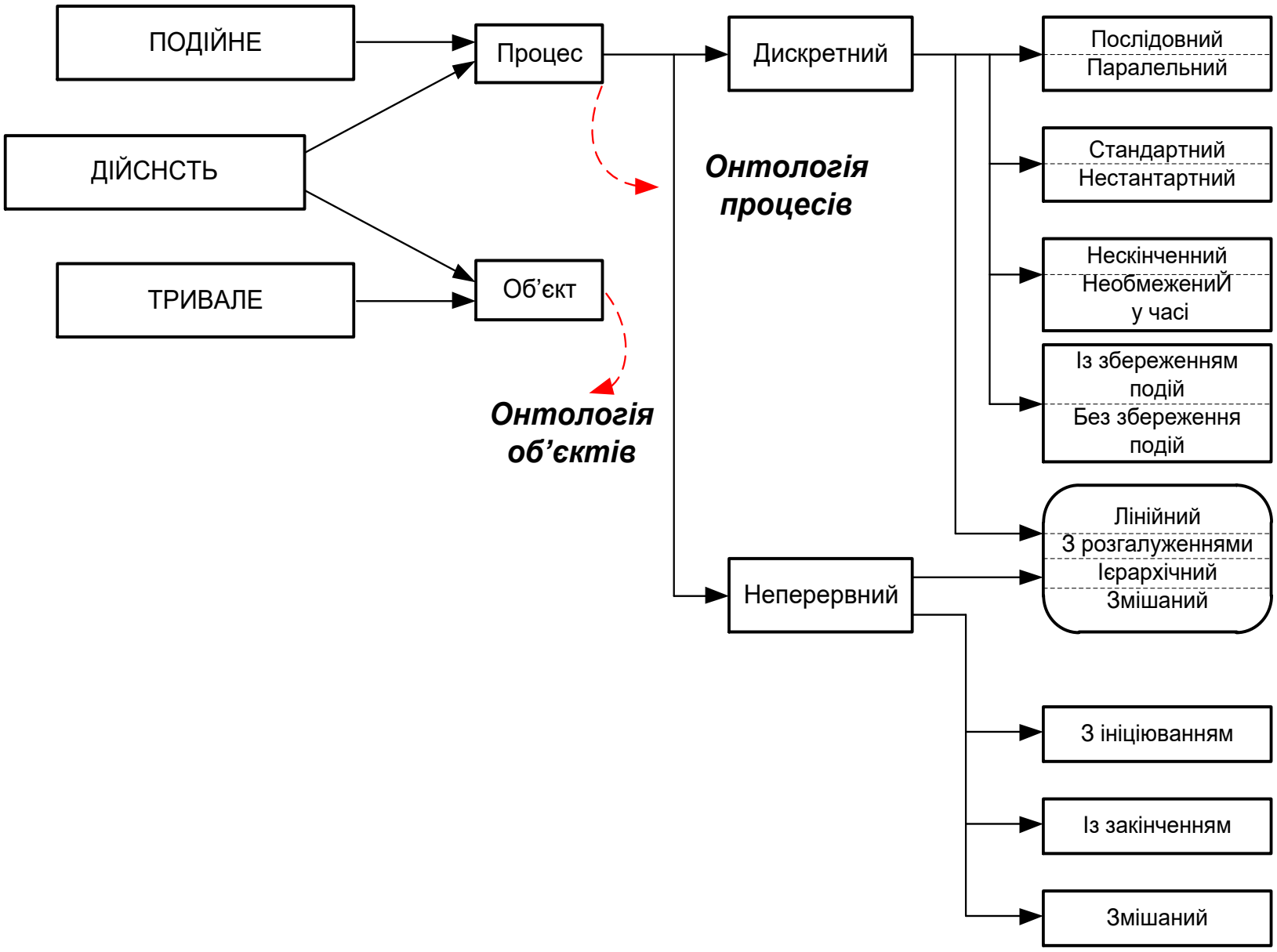

Рис. 8. Початкова ділянка онтології процесів

На рис. 9 представлена загальна схема онтології процесу ПдО, в якій категорія “Процес” представлена онтографом з $p$ рівнями і $n_p$

підпроцесами (Пп) на кожному рівні. Передостанній $(p-1)$ рівень представлений множиною дій (Д), на які розбивається кожен Пп попереднього рівня. В свою чергу, кожна дія на останньому ( $p$ -му) рівні розбивається на послідовність операцій $O_{i_j}^{p}$ .

Зв'язки між підпроцесами для сусідніх рівнів відповідають відношенням "Ціле-частина", а в середині кожного рівня – деякою змішаною формою організації з'єднань. На рис. 9 показаний окремий випадок такої організації – паралельний. Подальший розвиток (конкретизація) онтології процесів можливий, коли задана конкретна предметна область і відповідний проблемний простір, а в більш вузькому сенсі – конкретні ознаки розгалужень (умови ініціювання Пп, умови закінчення Пп та обмеження) в онтографі.

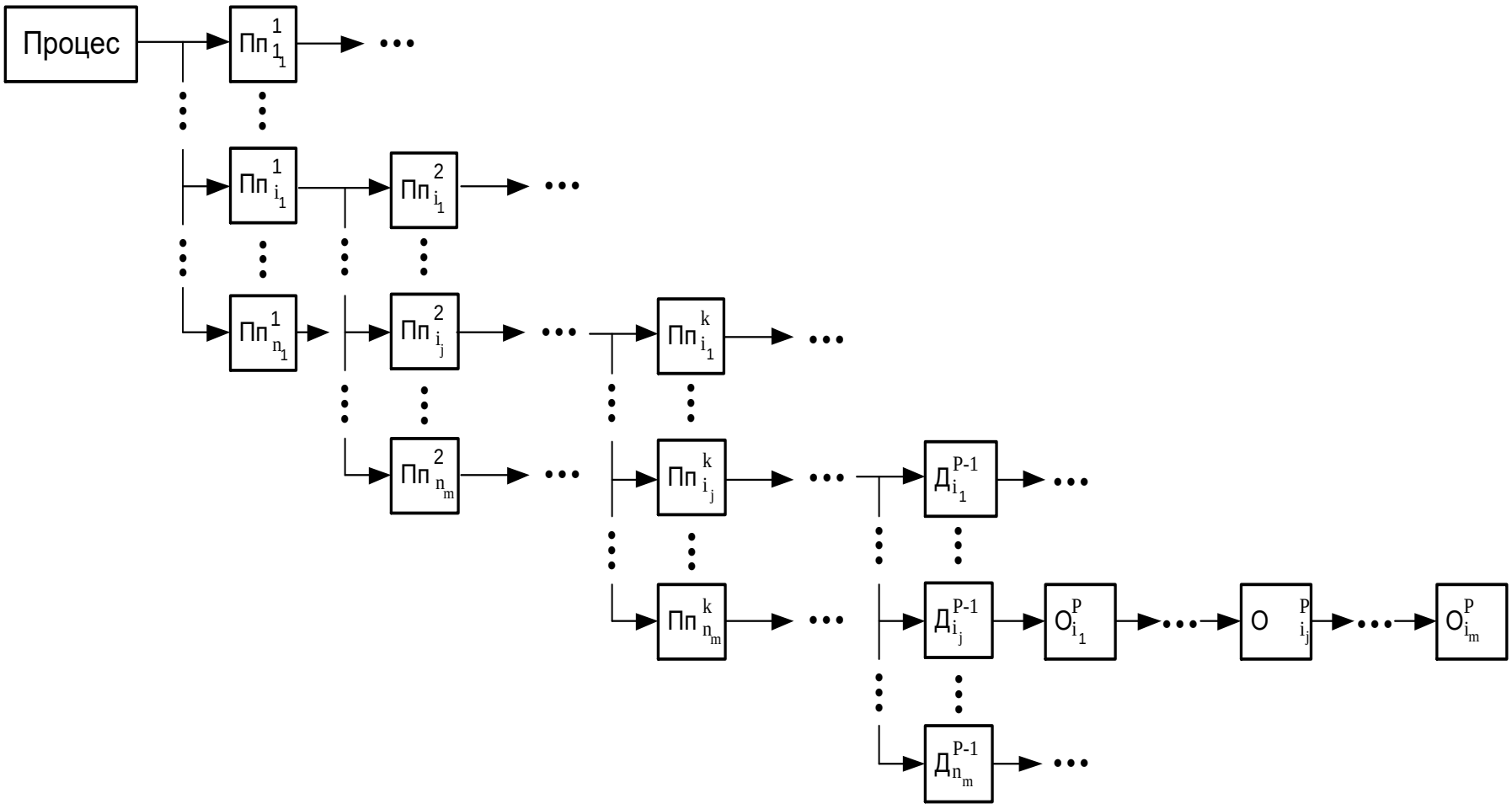

Рис. 9. Загальна схема онтології процесу

## 2.6.4 ***Онтологія задач проблемного простору***

З предметно-проблемних знань в загальному випадку виділяють поняття і процеси ПдО, поняття і класи задач проблемного простору (ПрП), а також методи рішення останніх. Сюди ж відносять і алгоритми, що реалізують відповідні методи. Перераховані типи знань з темпоральною ознакою поділяються на статичні і динамічні, а за онтологічною схемою на:

- онтологія об'єктів ПдО (в деяких публікаціях використовується термін "тезаурус";

– онтологія процесів ПдО;
– онтологія застосунків;
– онтологія класу задач ПрП;
– онтологія методів рішення задач.

Відомі різні схеми конструювання (синтезу) онтологічних структур з предметно-проблемних знань [35, 36]. В даній роботі в онтологію ПдО, включені об’єкти і процеси (статистичні знання), які згруповані в окремі онтології, а в онтологію задач включені класи задач, методи їх рішення та відповідні алгоритми. При такій схемі групування предметно-проблемних знань, останні строго розділені на статичні і динамічні. Якщо змінюється клас розв’язуваних задач, то онтології об’єктів і процесів вже готові до повторного використання, перепроєктується тільки онтологія задач [21, 22].

В текстових описах (специфікаціях) цільових задач виділяються набір об’єктів і набір процесів (методів), необхідних і достатніх для виконання конкретних цільових задач. Можна виділити деяку уніфіковану (поповнювану) множину базових задач (типових фрагментів задач), на основі яких за допомогою певних логічних послідовностей конструюються більш складні задачі.

Полем констрункта, що описує схеми типів задач, які вирішуються, є певна семантична вісь з протилежними сторонами “Вибір” і “Побудова”, концепти яких описують просту і складну схеми рішення задач. Під простою схемою (схема “Вибір”) розуміється така схема, в якій для рішення задачі онтологічній системі необхідно зробити вибір із сукупності відомих схем рішень. Коли вибір зроблено, то відомий метод, процес та його стани, тобто рішення задачі стає тривіальним. Протилежністю простій схемі є складна схема (схема “Побудова”), коли всі складові процесу рішення задачі невідомі. На практиці зустрічаються певні проміжні схеми рішення задач. Наприклад, схеми, в яких початкові і цільові стани, сукупності методів задаються в явному вигляді. Загальною стратегією рішення складних задач є багаторівнева декомпозиція вихідної задачі до того рівня, на якому отримані підзадачі є простими. При цьому онтологічні знання є активною основою процесу декомпозиції.

Онтологічні знання, що описують деякий ПрП, можна розділити на наступні компонентні знання: типи вхідних і вихідних даних, інструменти, оператори (людина або комп’ютерна програма) та операції (дії оператора або *Вирішувача задач*) [3].

Для реалізації підходу необхідно розробити уніфіковану мову представлення онтологічних знань та інструментальне середовище як набір спеціалізованих й універсальних базових операцій, які керують процесом рішення. Необхідно також розробити Вирішувач задач, який здійснює вибір засобів і методів формування структури задачі на основі базових операцій.

Схема моделі онтології задач описується трійкою [3]:

$$\mathrm{O}^3 = \left\langle ОЗ^{ПрП},\ M,\ РЗ \right\rangle, \tag{6}$$

де $ОЗ^{ПрП}$ – узагальнена задача проблемного простору, що складається з $p$ задач, які, в свою чергу, складаються з $w=\overline{1,W}$ фрагментів кожна. Кожен фрагмент представлений процедурою, реалізованою на множині $v=\overline{1,V}$ операцій кожна. Крім того, задача

$$З^p = \left\langle D_{in}^p,\ R^p, C^p,\ D_{out}^p \right\rangle \tag{7}$$

визначається множинами вхідних даних $D_{in}^p$, вимог (умов, обмежень) $R^p$, контексту задачі $C^p$ і вихідними даними (або метою рішення задачі) $D_{\text{out}}^p$;

$M$ – множина методів рішення задач. Визначається як відображення

$$M^p : \left(D_{in}^p,\ R^p, C^p\right) \to D_{out}^p, \tag{8}$$

компоненти якого визначені вище;

*ВЗ* – вирішувач задач.

На рис. 10 представлена схема онтології задач.

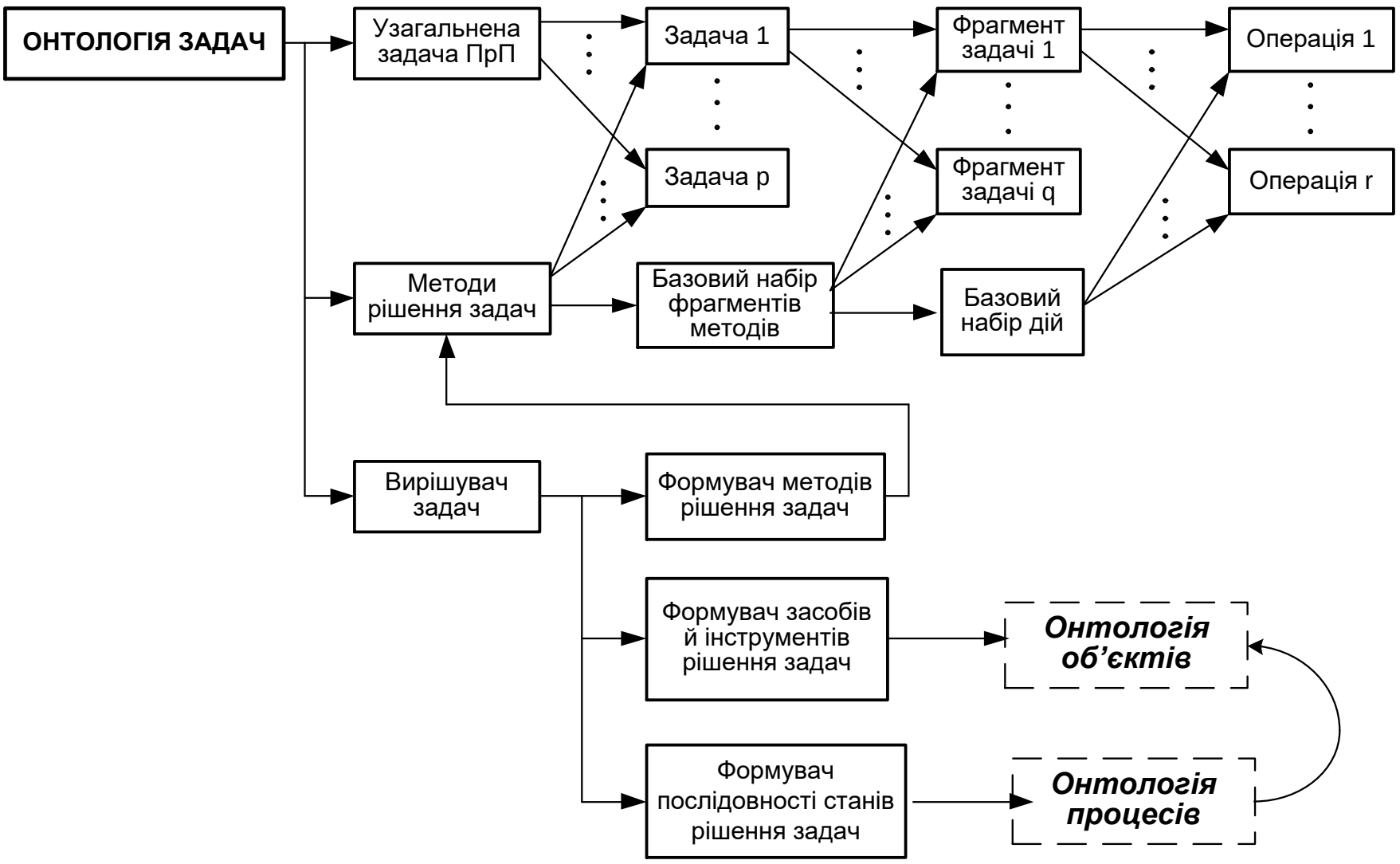

Рис. 10. Схема онтології задач

Одна з переваг онтологічного підходу, зокрема, ієрархічного подання полягає в тому, що складна задача великої розмірності розбивається на послідовно розв’язувані групи задач малої розмірності. Онтологія задач в якості понять містить типи розв’язуваних задач, а відношення цієї онтології, як правило, специфікують декомпозицію задач на підзадачі.

### 2.6.5 *Формалізовані аспекти інтеграції та її оцінка*

Розроблення і комерційне використання баз знань і відповідного інструментарію в багатьох прикладних областях (Knowledge based engineering systems) на основі онтолого-керованих інформаційних систем у багатьох науково-дослідних центрах і корпораціях привели до зростання теоретичних розробок формалізованих методологій проєкттування онтологічних структур. При цьому сутність зазначених методологій зводилася до формального обґрунтування структурування ієрархічного дерева онтології (формалізованої побудови наборів концептів і концептуальні відношення або категоризації, які їх пов’язують) і розробці формальних мов подання знань, які описують аксіоматизацію концептів предметної області. Наступним кроком розвитку теорії баз знань була необхідність теоретично обґрунтованого об’єднання (або системну інтеграцію) вже розроблених як загальнодоступних онтологій, так і комерційних баз знань для різноманітних прикладних задач, проблем, цілих предметних областей і трансдисциплінарних знань широкого призначення [37].

В загальному вигляді процес, що забезпечує системну інтеграцію множини онтологій, можна описати наступною формулою:

$$\bigcap_{KO} O_i, i = \overline{1, N}, \quad (9)$$

де $\bigcap_{KO}$ – знак концептуального об’єднання. Смисл цього знака полягає в системній інтеграції вихідних онтологічних графів з урахуванням областей визначень $O_i\ (i = \overline{1, N})$ та їх взаємозв’язку (взаємодії).

Об’єм знань $W$ в предметних областях можна оцінити через характеристики (параметри) їх формально-онтологічних представлень. Зокрема, при представленні онтологічним графом (без урахування типів відношень і складності функцій інтерпретації) величина $W$ може характеризуватися числом вершин ОГ. У разі простої деревовидної структури це число може бути виражене формулою:

$$W = \sum_i \sum_h \sum_l O_i \cdot S_{h,l}\,, \tag{10}$$

де $O_i$ – онтограф і-ої предметної області, $i = \overline{1,N}$, $S_{h,l}$ – ступінь вершини, що дорівнює числу ребер, що виходять з неї, $h = \overline{1,H}$ – кількість рівнів ОГ, $l = \overline{1,L_h}$ – номер вершини на відповідному $(h-му)$ рівні ОГ.

При рівномірній щільності розподілу ОГ, тобто при $S_{h,l} = S(H, l = 1,2,...)$ (7) зводиться до відомої формули суми геометричної прогресії:

$$W = \sum_i O_i \left( \frac{1-S^h}{1-S} \right), \tag{11}$$

Облік типів відношень і складність функцій інтерпретації призводить до ОГ зі зваженими вершинами і ребрами. Вираз (11) при цьому перетворюється в вид:

$$W = \sum_i \sum_h O_i \left( \alpha_l + \sum_j \beta_{l,j} \right), \tag{12}$$

де $\alpha_l$ і $\beta_{l,j}$ – значення вагових функцій відповідних відношень і функцій інтерпретації, приписані вершинам $(\alpha_l)$ і ребрам $(\beta_{i,j})$ ОГ. Вираз (12) дає повну оцінку складності ОГ, а відношення $\omega = W^0 / W$ характеризує середню щільність зваженого ОГ.

Розглянуті оцінки дозволяють порівнювати різні варіанти представлення знань про предметні області, а також відслідковувати процес еволюції наукових теорій.

Процес розвитку знань в будь-якої ПдО пов'язаний з її аналізом, концептуалізацією і побудовою формальної теорії. При цьому формалізація в загальному випадку відноситься до чотирьох основних видів представлення інформації:

$$I = I(V, A, T, G), \tag{13}$$

тобто до вербального (*V*), аналітичного (*A*), табличного (*T*) і графічного (*G*). Між ними існує взаємно-однозначна відповідність, не завжди на практиці реалізується строго і повністю. Тому всі вони знаходять своє, цілком певне місце при описі наукової теорії. В багатьох випадках коректним виявляється їх обмеження до двох: вербального і аналітичного.

Як правило, процес розвитку теорій супроводжується перерозподілом об'єму інформації про предметну область між вербальною і формальної компонентами, тобто між природномовним описом предмета дослідження і формально-аналітичним (формульним *Ф*, табличним *T*, графічним *G* представленням суті теорії). Очевидно, що формалізоване подання є більш компактним, а головне, – більш строгим і придатним для комп'ютерного оброблення [37].

*Перспективи дослідження проблем системної інтеграції онтологічних знань*

Дослідження проблем системної інтеграції онтологічних знань необхідно розглядати в контексті міждисциплінарних наукових досліджень (МНД), хоча вони мають і самостійне науково-практичне застосування. Концепція розвитку МНД викладена в [5], в якій методологія системної інтеграції онтологічних знань, побудова онтологізованих систем знань предметних галузей є одними з головних складових компонентів. При цьому основну роль відводиться формальним (комп'ютерним) онтологіям предметних галузей.

Онтології дозволяють формалізувати та компактно представляти накопичені знання, одночасно визначаючи та об'єднуючи термінологію різних предметних галузей, будувати єдину наукову картину світу як результат комплексних трансдисциплінарних досліджень [38].

## 2.7. **Основні технології та інструментарій Semantic Web**

### 2.7.1 ***Коротко про основні засади Semantic Web***

Вже тривалий час онтології застосовуються в мультиагентних технологіях в середовищі Semantic Web (SW). Остання – це напрямок розвитку Всесвітнього павутиння, основною метою якого є представлення інформації у вигляді, зручному для машинного оброблення на основі технологічних стандартів. SW передбачає запис інформації у вигляді семантичної мережі за участю онтологій, що дозволяє агентам безпосередньо добувати із SW факти та генерувати логікові наслідки з цих фактів у взаємодії з користувачем.

Розробка SW, як і онтологій, вимагає уніфікації представлення даних та мови маніпулювання цими даними. Ця уніфікація повинна як візуалізувати дані, так і виконувати перенесення правил виведення в SW з метою перетворення їх у систему семантичного рівня [39].

Онтологічний підхід до розробки SW передбачає розвиток засобів анотування документів, якими могли б скористатися Web-сервіси та спеціалізовані програми-агенти у процесі оброблення складних запитів користувачів. У зв'язку з цим консорціум W3C вирішив, що для практичного використання SW досить розробити універсальну мову подання знань з посиланнями на онтології (для чого слугує RDF – мова низького рівня подання метаданих), мова подання онтологій (OWL і OWL2), мова опису Web- сервісів (WSDL, OWL-S), мова запитів до бази знань (SPARQL), семантичні пошукові системи та програми-агенти.

Таким чином, SW є системою з елементами ШІ, а онтології в такій системі слугують допоміжним засобом концептуалізації та уніфікації даних і доступу до них. Для реалізації доступу до даних, що містяться в онтологіях, слугує RDF документ. На початку цього документа наведено список посилань на об'єкт із деякої онтології. Крім того, вузли цього списку можуть містити додаткову інформацію, що дозволяє будувати на основі RDF мови вищого рівня (RDF Schema, OWL). А це дозволить створювати спеціалізовані формати для представлення різноманітних типів об'єктів. На шляху вирішення цієї проблеми консорціум W3C розробив спеціалізовану мову запитів до RDF сховищ – SPARQL, яка дозволяє здійснити складні вибірки із масивів метаданих [39, 40].

На цьому рівні розвитку SW ґрунтується на трьох базисних компонентах: програмах-агентах, мові розмітки XML, що розширюється, і Web-онтологіях. Ці базисні принципи та компоненти використані у різних окремих проектах, серед яких виділяються два: Dublin Core (Дублінське ядро) та BDpedia. Перший з них реалізований ініціативною організацією Dublin Core Metadata Initiative (DCMI) та представляє відкритий проект розроблення метаданих, незалежних від платформ та придатних для використання у різних галузях. У рамках цього проекту здійснюється розроблення словників (тезаурусів) метаданих загального використання, які стандартизують представлення ресурсів у форматі RDF. Другий проект BDpedia реалізує добування структурованої інформації з проекту Wikipedia. При цьому користувач повинен запитувати інформацію, що базується на взаємозв'язках та властивостях ресурсів Wikipedia, а також з посиланнями на відповідні бази даних. Цей проект використовує RDF для представлення отриманої інформації.

### 2.7.2 *Короткий огляд інструментальних засобів Semantic Web*

Доступний цілий ряд різних SPARQL процесорів для виконання запитів, як до локальних, так і до віддалених даних (кінцевих точок). Нагадаємо, що *кінцева точка* – це ресурс, з яким процес може зв'язатися та використовувати як службу; кінцева точка SPARQL приймає SPARQL запити та повертає результати, використовуючи протокол SPARQL для RDF.

Для запитів до файлу даних на жорсткому диску комп'ютера доступна безкоштовна, заснована на Java, програма ARQ, що робить її досить простою. ARQ є частиною інфраструктури Apache Jena і для її отримання необхідно перейти за посиланням «Завантаження» з домашньої сторінки ARQ.

ARQ процесор зручний, коли вивчається SPARQL мова, тому що відразу після його встановлення можливо вказати йому файл даних і файл із запитом і відразу отримати результати виконання запиту для цих даних. Для встановлення та запуску сервера не потрібні жодні кроки з налаштування або конфігурації [41].

Коли необхідно розробити досить складний застосунок і налаштувати сервер, який має можливість приймати SPARQL запити і повертати результати, має сенс зібрати просту систему для відпрацювання майбутніх рішень. Для цього можна використовувати ARQ процесор, оскільки він є частиною проєкту Jena з відкритим вихідним кодом. Можливо використовувати його бібліотеки і вихідний код для його інтеграції як компонент механізму запитів для більшого проєкту на основі JVM, як це роблять сервери Jena Fuseki і Joseki SPARQL. Цей підхід не призведе до створення швидкого, надійного та масштабованого застосунку, але він добре підходить для створення прототипів при оцінці даних, з якими можна працювати [41].

Якщо в онтології є багато класів, властивостей та відношень між ними для визначення або зміни, то замість безпосереднього звернення до синтаксису RDF/XML або Turtle ваших моделей RDFS або OWL буде простіше використовувати графічний інструмент (або онторедактор), що дозволяє вказати необхідне, натиснувши, перетягнувши та заповнивши діалогові вікна, у тому числі сформувати OWL онтологію у необхідному синтаксисі. Для цього добре відомим інструментом з відкритим кодом є Protégé, розроблений в Стенфордському університеті. Найпопулярніший комерційний інструмент – TopBraid Composer. (Безкоштовна версія TopBraid Composer дозволяє виконувати все необхідне моделювання, а також виконувати SPARQL запити до даних; у комерційних версіях додані функції для розробки та розгортання застосунків). Обидва ці інструменти зберігають виконану роботу у стандартних синтаксисах [41].

*Логікове виведення*

У словнику New World College Dictionary Вебстера визначається "*зробити виведення*" як "зробити виведення або прийняти рішення з чогось відомого чи передбачуваного". Коли виконується RDF висновок, існуючі трійки є «чимось відомим», і інструменти виведення будуть виводити з них нові трійки. (Ці нові трійки дозволяють дійти будь-якого висновку чи рішення, залежно від того, що програма робить з ними).

OWL стандарт заснований на стандарті RDFS, тому OWL ризонер з відкритим вихідним кодом Pellet "розуміє", що свідчать властивості rdfs:domain і rdfs:range. Зазначимо, що Protégé 5 поставляється із різонером HermiT.

Будучи OWL різонером, Pellet також розуміє значення спеціальних класів та властивостей з простору імен OWL, таких як owl:SymmetricProperty та owl:inverseOf, що надає більшої виразності для SPARQL запитів, що виконуються за допомогою Pellet.

Комбінації логікового виведення та виконання SPARQL запитів зазвичай поділяються на дві категорії:

– запуск інструменту, що поєднує у собі етап логікового виведення та етап запитів. Деякі з цих інструментів дозволяють налаштувати, чи повинні вони виконувати будь-які відповідні виведення перед виконанням SPARQL запиту; наприклад, процесор Sesame включає прапорець «Включити виведені оператори» на екрані, де можна вводити SPARQL запити, а TopBraid Composer пропонує аналогічну кнопку на своїй панелі SPARQL запитів.
– використання одного інструменту для виконання логікового виведення, а іншого – для виконання SPARQL запиту на комбінацію вхідних та вихідних даних кроку логікового виведення.

З точки зору обслуговування людських потреб – ідея Semantic Web полягає у звільненні людини від обтяжливих рутинних завдань з добування, пошуку, обліку та індексування інформації, що міститься в Web. "Semantic Web – це бачення наступного покоління Інтернет, який дозволить веб-застосункам автоматично збирати веб-документи з різних джерел, враховувати та обробляти інформацію, а також взаємодіяти з іншими застосунками для виконання складних завдань" [41].

З точки зору поліпшення анотування – "ідея Semantic Web полягає в забезпеченні користувачів Web анотаціями, поданими у машинно-оброблюваній формі і пов’язаними між собою" [42].

З погляду поліпшення пошуку – реалізація пошуку як за ключовими словами, так і за контентом.

З точки зору веб-сервісів – "Semantic Web повинен забезпечити доступ не тільки до статичних документів, що містять корисну інформацію, але і до сервісів, які надають корисні послуги" [43].

Таким чином, задачі Semantic Web полягають у наступному [44]:

– індексація та пошук інформації;
– розроблення та підтримка метаданих;
– розроблення та підтримка методів анотування;
– подання Web у вигляді великої, інтероперабельної бази даних;
– організація машинного добування даних;
– виявлення (discovery) та надання веб-орієнтованих сервісів;
– дослідження в галузі інтелектуальних програмних агентів.

Розглянемо більш детально опис та алгоритм роботи з сервером *Apache Jena Fuseki*, який застосовується в онтологічній системі оброблення баз даних наукових публікацій (розділ 4).

*Формування та встановлення віддаленої прикінцевої точки «Apache Jena Fuseki»* [45]

Для зберігання та оброблення формальних описів наукових публікацій використовується застосунок *Apache Jena Fuseki*, який побудований на основі *Apache Web Server*. Він представляє собою SPARQL сервер, який можна запускати як сервіс операційної системи, Java web додаток, або автономний сервер. Сервер Fuseki доступний для завантаження у двох формах, як цільно-системний web застосунок, з графічним інтерфейсом для адміністрування та формування запитів, або як основний сервер, придатний для запуску як частина масштабнішого системного розгортання. Другий варіант можна запускати разом з Docker, або як вбудовану програму. Обидва варіанти використовують одну реалізацію протоколу та формат конфігураційних файлів. *Docker* – це відкрита платформа для розробки, постачання та виконання програм. Вона дозволяє розділити інфраструктуру та програми для пришвидшення постановки програмного забезпечення. Fuseki надає підтримку SPARQL протоколу версії 1.1 для запитів та оновлення даних. Також Fuseki підтримує SPARQL graph store protocol, тісно інтегрований з TBD (компонент Jena для зберігання RDF та запитів до RDF ресурсів) для надання надійної та стійкої підсистеми зберігання даних. TBD підтримує повний набір Jena API (Application Programming Interface) і може використовуватись як RDF сховище з швидкісним доступом на одному комп’ютері. Керувати та використовувати TDB сховище можливо за допомогою наборів команд командної строки та за допомогою Jena API. Ця підсистема також включає Jena text query(модуль для повного індексного текстового пошуку). Apache Jena Fuseki надає широкі функціональні можливості для використання, зокрема нами використано у вигляді віддаленої (або локальної) прикінцевої точки з доступом за HTTP протоколом.

На офіційній сторінці Apache Jena Fuseki (jena.apache.org) доступні декілька дзеркал для завантаження необхідного дистрибутиву. Також можливо завантажити попередні версії продукту. Нами застосовується версія 4.1.0. Для запуску Fuseki як автономного сервера достатньо виконати наступні кроки.

В Linux системі виконати команду "fuseki-server [–loc=DIR] [[--update] /NAME]", ця команда запустить сервер за адресою http://localhost:3030/NAME. NAME у даному випадку вказує на ім'я опублікованого набору даних. В цій команді використовується база даних типу TBD1, для того щоб використовувати TBD2 необхідно змінити команду на "fuseki-server –tbd2 [–loc=DIR] [[--update] /NAME]". Якщо необхідно створити набір даних у пам'яті можна використати команду "fuseki-server –mem /NAME". Для завантаження файлу з даними необхідно використати команду "fuseki-server –file=Data.ttl /NAME" де Data.ttl може бути будь яким RDF файлом. Якщо для запуску сервера використати одну з наведених вище команд адміністрування буде доступне тільки з локального фізичного сервера.

*Формування RDF-сховища наукових публікацій*

Вище було наведено декілька консольних команд для завантаження даних на сервер. Простішим варіантом є використання графічного інтерфейсу. Якщо перейти за адресою http://127.0.0.1:3030 в веб-браузері на комп'ютері з запущеним Fuseki, то на екрані з'явиться графічний інтерфейс для адміністрування серверу. Для завантаження наших RDF файлів необхідно натиснути кнопку "dataset" і вибрати вкладку "upload files". В полі "destination graph name" можна вказати назву графу в який будуть завантажені дані.

Після натискання на кнопку "upload all" дані будуть додані до RDF сховища і з ними можливо виконувати SPARQL запити.

У Розділі 4 наведено приклади виконання SPARQL запитів до ОнС та інструкцію користувача.

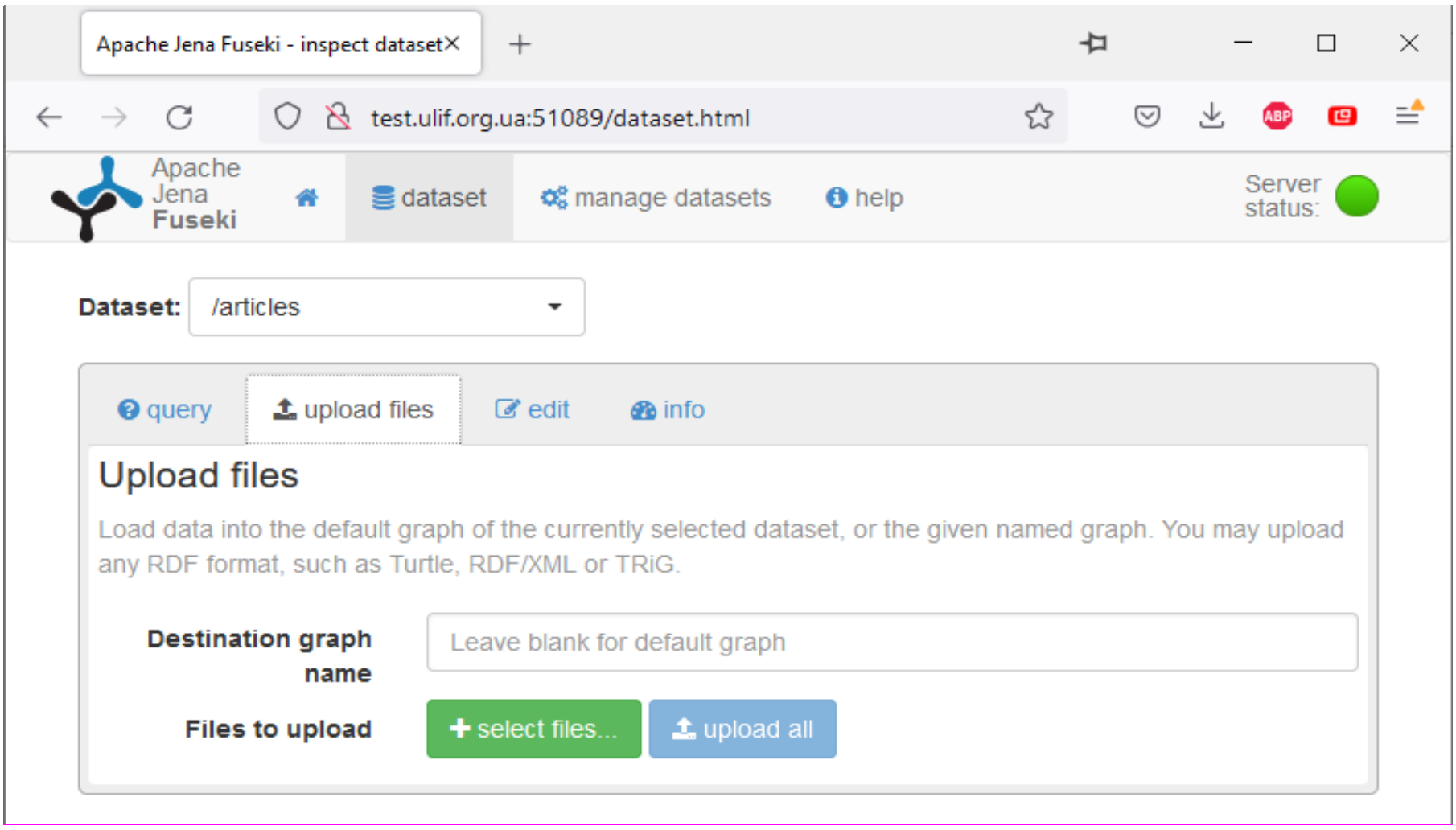

Рис. 11. Графічний інтерфейс адміністрування *Apache Jena Fuseki*

**Висновки за розділом**

1.Онтологічний підхід до представлення й інтеграції наукових знань дозволяє створити ефективні засоби побудови систем і технологічну базу системології трансдисциплінарної взаємодії й онтологічного інжинірингу.

2. Онтологічний інструментарій дозволяє побудувати мовно-онтологічну картину світу (різновидність лексикографічної системи), розглядаючи її як складову наукової картини світу, яка є основою трансдисциплінарної концепції наукових досліджень. У цьому випадку мовно-онтологічна картина світу виконує функції категоріальної надбудови баз знань в конкретних предметних областях й інтегрованих сховищах знань. Очевидно, що системна інтеграція знань виконується з врахуванням специфічних формально-методологічних вимог і критеріїв при формуванні достовірних тверджень і виведень, а категоріальний рівень з відповідною системою базових відношень представляє верхній рівень ієрархії мовно-онтологічної картини світу.

3. Наукова картина світу припускає багатовимірне його представлення, а тому наряду з понятійними компонентами онтограф повинен мати і їх *образні еквіваленти*. Тут доречно відмітити, що *онтологічна парадигма* почала і розвивалась практично одночасно з віртуальною. Сьогодні в повсякденну практику увійшли такі поняття, як віртуальний світ, віртуальна організація, віртуальна лабораторія, віртуальна система, віртуальна адресація тощо.

4. Теорія і практика створення та використання систем, заснованих на знаннях, – найбільш актуальний й такий, що інтенсивно розвивається, напрямок Computer Science, який дозволяє підвищити ефективність створення та застосування комп'ютерних технологій, прикладних систем й інструментальних засобів.

5. Особливість поточного періоду розвитку інформаційних техніки та технологій полягає в інтеграції результатів двох колись паралельно і незалежно сфер штучного інтелекту, що розвивалися: knowledge-engineering і комп'ютерної лінгвістики (когнітивної семантики), що відображає природну схему взаємодії людини з навколишнім світом. Свідомість у ній виступає як персоніфікований інструмент, що виробляє сукупність предметно, ситуаційно чи причинно-пов'язаних сутностей, які складають «*свідому*» картину світу.

6. Онтології – це суть понятійні системи, а понятійне мислення – найдосконаліша форма функціонування свідомості та інтелекту. Прототипом такої системи можуть слугувати знання-орієнтовані інформаційні системи з онтолого-керованою архітектурою, що активно розвиваються.

7. Загальне завдання онтології – компенсувати відсутність стандартів на подання знань при взаємодії користувача з інформаційними системами та останніх між собою, а також інтеграція знань предметних областей як головне завдання трансдисциплінарних наукових досліджень.

8. Перехід від недетермінованого режиму продукування та використання знань суб'єктами наукового процесу до режиму ефективного управління знаннями, представленими в уніфікованій формі на всіх етапах їх життєвого циклу, забезпечить зростання результативності та якості наукових досліджень. При цьому усталене знання стане інтелектуальним капіталом, а суб'єкти науки – безпосередніми учасниками економічної діяльності суспільства, що створить сприятливі умови для стимулювання розвитку, як самої науки, так і креативного суспільства.

9. Центральною ідеєю системно-онтологічного підходу є розробка онтологічних засобів підтримки рішення прикладних задач – поліфункціональної онтологічної системи. Така система (точніше, її концептуальна частина) описується двійкою, що включає онтологію предметної області (складається з онтології об'єктів і онтології процесів) і онтологію задач.

10. Процес розвитку теорій супроводжується перерозподілом об'єму інформації про предметну область між вербальною і формальною компонентами, тобто між природномовним описом предмета дослідження і формально-аналітичним представленням суті теорії. Очевидно, що формалізоване подання є більш компактним, а головне, – більш строгим і придатним для комп’ютерного оброблення.

11. Онтології застосовуються в мультиагентних технологіях в середовищі Semantic Web, яке є напрямком розвитку Всесвітнього павутиння, основною метою якого є представлення інформації у вигляді, зручному для машинного оброблення на основі технологічних стандартів. Semantic Web передбачає запис інформації у вигляді семантичної мережі за участю онтологій, що дозволяє агентам безпосередньо добувати із Semantic Web факти та генерувати логікові наслідки з цих фактів у взаємодії з користувачем.

12. Semantic Web – це динамічна концепція, яка постійно розвивається, а не набір комплексних працюючих систем.

13. Що до машинного оброблення даних, то Semantic Web – це ідея зберігання даних у Web таким чином, щоб вони були визначені і пов’язані між собою для подальшої можливості автоматизованого оброблення, інтеграції та повторного використання їх у різних застосунках.

14. Що до інтелектуальних агентів, то метою Semantic Web є зробити існуючий Web більш “машинно-читаємими” для того, щоб мати можливість використовувати інтелектуальних агентів для пошуку та оброблення відповідної інформації.

**Перелік посилань до розділу 2**

1. Palagin A. V. Transdisciplinarity Problems and the Role of Informatics. Cybernetics and Systems Analysis volume 49, pages 643–651 (2013). DOI: https://doi.org/10.1007/s10559-013-9551-y .
2. Palagin O.V. Transdisciplinarity, informatics and development of modern civilization. Visn. Nac. Akad. Nauk Ukr. 2014. (7): 25-33. DOI: http://doi.org/10.15407/visn2014.07.025/ .
3. Палагин А.В., Крывый С.Л., Петренко Н.Г. Онтологические методы и средства обработки предметных знаний. Луганск: изд-во ВНУ им. В. Даля, 2012. 323 с.
4. Palagin A. Information Technologies for Solution of Complex Science and Technical Problems // Розподілені комп'ютерні системи: Збірник праць Ювілейної міжнародної науково-практичної конференції. РКС, 2010. Том 1. С. 12–13.
5. Palagin A.V., Kurgayev A.F. Interdisciplinary scientific research: optimization of system-information support. Visn. Nac. Akad. Nauk Ukr.2009. (3): 14-15. [in Ukrainian]. URL: http://dspace.nbuv.gov.ua/bitstream/handle/123456789/3466/st3-statta.pdf/. (дата звернення: 26.06.2022).
6. Palagin, A.V. Architecture of ontology-controlled computer systems. Cybern Syst Anal. С. 111–124. DOI: https://doi.org/10.1007/s10559-006-0061-z.
7. Palagin, A.V. An Ontological Conception of Informatization of Scientific Investigations. Cybern Syst Anal 52, 1–7 (2016). DOI: https://doi.org/10.1007/s10559-016-9793-6.
8. Величко В.Ю., Попова М.А., Приходнюк В.В., Стрижак О.Є. ТОДОС– IT-платформа формування трансдисциплінарних інформаційних систем. *Системи озброєння і військова техніка*, 2017. № 1(49). С. 10–19.
9. Палагін О.В., Петренко М.Г. Тлумачний онтографічний словник з інженерії знань: Київ: Інтерсервіс, 2017. 478 с.
10. Войшвилло Е.К. Понятие: Изд-во Московского университета, 1967. 288 с.
11. Palagin A.V., Petrenko N.G. Methodological Foundations for Development, Formation and IT-support of Transdisciplinary Research. Journal of Automation and Information Sciences, Volume 50, 2018, Issue 10, pp. 1–17.

DOI: https://doi.org/10.1615/JAutomatInfScien.v50.i10.10.

12. Kurgaev A.F., Palagin A.V. Concerning the information support for research. Visn. Nac. Akad. Nauk Ukr. 2015. (8): 33–48. DOI: https://doi.org/10.15407/visn2015.08.033.
13. Голышев Л.К. Системный подход к анализу и проектированию сложных систем: Системный проект: науч. моногр. К.: ГП «Информационно-аналитическое агенство», 2011. 555 с.
14. Згуровский М.З., Панкратова Н.Д. Системный анализ: проблемы, методология, приложения. К.: Наук. Думка, 2005. 744 с.
15. Философско-методологические основания системных исследований: системный анализ и системное моделирование: сб. научн. Трудов. ВНИИ системных исследований. М.: Наука, 1983. 324 с.
16. Rumbaugh, J., Blaha, M., Premerlani, W., Eddy, F. and Lorensen, W. (1991).Object-oriented modeling and design. Englewood Cliffs, New Jersey: Prentice Hall.
17. Natalya F. Noy and Deborah L. McGuinness. Ontology Development 101: A Guide to Creating Your First Ontology. Stanford Knowledge Systems Laboratory Technical Report KSL-01-05 and Stanford Medical Informatics Technical Report SMI-2001-0880, March 2001. http://protege.stanford.edu/publications/ontology_ development/ontology101.html. (дата звернення: 19.02.2012).
18. Палагин А.В., Петренко Н.Г. Системно-онтологический анализ предметной области. УСиМ, 2009. № 4. С. 3–14. URL: http://usim.org.ua/arch/2009/4/2.pdf. (дата звернення: 19.02.2012).
19. Guarino N. Formal Ontology and Information Systems / In N. Guarino (ed.) Formal Ontology and Information Systems / FOIS'98, 6–8 June, 1998. Trento, Italy: IOS Press, Amsterdam, 1998. pp. 3–15.
20. T. A. Gavrilova and V. F. Khoroshevskii, Knowledge Bases of Intelligent Systems [in Russian], Piter, St. Petersburg (2001). 2001. 384 с.
21. Добров Б.В., Иванов В.В., Лукашевич Н.В., Соловьев В.Д. Онтологии и тезаурусы: модели, инструменты, приложения. Интернет-университет информационных технологий. ИНТУИТ.ру, БИНОМ. Лаборатория знаний, 2009. 176 с. ISBN: 978-5-9963-0007-5.

22. Добров Б.В., Лукашевич Н.В., Невзорова О.А., Федунов Б.Е. Методы и средства автоматизированного проектирования прикладной онтологии. *Изв. РАН. Теория и системы управления*. М., 2004. № 2. С. 58–68.

23. Невзорова О.А. Онтолингвистические системы: методологические основы построения // Научная сессия МИФИ-2007: сб. научн. трудов. Интеллектуальные системы и технологии. М., 2007. Том 3. С. 84–85.

24. Методический комплекс по дисциплине “Современные проблемы науки” / Методология анализа предметных знаний. URL: http://old.ulstu.ru./people/SOSNIN/umk/Modern_Scientific_Problems/belobo rodov/item_znan.htm. (дата звернення: 11.10.2007).

25. Palagin A. V. and Yakovlev Yu. S., “System integration of computer means,” [in Russian], Universum, Vinnitsa, (2005). 680 p.

26. Марка Д. А., МакГоуэн К. Методология структурного анализа и проектирования. М.: "МетаТехнология", 1993. 239 с.

27. Gomez-Perez A., Fernandez-Lopez O. M. Ontological engineering: With Examples from the Areas of Knowledge Management, E-commerce and the Semantic Web. O Corcho (2004). Springer, 2004. 403 p. DOI: https://doi.org/10.1007/b97353 .

28. Клещев А.С., Артемьева И.А. Отношения между онтологиями предметных областей. Ч.1. Информационный анализ. Выпуск 1, 2002. С. 4–9.

29. Андреев А.М., Березкин Д.В., Симаков К.В. Особенности проектирования модели и онтологии предметной области для поиска противоречий в правовых электронных библиотеках. URL: www.inteltec.ru/publish/articles/textan/ RCDL2004.shtml. (дата звернення: 12.12.2006).

30. Niles I., Pease A. Towards a Standard Upper Ontology. 2nd International Conference on Formal Ontology and Information Systems (FOIS-2001), Welty C., and Smith B., eds., Ogunquit, Maine. 17–19, October, 2001. 8pp. URL: http://home.earthlink.net/~ adampease/professional/FOIS.pdf. (дата звернення: 16.12.2008).

31. Sowa J. F. Conceptual Graphs as a universal knowledge representation. In: Semantic Networks in Artificial Intelligence. Spec. Issue of An International Journal Computers & Mathematics with Applications. (Ed. F. Lehmann),

Vol.23, Number 2–5, 1992. Part 1. pp. 75–95.

32. The Mikrokosmos Ontology. URL: http://crl.nmsu.edu/Research/Projects/mikro/htmls/asis.paper-htmls/node4.html. (дата звернення: 07.03.2007).

33. SUO, (2001), The IEEE Standard Upper Ontology web site. URL: http://suo.ieee.org. (дата звернення: 17.05.2006).

34. Палагін О.В., Петренко М.Г. Модель категоріального рівня мовно-онтологічної картини світу. Математичні машини і системи, 2006. – №3. – С. 91–104. URL: http://www.immsp.kiev.ua/publications/articles/2006/2006_3/Palagin_03_2006.pdf. (дата звернення: 07.03.2007).

35. Гуревич И.Б., Трусова И.Б. Тезаурус и онтология предметной области “Анализ изображений”. Всероссийская конф. с междунар. участием "Знания – Онтологии – Теории" (ЗОНТ–09). Новосибирск: Институт математики им. С.Л. Соболева СО РАН, 2009. 10 с. URL: http://www.math.nsc.ru/conference/ zont09/reports/95Gurevich-Trusova.pdf. (дата звернення: 17.12.2010).

36. Martins A.F., Falbo R.A. Models for Representing Task Ontologies. URL: http://nemo.inf.ufes.br/files/models_for_representing_task_ontologies_2008.pdf. (дата звернення: 23.06.2022).

37. Палагин А.В., Петренко Н.Г. К вопросу системно-онтологической интеграции знаний предметной области. Математические машины и системы, 2007. №3,4. С. 63–75. URL: http://www.immsp.kiev.ua/publications/articles/2007/2007_3,4/Palagin_034_2007.pdf. (дата звернення: 23.06.2022).

38. Палагин А.В., Риппа С.П., Саченко А.А. Концептуализация и проблематика онтологий. Искусственный интеллект, 2008. №3. С. 374–379.

39. Крывый С.Л. Формализованные онтологические модели в научных исследования. Управляющие системы и машины, 2016. № 3. С. 4–15. DOI: https://doi.org/10.15407/usim.2016.03.004.

40. Андон Ф.И., Гришанова И.Ю., Резниченко В.А. Semantic web как новая модель информационного пространства интернет. Проблеми програмування, 2008. № 2–3. С. 417-430.

URI: http://dspace.nbuv.gov.ua/handle/123456789/1414. (дата звернення: 23.06.2022).

41. DuCharme B. Learning SPARQL. Querying and Updating with SPARQL 1.1 (Second edition). O'Reilly Media. All rights reserved. August 2013: ISBN: 978-1-449-37143-2. 367p.
42. Tuttle M., Brown S., Campbell K., Carter J., Keck K., Lincoln M., Nelson S. Semantic Web As "Perfection Seeking:" A View from Drug Terminology. Stonebraker M. 2001.
43. Anutariya, Wuwongse, Akama, Wattanapailin. Semantic Web Modeling and Programming with XDD, In Proceedings of SWWS'2001.
44. Euzenat. Towards a principled approach to semantic interoperability, IJCAI 2001. Workshop on ontology and information sharing, 2001, Seattle (WA US).
45. Петренко М.Г., Палагін О.В., Бойко М.О., Матвєйшин С.М. Знання-орієнтований інструментальний комплекс оброблення баз даних наукових публікацій. II. Control systems and computers, 2022, № 3. С.11-28. DOI: https://doi.org/10.15407/csc.2022.03.011.

## Розділ 3. Системи дослідницького проєктування

### 3.1. **Інтелектуальні інформаційні системи дослідницького проєктування, які розвиваються**

Розвиток та застосування інтелектуальних інформаційних систем (ІІС) у різних галузях людської діяльності призвели до створення ІІС нового класу, що поєднують в собі властивості трансдисциплінарності, онтологічного керування, об’єднаних концепціями цілеспрямованого розвитку та віртуальності. Це – клас трансдисциплінарних онтолого-керованих систем дослідницького проєктування (ТРОСДП). Крім завдань інфраструктурної підтримки наукових досліджень, тут на перший план виходять завдання їх *методологічного супроводу* та забезпечення процесів інтеграції, конвергенції, уніфікованого подання трансдисциплінарних знань та операцій над ними. Істотну роль відіграє системологічна підготовка навичок та розширення діапазону світогляду наукових дослідників з метою забезпечення двоєдності концепцій поглиблення знань у конкретній предметній галузі, з одного боку, та розширення охоплення проблеми, виходячи з реальності єдності світу та необхідності формування єдиної системи знань про світ, – з іншого [1].

Одним важливим різновидом наукових досліджень є дослідницьке проєктування (ДП). Характерною рисою ДП є те, що його основні етапи пов’язані з процесом опису образу проєктованого об’єкта нової техніки (ОНТ) за відсутності самого об’єкта [1]. Сам процес проєктування будується як ряд інтерактивних процедур залучення додаткової інформації та формування проміжних гіпотетичних варіантів технічного рішення (ТР), починаючи із загальної концепції (концептуальний етап проєктування), технічної пропозиції та технічного завдання (ТЗ), (передпроєктний етап проєктування) і закінчуючи порівнянням цих варіантів та вибором найкращого.

Технічне рішення є системним описом проєктованого об’єкта, основними складовими якого є функціональна структура (або окремо – функціональний і структурний описи) і параметричні характеристики.

Кожна ітерація у пошуку ТР є, по суті, перетворення описів об’єкта проєктування із залученням додаткової інформації про об’єкт проєктування та предметної області загалом, знань та досвіду проєктувальника.

На передпроєктному етапі виділяють дві фази: концептуальне та парадигматичне проєктування [2]. Перша фаза зводиться до формування набору початкових понять і суджень на основі понятійної моделі реального світу. Друга фаза полягає у породженні природномовного опису проєктованого виробу у формі технічного завдання. Його призначення – формування первинного образу об’єкта проєктування, який створюється. За своєю сутністю цей образ метафоричний, тому що його опис будується шляхом наділення неіснуючого об’єкта властивостями, які мають існуючі об’єкти. Формування таких образів виконується шляхом побудови *метафоричних моделей*, на основі яких виконується синтез природномовного опису об’єкта проєктування і весь процес проєктування. Метафорична модель завжди є результатом порівняння предмета з іншими на підставі їх загальних ознак із залученням знань та досвіду проєктанта (суб’єктивний фактор). Роль останнього, проте, менш велика, оскільки його діяльність спирається на загальну понятійну модель реального світу та предметної області.

Виходячи з викладеного, актуальним є розроблення системи інформаційно-технологічної підтримки процесу ДП, яка успадковує одночасно як функції САНД (системи автоматизації наукових досліджень), так і функції сучасних САПР. Такі системи дослідницького проєктування (СДП) характеризуються високим ступенем інтелектуалізації на всіх рівнях:

– методологічному;

– знань предметної області (з процедурами логікового виведення);

– мов моделювання та опису проєкту;

– методу багатокритеріального вибору альтернативних варіантів;

– забезпечення роботи з не повністю визначеною вхідною інформацією на основі семантичного аналізу текстів.

Побудова ефективної архітектури знання-орієнтованих інформаційних систем дослідницького проєктування бачиться на шляху конструктивного використання таких розділів сучасної інформатики, як:

– трансдисциплінарність;

– опрацювання знань;

– онтологічна концепція;

– віртуальна парадигма та її застосунки [1].

Узагальнена схема функціонування інтелектуальної інформаційної комп’ютерної системи дослідницького проєктування може бути виражена продукційним ланцюжком: “вхідний сигнал → система знань → реакція → рішення”.

Основою предметної діяльності СДП є *система знань*, яка утворюється в процесі виконання проєкту та визначає технічні рішення на всіх етапах проєктування, і, яку можливо представити у вигляді підсистеми загальних знань предметної області, що взаємодіє з множиною підсистем знань, відповідних даному конкретному проєкту та конкретному етапу проєктування. Інтерес представляє новий підклас СДП, що тільки формується, – системи, що володіють функціями саморозвитку в on-line режимі.

### 3.1.1 *Онтолого-керовані системи дослідницького проєктування, які розвиваються*

Як випливає з викладеного вище, особливість системи дослідницького проєктування, зокрема, що розвивається (РСДП) полягає в тому, що процес проєктування заданого класу ОНТ з використанням цієї системи відбувається одночасно з супроводжуючим його процесом розвитку в реальному часі самої системи ДП та її функціональних підсистем, тобто процесу проєктування РСДП. Такі системи за своєю суттю є двопроцесними ($P_1$, $P_2$), в яких функції ведучої (*P1*) і такої, яка ведеться, (*P2*) підсистем міняються місцями [3]:

$$\text{а) } P_1 = F\left(P_2\right);\ \text{б) } P_2 = F\left(P_1\right), \qquad (1)$$

Функцію знання-орієнтованого розвитку РПДС на основі заздалегідь сформульованої стратегії розвитку слід розглядати як найбільш інтелектуальну у класі систем даного типу. У найпростішому випадку її можна звести до збільшення вже існуючих знань (відома формула Брукса) [4]:

$$K\left(S\right) + \Delta I = K\left(S + \Delta S\right), \qquad (2)$$

яка інтерпретується у такий спосіб. Якщо до вихідних знань $K\left(S\right)$ інформаційної системи, що розглядається, представленим деякою вихідною структурою $S$, додати порцію інформації $\Delta I$, то знання системи зміняться: $K\left(S\right) \rightarrow K\left(S + \Delta S\right)$. Випадок $\varphi\left(V\right) = \varphi_2\left(\varphi_1\left(V\right)\right) = ... = \varphi_1^*\varphi_2\left(V\right)$ означає, що в порції інформації $\Delta I$ міститься порція знань, що не змінює вихідну структуру $S$, тобто система раніше вже містила представлені в $\Delta I$ знання, можливо лише в іншій формі. Сам процес, що описується виразом (2), заснований на процедурі зіставлення вихідної структури знань $S$ зі структурою, що явно або неявно міститься в $\Delta I$. Стосовно онтологічної складової подання знань ця процедура завершується додаванням до множин *X* і *R* нових елементів.

У загальному випадку реалізація відображення $IS : \Delta I \rightarrow \Delta S$ є головною проблемою knowledge discovery. Вона, по суті, зводиться до проблеми управління всім процесом функціонування РСДП і вирішується за рахунок застосування методів семантичного аналізу текстів, data mining та ontology-driven system. Основу останніх складають формальні комп'ютерні онтології.

В якості основних онтолого-керованих функцій РСДП можна назвати [3]:

– ефективне компактне представлення системи знань конкретної ПдО на базі сучасних інформаційних технологій (специфікація, концептуалізація);

– прийом та оброблення сигналів зовнішнього світу;

– пошук інформації в системі знань ПдО (довідникові, навчальні функції);

– пошук необхідної інформації у просторі Інтернет;

– постановка та рішення прикладних задач у заданій ПдО (наукових досліджень, проєктування ОНТ та технологій, методів, методик, варіантів рішень);

– розвиток системи та отримання нових знань (або впорядкування існуючих, перевірка їх несуперечності, корекція категоріального дерева тощо).

Створення та впровадження онтолого-керованих інформаційних систем є однією з головних тенденцій розвитку галузі інформатики.

Особливість сучасного етапу розвитку науки і техніки полягає в тому, що з'явилося багато складних системних *трансдисциплінарних проєктів* як у плані управління процесом ТД-досліджень (підтримка всіх етапів життєвого циклу НДР), так і в плані управління знаннями (процедури формалізації, узагальнення, актуалізації та оцінки знань) та процесом проєктування ОНТ. Відсутні ефективні формальні методи інтеграції знань різних ПдО. Незважаючи на це, стихійно розпочався та продовжується процес кластеризації (формування *кластерів конвергенції*) наукових дисциплін та технологій, об'єднаних загальними цілями розвитку, факторами впливу та зворотними зв'язками [1].

Тут доречно зауважити, що онтологічна парадигма формувалася та розвивалася практично одночасно з парадигмами віртуальності та трансдисциплінарності. Сьогодні в повсякденну практику увійшли такі поняття, як віртуальний світ, віртуальна організація, віртуальна лабораторія, віртуальна система, віртуальна адресація тощо. Віртуальна парадигма при найближчому розгляді є не що інше, як різновид концепції систем, що розвиваються.

*Віртуальне дослідницьке середовище* – це комплекс мережевих інструментів, систем та процесів, що підтримують процес дослідницького проєкттування. Воно включає засоби підтримки: адміністрування проєктом, доступу до інформаційних ресурсів, створення та використання баз даних і знань, супроводу процесу впровадження результатів та захисту авторських прав, колективної on-line взаємодії учасників проєкту. Однією з прикладів перспективних віртуальних систем є Віртуальні Науково-інноваційні Центри (ВНІЦ), запропоновані ще в 90-ті роки [5].

Головний вектор діяльності ВНІЦ спрямований на реалізацію нової інноваційної стратегії виконання НДР та дослідницьких проєктів, яка виходить з поняття «новий продукт», чіткого формулювання науково-практичних проблем, створення оргструктур та визначення типу управління у поєднанні з розвиненими механізмами підтримки процесів формування та ефективного використання знань ПдО. Особливе місце у запропонованій інноваційній стратегії займає регламентація виконання етапів *життєвого циклу*:

– управління інноваційними проєктами;
– комерціалізації перспективних науково-дослідних робіт та проєктів;
– розвитку інфраструктури науки – виробництва – освіти.
– ефективної взаємоув'язки дослідницьких та інноваційних засад.

Стосовно СДП в [6] описаний метод, з абревіатурою назви КЕСПП – конструювання ефективної сукупності споживчих ознак. Сутність методу полягає в тому, що, використовуючи підходи трансформаційного і морфологічного синтезу, процес пошуку технічних рішень виконується як послідовність процедур пошуку кластерів функцій ОНТ, перспективних з позицій удосконалення архітектури проєктованого об'єкта ринкових вимог, ціноутворення та просування товару (технології).

Свого часу метод пройшов перевірку при створенні класу засобів обчислювальної техніки, який отримав назву «електронні комбайни».

Сучасне інформаційно-комунікаційне середовище ВНІЦ дозволяє формувати в ньому структури з різними функціями та конфігураціями, що ґрунтуються на використанні віртуальної парадигми. Типовим прикладом таких структур може бути *віртуальна дослідницька лабораторія* (ВДЛ), що є об'єднанням інтелектуальних, матеріальних і документальних ресурсів щодо виконання дослідницьких та проєктних робіт в єдиному інформаційному просторі для отримання швидких і якісних результатів, подальшого їх використання та комерціалізації отриманих результатів. Сучасний досвід ВДЛ дозволить виділити, окрім інших, три головні фактори, що свідчать про її переваги:

– забезпечення ефективної міждисциплінарної взаємодії;

– об’єднання у віртуальному середовищі унікального високовартісного обладнання [5];

– створення розподіленого комп’ютерного середовища з використанням сучасних Grid-мереж.

### 3.1.2 *Особливості інформаційних систем, що розвиваються*

Інформаційні системи, що розвиваються (саморозвиваються) сьогодні є переднім краєм інформатики. Їх головна особливість полягає в тому, що вони увібрали в себе властивості як складних природних, насамперед біологічних систем, так і сучасних систем штучного інтелекту. Цей процес взаємного запліднення двох цих систем не випадковий і підпорядковується загальному закону ноосферогенезу, відкритому століття тому нашим великим співвітчизником В.І. Вернадським [3]. Всі космічні, біологічні, соціальні, антропогенні та інформаційні системи підпорядковуються цьому закону і відносяться до класу складних систем, що розвиваються (еволюціонують). Розвиток таких систем пов’язане з надбанням і накопиченням нових якісних ознак і появою в реальному часі *нових рівнів* організації, які є результатом взаємодії системи із зовнішнім середовищем, заснованого на принципі *зворотного зв’язку*. Закони такої взаємодії, як правило, виходять за рамки цільової причинності і через ситуаційну невизначеність, виникнення атракторів, процесів диференціації-інтеграції призводять до зміни головної лінії розвитку, а іноді непрогнозованої втрати стійкості, що має бути предметом особливої уваги наукових дослідників [3].

Особливе місце у знаннях про системи, що розвиваються, в цілому та інформаційних, зокрема, займає *синергетична* парадигма. З одного боку, вона апелює до цілісності та інтегрального уявлення, системно визначаючи ефекти взаємодії об’єктів, процесів та суб’єктів, а з іншого, – акцентує увагу на нелінійностях, нестійкості та появі атракторів, що змінюють у результаті багаторівневу організацію та поведінку системи. В обох випадках вона виражається сукупністю формальних моделей самоорганізації та спрямована на відтворення наукової картини світу, що особливо важливо під час переходу до *трансдисциплінарного підходу* у наукових дослідженнях та втілення парадигми *глобального еволюціонізму*. Наукова картина світу при цьому може бути представлена як трансдисциплінарна онтологія, що увібрала в себе не тільки онтологію окремих дисциплін, але й методи останніх, включаючи варіанти їхнього перехресного впливу. Трансдисциплінарність на додаток до синергетики дозволяє побудувати єдину трансдисциплінарну методологію аналізу та синтезу, включивши її до загальнонаукової картини світу. З позицій, що розглядаються, всі інформаційні системи, що розвиваються, можна розділити на чотири взаємопов'язані класи:

– генетичні;

– з віртуальною архітектурою та реконфігурацією;

– знання-орієнтовані;

– трансдисциплінарні.

Крім того, вони включають два принципово різних підкласи: автономні і неавтономні системи. Останні призначені для активної людино-машинної взаємодії, а точніше – природного та штучного інтелекту [3].

### 3.1.3 *Персональні бази знань та онтологічне керування*

Центральним користувачем РСДП, як суто дослідних, так і проєктних СДП є дослідник в особі наукового співробітника чи інженера. Тому необхідно забезпечити комфортну інформаційно-технологічну та методологічну підтримку саме цьому класу користувачів. Однією з основних підсистем сучасної СДП є персональна база знань (ПБЗ) дослідника-проєктувальника. У ній можна виділити чотири типи інформаційних файлів [3]:

– документація, що відноситься до класу проєктованих об'єктів нової техніки;

– інформація про предметну область (статті, монографії, патенти та їх інтегральний опис на рівні наукових теорій);

– публікації автора (користувача);

– сховище ідей, гіпотез, нарисів.

Кожному типу відповідає свій функціональний блок ПБЗ, але вони обслуговуються єдиним механізмом онтологічного керування, контентно-семантичного аналізу й онтологічної репрезентації та розмітки текстів. Цей механізм реалізується комплексом інформаційно-програмних та методологічних засобів управління знаннями, що функціонує на основі трансдисциплінарного підходу, онтологічних методів аналізу та синтезу інформації, інтернет-ресурсів. Вміст ПБЗ є особистою власністю користувача, і доступ до її файлів у мережевому режимі роботи регламентується відповідно до установок користувача. Вміст ПБЗ можна використовувати як основний корпоративний актив – БЗ:

$$\text{БЗ} = \bigcup_{i} \text{БЗ}_i \ ,$$

де $i$ – порядковий індекс співробітника корпорації чи учасника проєкту, керованого підсистемою організації колективного взаємодії [3].

Основні функції ПБЗ:

– зручний інтерфейс для взаємодії з користувачем;

– формулювання запитів до вмісту ПБЗ і до зовнішнього інформаційного простору відповідно до задач, що цікавлять користувача, зокрема, на основі заздалегідь виділених онтологічних кластерів;

– підготовка матеріалів для наукових публікацій, звітів, патентних досліджень;

– формування опису предметної області;

– зіставлення результатів досліджень автора-користувача з публікаціями в заданій предметній області тощо [3].

Функціонування ПБЗ спирається на можливості підсистеми синтактико-семантичного аналізу текстів, поточної версією якої є система типу ТОДОС [7]. Головною функцією ТОДОС є інтеграція існуючих корпоративних інформаційних ресурсів шляхом оброблення природномовних текстових документів з подальшим виділенням поверхневої семантики та аналізом первинних знань на основі їх онтологічного подання з метою прийняття рішень.

Підсистеми ПБЗ і ТОДОС є складовими системи вищого рівня – системи інформаційно-когнітивної підтримки наукового дослідника широкого профілю, яка, у свою чергу, слугує основою систем дослідницького проєкттування, що мають можливості персоніфікації, тобто функціональної орієнтації на конкретного користувача, природно з функцією автентифікації. Винятково важливим як ПБЗ, так СДП загалом є *онтологічний підхід* до *подання результатів дослідницького проєктування*.

В [8] описаний варіант такого подання. Він включає такі компоненти:

– онтологічний опис функціонального фрагмента знань у вигляді онтографа та тезаурусу термінів;

– образну компоненту опису (просте зображення, 3D-графіка, мультимедійний відеоряд);

– подання знань на рівні формальної теорії;

– повний лінгвістичний корпус, що представляє задану предметну область або її фрагмент;

– підсистему сервісів, що надаються користувачеві.

У разі потреби передбачається можливість виведення зазначеної інформації на сайт або WEB-портал для взаємодії із зовнішніми користувачами.

Таким чином, СДП побудована на основі викладених вище концепцій (ноосферної парадигми, трансдисциплінарності, еволюціонізму, віртуальності, онтологічного керування, багаторівневого подання знань), є інструментом проєктування складних об'єктів нової техніки і технологій, перш за все в галузі інформаційних технічних і програмних засобів і технологій. З іншого боку, її проєкція на інтереси наукових дослідників та інженерів може значно розширити можливості останніх за рахунок інтелектуальної підтримки його професійних функцій та навичок, а також ефективного знання-орієнтованого інтерфейсу користувача. Можливість еволюціонування системи в часі, накопичення знань про навколишній світ, предметну область і самого користувача відкриває широкі перспективи для її застосування в наукових дослідженнях та наукомістких інженерних розробках, висуваючи на перший план питання проблемної орієнтації та тиражування.

## 3.2. **Інформаційна технологія й інструментальні засоби підтримки процесів ДП**

### 3.2.1 *Дослідницьке проєктування Smart-систем*

Рівень розвитку інтелектуальних інформаційних технологій значною мірою впливає на ефективність процесів, що відбуваються в економічній, науково-технічній, освітній та інших сферах діяльності людського суспільства. Процеси глобальної інформатизації світового співтовариства орієнтовані, перш за все, на побудову суспільства знань і носять все більш яскраво виражений трансдисциплінарний характер. Безперечним лідером при цьому є технології інженерії знань, у тому числі її нового напрямку – онтологічного інжинірингу. Ці технології реалізують процеси управління знаннями (Knowledge Management) і успіхи у цьому напрямку багато в чому визначаються рівнем інтелектуалізації та ефективності комп'ютерних систем [9].

Нині намітилася тенденція активізації наукових досліджень на стику різних предметних дисциплін (міждисциплінарні дослідження), так і в кластерах конвергенції (трансдисциплінарні дослідження). Для підтримки цих досліджень важливими факторами є побудова знання-орієнтованих інформаційних систем, удосконалення процесів дослідницького проєктування, розробка методів та інструментальних засобів онтологічного аналізу природномовних об'єктів з метою вилучення з них знань, прикладних аспектів застосування онтологій, зокрема при розробці електронних навчальних курсів (та інших аспектів електронної освіти), метаонтологій, систем інтеграції знань у трансдисциплінарних кластерах конвергенції тощо.

Далі розглядаються методологічні засади побудови систем дослідницького проєктування об'єктів нової техніки (як різновиду Smart-систем) з використанням онтологічної концепції та технології Smart Research & Development, що включають моделі, інформаційну технологію та інструментальні засоби когнітивних процесів оброблення інформації.

*Smart-системи*

Процес інтелектуалізації у галузі засобів інформатики та обчислювальної техніки вийшов на новий масштабний рівень, розширивши простір та функціональність їх застосунків у сучасному суспільстві, чому значною мірою сприяла стратегічна рамкова програма «Горизонт 2020». Вона відкрила епоху інтенсивного розвитку Smart-систем різного рівня та призначення [10]. Цей напрямок підтримано спільною технологічною ініціативою рядом європейських міжнародних організацій. Багаторічний стратегічний план реалізації програми досліджень та інновацій у галузі електронних компонент, систем та технологій (MASRIA) забезпечує розвиток наступних основних функціональних доменів: smart society, smart mobility, smart energy, smart health, smart production. По суті, йдеться про технології, що впливають на всі аспекти функціонування сучасного суспільства, такі як: взаємодія в реальному часі між машинами, людьми та об'єктами навколишнього світу; забезпечення безпеки особистості та суспільства; ефективне постачання та розподіл (вода, продукти харчування тощо); логістика; смарт-адміністрування та загалом підтримка сталого розвитку суспільства. Уся множина перерахованих задач та сфер діяльності є об'єктом дослідження та розробок стратегічного плану MASRIA. Слід також зазначити, що у "Концепції розвитку електронного урядування в Україні" та цифровій економіці передбачено застосування сучасних інноваційних підходів, методологій та технологій, зокрема Інтернет-речей, хмарної інфраструктури, Blockchain, Big Data та ін. [11]. У таблиці 3.1 подано класифікацію атрибутів, що характеризують Smart-системи та їх короткий опис [12].

Таблиця 1. Класифікація атрибутів Smart-систем

| Атрибут | Опис |
|---|---|
| Адаптація, пристосування | Здатність змінювати фізичні або поведінкові характеристики відповідно до змін у навколишньому середовищі або виживання в ньому |
| Сприйняття, зчитування, розпізнавання, розуміння | Здатність ідентифікувати, розпізнати, зрозуміти та/або усвідомити феномен, подію, об’єкт, вплив тощо |
| Логікове виведення | Здатність робити логікові виведення на основі вихідних даних, обробленої інформації, спостережень, доказів, припущень, правил та логічних міркувань |
| Навчання, освоєння | Здатність набуття нових чи зміни існуючих знань, досвіду, поведінки підвищення продуктивності, ефективності, навичок тощо |
| Антиципація (передбачення подій) | Здатність думати чи міркувати, щоб передбачати майбутні події чи свої подальші дії |
| Самоорганізація та реструктуризація (оптимізація) | Здатність системи змінювати свою внутрішню структуру (компоненти), самовідновлюватись та самопідтримуватись цілеспрямованим (невипадковим) способом за відповідних умов, але без зовнішнього агента/сутності |

Поняття "Smart-система" в онтологічній ієрархії тісно пов’язане на верхньому рівні з такими поняттями як кібер-фізичні системи та іншими розділами Smart-суспільства. На нижніх рівнях розташовані різні Smart-пристрої, що забезпечують реалізацію застосунків користувачів. Аналіз публікацій у зарубіжних та вітчизняних наукових фахових виданнях дозволив синтезувати кілька верхніх рівнів онтології ПдО “Smart-система” [10, 13].

На рис. 1 наведено онтографічне представлення поняття “Smart-система”.

Відомі наступні визначення кібер-фізичних систем та Smart-пристроїв [10].

*Кібер-фізичні системи* – це електронні системи, компоненти та програмне забезпечення, які тісно взаємодіють з фізичними системами та їх довкіллям: вбудований інтелект надає можливості для відстеження, моніторингу, аналізу та управління фізичними пристроями, компонентами та процесами у різних галузях застосування. Їх здатність підключатися і взаємодіяти через всі види мереж і протоколів (включаючи Інтернет, дротовий, бездротовий зв’язок) дозволяє їм координувати та оптимізувати функціональні можливості фізичних систем.

На сьогодення загальноприйнятого визначення *Smart-системи* немає. На наш погляд, коректним буде таке визначення.

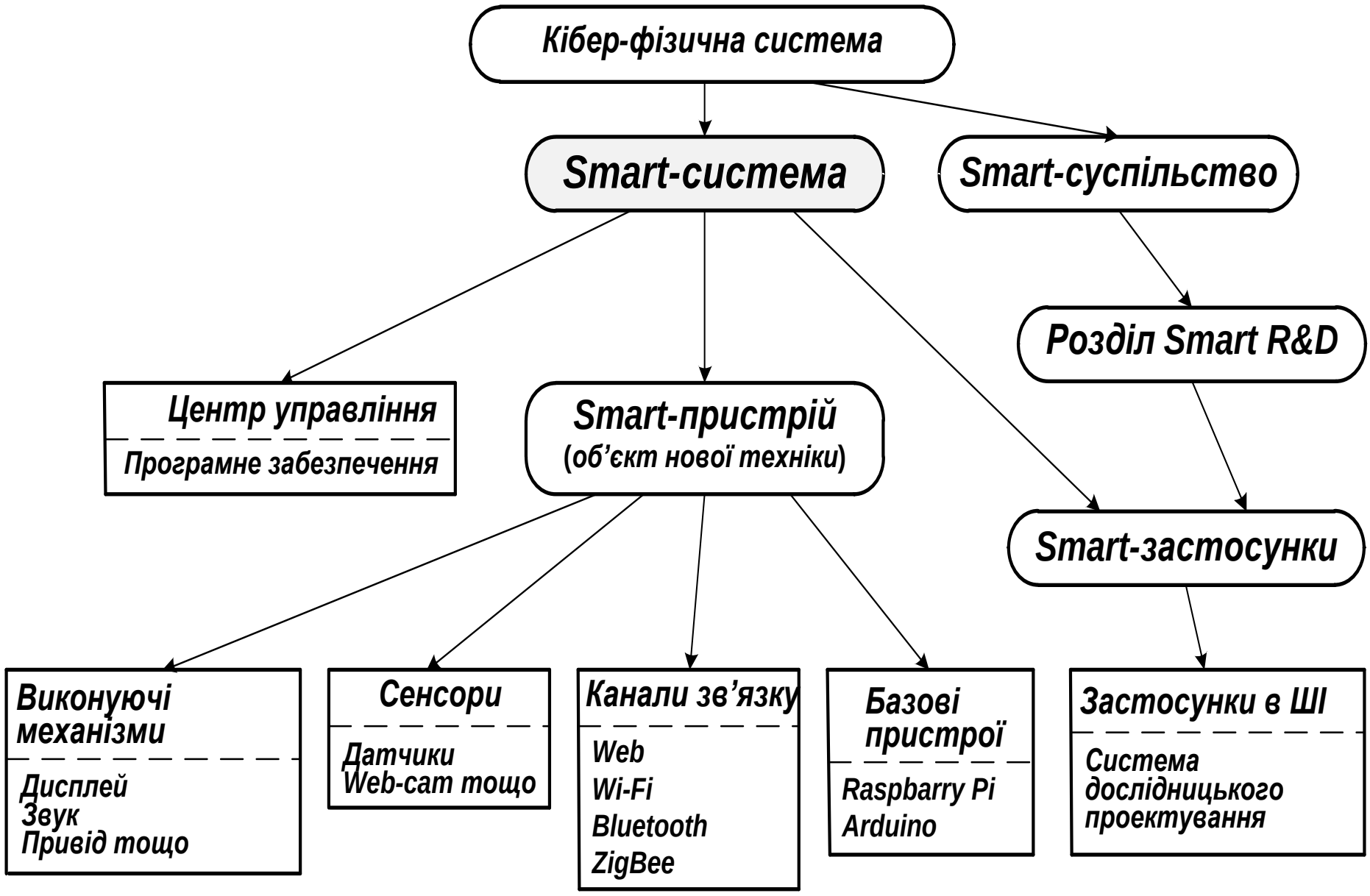

Рис. 1. Онтографічне представлення поняття "Smart-система"

*Smart-системи* – це системи, які можуть функціонувати як автономно, так і у складі складніших, кібер-фізичних систем, та здатні забезпечити:

- взаємну адресацію та ідентифікацію один одного;
- діагностування, опис та кваліфікацію навколишнього середовища;
- прогнозування подій, прийняття рішень та їх виконання;
- наявність більше однієї комунікаційної технології.

*Smart-пристрій* – це фізичний об'єкт, який у цифровому форматі взаємодіє з одним або декількома об'єктами:

- сенсори (температура, світло, рух тощо);
- виконавчі механізми (дисплеї, звук, двигуни тощо);
- кероване обчислення (може запускати програми та логіку);
- інтерфейси зв'язку.

Сформульовані в [9] основні принципи комплексного підходу до вирішення проблеми розроблення смарт-систем для потреб сучасного суспільства у конкретній реалізації та з урахуванням технологій Internet та Semantic Web трансформуються у наступні:

– (1) оброблення знань – при проєктуванні ОНТ необхідно мати знання кількох предметних областей, формально описаних однією з мов, характерних для Semantic Web (наприклад, RDF);

– (2) трансдисциплінарність – системна інтеграція знань задіяних предметних областей;

– (3) онтологічна концепція – структура знань ПдО представлена у вигляді онтологій з поділом на статичну та динамічну складові;

– (4) визначення проблемної ситуації та формулювання проблеми;

– (5) віртуалізація, уніфікація та стандартизація технічних рішень – комунікація "Smart-систем" між собою та з користувачами здійснюється через Web-середовище та/або "Хмара".

### 3.2.2 *Інструментальний комплекс підтримки процесів ДП*

Складність виконання сучасних наукових досліджень, у тому числі і Smart R&D, висуває все більш високі вимоги до інструментальних засобів їх підтримки, які повинні:

– оперувати великими обсягами текстової інформації з метою отримання з них предметних знань;

– формувати концептуальні структури знань (онтології) як на стику предметних дисциплін, так і в кластерах конвергенції;

– системно інтегрувати побудовані онтології з метою виявлення нових знань тощо. Такій проблемній орієнтації відповідає інструментальний комплекс онтологічного призначення (ІКОП), запропонований в [14].

ІКОП реалізує ряд компонентів єдиної інформаційної технології:

– пошук у мережі Internet та/або в інших електронних колекціях текстових документів, релевантних заданій ПдО, їх індексація та збереження в базі даних;

– автоматизоване оброблення природномовних текстів;

– добування з множини ТД-знань, релевантних заданій ПдО, їх системно-онтологічна структуризація та формально-логічне подання однією або кількома із загальноприйнятих мов опису онтологій (Knowledge Repressentation);

– створення, накопичення та використання великих структур онтологічних знань у відповідних бібліотеках;

– системна інтеграція онтологічних знань як один із основних компонентів методології міждисциплінарних та трансдисциплінарних наукових досліджень.

ІКОП складається з трьох підсистем і являє собою інтеграцію різного роду інформаційних ресурсів, програмних засобів оброблення і процедур користувача, які, взаємодіючи між собою, реалізують сукупність алгоритмів автоматизованої ітераційної побудови понятійних структур предметних знань, їх накопичення та/або системної інтеграції [14, 15].

**Висновки за розділом**

1. Система інформаційно-технологічної підтримки процесу дослідницького проєктування успадковує одночасно як функції системи автоматизації наукових досліджень, так і функції сучасних систем автоматизованого проєктування. Такі системи дослідницького проєктування характеризуються високим ступенем інтелектуалізації на всіх рівнях: методологічному; знань предметної області (з процедурами логікового виведення); мов моделювання та опису проєкту; методу багатокритеріального вибору альтернативних варіантів; забезпечення роботи з не повністю визначеною вхідною інформацією на основі семантичного аналізу текстів.

2. Створення та впровадження онтолого-керованих інформаційних систем є однією з головних тенденцій розвитку галузі інформатики.

3. Віртуальне дослідницьке середовище – це комплекс мережевих інструментів, систем та процесів, що підтримують процес дослідницького проєктування.

4. Система дослідницького проєктування, побудована на основі концепцій ноосферної парадигми, трансдисциплінарності, еволюціонізму, віртуальності, онтологічного управління, багаторівневого представлення знань є інструментом проєктування складних об’єктів нової техніки та технологій, перш за все в галузі інформаційних технічних та програмних засобів та технологій . З іншого боку, її проекція на інтереси наукових дослідників та інженерів може значно розширити можливості останніх за рахунок інтелектуальної підтримки його професійних функцій та навичок, а також ефективного знання-орієнтованого інтерфейсу користувача. Можливість еволюціонування системи в часі, накопичення знань про навколишній світ, предметну область і самого користувача відкриває широкі перспективи для її застосування в наукових дослідженнях та наукомістких інженерних розробках, висуваючи на перший план питання проблемної орієнтації та тиражування.

**Перелік посилань до розділу 3**

1. A. V. Palagin. Transdisciplinarity Problems and the Role of Informatics. Cybernetics and Systems Analysis volume 49, pages 643–651 (2013). DOI: https://doi.org/10.1007/s10559-013-9551-y .
2. Филиппович Ю.Н. Метафорическое проектирование информационных технологий и систем // Интеллектуальные технологии и системы. Сб. Статей. Вып. 2. М., 1999. С. 7–30.
3. Палагин А.В. Введение в класс трансдисциплинарных онтолого-управляемых систем исследовательского проектирования. Управляющие системы и машины. 2016. № 6. С. 3–11. URL: http://usim.org.ua/arch/2016/6/2.pdf. (дата звернення: 23.03.2023).
4. Palagin, A.V. Architecture of ontology-controlled computer systems. Cybern Syst Anal 42, 254–264 (2006). https://doi.org/10.1007/s10559-006-0061-z – С.111-124. DOI: https://doi.org/10.1007/s10559-006-0061-z .
5. A. V. Palagin and I. V. Sergienko, “Virtual scientific and innovational centers: creation concept and development prospects,” USiM, No. 3, 3–11 (2003).
6. Палагин А.В. Разработка информационного «электронного комбайна». Науковедение, 1995. №3–4. С. 6–22.
7. Величко В.Ю., Попова М.А., Приходнюк В.В., Стрижак О.Є. ТОДОС– IT-платформа формування трансдисциплінарних інформаційних систем. *Системи озброєння і військова техніка*, 2017. № 1(49). С. 10–19.
8. Palagin, A.V. An Ontological Conception of Informatization of Scientific Investigations. Cybern Syst Anal 52, 1–7 (2016). DOI: https://doi.org/10.1007/s10559-016-9793-6 .
9. Палагин А.В. Функционально-ориентированный подход в исследовательском проектировании // Кибернетика и системный анализ. – 2017. – № 6. – С. 185–192. URL: http://www.kibernetika.org/volumes/2017/numbers/06/articles/17/17.pdf. (дата звернення: 23.03.2023).
10. Multi Annual Strategic Research and Innovation Agenda for ECSEL Joint

Undertaking 2017. 118 p. URL: https://artemis–ia.eu/publication/download/masria. (дата звернення: 15.03.2019).

11. Концепція розвитку електронного урядування: Розпорядження Кабінету Міністрів України від 20 вересня 2017 р. № 649-р. URL: https://zakon.rada.gov.ua/laws/show/649-2017-%D1%80#Text. (дата звернення: 19.03.2018).
12. Палагин А.В., Петренко Н.Г., Малахов К.С. Информационная технология и инструментальные средства поддержки процессов исследовательского проектирования Smart-систем. УСиМ. – 2018. – № 2. – С. 19–30. URL: http://usim.org.ua/arch/2018/2/4.pdf. . (дата звернення: 23.03.2023).
13. Guinard D.D., Trifa V.M. Building the Web of Things. With examples in Node.JS and Raspberry PI // Manning Publications Co. Shelter Island. NY, 2016. 346 p.
14. Палагин А.В., Крывый С.Л., Петренко Н.Г. Онтологические методы и средства обработки предметных знаний. Луганск: изд-во ВНУ им. В. Даля, 2012. 323 с.
15. Палагин А.В., Петренко Н.Г., Величко В.Ю., Малахов К.С., Тихонов Ю.Л. К вопросу разработки инструментального комплекса онтологического назначения. Проблеми програмування, 2012. № 2–3. С. 289–298.
URL: http://dspace.nbuv.gov.ua/bitstream/handle/123456789/86614/34-Palagin.pdf?sequence=1. (дата звернення: 20.06.2022).

# Розділ 4. Онтологічна система оброблення баз даних наукових публікацій

## 4.1. Призначення та архітектура онтологічної системи оброблення баз даних наукових публікацій

Широко відомі численні застосунки й інструментарій, що реалізують технології пошуку інформації у різних текстових джерелах у відповідності з заданими параметрами. Причому результати пошуку надаються користувачу тільки по кожному параметру пошуку окремо і не пов'язані між собою. А застосування технологій Semantic Web з метою багатопараметричного і пов'язаного пошуку інформації у різних джерелах в Україні находиться на початковій стадії розвитку. Окремою проблемою є мультимедійне (образно-понятійне) подання результатів пошуку та їх порівняння з понятійною структурою предметної галузі (Knowledge Domain), яка представляє зацікавленість, з метою добування нових знань. Під таким кутом зору для наукових досліджень оброблення наукових публікацій одного автора, авторів наукового підрозділу чи академічного інституту загалом із застосуванням технологій Semantic Web, мультимедійним поданням інформації й ефективною підтримкою процесу добування нових знань є актуальним.

Йдучи за онтологічною концепцією візуального моделювання тематики наукових досліджень предметних галузей (ПдГ) на рис. 1 представлено онтограф категорій і відповідних понять інтелектуального інструментального комплексу оброблення наукової інформації [1], базова структура концептів якого взята за основу розроблення онтологічної системи оброблення бази даних наукових публікацій (ОнС БДНП), яка в певному сенсі є семантичною електронною бібліотекою наукового співробітника.

Онтологічна система оброблення баз даних наукових публікацій (далі просто ОнС) є системою, яка реалізує технології Information Retrieval і Knowledge Discovery in Databases з акцентом на технології й інструментарій Semantic Web та когнітивної графіки. Остання технологія та відповідний інструментарій дозволяють створювати мультимедійне подання образно-понятійних структур, які описуються у наукових документах. Технології Semantic Web включають процеси створення й оброблення сховищ RDF документів наукових публікацій (НП), розробки локальних і/або віддалених прикінцевих точок, формування та виконання SPARQL запитів користувача. Із всієї широкої множини технологій Semantic Web слід виділити SPARQL технологію, яка надає науковому робітнику (НР) можливість створювати запити довільного рівня складності й отримувати відповіді на них, включаючи різного роду інформацію для:

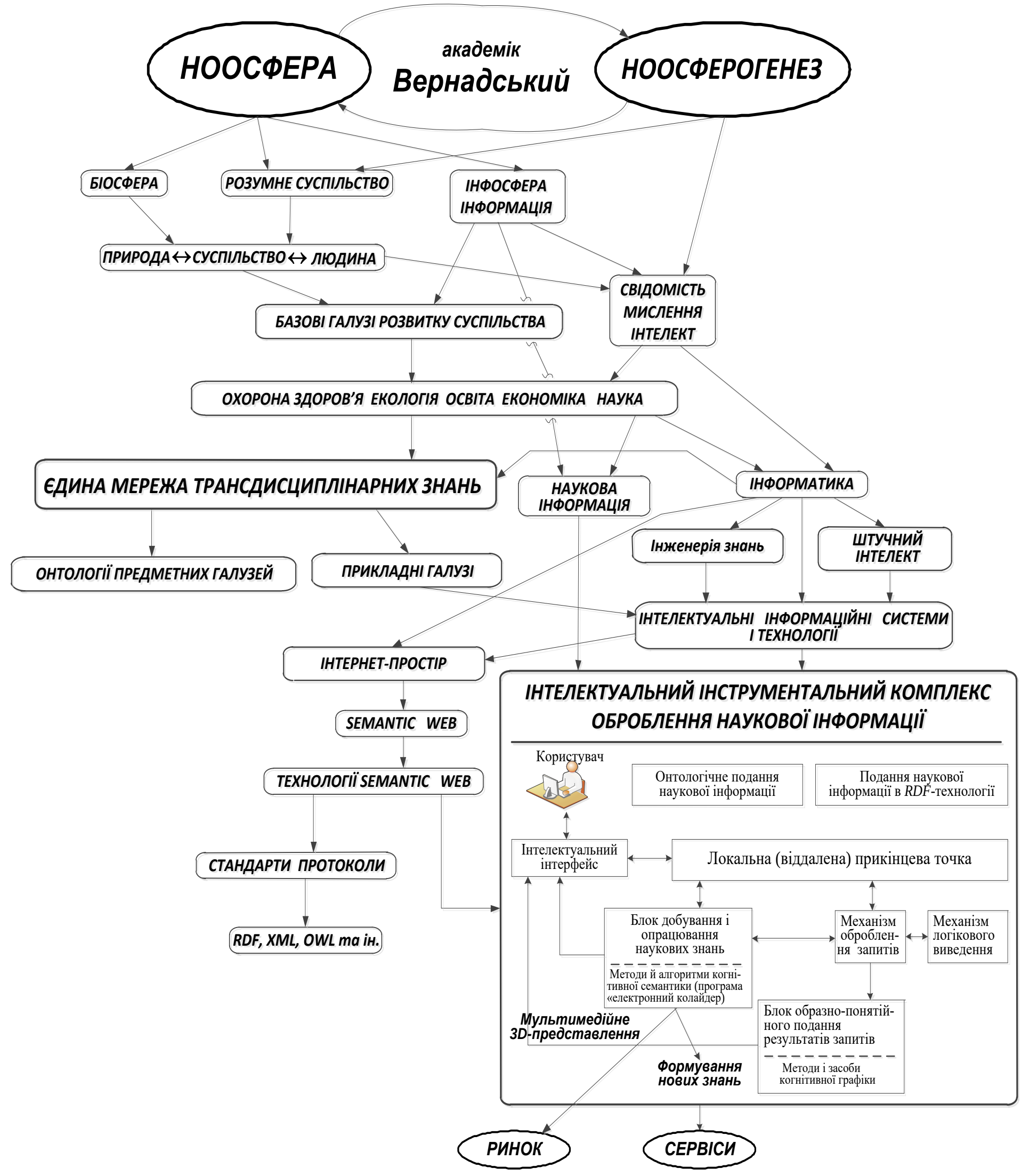

Рис. 1. Онтограф інтелектуального інструментального комплексу оброблення наукової інформації

– аналізу охоплення науковими дослідженнями області наукових інтересів НР, наукового відділу, академічного інституту (у відповідності з опублікованими науковими матеріалами);

– аналізу поточного стану виконаних наукових досліджень (НД) та опублікованих результатів;

– аналізу статистичних показників НП;

– підтримки створення нових наукових документів (статей, доповідей, звітів з науково-дослідних робіт, патентів тощо);

– мультипараметричного пошуку структурної та семантичної інформації;

– мультимедійного представлення результатів запитів [1, 2].

Архітектура ОнС спроєктована на основі онтологічного підходу [3] і представлена на рис. 2 [2]. Вона складається із трьох підсистем і блоку зв'язку із зовнішнім середовищем, який забезпечує користувачу можливість завантажувати в базу даних нові наукові публікації. При цьому останні повинні пройти попередню обробку на відповідність прийнятим в системі вимогам подання НП в БД. ОнС представляє собою інтеграцію різного роду інформаційних ресурсів, оригінальних і вільно доступних програмних засобів, які, взаємодіючи між собою, реалізують сукупність алгоритмів автоматизованої побудови й використання БДНП. Попередньо користувач формує масив текстоsвих документів (за замовчуванням – це електронні файли наукових публікацій з можливими розширеннями *doc*, *pdf* або *txt*) і передає їх інженеру зі знань (ІЗ).

Підсистема *Програмна оболонка й інтерфейс користувача* включає інтерфейс (головне меню) користувача/інженера зі знань, блок онтологічних структур галузі наукових досліджень (ГНД), блок створення та підтримки БДНП, блок формування машинно-читаємих результатів виконання запитів і блок представлення та візуалізації людино-читаємих результатів виконання запитів користувачів.

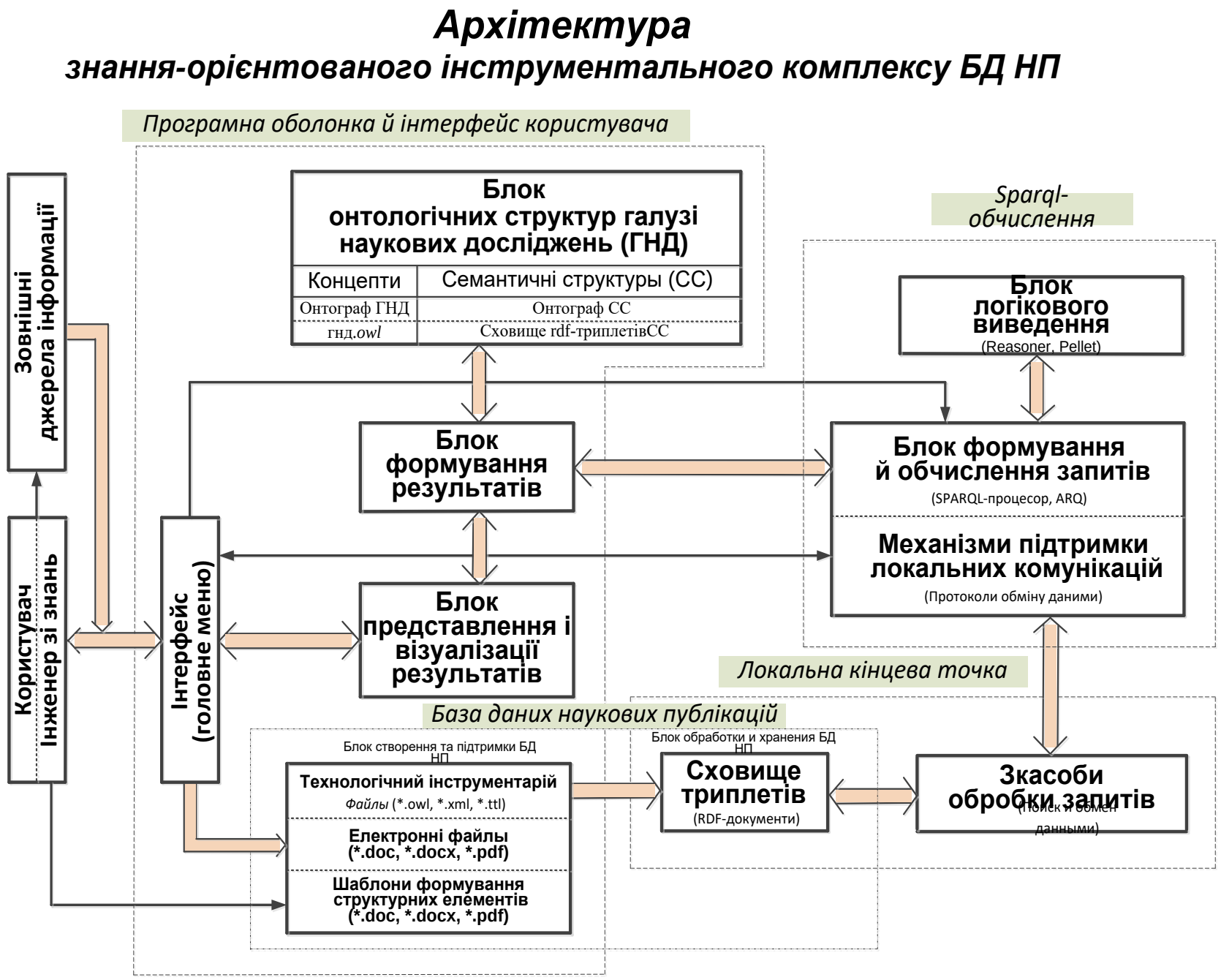

Рис. 2. Архітектура ОнС

*Перший* компонент забезпечує взаємодію користувача та ІЗ з системою.

*Другий* компонент є інформаційним ресурсом, який створюється інженером зі знань і представляє собою набір онтографів галузі наукових досліджень і семантичних структур (СмС), а також відповідні файли їх формальних описів.

Особливістю представлення й обробки наукових публікацій в ОнС є узагальнення понятійних структур множини НП у вигляді онтографічного представлення так званої галузі НД автора і деталізація кожної НП до рівня сукупності простих понятійних структур (або елементарних сенсів (ЕС)) і їх онтографічного представлення. ЕС – це монада знань, неподільна одиниця. У лінгвістиці вона відповідає судженню чи двоскладному простому реченню. В основі ЕС лежить поняття (об’єкт), навколо якого групуються предикати, атрибути об’єктів та атрибути дій. Будь-яка система знань може бути представлена “згустком” ЕС, поданих в деякій просторовій формі.

*Третій* компонент також створюється ІЗ і призначений для формування та підтримки в актуальному стані сховища триплетів (*rdf* документів) наукових публікацій. При цьому всі НП розбиті на види, такі як статті, монографії, енциклопедії, словники, препринти, доповіді, тези конференцій тощо. Для кожного виду НП розробляється свій шаблон, або метарівень онтографа. На рис. 3 представлена блок-схема шаблона для всіх видів публікацій.

Так як статті є найбільшим за кількістю видом НП, далі буде розглядатися тільки цей вид публікацій (на першому етапі розробки).

Важливою інформаційною компонентою в процесі перетворення електронної версії статті в її формальний опис в онторедакторі Protégé є файл “Опис_статті_№ N.docx”. В ньому відображені текстові структури статті у відповідності з онтографом цієї статті.

Взаємозв’язки між відповідними вершинами онтографів шаблонів видів НП і самих НП (на рисунках показані онтографи тільки для статей) наступні:

1) всі онтографи мають наскрізну нумерацію;

2) онтограф шаблона для всіх видів НП представлений на листі 1 (рис. 3);

3) онтограф шаблона для статей представлений на листі 2 (рис. 4);

4) всі атрибути статті $S_i$ включені в перехід з листа 1 “Стаття $S_i$ Лист 2 – 9”;

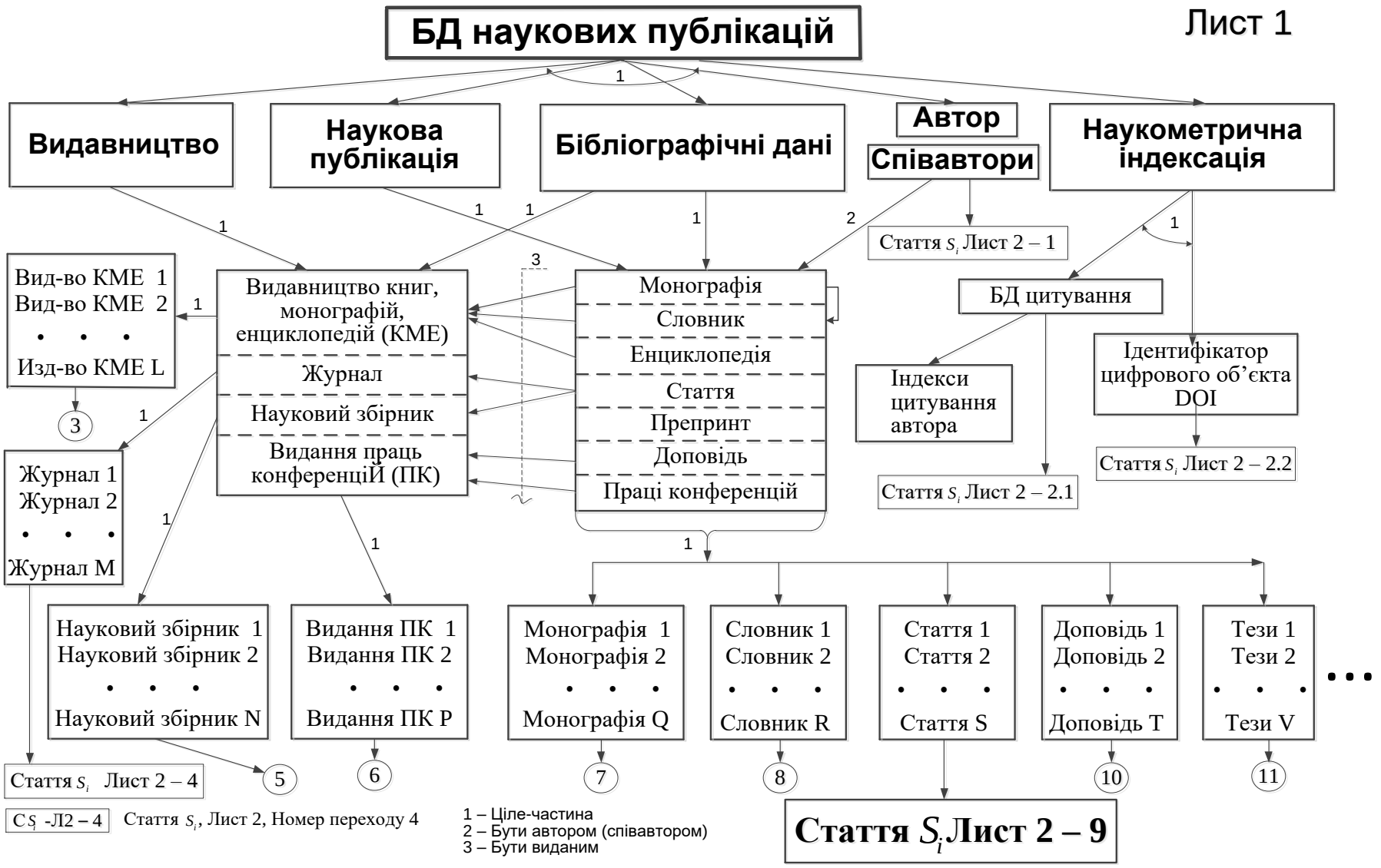

Рис. 3. Блок-схема шаблона для всіх видів НП

5) інші атрибути, безпосередньо не пов’язані із статтею $S_i$, пронумеровані як:

– “Автор, співавтор – Стаття $S_i$ Лист 2 – 1”;

– “База даних цитування – Стаття $S_i$ Лист 2 – 2.1”;

– “Ідентифікатор цифрового об’єкта DOI – Стаття $S_i$ Лист 2 – 2.2”;

– “Журнал – Стаття $S_i$ Лист 2 – 4”;

6) для видів НП онтографи нумеруються як “Наукова публікація № Лист 1”, “Наукова публікація № Лист 2 (продовження)” тощо. Наприклад, для статей онтографи пронумеровані як “Стаття № Лист 1”, “Стаття № Лист 2 (продовження)”.

Слід також зазначити, якщо у НП який-небудь атрибут відсутній, то відповідне поняття (клас) в онтографі не буде мати індивідів.

*Четвертий* блок формує машинно-читаємі результати виконання запитів користувача сумісно з онтологічними структурами ГНД та їх формальними описами.

І *п’ятий* блок призначений для людино-машинного представлення обчислених результатів користувачу, в тому числі і їх візуалізації в графічній формі.

Вільно доступними засобами реалізації ОнС є поліфункціональний інструментарій Protégé 5.5.0, який включає онторедактор, *Sparql*-процесор і різонер (або механізм логікового виведення). До нього можливо підключити різного роду *plag in* для розширення функціональних можливостей. Зокрема, в ОнС в якості *Sparql*-процесора підключений *ARQ*-процесор, а в якості різонера – *Pellet*. Останні є широко розповсюдженими інструментальними засобами (в тому числі для створення локальних прикінцевих точок). Вони підтримують версію рекомендацій W3C SPARQL 1.1.

В підсистемі *Програмна оболонка та інтерфейс користувача* блоки онтологічних структур ГНД і база даних НП реалізовані в Protégé 5.5.0, а блоки формування машинно-читаємих результатів виконання запитів і представлення і візуалізації людино-читаємих результатів є оригінальними програмами.

Підсистема *Sparql-обчислення* представляє собою *Sparql*-процесор і різонер Hermit, вбудовані в Protégé 5.5.0, а також *plag-in* – процесор *ARQ* и різонер *Pellet*. Також в цю підсистему включений блок механізмів підтримки локальних комунікацій, який забезпечує передачу різного роду інформації у відповідності із стандартизованими протоколами консорціуму W3C.

Підсистема *Локальна прикінцева точка* підтримує протокол *Sparql*-запитів і дозволяє користувачу звертатися до бази даних НП й отримувати від неї інформацію. Включає сховище *rdf* документів НП, створене інженером зі знань, і засоби оброблення запитів. Останні також виконують стандартизовані функції передачі та перетворення інформації.

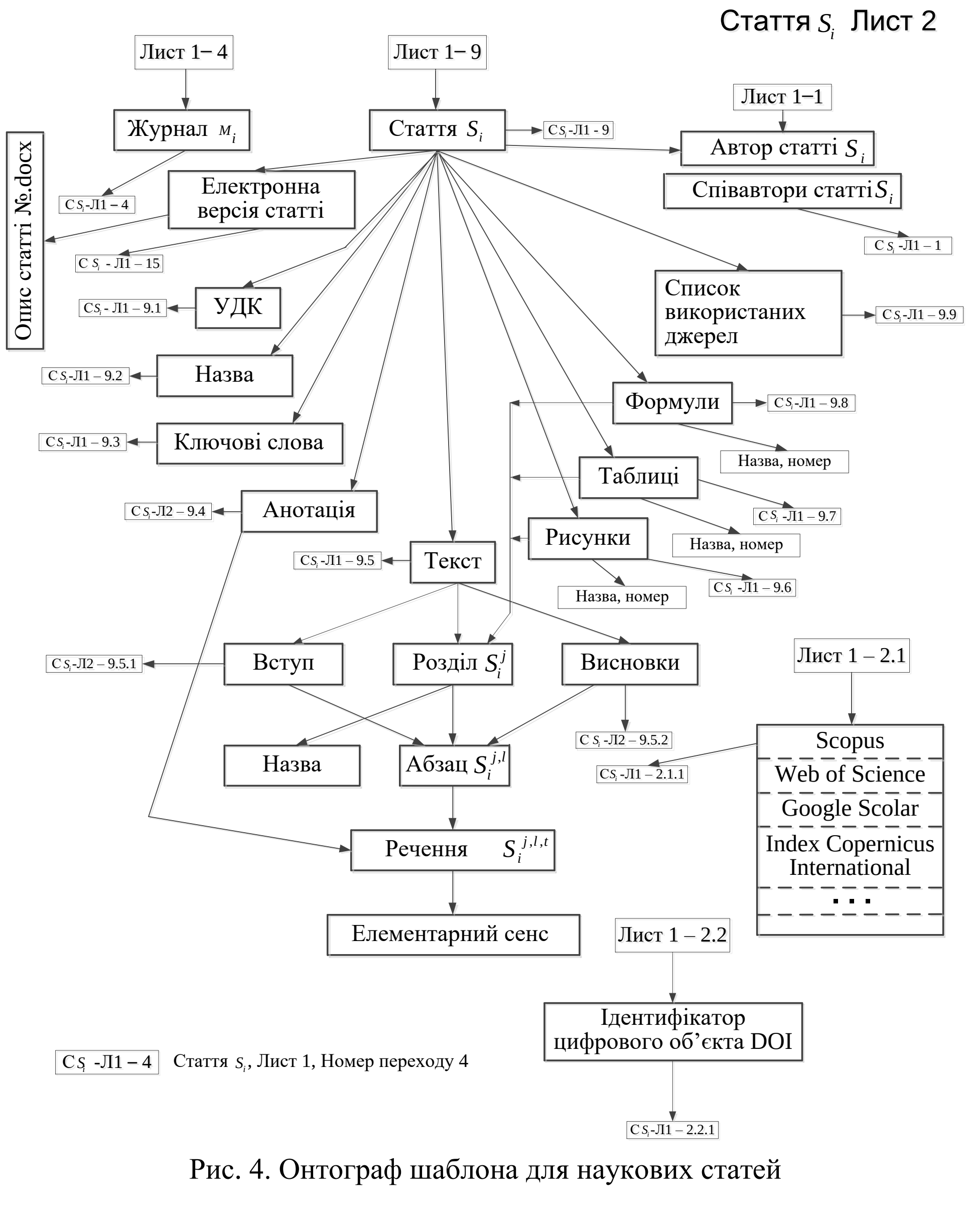

Рис. 4. Онтограф шаблона для наукових статей

## 4.2. **Підготовчий етап та його UML-діаграми**

Узагальнена блок-схема розроблення ОнС представлена на рис. 5. Вона включає: блок підготовчого етапу та блоки основного етапу з варіантами А і В.

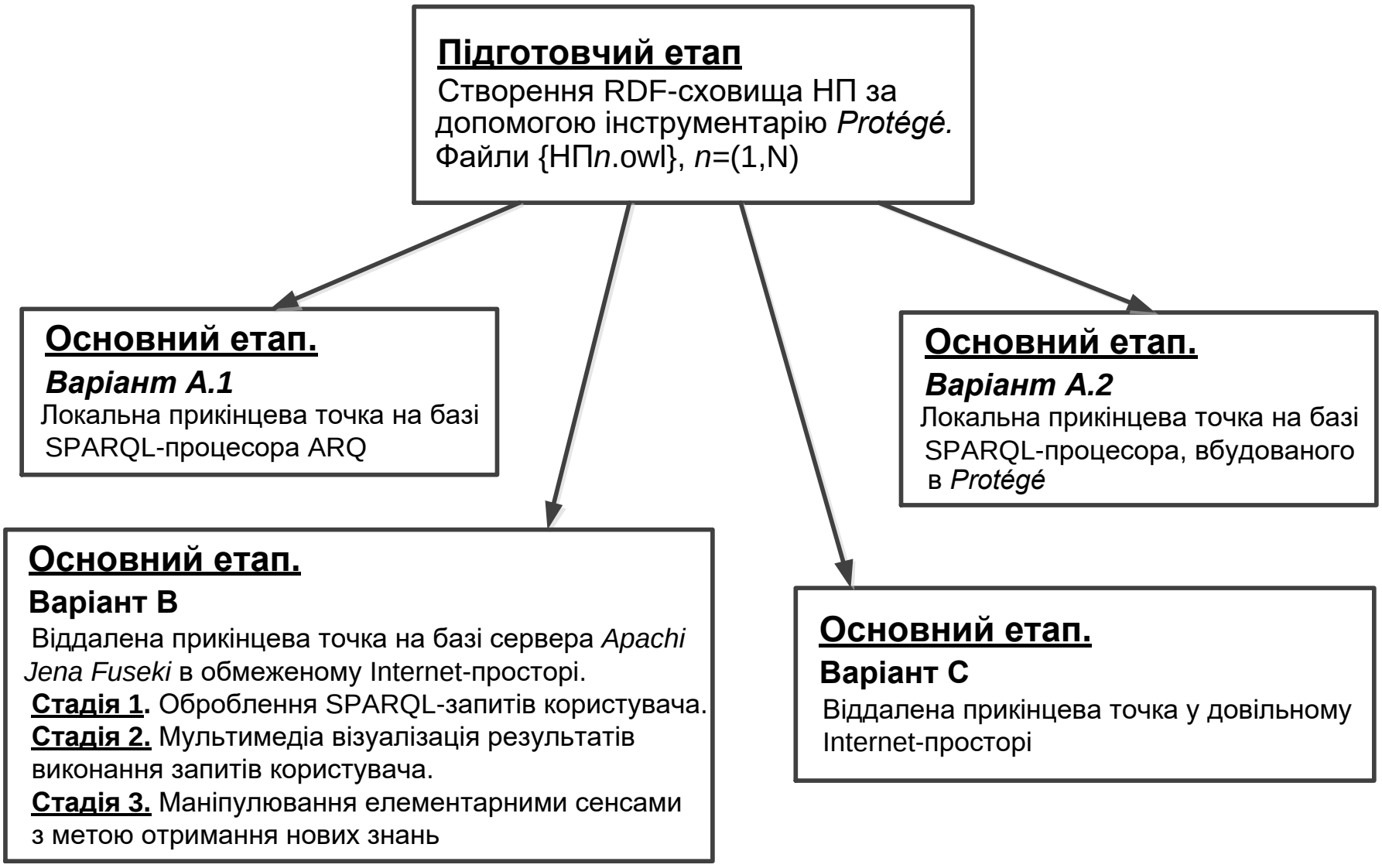

Рис. 5. Узагальнена блок-схема розроблення ОнС

На *підготовчому етапі* ІЗ необхідно виконати наступні дії:

1) розробити множину шаблонів формування структурних елементів НП. При цьому в Microsoft Word формуються файли: “Шаблон_стаття.*docx*”, “Шаблон_доповідь.*docx*”, “Шаблон_звіт.*docx*”, “Шаблон_патент.*docx*” тощо. Вихідною інформацією для розробки є вимоги та рекомендації відповідних видавництв наукової продукції;

2) сформувати базу електронних файлів НП, які надаються користувачем (автором статей), у відповідності з розробленими шаблонами;

3) сформувати базу даних НП у відповідності з онтографом, верхній рівень якого (загальний для всіх видів наукових публікацій) наведений на рис. 3. При цьому для кожної НП в онторедакторі Protégé формується файл з розширенням **.owl*;

4) сформувати сховище триплетів *rdf* документів НП;

5) створити локальну прикінцеву точку збереження й оброблення *rdf* документів НП;

6) поповнювати сховище триплетів в режимі надходження нових НП;

7) у взаємодії з користувачами розширювати функціональні можливості ОнС (створювати нові запити);

8) розробити онтологічну структуру концептів галузі наукових досліджень. При цьому для ГНД в онторедакторі Protégé розробляється онтограф і формується файл з розширенням "гнд.*owl*". Передбачається, що вказана онтологічна структура концептів формується на основі множини електронних версій НП у відповідності з методикою системної інтеграції й інструментарієм, описаних в [3];

9) сформувати сукупність семантичних структур ГНД і в онторедакторі Protégé розробити онтограф СмС галузі НД і сховище *rdf* триплетів СмС.

Далі наведемо *UML*-діаграми підготовчого етапу, які значно деталізують процеси та виконання відповідних процедур. Варіант А1 основного етапу інженером зі знань розглядається як етап відпрацювання його творчих задумів. Тому цей етап не розглядається.

Для етапу А2 буде наведено короткий опис та основні *UML*-діаграми. Він розглядається як експериментальна перевірка функціонування ОнС в локальній мережі, напрацювання списку запитів природною мовою (ПМ), узгодження роботи інтерфейсів користувачів та ІЗ, а також вирішення питань автоматизації створення *owl*-файлів НП. А основну увагу далі зосереджено на описі етапу В, стадія 1 (В1).

Наявна персоніфікована база знань науковця-дослідника, в якій задекларовано сукупність функціональних можливостей, що підтримують процеси наукової та творчої діяльності [4]. Така персоніфікована база знань є:

– засобом підтримки наукових досліджень і одним з центральних напрямів розвитку практичної інформатики [4, 5];

– розвитком системи знань НР й отримання нових знань (або впорядкування наявних, перевірка їх на коректність і суперечності тощо) [6–8];

– однією з основних підсистем сучасної системи дослідного проєктування [9], автоматизованого робочого місця НР [3];

– одним з основних елементів підтримки функціонування знання-орієнтованої інформаційної системи [9] і конструювання сталого канонічного знання [10].

Відомо, що є тісний зв'язок між технологіями *Semantic Web* і *UML*-технологіями, зокрема, зв'язок між *OWL*-синтаксисом і візуальним моделюванням *UML*-діаграм. Мова *UML* подається як загальноцільова мова візуального моделювання, яку розроблено для специфікації, візуалізації, проєктування та документування компонентів програмного забезпечення, бізнес-процесів та інших систем.

Мова *UML* одночасно слугує простим і потужним засобом моделювання, який може бути ефективно використаний для побудови концептуальних, логікових і графічних моделей складних систем найрізнішого цільового призначення. Ця мова увібрала в себе найкращі якості методів програмної інженерії, які з успіхом використовувалися протягом багатьох років при моделюванні великих і складних систем [11].

Візуальне моделювання в *UML* можна представити як певний процес порівневого спуску від найзагальнішої та найабстрактнішої концептуальної моделі вихідної системи до логікової, а потім і до фізичної моделі відповідної програмної системи. Для досягнення цих цілей спочатку будується модель у формі так званої діаграми варіантів використання, що описує функціональне призначення системи або, інакше кажучи, те, що система виконуватиме в процесі свого функціонування. Діаграма варіантів використання слугує вихідним концептуальним поданням або концептуальною моделлю системи в процесі її проєктування та розроблення [12, 13].

В основному режимі функціонування ОнС користувачу на основі запитів (список яких сформовано в головному меню інтерфейсу зв'язку) надається можливість рішення наступних задач (нижче буде наведено розширення списку запитів і кількісно, і якісно).

I. Задачі інформаційного пошуку в наукових публікаціях:

- формування списку публікацій (за назвами) в заданому журналі (збірнику);
- формування списку статей за певною тематикою;
- формування списку статей за роками і датами;
- формування списку статей автора із заданим співавтором;
- формування списку статей за бібліографічними і наукометричними показниками;
- формування тексту (рисунків, формул, таблиць, списку літературних джерел) заданої статті;
- формування списку анотацій (висновків, назв розділів тощо) заданих статей;
- формування списку видань (видавництв, журналів тощо) за заданими критеріями;
- пошук у НП інформації за заданими критеріями:

– за ключовим словом;

– за групою ключових слів;

– за фрагментом тексту;

– комбінований пошук інформації у декількох базах даних НП.

II. Задачі побудови онтографів і візуалізації інформації, яка міститься у відповідях на запити:

- побудова та візуалізація онтографа наукової дисципліни;
- побудова та візуалізація онтографа наукової публікації;
- побудова та візуалізація онтографа заданого фрагмента тексту із НП;
- побудова та візуалізація онтографа за ключовими словами;
- обчислення та візуалізація у графічному вигляді статистичних показників НП.

III. Задачі роботи із семантичними структурами (елементарними сенсами):

- побудова та візуалізація онтографа ЕС наукової публікації (фрагмента НП).

Задачі користувача II, III потребують спільного оброблення результатів виконання запитів у підсистемах *Локальна прикінцева точка* та *Sparql-обчислення* й інформації, яка міститься в блоці онтологічних структур ГНД підсистеми *Програмна оболонка й інтерфейс користувача*.

*UML-діаграми функціонування ОнС для реалізації процесів підготовчого етапу*

На підготовчому етапі всі процеси та процедури виконує інженер зі знань, користувач тільки надає йому масив текстових документів НП. Створення сховища *rdf* документів НП виконується на персональному комп'ютері (ПК) ІЗ, за його проєктом (див. рис. 6) та у відповідності з методичними матеріалами [14, 15]. Також слід додати, що основний варіант А припускає оброблення за запитом інформації, взятої тільки з однієї НП, тобто обчислення можливе тільки для однієї наукової статті. Як буде показано далі, варіант В припускає обчислення з будь-якої кількості НП.

На рис. 6 наведено *UML*-діаграму використання ОнС для реалізації підготовчого етапу, на рис. 7 – *UML*-діаграму компонентів, на рис. 8 – *UML*-діаграму послідовності і на рис. 9 – *UML*-діаграму діяльності. Це основні *UML*-діаграми підготовчої фази етапу А2 функціонування ОнС. Як видно з рисунку, програмні блоки у підсистемі створення/підтримки БДНП реалізовано в онторедакторі *Protege*, а базу даних НП – на жорсткому диску ПК ІЗ.

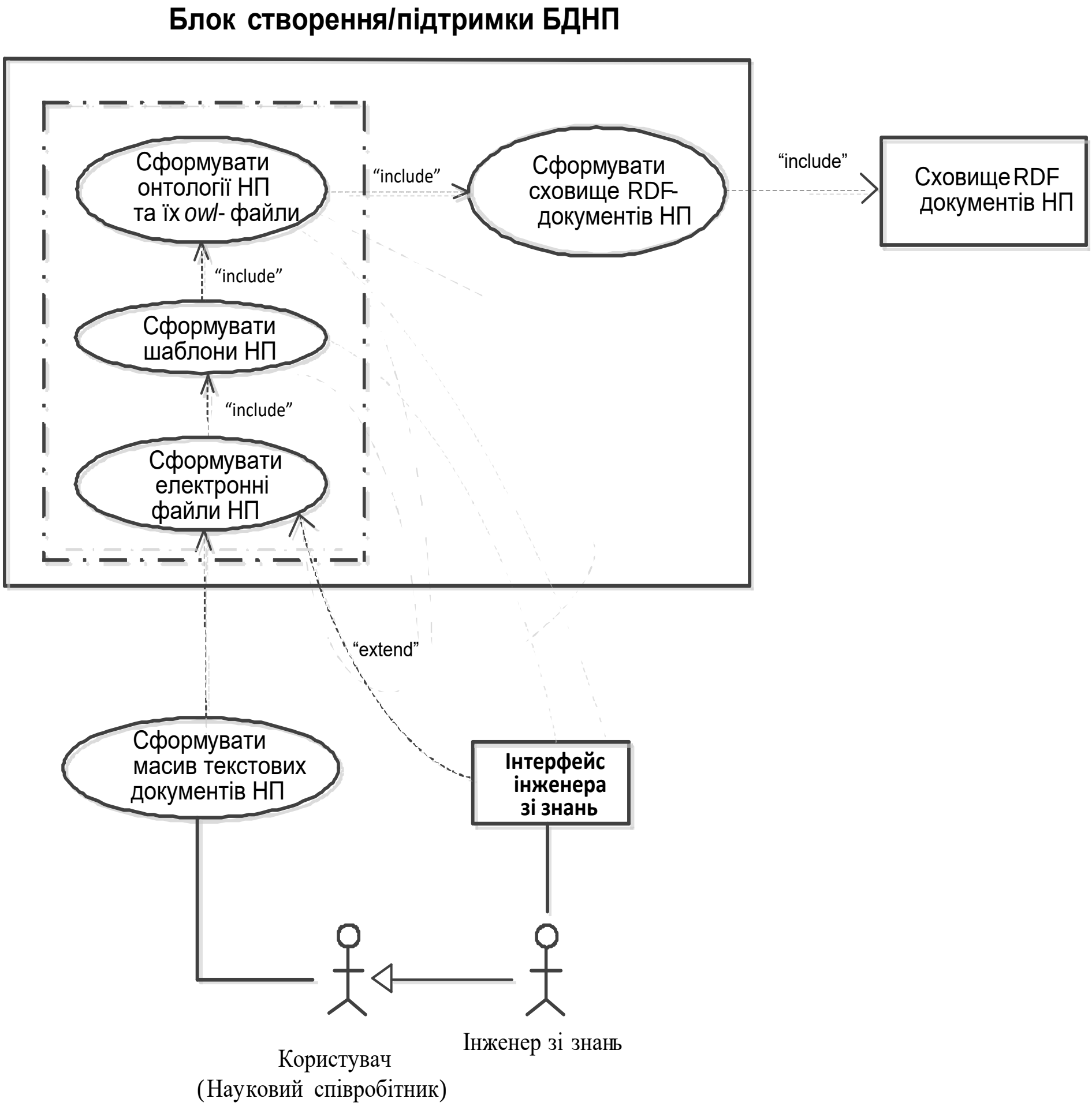

Рис. 6 Діаграма варіантів використання ОнС на підготовчому етапі

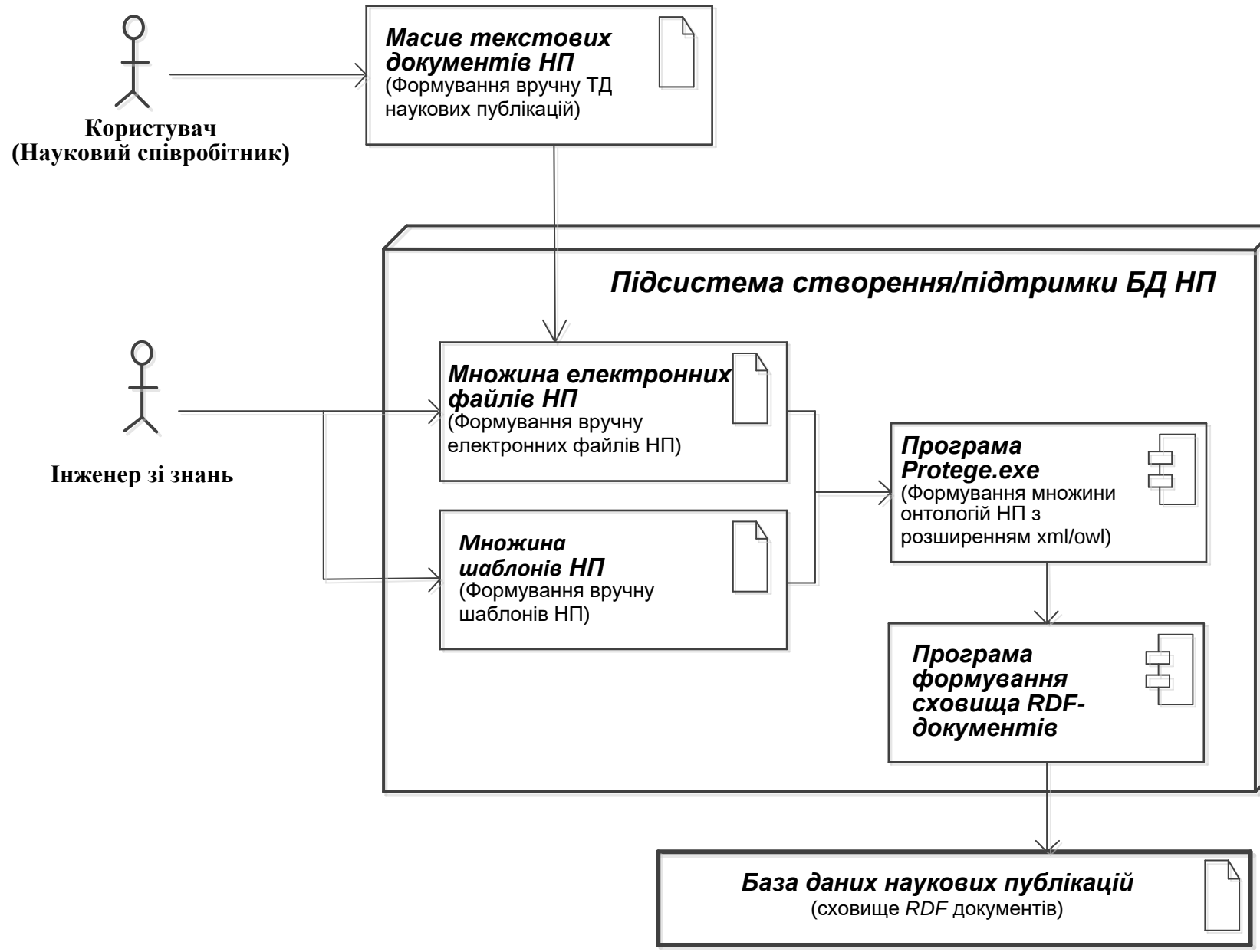

Рис. 7 Діаграма компонентів (підготовчий етап)

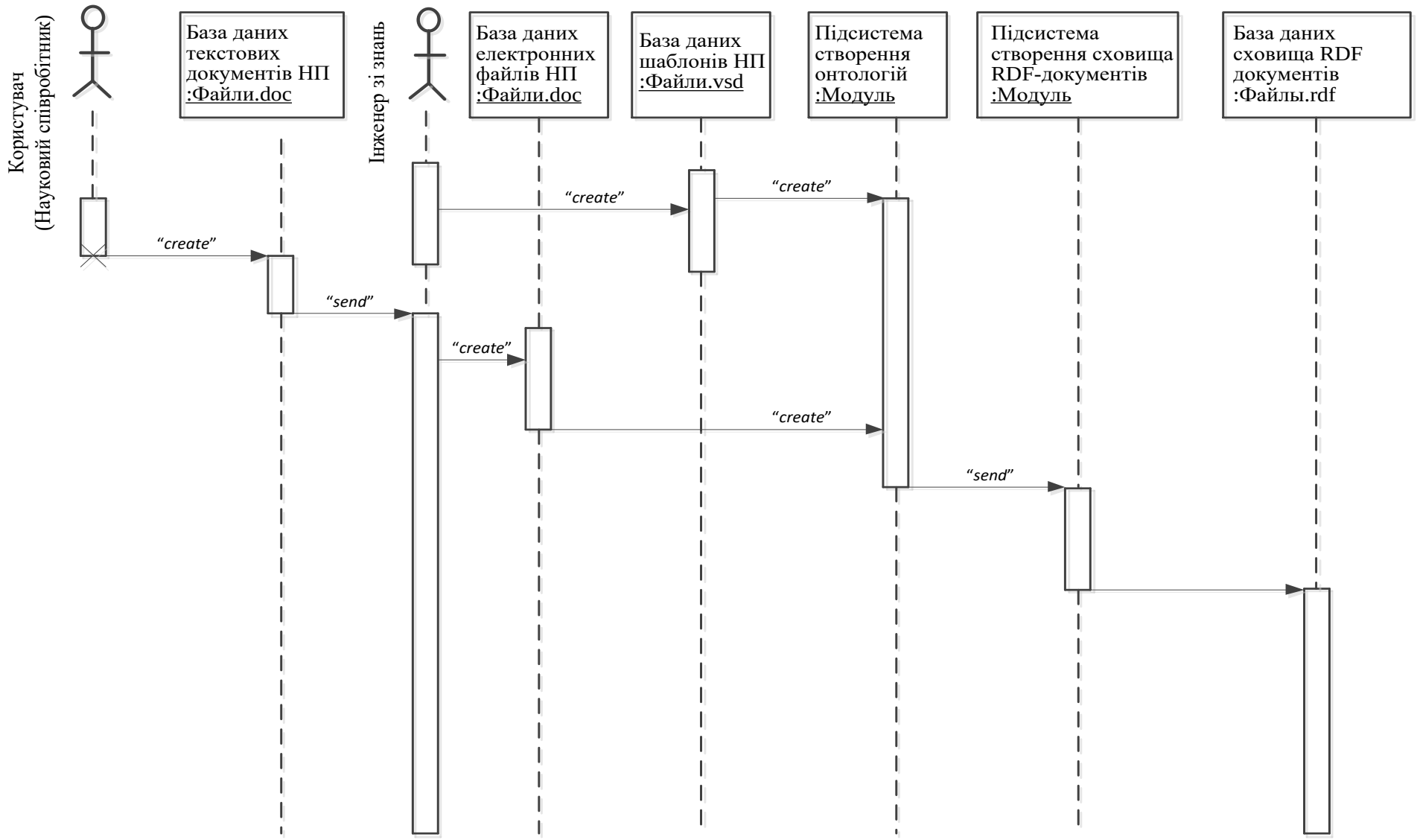

Рис. 8. Діаграма послідовності (підготовчий етап)

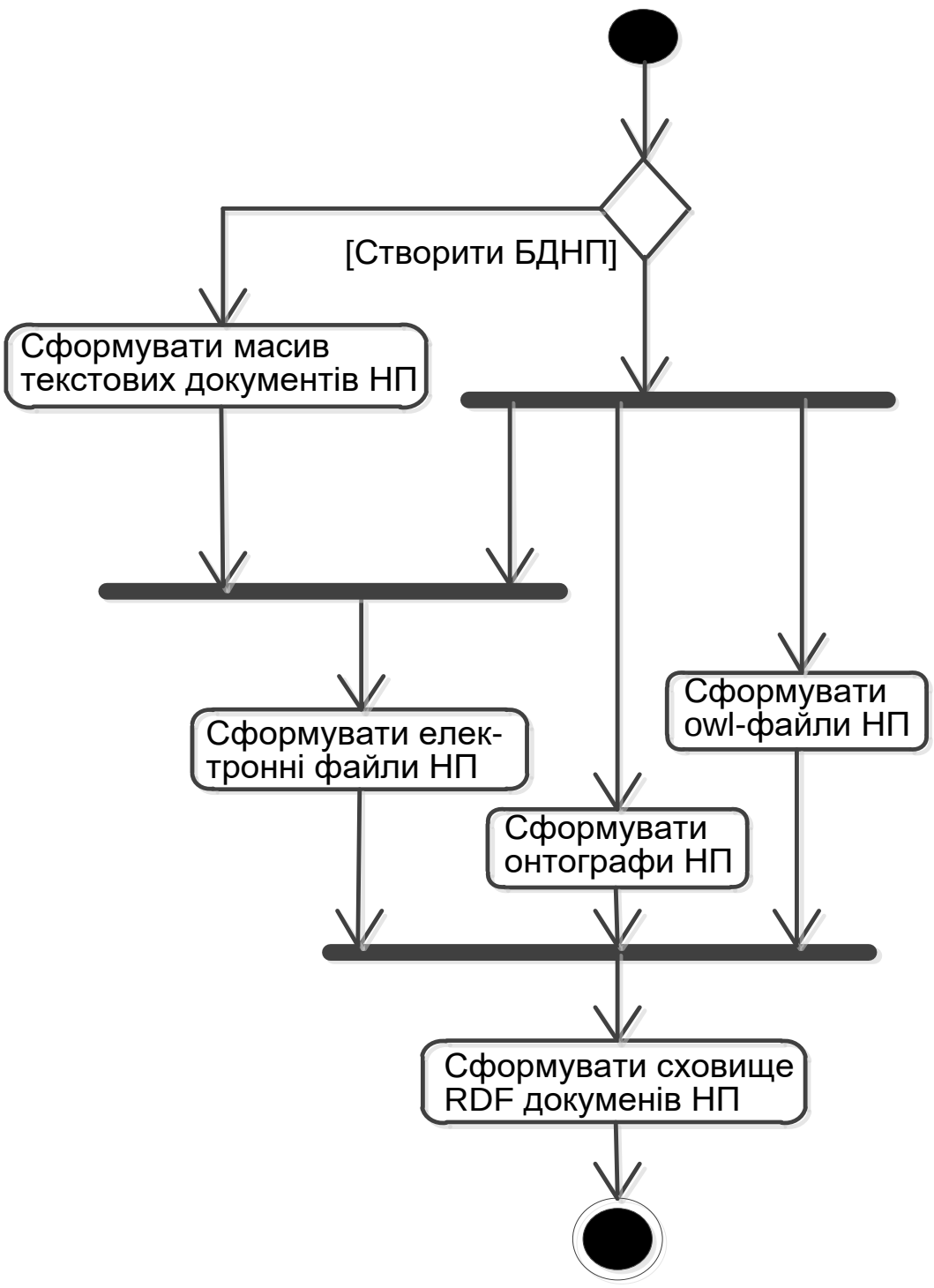

Рис. 9. Діаграма діяльності (підготовчий етап)

## 4.3. **UML-діаграми функціонування ОнС на основному етапі А2**

На рис. 10 наведено блок-схему функціонування ОнС. Як видно з рисунка, вона функціонує у локальній мережі “Інженер зі знань_Користувач (один або декілька)”. На ПК користувача відображається список запитів на ПМ з можливістю вибору відповідних аргументів та візуалізується результат обчисленого запиту. Ця інформація передається до ПК інженера зі знань. Як видно з рис. 10, у ньому функціонують блоки інтерфейсу ІЗ та інструментарію *Protege*, в якому встановлено відповідні *SPARQL* процесор та *Reasoner*. Опис відповідних процесів і процедур наведено на рис. 11–14.

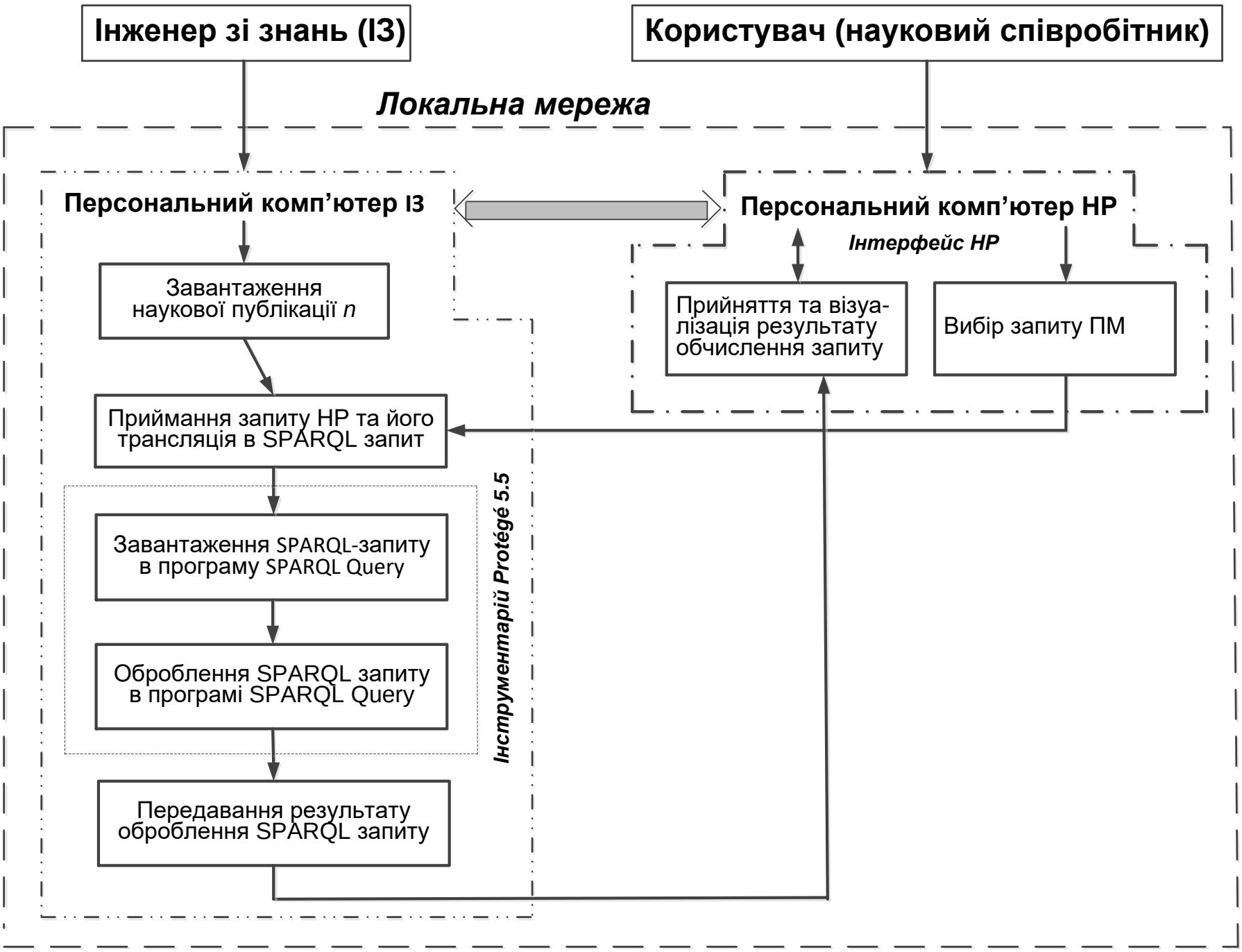

Рис. 10. Блок-схема ОнС (основний етап А2)

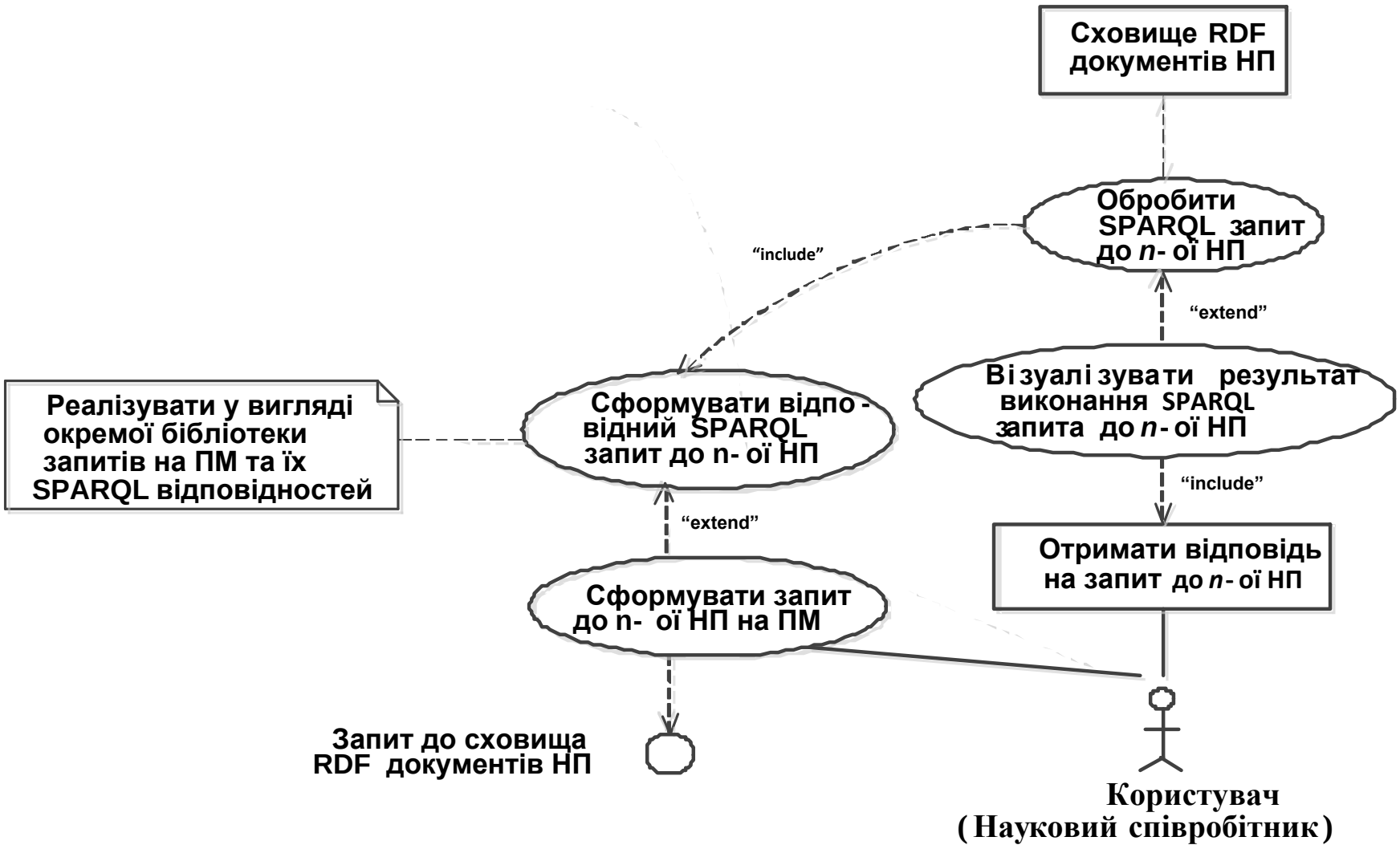

Рис. 11. Діаграма варіантів використання ОнС (основний етап А2)

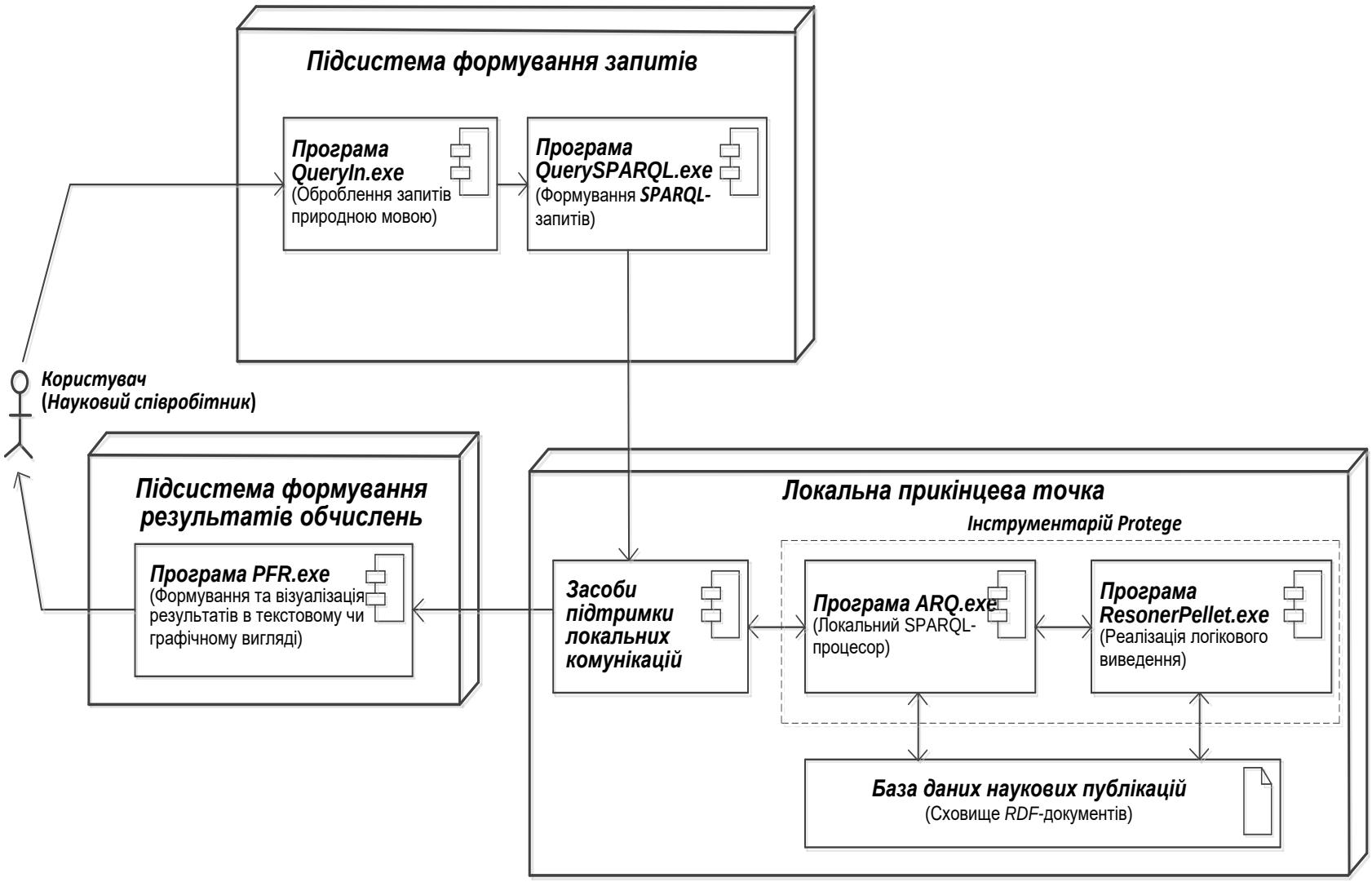

Рис. 12. Діаграма компонентів використання ОнС (основний етап А2)

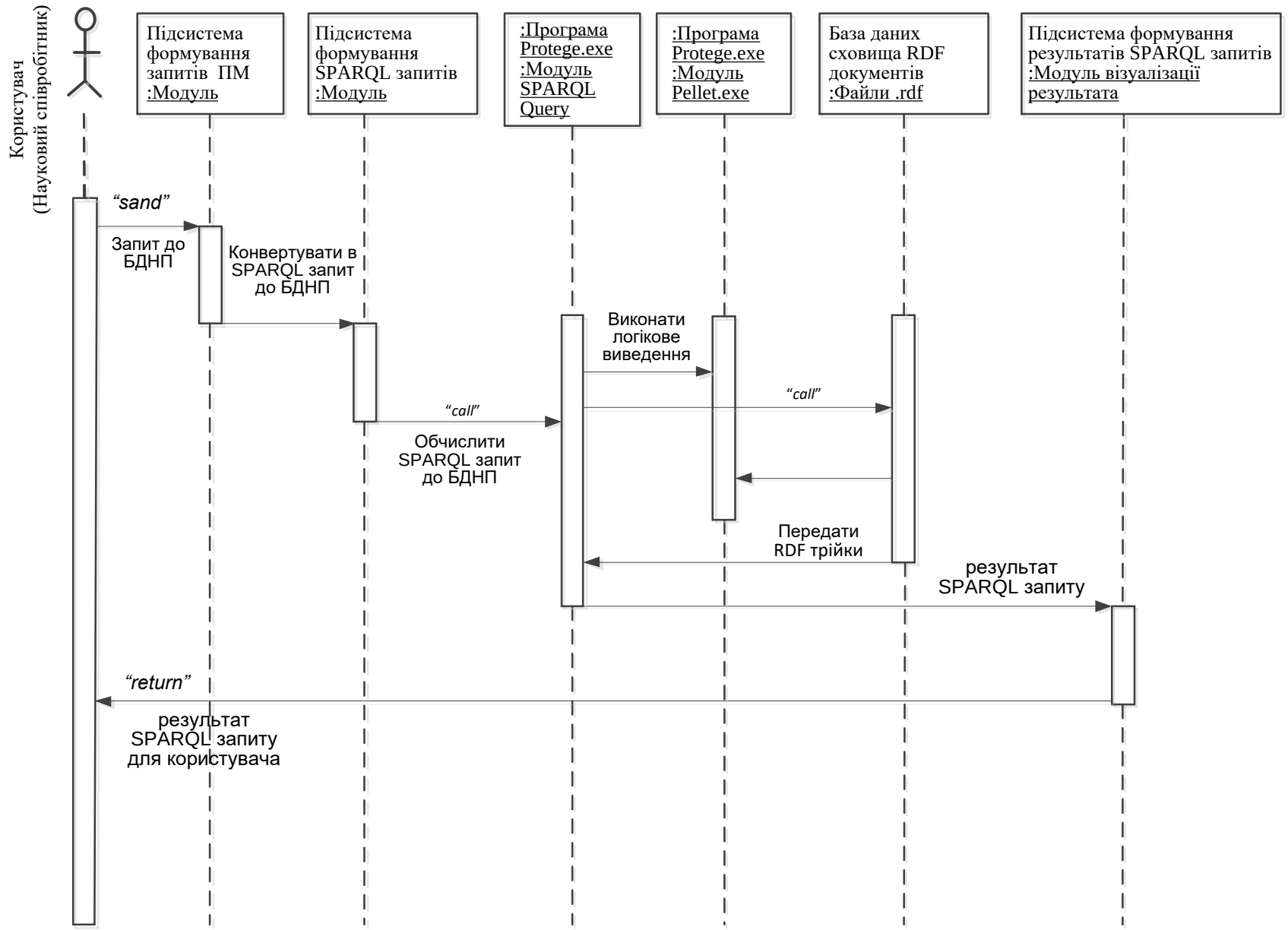

Рис. 13. Діаграма послідовності використання ОнС (основний етап А2)

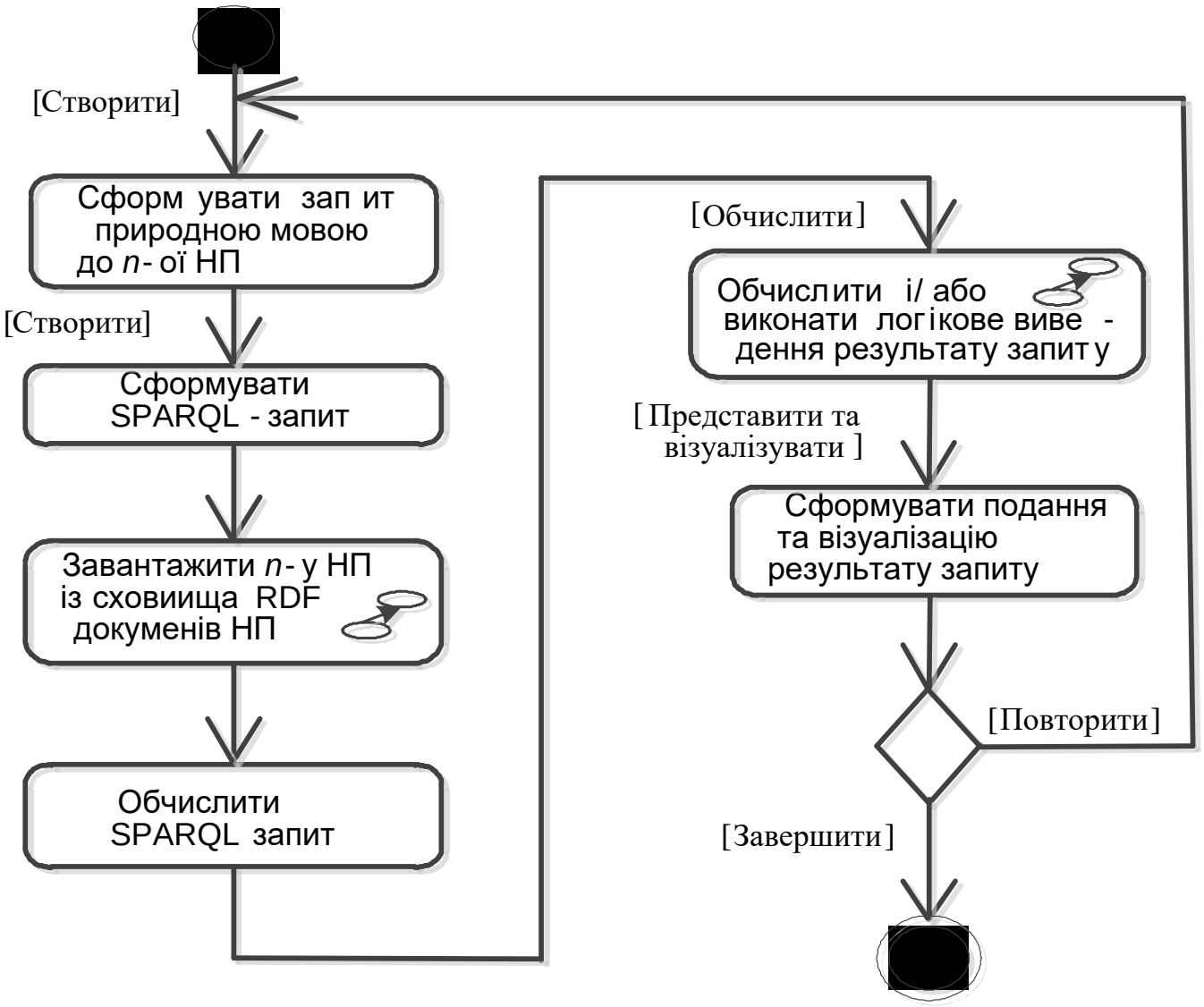

Рис 14. Діаграма діяльності ОнС (основний етап А2)

### 4.4. **UML-діаграми функціонування ОнС на основному етапі, варіант B1**

ОнС для цього варіанта функціонує як віддалена прикінцева точка на базі сервера *Apachi Jena Fuseki* і складається з трьох стадій: стадія 1 – оброблення *SPARQL* запитів користувача; стадія 2 – мультимедіа візуалізація результатів виконання запитів користувача або побудова та використання образно-понятійних структур галузі наукових досліджень; стадія 3 – маніпулювання елементарними сенсами з метою отримання нових знань. На рис. 15 наведено блок-схему ОнС для варіанта В1.

Спочатку інженеру зі знань необхідно розгорнути сервер *Apachi Jena Fuseki* як віддалену прикінцеву точку (див. 2.7.2 і [16–18]) і завантажити у сховище наукових публікацій файли *RDF* графів наукових статей, що сформовані на попередньому етапі.

Користувачу в його інтерфейсі надається список можливих запитів природною мовою до його БДНП. Із цього списку він поступово вибирає по одному запиту, який передається через локальну мережу в персональний комп'ютер ІЗ, тобто крок за кроком деталізує необхідну йому інформацію. Причому, в одному запиті користувач має можливість вказати декілька статей, з яких йому потрібно отримати інформацію.

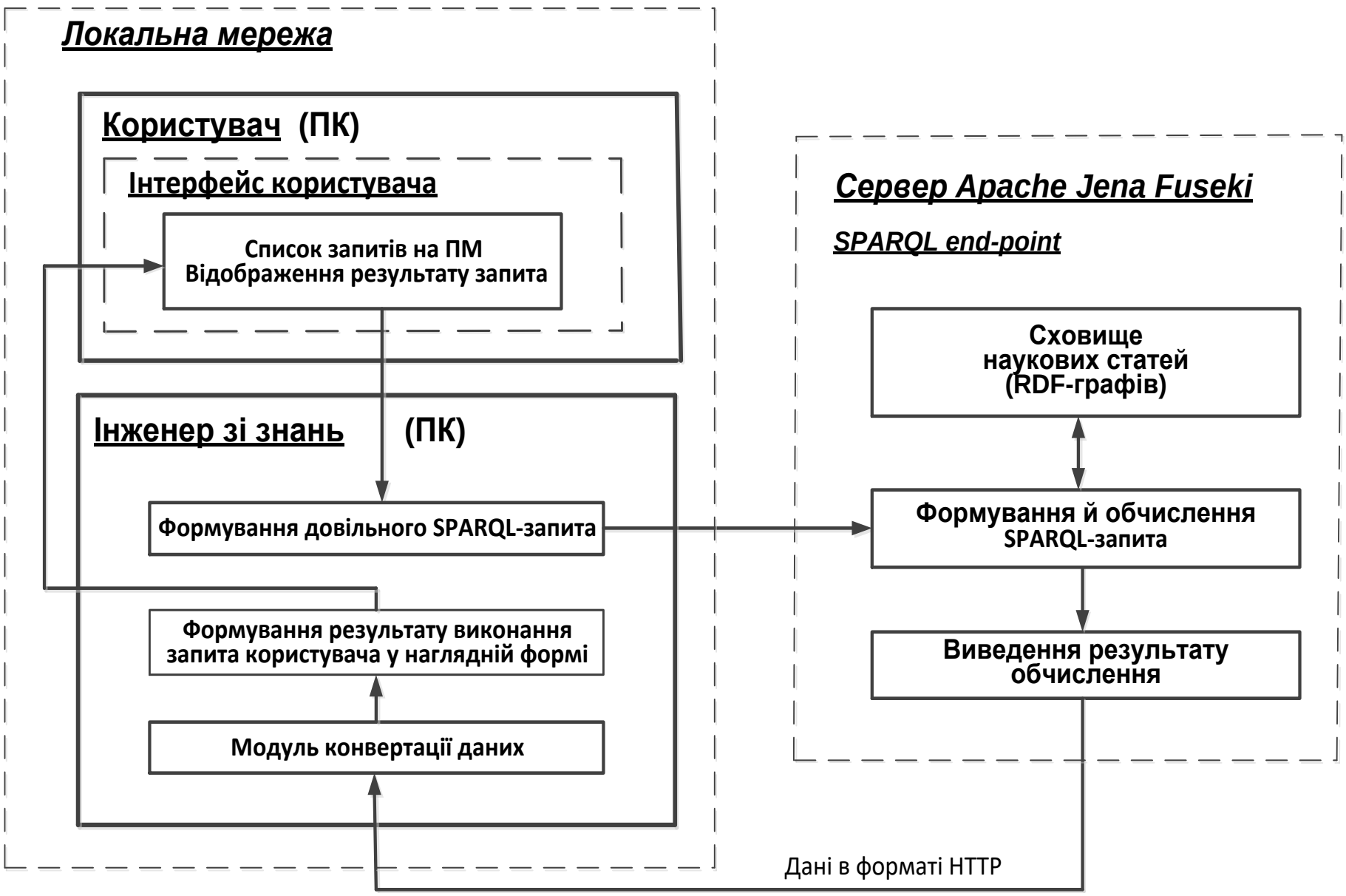

Рис. 15. Узагальнена блок-схема ОнС

На підготовчому етапі інженер зі знань (враховуючи побажання користувачів) формує початковий список запитів природною мовою. Далі наведено варіант такого початкового списку.

*Запити користувача початкові*

Вихідна інформація.

Науковий співробітник у своїй базі даних має опублікованими у відомих наукових журналах *N* наукових статей. Множина *N* описується у такий спосіб:

$$N = 1, 2, \ldots, m1, \ldots, m1+1, \ldots, m2, \ldots, m2+1, \ldots, mk, mk+1, \ldots, N–1.$$

Припускається, що автором є перший у публікації співавтор, або той, кому належить база даних НП.

1 *Вивести* назви всіх *N* статей:

1.1 у порядку зростання дати публікації;

1.2 після ___ року публікації;

1.3 без співавторів.

2. *Вивести* назви статей *m*1, *m*1+b… *m*2, …, *m*2+*c*, *mk*, …, *mk*+*d*, …

змінні параметри *m*1, *b*, *m*2, *c*, *mk*, *d* … задаються користувачем.

3. *Вивести* ПІБ співавторів статей *m*1, *m*1+*b*…*m*2, …, *m*2+*c, mk*, …, *mk*+*d*, …

……..

7. *Вивести* анотації статей *m*1, *m*1+*b*… *m*2, …, *m*2+*c, mk*, …, *mk*+*d*, …

…….

13. *Вивести* назви статей на тему «*трансдисциплінарність*».

15. *Вивести* назви розділів статей *m*1, *m*1+*b*…*m*2, …, *m*2+*c*, *mk*, …, *mk*+*d*, …

16. *Вивести* текст розділу *n* статті *m*.

……..

20. *Вивести* назви рисунків у статтях *m*1, *m*1+*b*… *m*2, …, *m*2+*c, mk*, …, *mk*+*d*, …

……..

23. *Вивести* тексти абзаців в статтях *m*1, *m*1+*b*… *m*2, …, *m*2+*c, mk*, …, *mk*+*d*, …, у яких

є фраза {*fi*}. Фраза для перевірки запиту – «*трансдисциплінарні наукові дослідження*», «*онтологічний*» та інші.

…

25. *Вивести* назви статей, в яких автором чи співавтором є «ПІБ».

26. *Вивести* назви статей, які опубліковані в журналі {*gк*}, де *gк* – назва журналу.

Слід зазначити, що у ПМ-запиті слово «*Вивести*» не слід трактувати як ключове, воно може бути замінене будь-яким його синонімом.

*UML-діаграми функціонування ОнС для варіанта В1*

Розгляньмо *UML*-діаграми, які розкривають суть функціонування ОнС для варіанта В1.

На рис. 16 представлено діаграму варіантів використання, на рис. 17 – діаграму класів, на рис. 18 – діаграму компонентів і на рис. 19 – діаграму послідовності.

До локальної мережі (ЛМ), адміністратором якої є інженер зі знань, підключено певну кількість персональних комп’ютерів НР. Роботу мережі ми розглянемо для одного користувача, для декількох користувачів запит виконується в аналогічний спосіб.

Персональний комп’ютер користувача функціонує як модуль загального інтерфейсу ЛМ. В ньому, з однієї сторони відображаються всі запити на ПМ, з яких НР може вибрати один з відповідними аргументами, а з іншої сторони – відображається результат виконання запиту у наглядній формі.

Другу частину інтерфейсу складає модуль ІЗ. У ньому формується відповідний SPARQL запит і передається за HTTP-протоколом до end-point. На сервері Apachi Jena Fuseki виконується SPARQL запит і відповідь на нього за HTTP-протоколом передається в ЛМ, інтерфейс, модуль ІЗ.

Процеси формування, оброблення запитів користувача та отримання ним відповідей детально показано на основних UML-діаграмах на рис. 13–16.

Слід зазначити, що: на діаграмах не показано процедури вибору аргументів НР та перетворення їх у номери статей у БД; ОнС зорієнтовано на оброблення НП, поданих українською чи російською мовами, а тому у SPARQL запитах ключі пошуку надаються двома мовами.

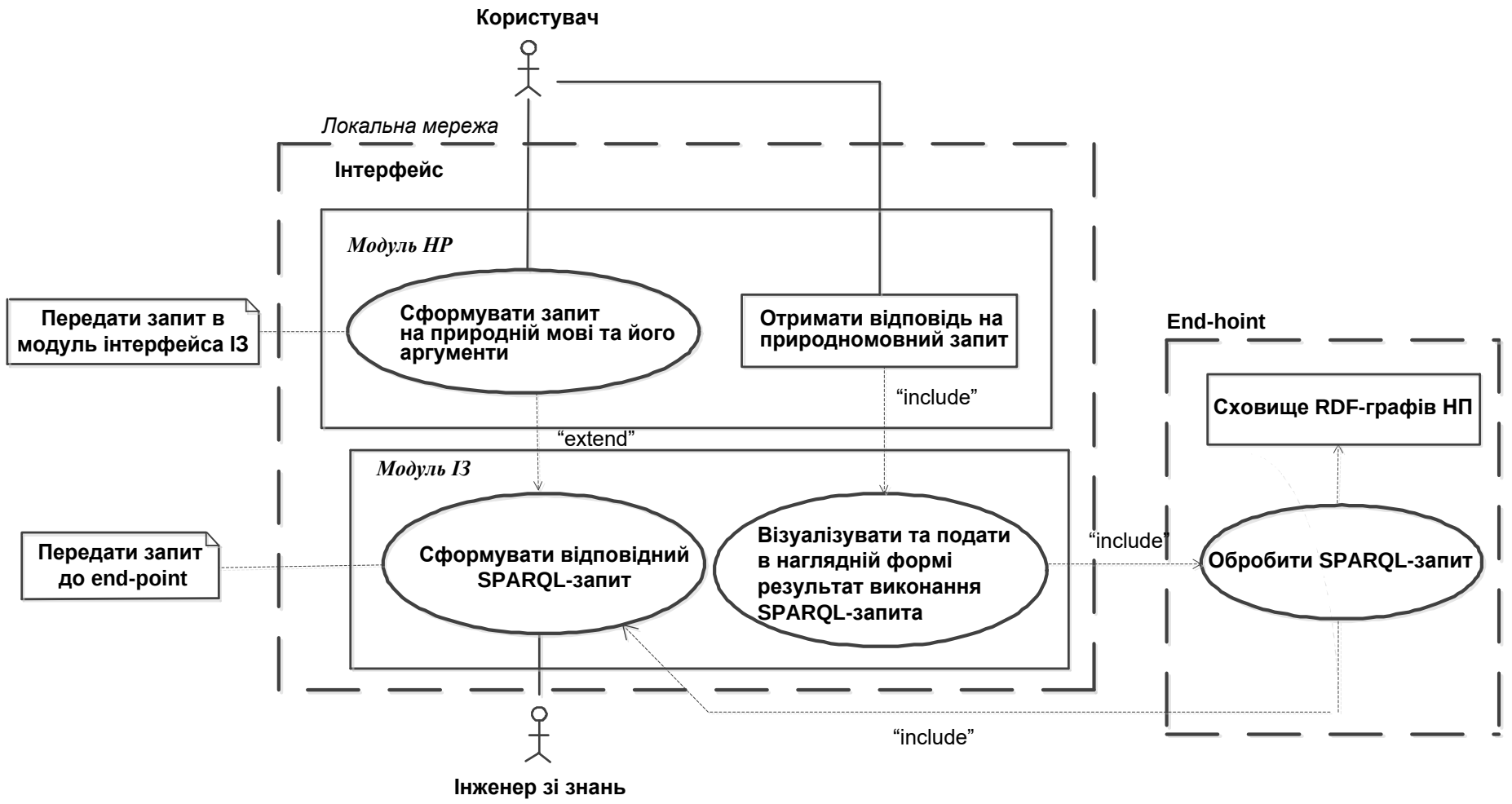

Рис. 16. Діаграма варіантів використання ОнС

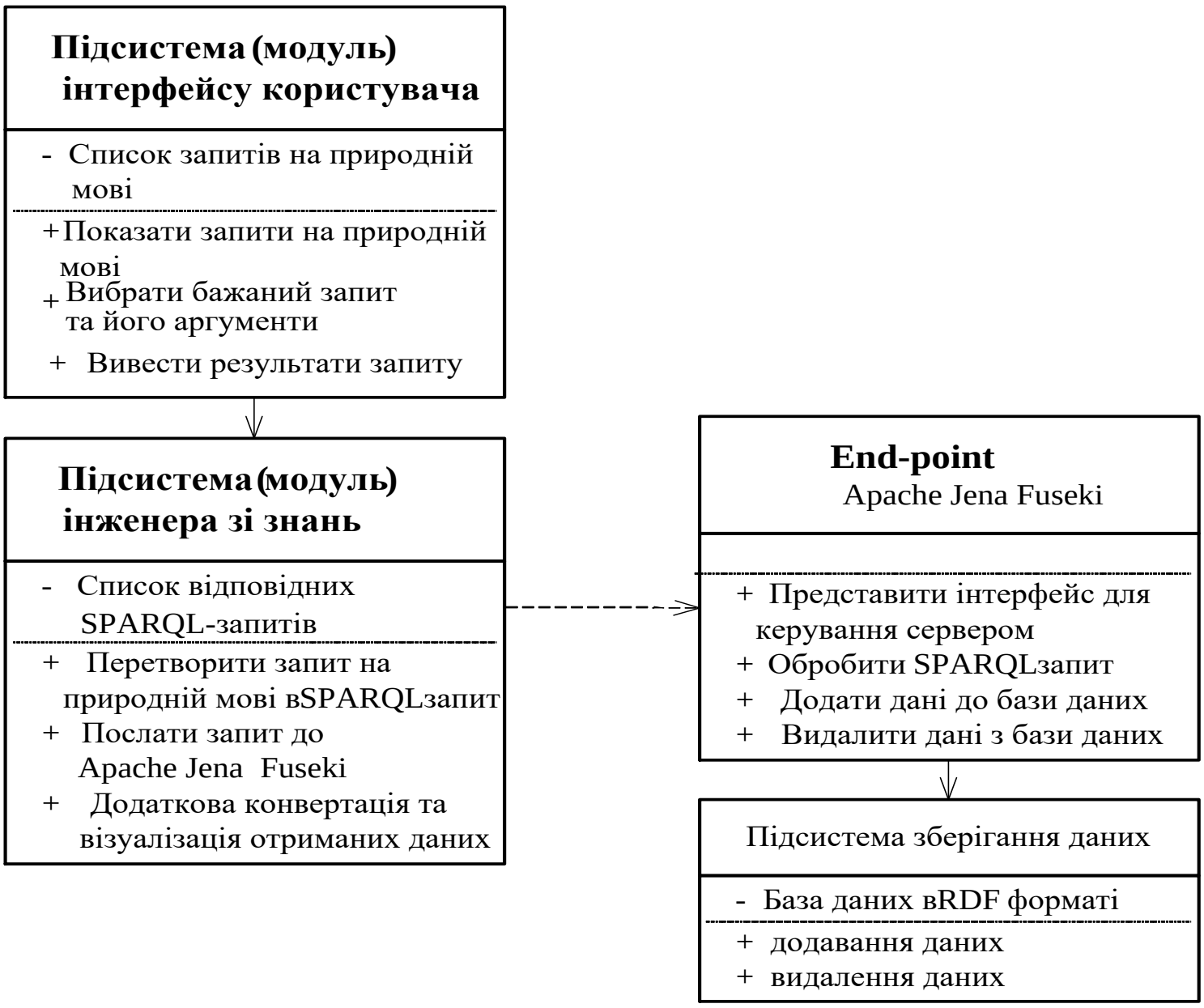

Рис. 17. Діаграма класів ОнС

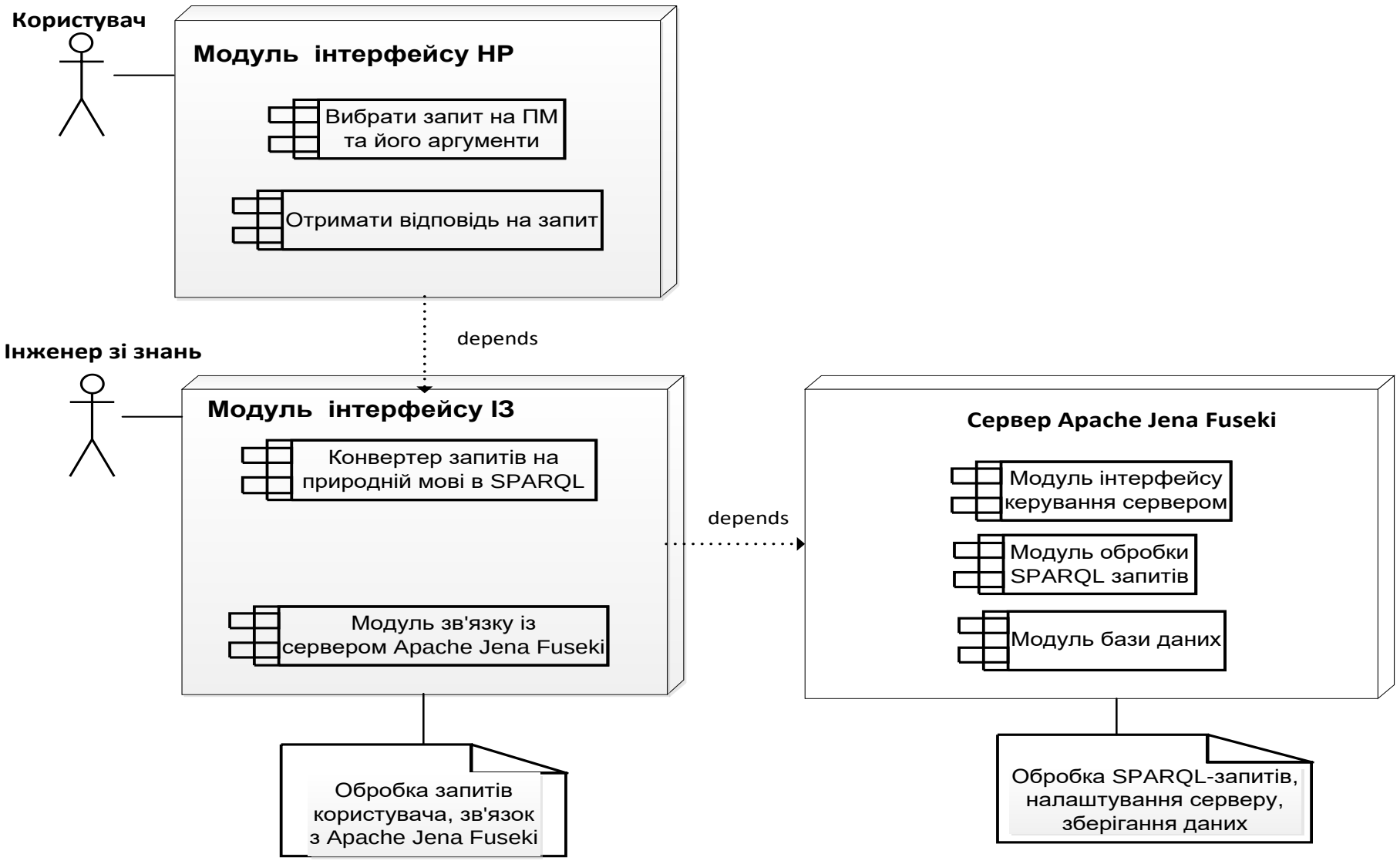

Рис. 18. Діаграма компонентів ОнС

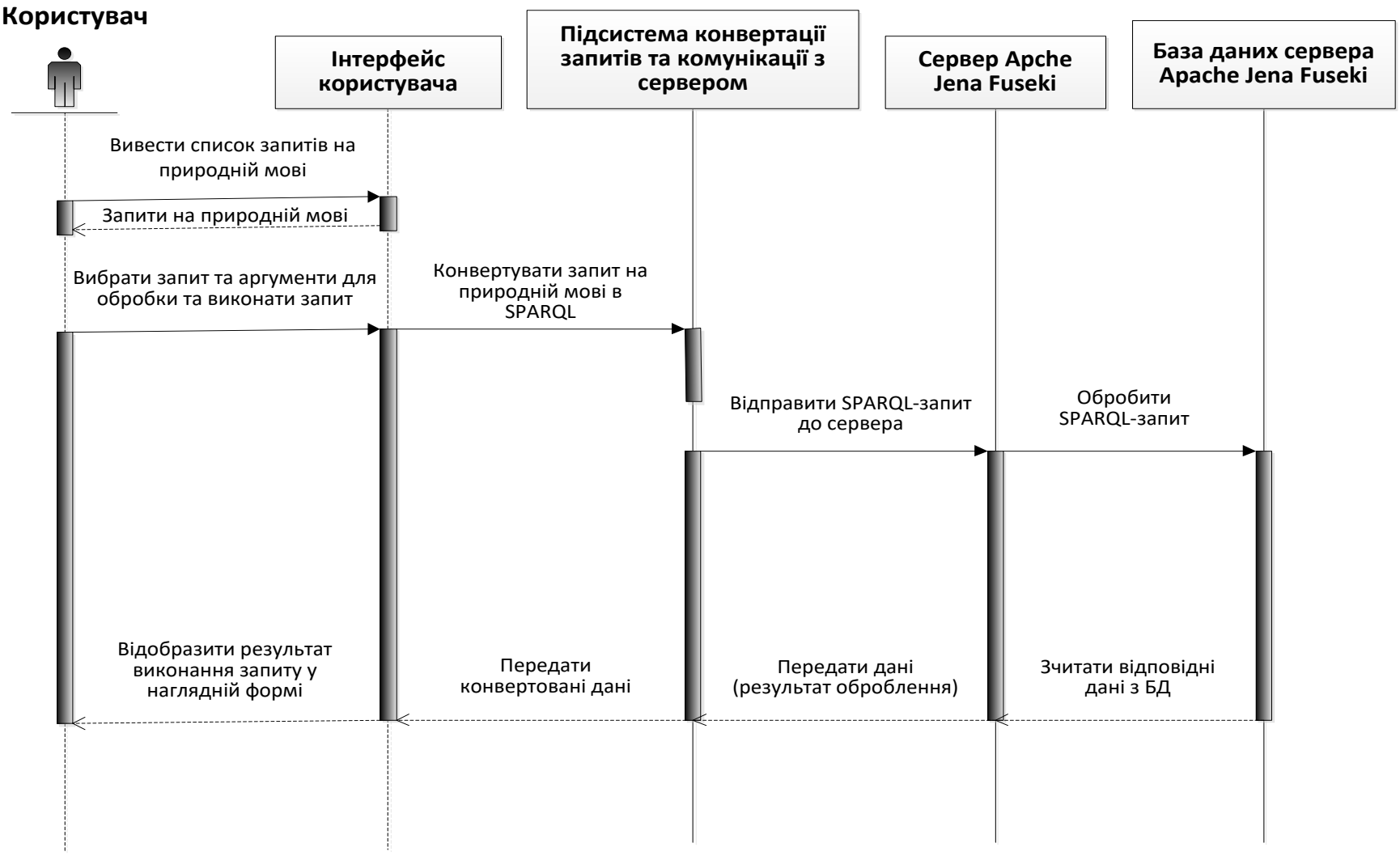

Рис. 19. Діаграма послідовності ОнС

## 4.5. **Приклади виконання SPARQL запитів та їхні результати**

Слід зазначити, що наукові публікації в БД пронумеровані по порядку, а їхній список з назвами доступний користувачу в його інтерфейсі.

1. Запит на ПМ.

*Вивести назви статей на тему “онтологічний”.*

Відповідний *SPARQL*-запит.

*PREFIX* : <http://www.semanticweb.org/николай/ontologies/2020/5/19/untitled-ontology36#>

*SELECT DISTINCT* ?номер_статті ?назва_статті

{

*GRAPH* ?номер_статті {?s1 :Название_статьи?назва_статті.

*FILTER REGEX*(?назва_статті, "онтолог", "i")}

}

Результат виконання.

1. Номер статті http://test.ulif.org.ua:51089/articles/data/**article5**

Назва статті «Про деякі особливості побудови онтологічних моделей предметних областей»

2. Номер статті <u>http://test.ulif.org.ua:51089/articles/data/**article7**</u>

Назва статті >«Введение в класс трансдисциплинарных онтолого-управляемых систем исследовательского проєкирования»

3. Номер статті <u>http://test.ulif.org.ua:51089/articles/data/**article8**</u>

Назва статті «Онтологическая концепция информатизации научных исследований»

4. Номер статті <u>http://test.ulif.org.ua:51089/articles/data/**article10**</u>

Назва статті «Архитектура онтолого-управляемых компьютерных систем»

5. Номер статті <u>http://test.ulif.org.ua:51089/articles/data/**article16**</u>

Назва статті «К вопросу системно-онтологической интеграции знаний предметной области»

6. Номер статті <u>http://test.ulif.org.ua:51089/articles/data/**article19**</u>

Назва статті «Знание-ориентированные информационные системы с обработкой естественно-языковых объектов: онтологический подход»

7. Номер статті <u>http://test.ulif.org.ua:51089/articles/data/**article21**</u>

Назва статті «Системно-онтологический анализ предметной области»

2. Запит на ПМ.

*Вивести назви статей на тему «трансдисциплінарність»*.

Відповідний *SPARQL*-запит.

*PREFIX* :
<http://www.semanticweb.org/николай/ontologies/2020/5/19/untitled-ontology36#>

*SELECT* ?номер_статті ?назва_статті

{

*GRAPH* ?номер_статті {?s1 :Название_статьи?назва_статті.

*FILTER REGEX* (?назва_статті, «трансдисципл», «i»)}

}

Результат виконання.

1. Номер статті <u>http://test.ulif.org.ua:51089/articles/data/**article1**</u>

Назва статті «Методологические основы развития, становления и IT-поддержки трансдисциплинарных исследований»

2. Номер статті <u>http://test.ulif.org.ua:51089/articles/data/**article2**</u>

Назва статті «Трансдисциплинарность, информатика и развитие

современной цивилизации»

3. Номер статті http://test.ulif.org.ua:51089/articles/data/**article6**

Назва статті «Проблемы трансдисциплинарности и роль информатики»

4. Номер статті http://test.ulif.org.ua:51089/articles/data/**article7**

Назва статті «Введение в класс трансдисциплинарных онтолого-управляемых систем исследовательского проєкирования»

3. Запит на ПМ.

*Вивести абзаци в розділах статей, у яких розглядається «формальна комп'ютерна онтологія ≫.*

Відповідний *SPARQL*-запит.

*PREFIX* : <http://www.semanticweb.org/николай/ontologies/2020/5/19/untitled-ontology36#>

*SELECTDISTINCT* ?номер_статті ?назва_статті ?paragraph ?текст_речення

*FROM NAMED* <http://test.ulif.org.ua:51089/articles/data/article1>

*FROM NAMED* <http://test.ulif.org.ua:51089/articles/data/article2>

*FROM NAMED* <http://test.ulif.org.ua:51089/articles/data/article4>

{

*GRAPH* ?номер_статті {?s1 :Название_статьи?назва_статті

{?paragraph :Иметь_Предложение?речення_абзацу}

{?речення_абзацу :Иметь_Текст?текст_речення}

FILTERREGEX (?текст_речення, «формальн.*онтолог.*», «i»)

}}

Результат виконання.

1. Номер статті http://test.ulif.org.ua:51089/articles/data/**article2**

Назва статті «Методологические основы развития, становления и IT-поддержки трансдисциплинарных исследований»

Абзац 4.11.1.17

Текст речення «Во-первых, формальные компьютерные онтологии являются одновременно и результатом развития, и инструментом knowledge engineering»

2. Номер статті http://test.ulif.org.ua:51089/articles/data/**article2**

Назва статті «Методологические основы развития, становления и IT-поддержки трансдисциплинарных исследований»

Абзац 4.11.1.15

Текст речення «Таким каркасом может служить совокупность формальных компьютерных онтологий конкретних предметных областей [5]»

3. Номер статті http://test.ulif.org.ua:51089/articles/data/**article2**

Назва статті «Методологические основы развития, становления и IT-поддержки трансдисциплинарных исследований»

Абзац 4.11.1.15

Текст речення «Формально онтологию можно представить четверкой множеств:»

4. Номер статті http://test.ulif.org.ua:51089/articles/data/**article4**

Назва статті «Информационная технология и инструментальные средства поддержки процессов исследовательского проєкирования *Smart*-систем»

Абзац 4.11.2.2

Текст речення «…извлечение из множества ТД знаний, релевантных заданной ПдО, их системно-онтологическая структуризация и формально-логическое представление на одном или нескольких из общепринятых языков описания онтологий (*Knowledge Representation*);»

5. Номер статті http://test.ulif.org.ua:51089/articles/data/**article1**

Назва статті «Методологические основы развития, становления и IT-поддержки трансдисциплинарных исследований»

Абзац 3.15.3

Текст речення «В ее недрах зародилась онтологическая концепция, сущность которой состоит в формальном онтологическом описании предметных областей и научной картины мира в целом (процесс только начат)»

6. Номер статті http://test.ulif.org.ua:51089/articles/data/**article1**

Назва статті «Методологические основы развития, становления и IT-поддержки трансдисциплинарных исследований»

Абзац 4.11.3.1

Текст речення ««Формально онтологию можно представить четверкой множеств:»

4. Запит на ПМ.

*Вивести анотації статей* 2, 6, 13.

Відповідний *SPARQL*-запит.

*PREFIX* rdfs: <http://www.w3.org/2000/01/rdfschema#>

*PREFIX* rdf: <http://www.w3.org/1999/02/22-rdf-syntax-ns#>

*PREFIX* : <http://www.semanticweb.org/николай/ontologies/2020/5/19/untitled-ontology36#>

*SELECT* ?номер_статті ?назва_статті (group_concat(?анотація) as ?анотація_повна)

*FROM NAMED* <http://test.ulif.org.ua:51089/articles/data/article2>

*FROM NAMED* <http://test.ulif.org.ua:51089/articles/data/article6>

*FROM NAMED* <http://test.ulif.org.ua:51089/articles/data/article13>

{

*GRAPH* ?номер_статті {?s1 :Название_статьи?назва_статті.

?subclass rdfs:subClassOf :Аннотация{?shortname rdf:type ?subclass}{?shortname :Иметь_Текст ?анотація}

}}

*groupby* ?номер_статті ?назва_статті

Результат виконання.

1. Номер статті <http://test.ulif.org.ua:51089/articles/data/**article6**>

Назва статті «Проблемы трансдисциплинарности и роль информатики»

Анотація повна «Предложены и проанализированы некоторые методы и средства информационно-технологической поддержи трансдисциплинарных научных исследований в их становлении и развитии»

2. Номер статті <http://test.ulif.org.ua:51089/articles/data/**article2**>

Назва статті «Трансдисциплинарность, информатика и развитие современной цивилизации»

Анотація повна «Перспективы и проблемы развития человеческой цивилизации всегда волновали общество. Особенно острый интерес к ним возникает, как правило, на крутых виражах истории, в периоды общественных потрясений и техногенних катаклизмов. …»

3. Номер статті <http://test.ulif.org.ua:51089/articles/data/**article13**>

Назва статті «Міждисциплінарні наукові дослідження: оптимізація системно-нформаційної підтримки»

Анотація повна «Епоху аналітизму й властиву йому диференціацію науки завершено. Реальні проблеми, що постають перед людством, набагато складніші за наукові. Сучасна наука не в змозі їх кардинально вирішити. Одна з причин – роз'єднаність наукових дисциплін, а отже, нескоординованість роботи наукових установ, їхньої тематики. …»

5. Запит на ПМ.

*Вивести назви* статей із співавтором «*Кургаєв*».

Відповідний *SPARQL*-запит.

*PREFIX* : <http://www.semanticweb.org/николай/ontologies/2020/5/19/untitled-ontology36#>

*SELECT* ?номер_статті ?назва_статті ?ПІБ_співавтора

{

*GRAPH* ?номер_статті {?s :Название_статьи?назва_статті.

?s1 :Иметь_Соавтора ?співавтор_статті.

?співавтор_статті :Иметь_ФИО ?ПІБ_співавтора

*FILTER REGEX*(?ПІБ_співавтора, "Кургаєв

О.П.", "i")}

}

Результат виконання.

1. Номер статті <http://test.ulif.org.ua:51089/articles/data/article9>

Назва статті «Міждисциплінарні наукові дослідження: оптимізація системно-інформаційної підтримки»

ПІБ співавтора «Кургаєв О.П.»

2. Номер статті <http://test.ulif.org.ua:51089/articles/data/article13>

Назва статті «Ноосферна парадигма розвитку науки та штучний інтелект»

ПІБ співавтора «Кургаєв О.П.»

## 4.6. **Інструкція користувача роботи з ОнС БДНП**

Нагадаємо, що на підготовчому етапі всі процеси та процедури виконує інженер зі знань, користувач тільки надає йому масив електронних текстових документів НП. ІЗ створює сховище *rdf* документів НП у відповідності з методичними матеріалами [14, 15, 18]. Для кожної накової статті формується файл “Стаття N.owl” в синтаксисі RDF/XML і додається до сховища *rdf* документів з відповідним порядковим номером. Інтерфейс користувача має декілька сторінок. На рис. 20 показано сторінку для основного етапу, варіант В, стадія 1 для одного з початкових запитів. На рис. 21 показано сторінку, на якій відображено фрагмент реалізації стадії В3, запит до статей 1, 2, 6 і 7 на виведення із них простих знань з їхніх анотацій.

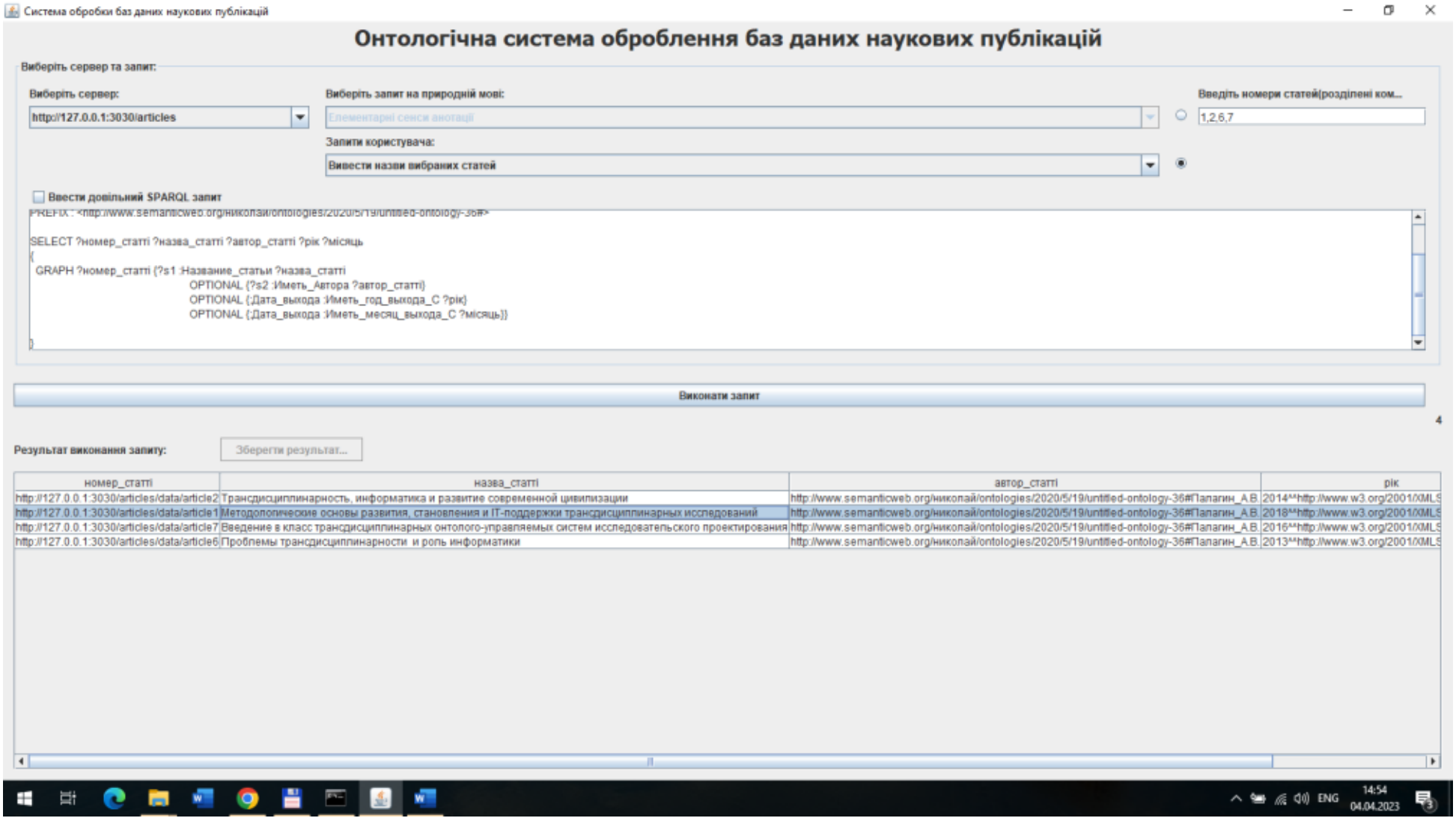

Рис. 20. Інтерфейс користувача ОнС (стадія 1)

Іншими словами, користувачу необхідно просто вибрати із спадаючого меню аргументи до запиту, який його цікавить.

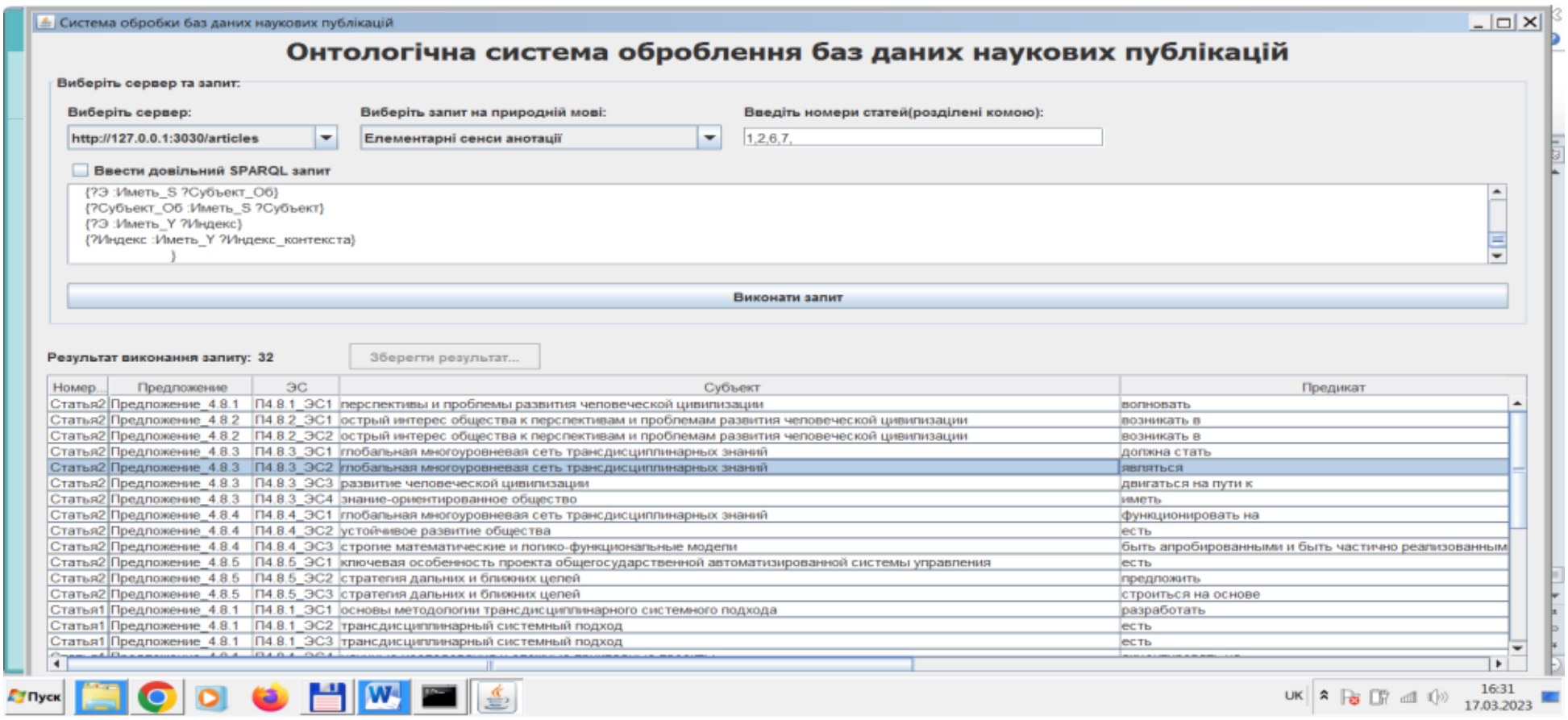

Рис. 21. Інтерфейс користувача ОнС (стадія 3)

## Висновки за розділом

1. Онтологічна система оброблення баз даних наукових публікацій орієнтована на пришвидшення отримання науковцем необхідної йому інформації з акцентом на технології й інструментарій Semantic Web та когнітивної графіки.

2. Існує тісний зв’язок між технологіями *Semantic Web* і *UML*-технологіями, зокрема, зв’язок між *OWL*-синтаксисом і візуальним моделюванням *UML*-діаграм.

3. У розділі розглянуто архітектуру онтологічної системи та алгоритми її функціонування на підготовчому й основному етапах. Детально розглянуто кроки підготовчого етапу, які реалізуються інженером зі знань. При цьому зазначено створення двох моделей онтології наукової статті із поданням відповідних онтографів: *CRF*-модель описує поняття, які містяться в статті, і *OWL*-модель описує структурні та семантичні компоненти статті. Наведено приклади запитів до бази даних наукових публікацій, які демонструють працездатність онтологічної системи оброблення баз даних наукових публікацій.

4. Розроблено й описано архітектурно-структурну організацію онтологічної системи оброблення баз даних наукових публікацій, яка включає локальну мережу із персональних комп’ютерів користувачів, персонального комп’ютера адміністратора-інженера зі знань та віддалену прикінцеву точку на базі сервера *Apachi Jena Fuseki*, основні *UML*-діаграми функціонування онтологічної системи та приклади виконання запитів користувача.

5. Надалі слід розширити використання в розробці онтологічної системи технологій, таких як когнітивні семантика та графіка, мультимедійне подання інформації, орієнтовані на ефективну підтримку процесів екстракції і/або генерації нових знань.

## Перелік посилань до розділу 4

1. Petrenko, M.G., Palagin, A.V., *Boyko M.O., Matveyshyn S.M.* Knowledge-Oriented Tool Complex for Developing Databases of Scientific Publications and Taking into account Semantic Web Technology II. Control systems and computers, 2022, № 3. С.11–28. DOI: https://doi.org/10.15407/csc.2022.03.011 (In Ukrainian).
2. Palagin, A.V., Petrenko, M.G., 2020. "Knowledge-Oriented Tool Complex Processing Data-bases of Scientific Publications". Control systems and computers. N 5, pp. 17–33. https://doi.org/10.15407/csc.2020.05.017 (In Ukrainian).
3. Палагин А.В., Крывый С.Л., Петренко Н.Г. Онтологические методы и средства обработки предметных знаний. Луганск: изд-во ВНУ им. В. Даля, 2012. 323 с.
4. .Палагін О. В., Малахов К. С., Величко В. Ю., Щуров О. С. Проєктування та програмна реалізація підсистеми створення та використання онтологічної бази знань публікацій наукового дослідника. Проблеми програмування. 2017. № 2. С. 72–81. URI: http://dspace.nbuv.gov.ua/handle/123456789/144477. (дата звернення: 20.06.2022).
5. Palagin A.V., Petrenko N.G., Velychko V.Yu., Malakhov K.S., 2014. Development of formal models, algorithms, procedures, engineering and functioning of the software system "Instrumental complex for ontological engineering purpose". In: Proceedings of the 9th International Conference of Programming UkrPROG. CEUR Workshop Proceedings 1843. Kyiv, Ukraine, May 20–22, 2014.URL: http://ceur-ws.org/Vol-1843/221-232.pdf. (дата звернення: 20.06.2022).
6. Palagin O.V., Velychko V.Yu., Malakhov K.S., Shchurov O.S., 2020. Distributional seman-tic modeling: a revised technique to train term/word vector space models applying the ontology-related approach. In: Proceedings of the 12th International Scientific and Practical Conference of Programming UkrPROG 2020. CEUR Workshop Proceedings 2866. Kyiv, Ukraine, September 15-16, 2020. DOI: https://doi.org/10.15407/pp2020.02-03.341.
7. Palagin A. V., Petrenko N. G. Methodological foundations for development, formation and it-support of transdisciplinary research. Journal of automation and information sciences. 2018. Vol. 50, no. 10. P. 1–17. DOI: https://doi.org/10.1615/jautomatinfscien.v50.i10.10.

8. A. V. Palagin. Transdisciplinarity Problems and the Role of Informatics. Cybernetics and Systems Analysis volume 49, pages 643–651 (2013). DOI: https://doi.org/10.1007/s10559-013-9551-y.
9. Петренко Н.Г., Зеленцов Д.Г. О практическом использовании онтологических моделей предметных областей. Комп'ютерне моделювання: аналіз, управління, оптимізація. 2019. №2(6). С. 58–73. DOI: 0.32434/2521-6406-2019-6-2-58-73.
10. Палагин А.В., Петренко Н.Г., Крывый С.Л. О построении знание-ориентированных компьютерных систем для научных исследований. УСиМ. 2015. № 2. С. 64–73. URL: http://usim.org.ua/arch/2015/2/7.pdf. (дата звернення: 20.06.2022).
11. Palagin A.V., Kurgayev A.F. Interdisciplinary scientific research: optimization of system-information support. Visn. Nac. Akad. Nauk Ukr.2009. (3): 14-15. [in Ukrainian]..
12. Booch G., Rumbaugh J. , Jacobson I. The Unified Modeling Language User Guide. Addison-Wesley. Reading. MA, 2005. 475 p.
13. Шмуллер Д. Освой самостоятельно UML 2 за 24 часа. Практическое руководство. Sams Teach Yourself UML in 24 Hours, Complete Starter Kit. М.: Вильямс, 2005. 416 с. ISBN 0-672-32640-X.
14. Леоненков А. В. Самоучитель UML 2. СПб.: БХВ-Петербург, 2007. 576 с.: ил. ISBN 978-5-94157-878-8.
15. OWL 2 Web Ontology Language Primer (Second Edition). W3C Recommendation. URL: http://www.w3.org/TR/2012/REC-owl2-primer-20121211/ (дата звернення: 23.06.2022).
16. Matthew Horridge. A Practical Guide To Building OWL Ontologies Using Protege 4 and COODE Tools. Edition 1.3. Copyright The University Of Manchester. March 24, 2011. 107p.
17. Apache Jena Fuseki. URL: https://jena.apache.org/documentation/fuseki2/ (дата звернення: 23.06.2022).
18. DuCharme B. Learning SPARQL. Querying and Updating with SPARQL 1.1 (Second edition). O'Reilly Media. All rights reserved. August 2013: ISBN: 978-1-449-37143-2. 367p.

# Розділ 5. Методологічні засади розроблення апаратного лінгвістичного процесора

## 5.1. Загальний підхід до проблеми аналізу природномовного тексту

Ступінь розвитку і впровадження комп'ютерних технологій значною мірою визначається не стільки можливостями традиційної обчислювальної техніки, скільки особливостями ПдО, і успіх ефективної комп'ютеризації останніх істотно залежить від глибини дослідження і розуміння явища, яке моделюється. Відповідно до цього етап узагальненого підходу до формування інформаційних технологій виключно з позицій потенційних можливостей обчислювальної техніки (ОТ) поступово зміщується у бік диференціації окремих напрямків інформаційних технологій (ІТ), де на першому плані виступають особливості (домена) конкретних ПдО. До такого напрямку слід віднести ІТ, які тим чи іншим чином пов'язані з мовним моделюванням поведінки людини, або системи, зорієнтовані на опрацювання природномовної інформації. Причому під опрацюванням природномовного тексту (ПМТ) розуміється реалізація низки ІТ, кінцевою метою яких є комп'ютерне оброблення знань як вищої форми розумової діяльності.

Розроблення формальної теорії комп'ютерного опрацювання знань становить одну з нагальних проблем загальної теорії ШІ. Складність зазначеної проблеми визначається, зокрема, необхідністю залучення цілого ряду наукових теорій (інформатики, математичної логіки, комп'ютерної та психологічної лінгвістики, нейрофізіології, нейрокібернетики, теорії семантичних мереж тощо), які в сукупності, будучи застосованими до вирішення проблеми формального подання та оброблення знань, склали б концептуально єдину (міждисциплінарну) формальну теорію. Складові зазначеної теорії повинні враховувати сутність етапів мовного та предметного оброблення інформаційних об'єктів (для першого – ПМТ, а для другого – знання, що добуті з ПМТ) і формального зв'язку між ними. Зрештою необхідна така теорія, яка дозволила б розробити комплекс інформаційних технологій комп'ютерного опрацювання ПМТ, подання знань та комп'ютерного оброблення знань.

При розробці зазначеної формальної теорії важливим завданням є побудова природномовних лінгвістичних моделей та створення на їх основі ефективних лінгвістичних процесорів, які разом з МОКС представляють мовно-онтологічну інформаційну систему.

В існуючих ІС виділяють п'ять основних рівнів розуміння ПМТ [1].

*Перший рівень* характеризується схемою, яка показує, що відповіді на питання система формує лише на основі прямого змісту, що випливає з тексту.

*Другий рівень*. На другому рівні додаються способи логікового виведення, що базуються на інформації, яка міститься в тексті. Це різні логіки тексту (тимчасова, просторова, каузальна тощо), які породжують інформацію, що явно відсутня в тексті.

*Третій рів*ень. До засобів другого рівня додаються правила поповнення тексту знаннями про середовище. Ці знання в ІС, як правило, мають логічний характер і фіксуються у вигляді сценаріїв або процедур іншого типу.

Три перелічені рівня розуміння повністю або частково реалізовані практично у всіх діючих ІС.

*Четвертий рівень*. На цьому рівні відбувається зміна вмісту бази знань. Вона доповнюється фактами, відомими системі та які містяться у тих текстах, які введені в систему. Різні ІС відрізняються одна від одної характером правил породження фактів із знань, спираючись на методи дедуктивного виведення та розпізнавання образів. Правила можуть бути засновані на принципах ймовірностей, розмитих висновків тощо. Але у всіх випадках база знань виявляється апріорно неповною. В ІС виникають складнощі з пошуком відповіді на запити. Зокрема, у базах знань стає необхідним немонотонне виведення.

*П'ятий рівень*. На цьому рівні відбувається породження метафоричного знання. Правила породження знань метафоричного рівня, що використовуються для цих цілей, є особливими процедурами, які спираються на виведення за аналогією та асоціацією [2].

Існують й інші інтерпретації феномена розуміння. Можливо, наприклад, оцінювати рівень розуміння щодо здатності системи до пояснення отриманого результату. Тут можливий не тільки рівень пояснення, коли система пояснює, що вона зробила, наприклад, на основі введеного в неї тексту, а й рівень обґрунтування (аргументації), коли система обґрунтовує свій результат, показуючи, що він не суперечить системі знань і даних, якими вона володіє. На відміну від пояснення, обґрунтування завжди пов'язане із сумою фактів та знань, що визначаються поточним моментом існування системи. І введений для розуміння текст в одних станах можливо буде сприйнятий системою як істинний, а в інших – як хибний. Існуючі ІС типу експертних систем, як правило, здатні давати пояснення і лише частково обґрунтування [3].

Узагальнений критерій розуміння ПМТ науково-технічного профілю полягає у здатності вирішувати на основі знань, що містяться в них, прикладні задачі.

5.1.1 ***Формальна постановка задачі аналізу ПМТ*** [4]

Нехай $T = t_1, t_2, ..., t_n$ – природномовний текст в алфавіті *X*, тобто $T \in L(X)$, де $L(X)$ – мова над алфавітом *X*, а $t_i \in T$ – речення, $i = \overline{1, n}$, *n* – потужність множини $T$.

Кожне речення $t_i \in T$, у свою чергу, має структуру $t_{i_1}, t_{i_2}, ..., t_{i_m}$, де $t_{i_j}$ змістовно означають граматичні одиниці, із яких побудовано речення $t_i$. Якщо $t_{i_j} \in t_i$, то $C_L(t_{i_j}) = t_{i_1}, t_{i_2}, ..., t_{i_{(j-1)}}$ и $C_R(t_{i_j}) = t_{i_{(j+1)}}, t_{i_{(j+2)}}, ..., t_{i_m}$ будемо називати *лівим* і *правим контекстом* $t_{i_j}$ відповідно.

З текстом $T$ зв'яжемо такі об'єкти:

*S* – словник мови $L(X)$, де містяться слова $t_{i_j}$ із своїми визначниками (зокрема, лінгво-семантичними характеристиками одиниць словника);

$\gamma \subseteq T \times S$ – відношення, яке визначає можливі значення і типи слів у словнику $S$;

– $A = (D, \Pi)$ – предметна модель, на якій інтерпретується текст $T$;

$\phi \subseteq T \times A$ – відношення інтерпретації тексту $T$ на області *D*.

Сигнатура предикатів $\Pi = \left\{\pi_1^{k_1}, ..., \pi_r^{k_r}\right\}$ містить атомарні предикати, із яких можливо будувати складні формули. Зафіксуємо цю сигнатуру, оскільки вона залежить від предметної моделі. Так як модель не уточнюється, то і сигнатуру уточнити неможливо. Відзначимо тільки, що кожен атомарний предикат має тип (тобто мова йде про деяку типізовану сигнатуру).

Визначимо правила обчислення відношень $\gamma$ і $\phi$.

Відношення $\gamma$ має достатньо простий спосіб обчислення:

$$\gamma(t_{i_j}) = \{(a_1, \tau_1), (a_2, \tau_2), ..., (a_s, \tau_s)\}, \quad (1)$$

де $a_i$ – можливі значення слова $t_{i_j}$, а $\tau_i$ – його можливі типи. Можливий випадок, коли $\gamma\left(t_{i_j}\right)=\varnothing$. У цьому випадку значення цього слова рахується невизначеним, і це потребує поповнення словника *S*.

Відношення $\phi$ визначається формулою:

$$\phi\,(T) = \phi\,(t_1),\ldots,\phi\,(t_n), \qquad (2)$$

де

$$\phi\,(t_i) = \left\{\phi\left(\gamma\left(t_{i_1}\right)\gamma\left(C_R\left(t_{i_1}\right)\right)\right), \phi\left(\gamma\left(C_L\left(t_{i_2}\right)\right)\gamma\left(t_{i_2}\right)\gamma\left(C_R\left(t_{i_2}\right)\right)\right),\ldots,\phi\left(\gamma\left(C_L\left(t_{i_n}\right)\right)\gamma\left(t_{i_n}\right)\right)\right\},$$

при цьому:

$$\phi\left(\gamma\left(t_{i_j}\right)\right) = \gamma\left(\phi\left(t_{i_j}\right)\right);$$

$$\phi\left(\gamma\left(C_L\left(t_{i_j}\right)\right)\right) = C_L\left(\phi\left(\gamma\left(t_{i_j}\right)\right)\right);$$

$$\phi\left(\gamma\left(\pi_r^k\left(p_1,\ldots,p_k\right)\right)\right) = \gamma\left(\phi\left(\pi_r^k\right)\right)\left(\phi\left(\gamma\left(p_1\right),\ldots,\gamma\left(p_k\right)\right)\right),$$

де $\gamma\left(\phi\left(\pi_r^k\right)\right)$ – ім’я предиката, тип якого узгоджений з аргументами $\gamma\left(p_1\right),\ldots,\gamma\left(p_k\right)$.

Із цієї формальної постановки проблеми аналізу ПМТ випливає, що основні задачі зводяться до наступних:

– конкретизувати предметну модель *A*;

– показати обчислювальність відношень $\gamma$ (1) і $\phi$ (2) на предметній моделі *A*;

– побудувати алгоритми обчислення відношень $\gamma$ і $\phi$;

– при обчисленні відношень $\gamma$ і $\phi$ контролювати відповідність типів аргументів і предикатів;

– визначити взаємодію алгоритмів обчислення $\gamma$ і $\phi$ із системами лінгвістичного аналізу тексту.

Вищенаведений процес аналізу ПМТ представлено в достатньо загальному вигляді та потребує уточнення. Виконаємо деяку конкретизацію цього визначення стосовно до предметної області “*Аналіз ПМТ*”.

### 5.1.2 *Лексико-граматичний аналіз*

Нехай $L$ – мова відношень, які представляють знання, $V$ – множина граматичних характеристик, включаючи граматичні розряди ПМ, а $D$ – область інтерпретації.

Лексико-граматичний аналіз приводить до конкретизації інтерпретації $\varphi: V \to D$ і відношень $R_i \in L$. Інтерпретація $\varphi$ в даному випадку представляє собою суперпозицію двох функцій $\varphi_1$ і $\varphi_2$, тобто $\varphi(V) = \varphi_2(\varphi_1(V)) = ... = \varphi_1^* \varphi_2(V)$, де * позначає суперпозицію функцій. Функції $\varphi_1$ і $\varphi_2$ реалізують процес синтаксичного і семантичного аналізу речень тексту *T*, а відношення $R_1$ і $R_2$ – це синтаксичні (правила мови, на якій написано текст *T*) і семантичні обмеження.

Одним з подальших можливих уточнень є уточнення відображення $\varphi_1$. Це відображення, у свою чергу, можливо розглядати як суперпозицію двох відображень, що реалізують морфологічний і синтаксичний аналіз речень ПМТ і разом з відображенням $\varphi_2$ утворюючих цілісну систему класичного типу, схема якої показана на рис. 1 [5].

Для перевірки коректності виконаного аналізу передбачається зворотний синтез речення до його запису у звичайному орфографічному вигляді. Така перевірка може виконуватися *у діалоговому режимі роботи системи з користувачем. Можлива структура словників, які* використовуються у наведеній схемі, і деяке її обгрунтування описані в роботах [6–8].

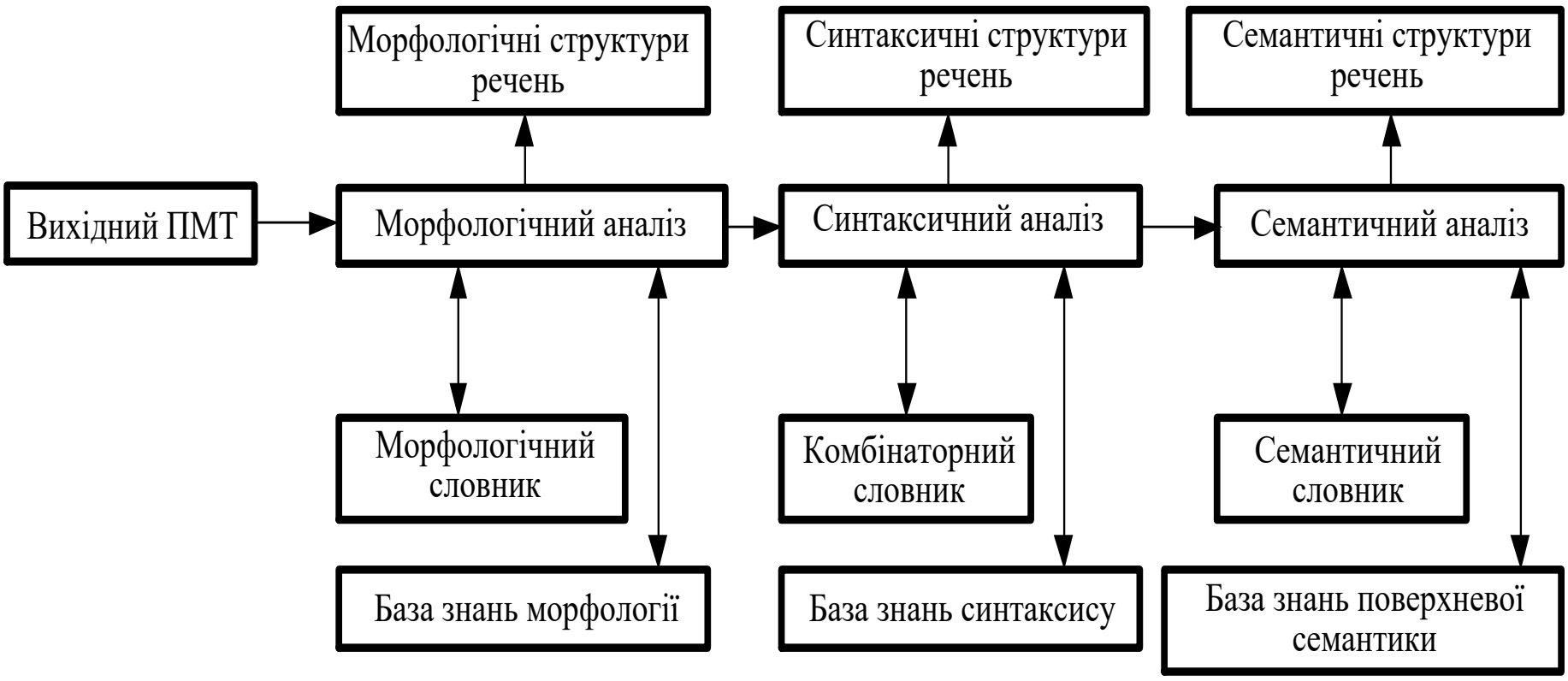

Рис. 1 Схема лінгвістичного аналізу класичного типу

Якщо детально проаналізувати типи загальнозначимих відношень, за допомогою яких будується ієрархія понять, то можливо помітити, що вони асоціюються, перш за все, з *відношенням часткового порядку*. А такого типу відношення складають дистрибутивну гратку, яка має ряд корисних властивостей, і ці властивості можливо використати для генерації відповідних наслідків, тобто для пошуку (генерації) нових знань [1].

## 5.2. **Онто-логічний підхід до побудови апаратних засобів лінгвістичного ана-лізу ПМО**

Оброблення мовної чи текстової інформації забезпечується лінгвістичним процесором, чи то на рівні "мовної" свідомості людини чи комп'ютерної системи. У КС він є основною компонентою, що реалізує розпізнавання та розуміння вхідної природномовної інформації, добування з неї первинних знань з їх подальшим формально-логічним поданням. Отриманої інформаційної структури вже достатньо для реалізації (знання-орієнтованих) процедур для рішення прикладних задач, прийняття рішень тощо.

### 5.2.1 ***Узагальнена модель апаратних засобів лінгвістичного аналізу ПМО***

Однією з важливих задач на шляху розроблення загальної теорії комп'ютерного оброблення предметних знань, поданих у природномовної формі, є побудова ефективних лінгвістичних процесорів. Це завдання особливо актуальне для додатків оброблення ЛКТ надвеликих об'ємів (і в реальному часі). Це пов'язано з тим, що сучасні персональні комп'ютери програмним способом виконують лінгвістичний аналіз одного слова вхідного тексту середньої довжини приблизно за одну мілісекунду (діаграма залежності часу оброблення вхідного слова від його довжини наведена на рис. 2), а такий аналіз займає значну частину комп'ютерного часу загального лінгвістичного оброблення. При цьому час оброблення вхідного тексту навіть щодо невеликого об'єму займає від кількох до десятків хвилин. В результаті для застосунків, які працюють у реальному часі, частина інформації можливо буде втрачена.

Тому задача суттєвого (на 2 порядки і більше) підвищення швидкодії лінгвістичного аналізу є актуальною.

Слід зазначити, що вказане вище підвищення швидкодії може бути досягнуто за допомогою додаткових апаратурних витрат як стандартної, так і спеціальної розробки. Апаратні засоби (АЗ) першого типу є продуктом відомих фірм, доступних на ринку та до них додається програмне забезпечення САПР. Безперечним лідером таких АЗ на ринку є плати із встановленими на них програмовними логіковими інтегральними схемами (ПЛІС) (в яких є надшвидка пам'ять) і швидкодіюча пам'ять великого об'єму [9]. АЗ другого типу є спеціалізованою розробкою, для них необхідно спроєктувати архітектурно-структурну організацію процесора, електричну схему або граф-схеми алгоритмів, спеціальне програмне забезпечення керування ними та драйвери суміщення з операційною системою комп'ютера. З точки зору реалізації лінгвістичного процесора обидва ці варіанти АЗ мають свої переваги та недоліки. Для АЗ першого типу перевагою є те, що вони доступні на ринку, їхня обчислювальна потужність постійно збільшується розробниками, до них вже додається програмне забезпечення, а проєкт апаратного лінгвістичного процесора (АЛП) може бути розроблений за час від 2-х місяців. Недоліком цих АЗ є мінімальний відсоток використання встановленого на платі устаткування. До переваг АЗ другого типу слід віднести підвищення швидкодії на 1–2 порядки у порівнянні з АЗ першого типу, що є головним критерієм при розробці АЛП. А до недоліків – необхідний колектив розробників (системотехніків та програмістів), і час розробки проєкту оцінюється від 1 року [10–11].

Підвищення швидкодії реалізації алгоритму лінгвістичного аналізу для обох типів АЗ досягається за рахунок перекладу операторів алгоритмічного та програмного рівня (реалізація лінгвістичного аналізу програмним способом) на нижні рівні інтерпретації (згідно з логіко-інформаційною моделлю [9]): для АЗ першого типу – на мікропрограмний рівень, для АЗ другого типу – на мікропрограмний та частково на фізичний рівні.

В [1] наведено додаткові докази доцільності реалізації ЛП загалом, та морфологічного процесора зокрема, апаратними засобами. Наприклад, апаратна реалізація надає можливість паралельного оброблення всіх слів одного речення одночасно. При цьому спрощуються алгоритми синтаксичного та семантичного аналізу.

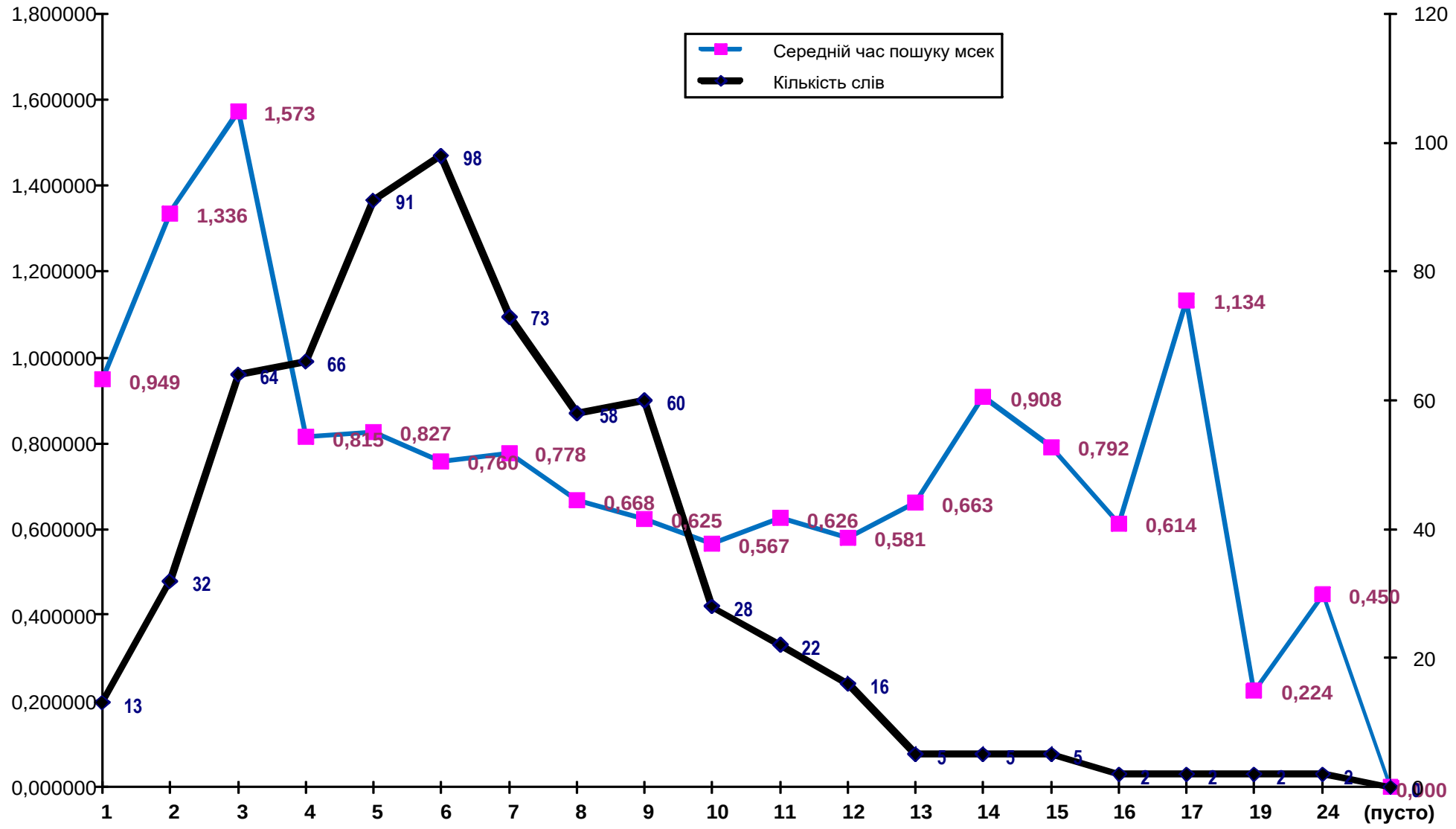

Рис. 2. Діаграма залежності часу оброблення вхідного слова від його довжини програмним способом

### 5.2.2 ***Метод багаторівневих проєкцій проєктування АЛП*** [10]

Метод багаторівневих проєкцій побудови апаратних лінгвістичних процесорів заснований на застосуванні онто-логічного підходу [12, 13] до проєктування архітектурної й інформаційної компонент онтолого-керованої інформаційної системи. При цьому схема логіко-інформаційної моделі перетворюється в схему онто-логічної моделі АЛП наступного виду (рис. 3): де $\tau_0$ – фізичний рівень; $\tau_1$ – мікропрограмний рівень; $\tau_2$ – програмний рівень; $\tau_3$ – алгоритмічний рівень; $\tau_4$ – предметний рівень.

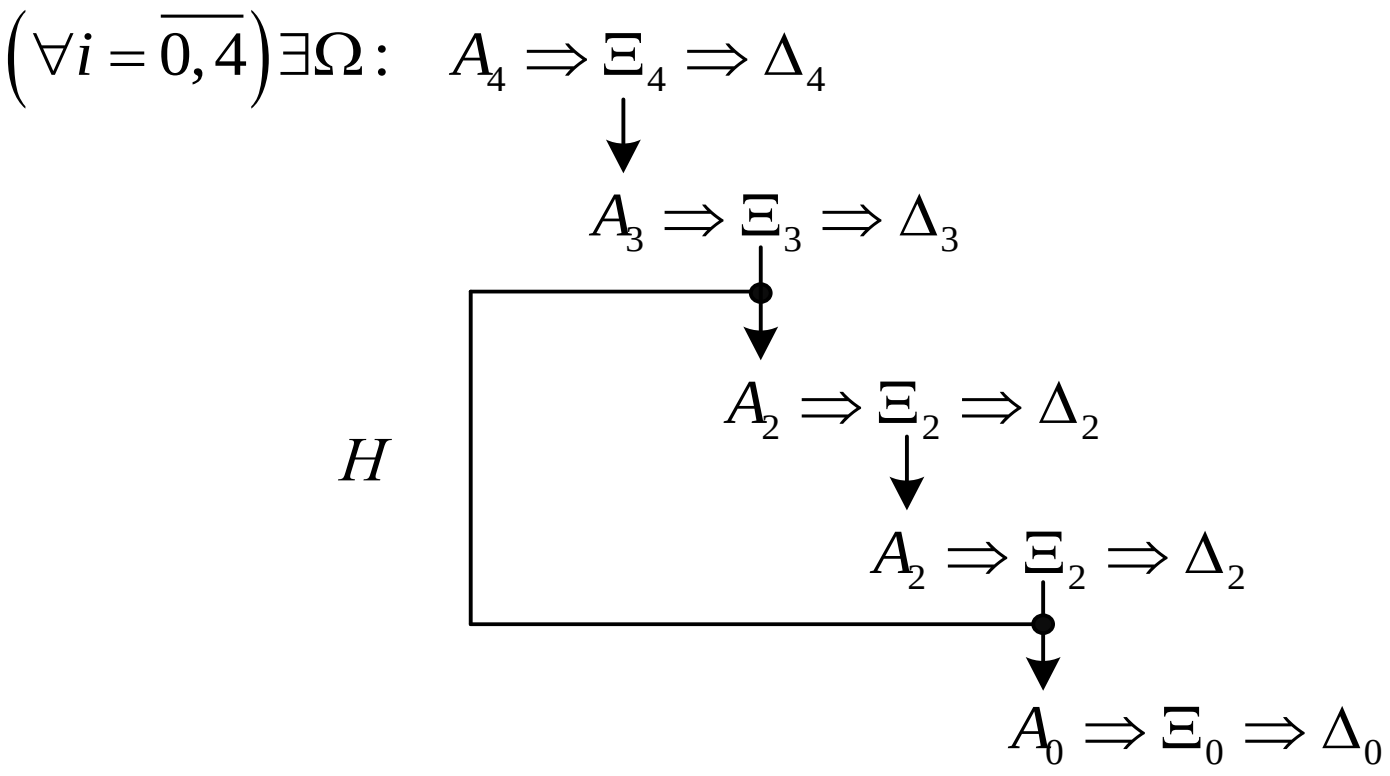

(3)

Рис. 3. Онто-логічна модель АЛП

В моделі (3) оператори алгоритмічного рівня за посередництва відображення $H$ переводяться на “нульовий” або фізичний рівень інтерпретації.

Модель (3) можливо описати мережею скінченних автоматів:

де $X$ – вхідний алфавіт мережі автоматів;

$\{M_n\}$ – множина структурних автоматів;

$Y$ – вихідний алфавіт мережі автоматів;

$\{\psi_i\}$ – множина функцій переходів в $\{M_n\}$;

$S$ – множина зв’язків між структурними автоматами;

$\varphi$ – початковий стан мережі;

$\mu$ – кінцевий стан мережі.

Враховуючи сучасні технічні характеристики мікроелектронної бази, можна реалізувати всі чотири етапи лінгвістичного аналізу на апаратно-мікропрограмному рівні у вигляді мережі комбінаційних автоматів з пам’яттю. При цьому виконується аналіз вхідної ПМ-інформації не послівно, а по реченням. Речення є мінімальною предикативною синтаксичною одиницею, для якої може бути виконаний повний синтаксичний та поверхнево-семантичний аналіз.

Суть методу багаторівневих проєкцій полягає в наступному.

На рис. 4 представлена структурна модель АЛП.

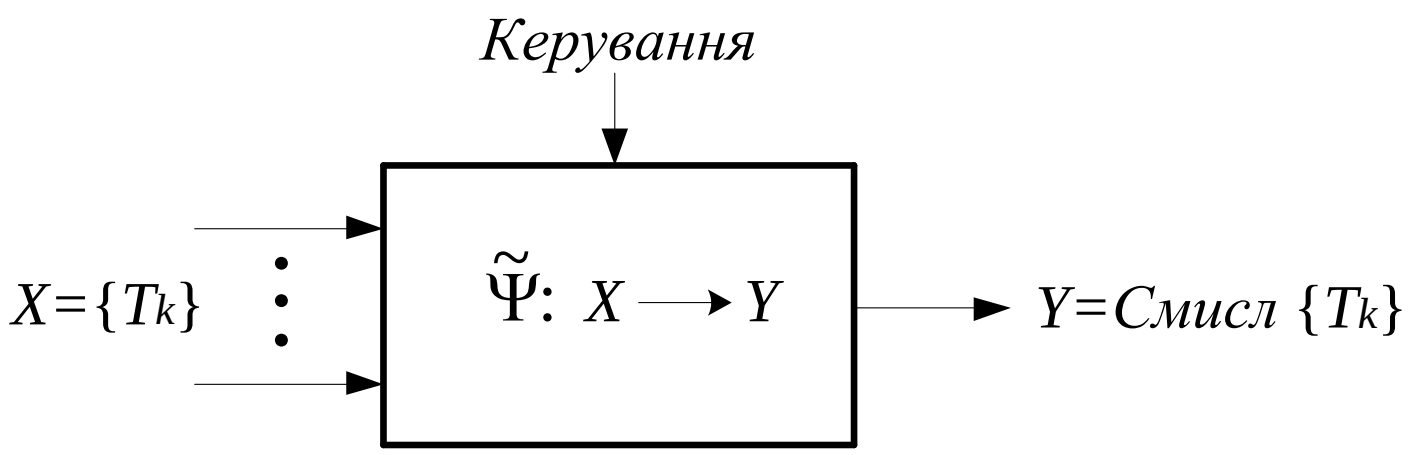

Рис. 4. Структурна модель АЛП

Вхідними даними моделі є множина текстів $\{T_k\}$, вихідними даними у загальному випадку є сенс обробленого тексту чи, по крайній мірі, його семантична структура. Усередині моделі реалізується відображення $\tilde{\psi} : X \to Y$ , яке поетапно переводить текстове подання вхідної інформації (текст, графічна структура тексту, морфологічна структура, синтаксична структура, семантична структура) у відповідні структури мережі операційних автоматів. При цьому мікропрограмні автомати природно налаштовуються на управління функціонуванням відповідних операційних автоматів.

Слід зазначити важливу особливість реалізації АЛП із застосуванням ПЛІС-технології, для якої існує можливість реконфігурації структури АЛП на фізичному рівні як інструменту налаштування (перебудови) на оброблення ЛКТ заданої ПдО або розв’язання задачі предметно-проблемної орієнтації апаратних засобів.

### 5.2.3 ***Оцінка складності реалізації алгоритму лінгвістичного аналізу*** [12]

Нижче пропонується формалізований підхід до оцінки складності ієрархічної системи оброблення інформації, що ґрунтується на поняттях абсолютної та відносної оцінок складності операторів, які складають мови різних рівнів ієрархії.

*Визначення* 1. Словом оператора називатимемо скінченний рядок літер із заданого алфавіту, що відзначає даний оператор при поданні його в пам’яті комп’ютерної системи.

*Визначення* 2. Довжиною слова називатимемо число букв у слові.

*Визначення* 3. Довжину опису алгоритму визначатимемо як суму довжин слів операторів мови, в якій представлений алгоритм.

Якщо множині операторів мови $\Lambda = \{\lambda_1, \lambda_2, ..., \lambda_q\}$ поставити у відповідність множину їх інформаційно-кодових представлень $U=\left\{u_1, u_2, ..., u_q\right\}$, де $u_i \in U\left(i=\overline{1,q}\right)$ – кодове слово, що відмічає оператор $\lambda_i \in \Lambda\left(i=\overline{1,q}\right)$ $\sum_{i=1}^{q} r_i = N_A$, то складність алгоритму А у відповідності з визначенням 3 дорівнює:

$$\lambda_i \in \Lambda\left(i=\overline{1,q}\right) \quad Q_A = \sum_{i=1}^{q} r_i \left|u_i\right|, \tag{4}$$

де $\left|u_i\right|$ – розрядність кодового слова $u_i$, $r_i$ – число операторів $i$-го типу в алгоритмі А, $\sum_{i=1}^{q} r_i = N_A$ – довжина алгоритму, тобто число кроків, на кожному з яких реалізується один із операторів $\lambda_i \in \Lambda\left(i=\overline{1,q}\right)$.

Якщо аналогічним чином представляти оператори кожного рівня ієрархічної системи, то вираз (4) прийме вигляд:

$$Q_A = \sum_{j=1}^{h} Q_j = \sum_{j=0}^{h} \sum_{i=1}^{q_i} r_{ij} \left|u_{ij}\right|, \tag{5}$$

де $Q_j\left(j=\overline{0,h}\right)$ – складність алгоритму, представленого на $j$–ому рівні; $j$=0 відповідає рівню мікрооперацій; $\underset{j=\overline{1,h}}{\forall} \sum_{i=1}^{q_i} r_{ij} = N_j$ – довжина алгоритму $j$–го рівня.

Такий підхід завжди має на увазі оцінку відносної складності алгоритмів та операторів. Дійсно, той самий об’єкт (процес, функція) може бути визначений алгоритмічно в термінах операторів будь-якої складності. При цьому очевидно, що чим вище складність операторів мови, тим менша довжина опису даного алгоритму в термінах мови цього рівня. Звідси

*Твердження* 1. Із збільшенням складності операторів мови складність алгоритму, поданого у цій мові, зменшується.

*Визначення* 4. Відносною складністю алгоритму $\left(\tilde{Q}_A\right)$ є відношення складності алгоритму $\left(\tilde{Q}_A\right)$ до складності операторів $\left(Q_\Lambda\right)$ мови, в якій він представлений:

$$\tilde{Q}_A = \frac{Q_A}{Q_\Lambda} \qquad (6)$$

В термінах теорії інформації вираз (6) можливо інтерпретувати наступним чином: кількість інформації, необхідна для завдання опису будь-якого процесу, обернено пропорційна потужності його алфавіту.

Скориставшись виразом (6), для ієрархічної системи можна записати наступне відношення, що визначає відносну складність алгоритму, представленого в операторах l-го мовного рівня:

$$\tilde{Q}_A(l) = \frac{\sum_{i=1}^{q_l} r_{il} \left| u_{il} \right|}{\sum_{j=0}^{l-1} \sum_{i=1}^{q_i} r_{ij} \left| u_{ij} \right|}, \quad l = \overline{1,h}. \qquad (7)$$

Знаменник у виразі (5) характеризує складність реалізації операторів *l*-го рівня і, таким чином, є нічим іншим, як мірою складності обчислень. Таким чином, вираз (7) встановлює взаємозв'язок між складністю алгоритму та складністю обчислень.

Поряд із поняттям відносної складності можливо визначити поняття абсолютної складності. Його доцільно застосовувати, насамперед, для елементарних операторів, які, з їхнього визначення, не можна подати через будь-які інші оператори. При цьому складність елементарного оператора може бути як завгодно велика. Прикладом може бути таблично реалізований елементарний оператор обчислення функцій однієї і більше змінних. Абсолютну складність елементарного оператора можна оцінювати за допомогою ентропійної міри, яку можна трактувати як ємність пам'яті (у бітах), необхідну для зберігання гіпотетичної таблиці, за допомогою якої цей елементарний оператор реалізується як оператор відображення:

$$F : X \to Y,$$

де $X$ і $Y$ – множини значень аргументу та функції відповідно.

Якщо під елементарним оператором розуміти мікрооперацію, то абсолютна складність мікрокоманди:

$$Q_l = \sum_{i=1}^{l} Z_i(F),$$

де $l$ – потужність мікрокоманди; $Z_i(F)$ – абсолютна складність і-ої мікрооперації.

У загальному випадку абсолютна складність оператора визначається: числом операндів; розрядністю операндів; основою системи числення. Підкреслимо, що відповідно до прийнятої концепції, ієрархічна система програмовних автоматів дозволяє представляти алгоритми на будь-якому рівні ієрархії. У зв'язку з цим для даного класу автоматів поняття складності алгоритмів та складності обчислень взаємопов'язані.

Задача оцінки складності алгоритму, в такий спосіб, зводиться до задачі оцінки складності мовних операторів та їх оптимальної ієрархії у системі.

## 5.3. **Розроблення функціональної схеми АЛП** [10]

Лінгвістичний процесор (апаратний чи програмний) інтерпретує текстову інформацію (деякий ПМО – документ, стаття, монографія чи ЛКТ) відповідно до етапів лінгвістичного аналізу: графематичного, морфологічного, синтаксичного та семантичного (точніше, поверхнево-семантичного). Результат роботи АЛП – інформаційна структура, призначена для проведення глибинно-семантичного аналізу в підсистемі екстралінгвістичного оброблення, завданням якої є добування та формування структури понять, тобто автоматичне добування з ПМО знань, реалізація їх прагматичної інтерпретації в термінах прикладної задачі або відповідна реакція, яка властива людині.

На рис. 5 представлено функціональну схему такого АЛП, яка включає відповідні підсистеми реалізації етапів лінгвістичного аналізу та мовно-онтологічну картину світу, використання якої є однією з основних відмінностей АЛП від класичного лінгвістичного процесора [5], а її (МОКС) проєктування розглянуто в [1].

На ньому прийнято такі скорочення:

АСС – загальноприйняті абревіатури, скорочення і спеціальні символи;

БПОС – блок порівняння основи;

БПОК – блок порівняння закінчення;

О – об'єкт, Д – дія, ХО – характеристика об'єкта, ХД – характеристика дії.

Вихідною інформацією для АЛП є ЛКТ заданої ПдО.

Лінгвістичний корпус текстів вміщує скінченну множину текстів $\{T_k\}, k = \overline{1, K}, K$ – кількість текстів в ЛКТ, які послідовно поступають на вхід підсистеми графематичного аналізу.

В процесі виконання загального алгоритму лінгвістичного аналізу текст поетапно перетворюється на графематичну, морфологічну, синтаксичну та семантичну структури, кожна з яких має свою модель подання. Розглянемо процес лінгвістичного аналізу у підсистемах АЛП.

### 5.3.1 ***Розроблення підсистем АЛП***

*Підсистема графематичного аналізу*

На рис. 6 представлено блок-схему графематичного аналізу (де $\Psi$ – процедура відображення граф-схеми алгоритму реалізації відповідної підсистеми лінгвістичного аналізу у відповідну мережу комбінаційних автоматів з пам'яттю (ТпМ) у термінах САПР ПЛІС, ГфА – графематичний аналіз), а на рис. 7 – діаграма станів (або граф-схема алгоритму) з описами виконуваних процедур та аналізованими умовами:

p11 – початок роботи АЛП. Запис у буфер-формувач речення прийнятої словоформи;

p12 – передача із буфера-формувача сформованого речення на вхід процедури p2;

p21 – індексація словоформ $S_l^m$ і речення $P_k^l$ в цілому;

p22 – запис речення $P_k^l$ в базу даних індексованих текстів;

p23 – передача із буфера-формувача речення $P_k^l$ на вхід процедури p31;

p24 – індексація $P_k^l \in T_k$ $\left\{\left(ЛЕ_l^1, МХ_l^1\right),\ldots,\left(ЛЕ_l^m, МХ_l^m\right),\ldots,\left(ЛЕ_l^M, МХ_l^M\right)\right\}$ і його запис в базу даних індексованих текстів;

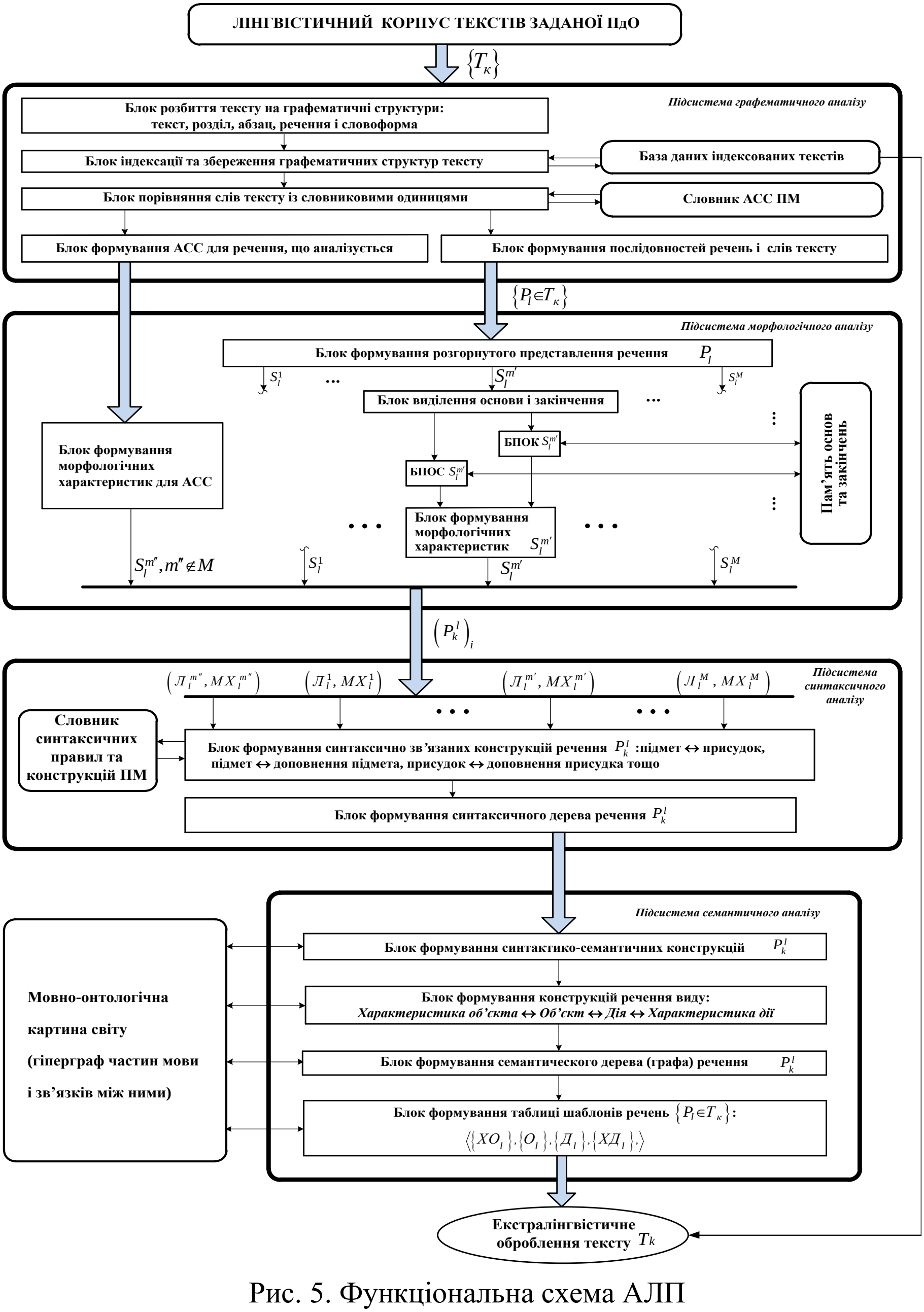

Рис. 5. Функціональна схема АЛП

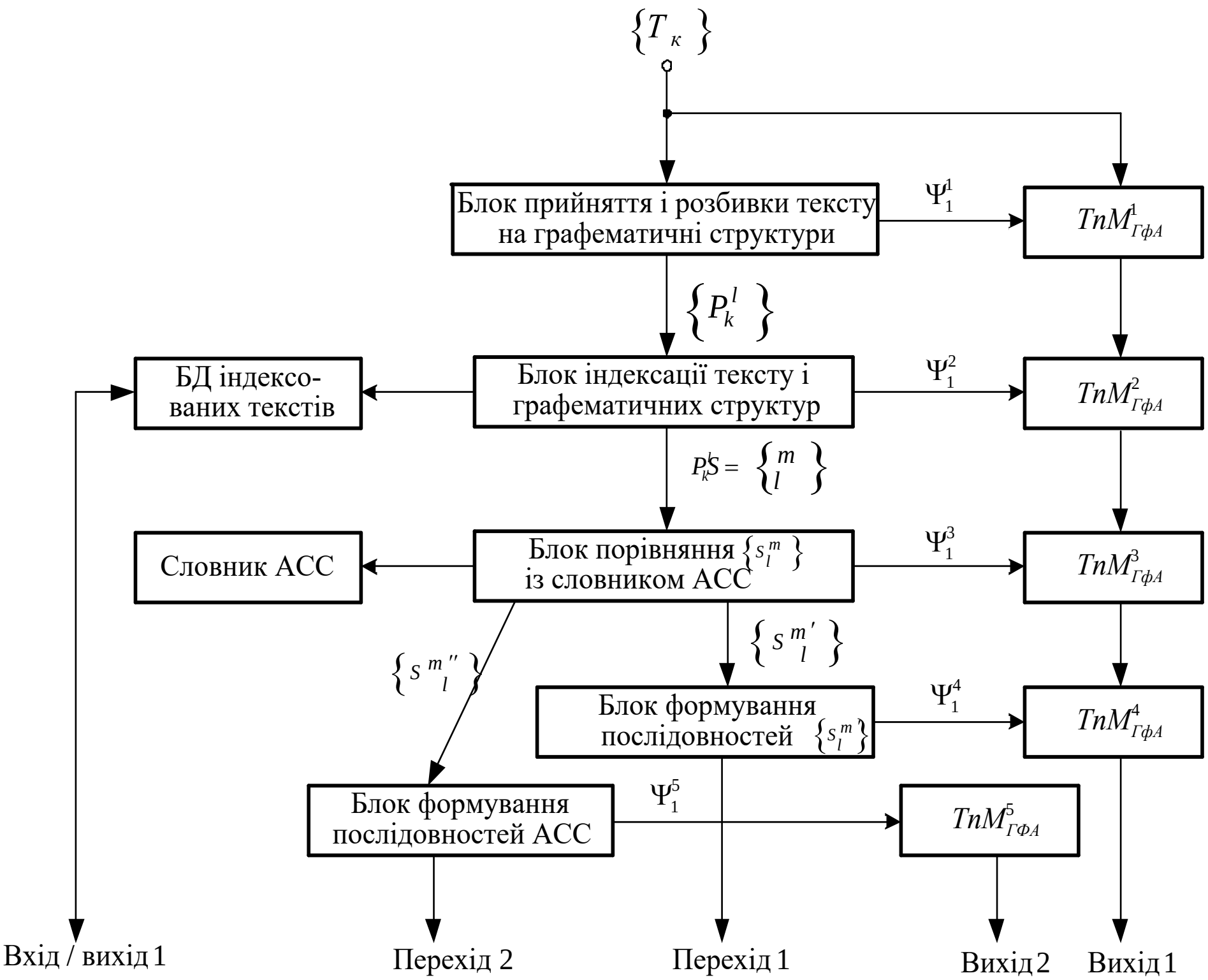

Рис. 6. Функціональна схема підсистеми ГфА

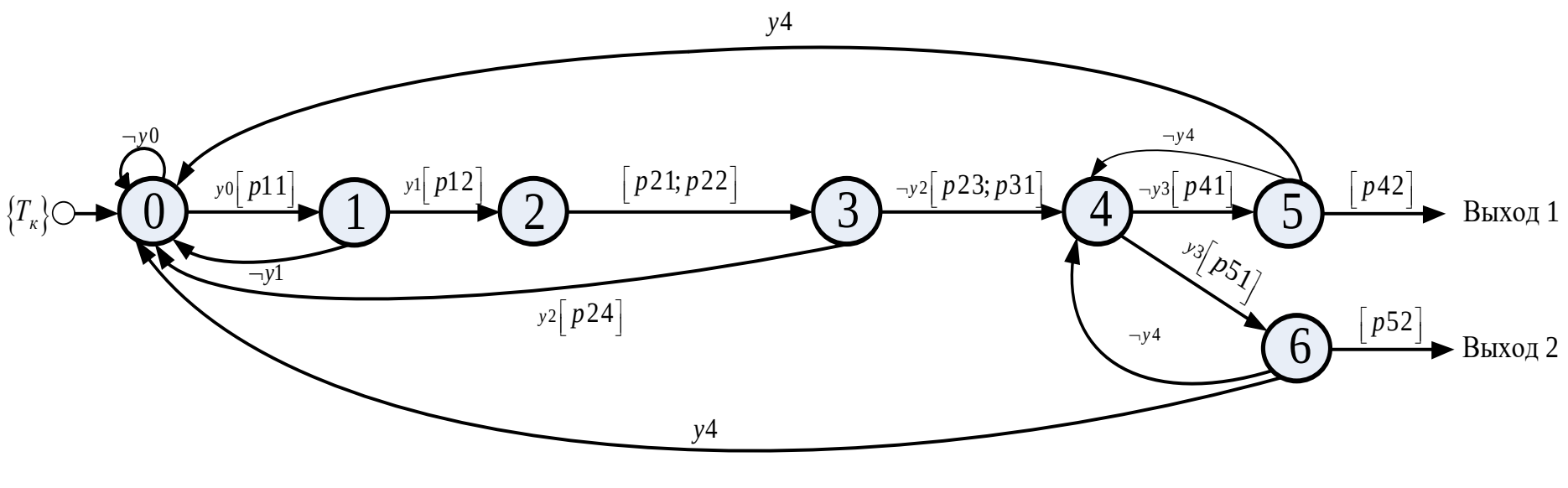

Рис. 7 Діаграма станів підсистеми ГфА

p31 – порівняння $S_l^m$ із словником АСС;

p41 – формування послідовності словоформ $S_l^m$ ;

p42 – передача сформованої послідовності словоформ $S_l^m$ на “Вихід 1” (рис. 6);

p51 – формування послідовності словоформ $S_l^m$ ;

p52 – передача сформованої послідовності словоформ $S_l^m$ на “Вихід 2” (рис. 6).

y0 – умова початку прийняття словоформи;

y1 – умова закінчення прийняття словоформи;

y2 – умова закінчення прийняття тексту $T_k$ ;

y3 – умова порівняння словоформи, що аналізується, з АСС;

y4 – умова закінчення порівняння словоформи $S_l^m$ з АСС.

Текст, що надійшов на оброблення, в блоці розбиття тексту на графематичні структури розбивається і структурується на множину розділів, абзаців, речень і словоформ. Далі структурні одиниці тексту індексуються і зберігаються в базі даних індексованих текстів, інформація в якій використовується в процесі виконання алгоритму всього лінгвістичного аналізу, а також при екстралінгвістичній обробці. Потім, для кожного речення тексту в блоці порівняння слів тексту зі словниковими одиницями виділяються АСС природної мови. Іншими словами, кожне речення розбивається на дві частини, причому для першої з них необхідно обчислити морфологічні характеристики (і вона надходить на оброблення до блоку формування послідовностей речень і слів тексту), а для другої вони добуваються з відповідного словника (і вона надходить до блоку формування АСС). На останньому етапі графематичного аналізу формуються дві паралельні послідовності фрагментів речень, що надходять до підсистеми морфологічного аналізу (МА), перша з них – до блоку формування морфологічних характеристик (БФМХ) для АСС, а друга – до блоку формування розгорнутого подання речення $P_k^l$ . Власне, друга послідовність є вхідною інформацією, що потребує оброблення в підсистемі МА.

*Підсистема морфологічного аналізу*

На рис. 8 представлена блок-схема підсистеми морфологічного аналізу, а на рис. 9 – діаграма станів з описами виконуваних процедур й умовами, що аналізуються.

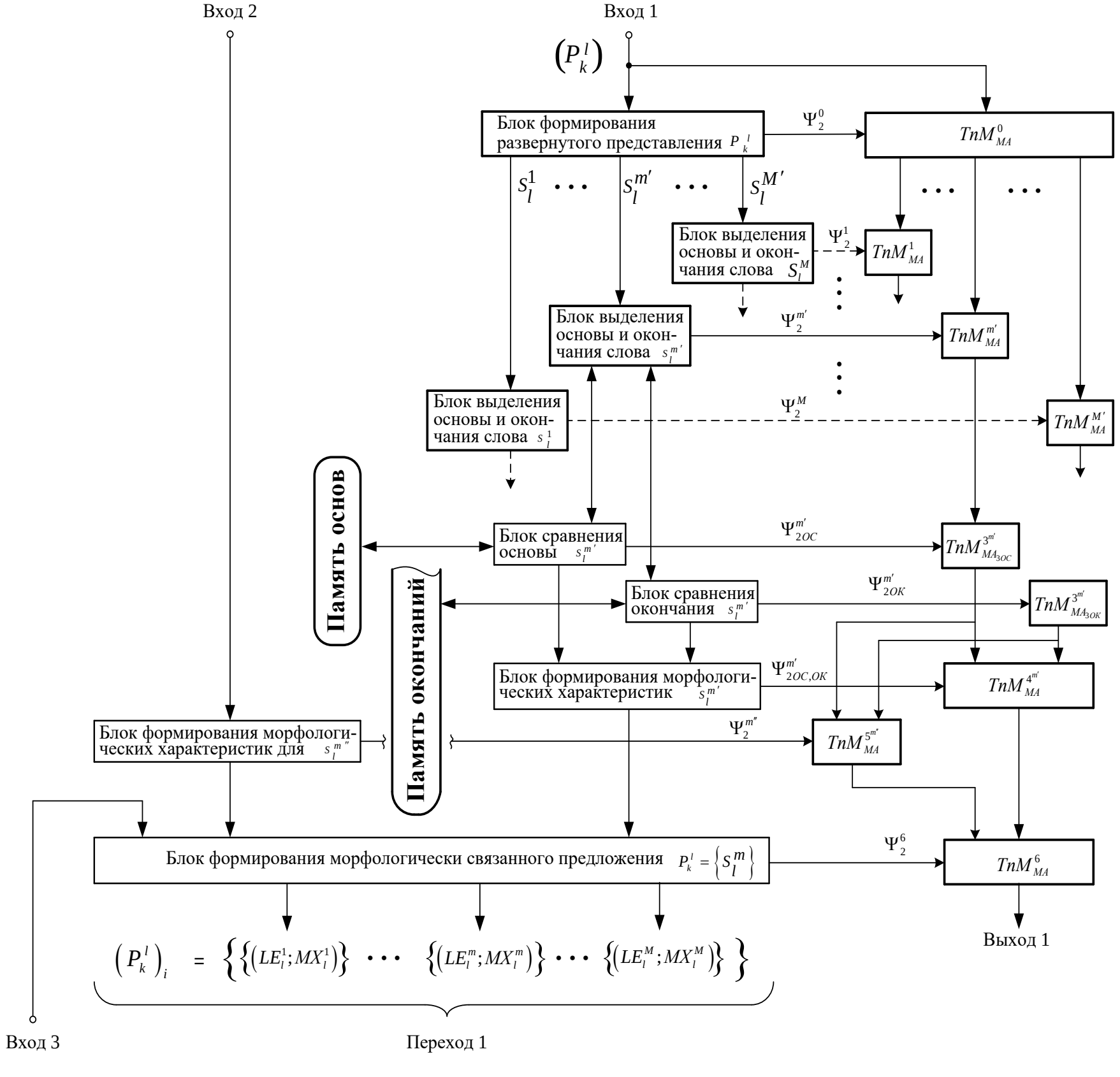

Рис. 8 Функціональна схема підсистеми МА

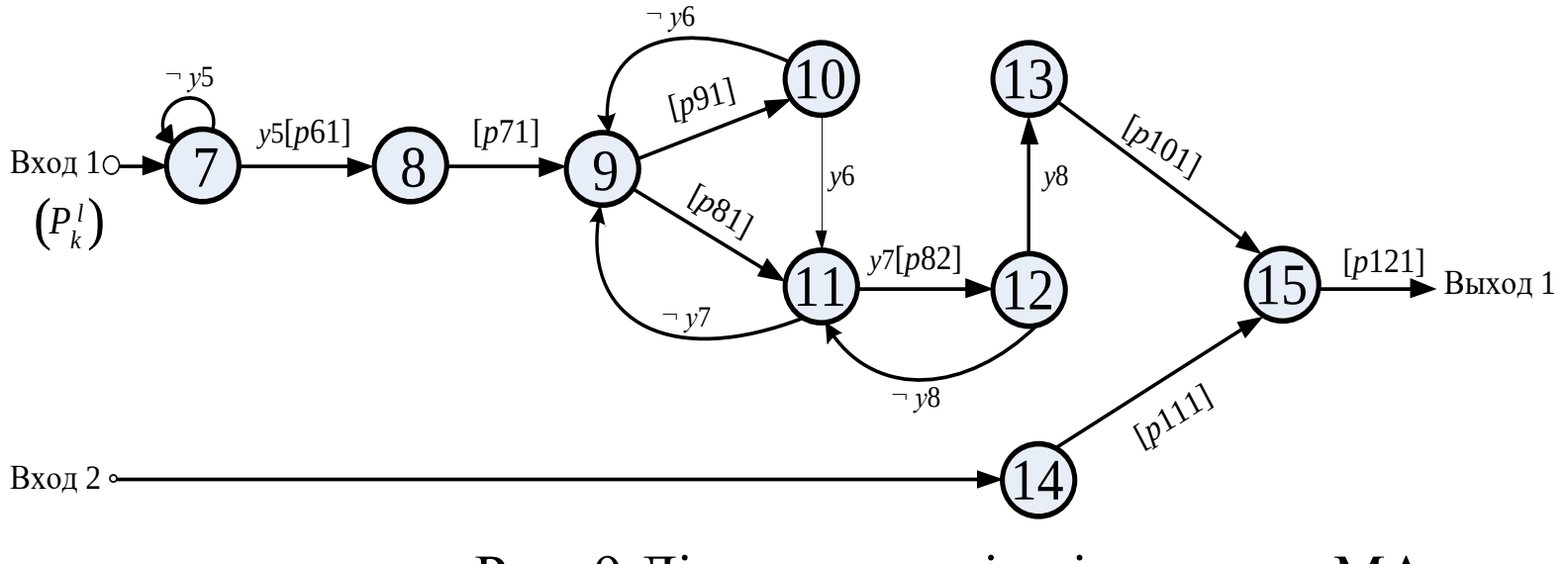

Рис. 9 Діаграма станів підсистеми МА

p61 – перша процедура для словоформ $P_k^l$, що не є АСС. Формує розгорнуте представлення речення $P_k^l$;

p71 – сумісно з процедурами p8 і p9 формує роздільне представлення основи і закінчення словоформи $S_l^m$. Виконує попередні установки інформаційних і керуючих регістрів;

p81 – виконує порівняння основи словоформи $S_l^m$ з вмістом пам'яті основ;

p82 – формує список основ-омонімів;

p91 – виконує порівняння закінчення словоформи $S_l^m$ з вмістом пам'яті закінчень;

p101 – формує лексеми і морфологічні характеристики $S_l^m$, в тому числі й омонімів;

p111 – формує морфологічні характеристики $S_l^m$;

p121 – формує морфологічно зв'язані речення $P_k^l$, роздільні для всіх неоднозначних словоформ;

y5 – умова початку прийняття послідовності словоформ речення $P_k^l$;

y6 – умова порівняння закінчення словоформи $S_l^m$;

y7 – умова порівняння основи словоформи $S_l^m$;

y8 – умова співставлення повного списку основ-омонімів.

В апаратному морфологічному процесорі (АМП) для оброблення кожної словоформи $S_l^m$ речення $P_k^l$ виділено окремий апаратний блок, в якому із словоформи виділяються основа і закінчення, причому принципи такого виділення відрізняються від традиційних [5] і описані в [14]. Сукупність таких блоків, які паралельно обробляють всі словоформи речення $P_k^l$, складають одну з основних компонент підсистеми морфологічного аналізу. Максимальна кількість блоків визначається на основі статистичних характеристик заданого ЛКТ, зокрема, параметра максимальної кількості входжень словоформ в речення.

Виділені основа і закінчення словоформи $S_l^m$ поступають відповідно в БПОС и БПОК, в яких за асоціативним принципом формується адреса фрагмента комірок пам'яті основ і закінчень, в якому зберігається морфологічна структура словоформи $S_l^m$.

Аналогічним чином формуються морфологічні характеристики для всіх словоформ $\left\{S_l^{m'}\right\}$ речення $P_k^l$. Відмітимо, що вказані морфологічні характеристики враховують особливості тільки текстів науково-технічного та ділового стилів.

В підсистемі МА виконується паралельний аналіз словоформ $\left\{S_l^{m'}\right\}$, що належать реченню $P_k^l \in T_k$, де $m' = \overline{1,M}, M$ – кількість словоформ в реченні $P_k^l$, $l = \overline{1,L}, L$ – кількість речень в тексті $T_k$. На виході підсистеми в блоках формування морфологічних характеристик формується вихідна інформація $\left\{Л_l^{m'}, МХ_l^{m'}\right\}$, до якої приєднується морфологічна інформація про присутніх в реченні, що аналізується, АСС $S_l^{m''}, m'' \notin M$. Причому остання множина може бути і пустою. На виході підсистеми МА формується морфологічна структура виду $\left\{\left(ЛЕ_l^1, МХ_l^1\right),...,\left(ЛЕ_l^m, МХ_l^m\right),...,\left(ЛЕ_l^M, МХ_l^M\right)\right\}$, яка є вхідною інформацією для підсистеми синтаксичного аналізу.

*Підсистема синтаксичного аналізу*

На рис. 10 представлена блок-схема синтаксичного аналізу (де СнА – синтаксичний аналіз), а на рис. 11 – діаграма станів з описами процедур, що виконуються, й умовами, що аналізуються.

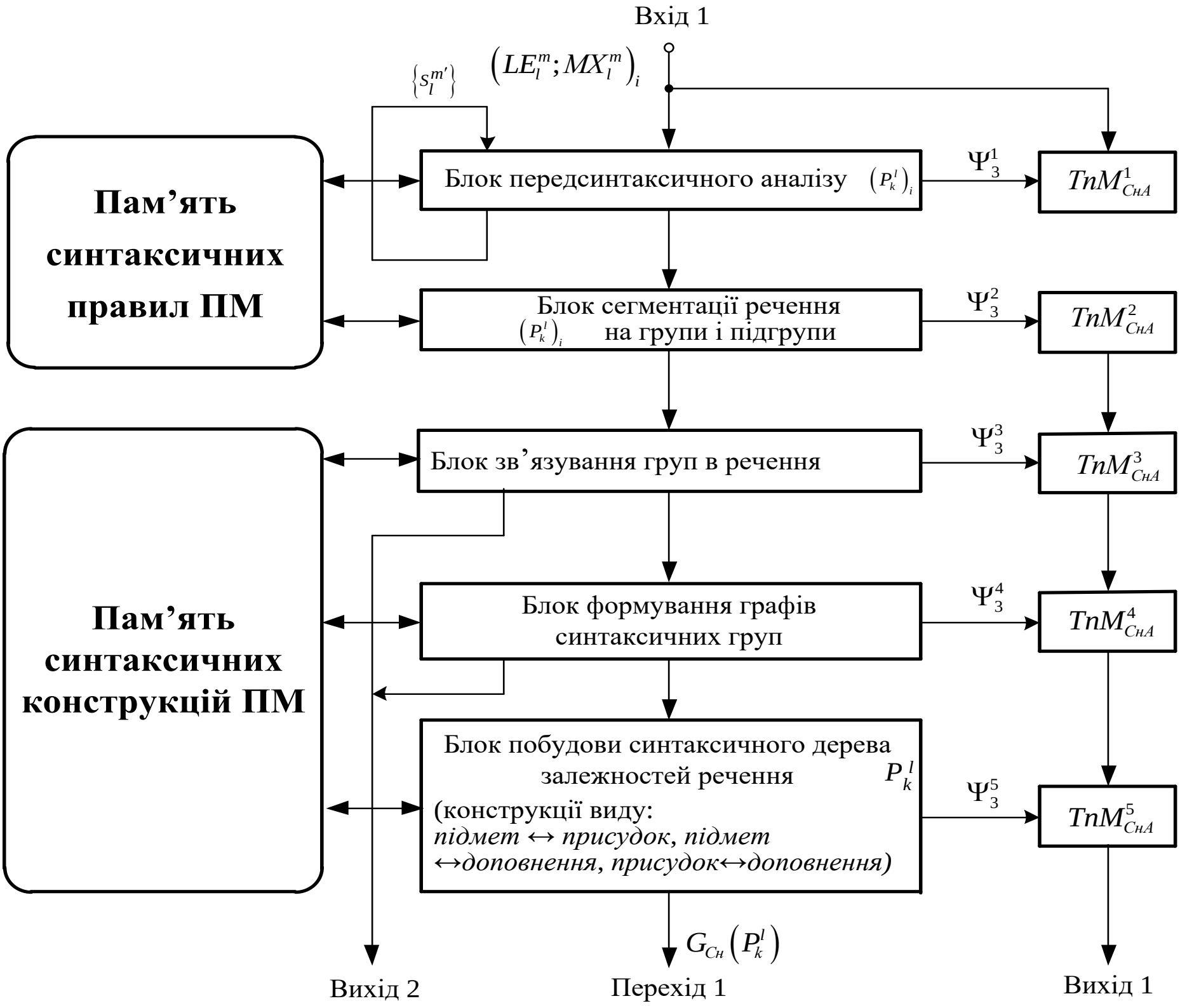

Рис. 10. Функціональна схема підсистеми СнА

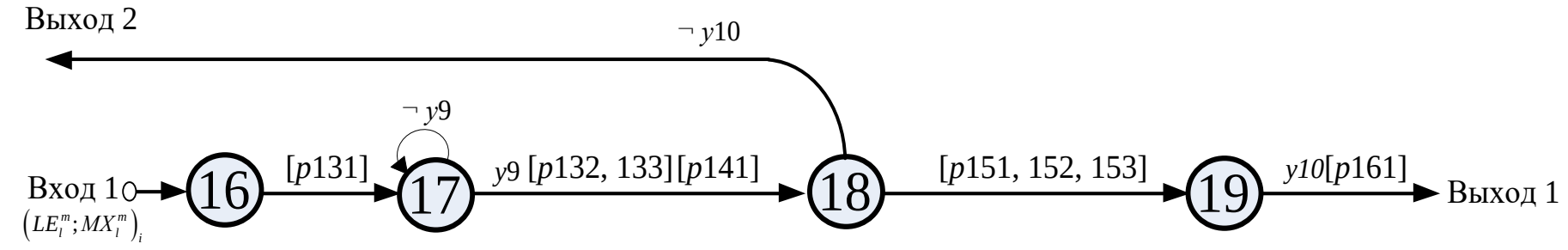

Рис. 11. Діаграма станів підсистеми СнА

p131 – виконує приймання від підсистеми МА послідовності речень $\left(P_k^l\right)_i$, де $i = \overline{1, I}$, $I$ – максимальна кількість неоднозначних речень;

p132 – виконує передсинтаксичний аналіз речення $P_k^l$ (вирішення граматичної та лексичної омонімії, зв'язування слів у реченні синтаксичними зв'язками);

p133 – виконує сегментацію речення на синтаксичні групи та підгрупи;

p141 – виконує почергове зв'язування груп у реченні. У цьому виконується послідовне оброблення неоднозначних речень. Після закінчення оброблення $i$-го речення на "Вихід 2" підсистеми видається повідомлення про завантаження ($i$+1)-го речення;

p151 – виконує формування вузлів графів груп та підгруп;

p152 – виконує побудову орієнтованих графів груп та підгруп;

p153 – виконує перетворення графів у синтаксичні дерева залежностей (конструкції виду: підмет↔присудок, підмет ↔ доповнення підмета, присудок ↔ доповнення присудка тощо).

p161 – виконує побудову спільного синтаксичного дерева речення;

y9 – умова відсутності неоднозначних словоформ у реченні;

y10 – умова перебору всіх неоднозначних речень.

На вхід підсистеми синтаксичного аналізу надходить морфологічна структура речення $P_k^l$, яка є лінійним об'єктом, на виході формується синтаксичне дерево залежностей, що є нелінійним об'єктом, складність якого суттєво вища порівняно з морфологічною структурою.

Під синтаксичною структурою речення $P_k^l$ розуміється розмічене дерево залежностей, таке, що: множина його вузлів утворює імена лексем, що входять до $P_k^l$; кожна дуга позначена ім'ям відповідного синтаксичного відношення, специфічного для заданої ПМ. Наприклад, для російської та української мов виділяють понад 50 синтаксичних відношень. У свою чергу, всі синтаксичні відношення поділяються на типи – актантні, атрибутивні, твірні та службові. Їх повний перелік наведено в [5].

Алгоритм синтаксичного аналізу розбивається на два блоки: передсинтаксичний аналіз та власне синтаксичний аналіз.

Основне призначення блоку передсинтаксичного аналізу полягає в усуненні лексичної та граматичної омонімії словоформ за лінійним контекстом. Крім того, цей блок для деяких речень встановлює конкретні синтаксичні зв'язки між тими словами, лексичні та синтаксичні властивості та відносне лінійне розташування яких дозволяють однозначно зробити висновок про наявність між ними таких зв'язків.

Вихідне речення розбивається на синтаксичні групи, які, своєю чергою, можуть складатися з підгруп. Виділяють групи підмета, присудка тощо. Далі алгоритм виконує зв'язування побудованих груп в один граф синтаксичних відношень.

Таким чином, в результаті застосування правил передсинтаксичного аналізу морфологічна структура речення перетворюється на деякий проміжний об'єкт, в якому, порівняно з вихідною структурою, суттєво скорочено лексико-граматичну омонімію та проведено деякі синтаксичні зв'язки.

В підсистемі СнА одними з основних компонентів є словники синтаксичних правил і конструкцій природної мови. За допомогою правил СнА формується набір гіпотез про можливі синтаксичні зв'язки між складовими елементами $P_k^l$. Основним критерієм, що лежить в основі правил, що формують синтаксичні гіпотези, є критерій максимальної узгодженості пов'язаних лексем за всіма типами приписаної комбінаторної інформації – як морфологічної, так і лексикографічної. Інформація першого типу добувається з морфологічної структури, у якій вона була вироблена у процесі морфологічного аналізу; лексикографічна інформація (включаючи синтаксичну, семантичну, сполучувану) добувається з відповідних статей комбінаторного словника синтаксичних правил та конструкцій ПМ.

Після того як формування множини синтаксичних гіпотез закінчено, алгоритм синтаксичного аналізу приступає до його оптимізації, усуваючи з цієї множини помилкові гіпотези на основі деяких універсальних і локальних вимог до правильної синтаксичної структури речення.

Синтаксичні правила – складні формальні об'єкти, які складаються з двох основних зон – зони умов і зони дій.

Умови в правилах являють собою логічні вирази, в записі яких можуть бути використані як елементарні, так і складові предикати, за допомогою яких можна описати наявність (або відсутність) у слові певних характеристик, узгоджені між виділеними словами фрази, необхідний (або неможливий) лінійний чи деревовидний контекст виділених слів фрази тощо.

Зона дій у правилах – це перелік інструкцій, послідовне виконання яких здійснює необхідне за правилом перетворення об'єкта, що розглядається. У переважній більшості правил синтаксичного аналізу об'єктом перетворення є морфологічна структура.

Нижче описано алгоритм синтаксичного аналізу [5].

1. Побудова набору гіпотетичних синтаксичних відношень для речення $P_k^l$. В результаті отримаємо орієнтований граф гіпотетичних синтаксичних відношень між словами речення. Він складається з $N$ вузлів, де $N$ – число слів у реченні, що аналізується.

2. Визначення вершини синтаксичної структури.

Необхідно з графа виділити дерево – синтаксичну структуру вихідного речення. Цей процес відбувається значно швидше, якщо одразу вдається визначити вершину дерева залежностей. Тому після побудови графа починає працювати блок, завданням якого є виявлення слів, які є кандидатами на роль вершини дерева.

Опис властивостей слів, які можуть бути вершинами в синтаксичної структури вихідної речення, задано у вигляді спеціального правила. Правило переглядає всі омоніми всіх слів речення та позначає ті з них, які за своїми характеристиками можуть бути вершиною. Крім того, за допомогою приписування певної ваги кандидати на роль вершини впорядковуються в порядку зменшення ймовірності того, що це слово буде обрано вершиною синтаксичної структури.

3. Застосування шаблонів синтаксичних конструкцій та перевірка вмісту побудованого дерева.

4. Застосування правил переваг. Якщо граф, залишаючись зв'язаним, не є деревом, відбувається звернення до правил переваг, які дозволяють окремих випадках шляхом порівняння гіпотетичних синтаксичних хазяїнів і слуг кожного слова надати перевагу одним гіпотетичним зв'язкам, знищивши інші.

5. Перегляд альтернативних дерев.

Описаний алгоритм призводить до побудови, принаймні, однієї правильної синтаксичної структури для речення, що обробляється.

Використовуючи інформацію, що зберігається у словнику, блок формування синтаксично пов'язаних конструкцій інтерпретує побудовану синтаксичну структуру і формує зв'язки типу "підмет ↔ присудок", " підмет ↔ доповнення підмета" (як правило, це прикметник, узгоджений з іменником у роді, числі та відмінку) "присудок ↔ доповнення присудка" тощо.

*Підсистема семантичного аналізу*

Підсистема семантичного аналізу (СА) будує послідовно семантичні структури кожного речення вхідного тексту. Семантична структура складається із семантичних вузлів та семантичних відношень. Семантичний вузол – це такий об'єкт текстової семантики, у якого заповнені всі валентності як експліцитно виражені в тексті, так і імпліцитні – ті, які добуваються з МОКС. Застосування МОКС у підсистемі семантичного аналізу є основною відмінністю від класичних реалізацій семантичного аналізу. Вона містить крім традиційної семантичної інформації (яка міститься в словниках і тезаурусах заданої ПМ) багаторівневі проєкції як узагальнених, так і конкретизованих шаблонів представлень семантичних конструкцій простих речень.

З визначення семантичного вузла також випливає, що в процесі семантичного аналізу виділяються семантичні вузли та їх атрибути, що входять до семантичних вузлів. Як і на всіх етапах аналізу, семантичні вузли утворюються зі слів вихідного речення. Головне джерело гіпотез про склад семантичного вузла дає синтаксичний аналіз. Багато синтаксичних груп можуть перейти в семантичні вузли, інші повинні перетворитися на атрибути вузлів. Іншим джерелом формування семантичних вузлів є МОКС.

Крім слів, семантичні вузли можуть включати: розділові знаки, стійкі обороти, стійкі словосполучення, жорсткі синтаксичні групи тощо.

Кожному вузлу приписано множину атрибутів, таких як набір графематичних слів, із яких складається даний вузол; номер семантично головного слова у вузлі; граматична інтерпретація вузла (зовнішня синтаксична характеристика); номер фрагмента, якому належить вузол; прийменник, який у синтаксисі керував цим вузлом; посилання на словникову статтю в МОКС тощо.

Загальна схема роботи алгоритму може бути наступною послідовністю кроків.

1. Ініціалізація семантичних вузлів.
2. Побудова множини словникових інтерпретацій вузлів.
3. Побудова груп часу.
4. Побудова вузлів із словосполучень у лапках.
5. Побудова вузлів типу "один одного".
6. Побудова стійких словосполучень.
7. Побудова лексичних функцій-параметрів.
8. Встановлення відношень між локативними вузлами.
9. Побудова графа гіпотетичних зв’язків.
10. Побудова множинних актантів.
11. Перевірка семантичних вузлів за семантичними характеристиками.
12. Перевірка проєктивності побудованої семантичної конструкції.
13. Побудова відношень за умовчанням.
14. Побудова міжфразних зв’язків.
15. Побудова анафоричних зв’язків.

На рис. 12 представлена блок-схема семантичного аналізу, а на рис. 13 – діаграма станів з описами виконуваних процедур і умовами, що аналізуються.

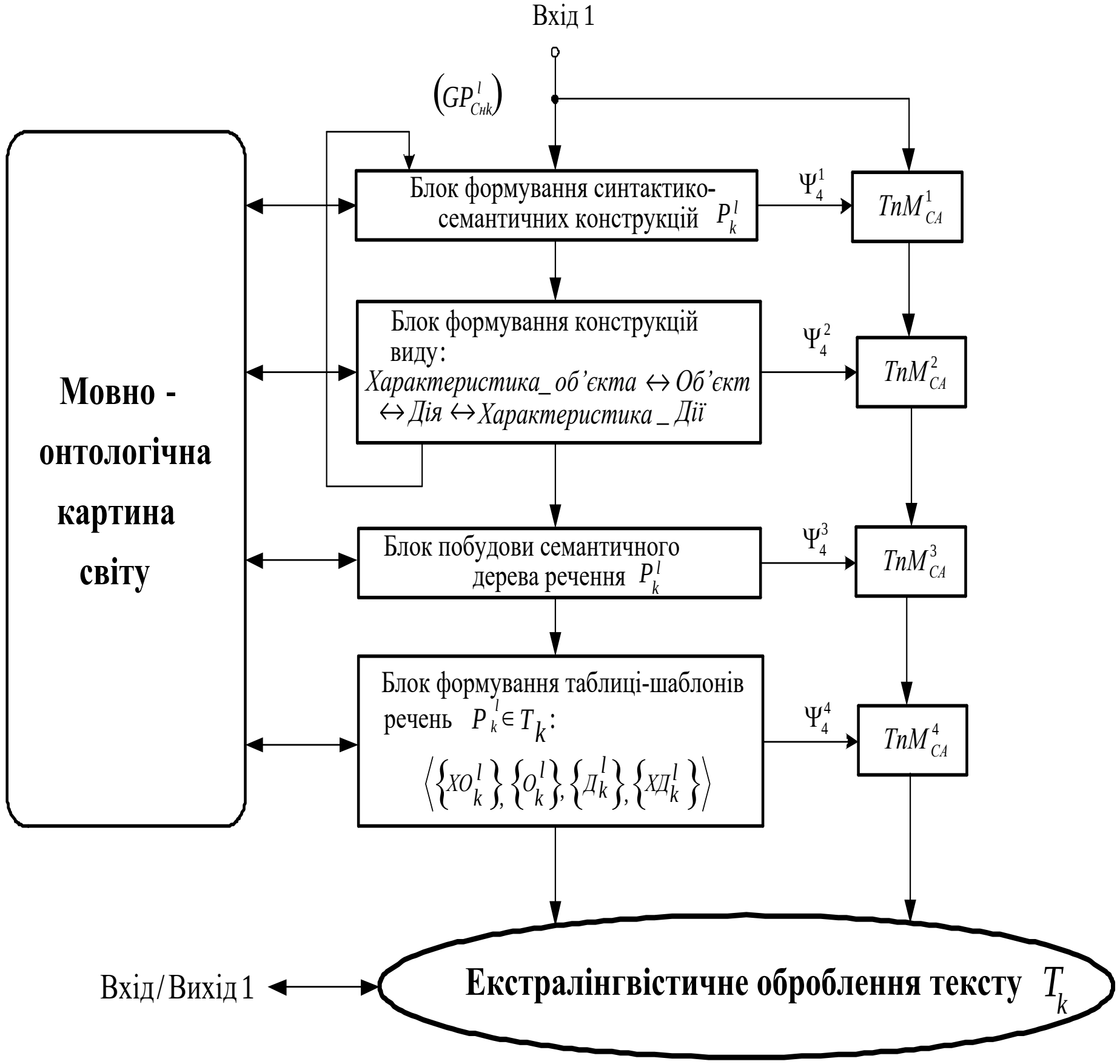

Рис. 12. Функціональна схема підсистеми СА

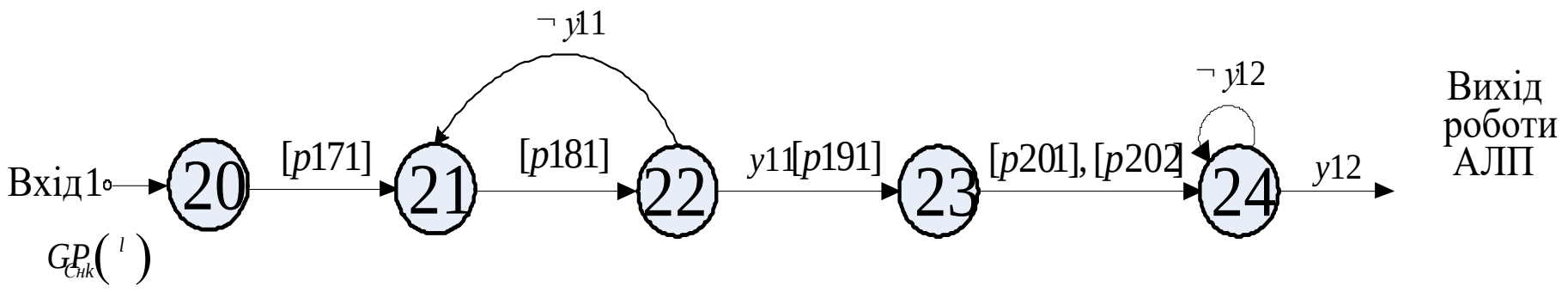

Рис. 13. Діаграма станів підсистеми СА

р171 – виконує перетворення синтаксичного дерева залежностей речення $P_k^l$ в синтактико-семантичну структуру. При цьому виконуються п.п. 1–2 загального алгоритму семантичного аналізу, наведеного вище;

р181 – формує шаблон-конструкцію для навігації по МОКС й ітераційно знаходить найбільш підходящий варіант. При цьому процедура може звернутися за інформацією, отриманою на будь-якій стадії оброблення. Виконуються п. п. 3–8 загального алгоритму;

р191 – виконує побудову семантичного дерева речення $P_k^l$. При цьому до сформованого попередньою процедурою шаблону додається смислові відтінки. Виконуються п. п. 9–12 загального алгоритму;

р201 – до сформованого шаблону речення додаються відсутні зв'язки. Виконує остаточне формування таблиці шаблонів тексту $T_k$;

р202 – процедура завершення роботи АЛП. По запиту блоку екстралінгвістичного оброблення передає вміст таблиці шаблонів речень тексту $T_k$ в цілому.

Умови на діаграмах станів інтерпретуються наступним чином:

у11 – умова знаходження підходящого варіанту шаблона-конструкції для речення $\left(P_k^l\right)_i$;

у12 – умова завершення роботи АЛП.

Як було вказано вище, на вхід підсистеми семантичного аналізу поступає синтаксичне дерево залежностей речення $P_k^l$.

Її перші два блоки, взаємодіючи з МОКС, реалізують переклад синтаксичних конструкцій речення $P_k^l$ в семантичні у відповідності з узагальненим шаблоном: “Характеристика об'єкта ↔ Об'єкт ↔ Дія ↔ Характеристика дії”.

Далі у блоці формування семантичного дерева виконується усунення синтаксичної омонімії та усунення анафоричних зв'язків.

Алгоритм усунення анафоричних зв'язків виконує заміщення анафор відповідними лексемами та описаний у [15].

Така конкретизація узагальненого шаблону вже достатня для побудови семантичного графа речення $P_k^l$. У цьому графі вершини є сутності (об'єкти), а дуги – відношення (зв'язку) між ними. Імена вершин і дуг зазвичай збігаються з іменами відповідних сутностей та відношень, що використовуються у природній мові. Дуга та дві пов'язані з нею вершини становлять мінімальну значущу інформацію – факт наявності зв'язку певного типу між відповідними об'єктами.

На закінчення роботи АЛП буде сформовано таблицю шаблонів речень тексту $T_k$, в якій зберігається інформаційна структура для всього тексту в цілому і яка є вхідною інформацією для підсистеми екстралінгвістичного оброблення тексту. В ньому виконується формально-логічне представлення (переклад) речень і тексту загалом у відповідній формальної теорії першого порядку, наприклад, спочатку модифіковані концептуальні графи і потім у логіку предикатів першого порядку.

### 5.3.2 ***Задача оптимального синтезу АЛП*** [11]

Введення додаткових апаратних витрат з метою суттєвого виграшу у швидкодії реалізації алгоритмів лінгвістичного аналізу вимагає розгляду питання оптимізації співвідношення "Апаратні витрати ($Q$) $\leftrightarrow$ Час виконання ($T$)". Таку оптимізацію можливо виконати на основі відомого методу Парето.

Цільова функція в аналітичному вигляді знаходиться одним з наближених методів, наприклад, лінійної або нелінійної інтерполяції або екстраполяції, за декількома опорними точками (структурні реалізації алгоритму), які отримуються шляхом попереднього формування варіантів реалізацій алгоритму. З множини цих точок, де кожній *r*-ій точці ($r = 1 \div m$) відповідає реалізація з параметрами $\langle T_r, Q_r \rangle$, формується множина Парето на площині $T - Q$ (рис. 5) з урахуванням впорядкування:

$$T_1 \le T_2 \le \ldots \le T_r \le \ldots \le T_m;$$
$$Q_1 \ge Q_2 \ge \ldots \ge Q_r \ge \ldots \ge Q_m.$$

У загальному випадку задача вибору (квазі) оптимального варіанта реалізації алгоритму зводиться до вибору певних граничних значень (граничних параметрів Tr, Qr) і мінімізації функціоналу:

$$F = \min_r \frac{c_r Q_r + b_r T_r}{c_r Q_0 + b_r T_0}, \tag{9}$$

де $c_r = (c_1, c_2, \ldots, c_m), b_r = (b_1, b_2, \ldots, b_m)$ – нормативні коефіцієнти (які можуть бути визначені, наприклад, методом експертних оцінок), такі, що $\sum_{r=1}^{m} c_r = 1, \sum_{r=1}^{m} b_r = 1$, і $c_r \bullet b_r = k$, де $k$ – коефіцієнт оберненої пропорціональності. Суть цього коефіцієнта полягає в тому, що чим менший час $T_r$ реалізації алгоритму, тим необхідно більше витрат обладнання $Q_r$ і навпаки.

Якщо функціонал (9) знайдено і знайдено мінімальне значення $c = \min_r \frac{c_r Q_r + b_r T_r}{c_r Q_0 + b_r T_0}, \text{то } \forall c' \geq c$ отримаємо нерівність $c'\left(c_r Q_0 + b_r T_0\right) \geq c_r Q_r + b_r T$ .

Остання нерівність означає, що який би не був параметр $c'$, реалізація не буде перевершувати значення нормативної суми $c_r Q_0 + b_r T$ .

### 5.3.3 *Структурна організація та проєктування АМП*

Нижче описано апаратну реалізацію підсистеми морфологічного аналізу (або апаратного морфологічного процесора), причому лише послідовного аналізу словоформ вхідного речення. Як зазначалося вище, для реалізації паралельного оброблення всіх словоформ речення потрібно *К* блоків морфологічного аналізу, де *К* – максимальна кількість входжень словоформ у речення.

Загальна схема реалізації морфологічного аналізу (МА) (незалежно від способу реалізації) зводиться до прийому послідовності слів, що становлять вхідний текст, розпізнавання або дешифрації аналізованого слова та знаходження відповідної йому так званої “точки в гіперпросторі” (або реалізації табличного методу аналізу), в якій слову, що аналізується, приписані всі необхідні морфологічні характеристики. Цей простір є віссю $X_i$ частини мови заданої ПМ, де $i = \overline{1,n}$, *n* – кількість частин мови, а по вісі $Y_i$ – послідовність словоформ і-ої частини мови.

Описана вище послідовність кроків МА є "ідеальною" і практично нереалізованою для сучасних мікроелектронних технологій, а наближення до неї можливе лише для апаратної реалізації алгоритму МА. Для “ідеальної” реалізації знадобився дешифратор (або пам’ять) з адресацією 256 розрядів. Цей параметр визначено з того, що для кодування однієї літери (символу) слова потрібно 8 біт, а максимальна кількість символів найдовших словоформ у лексикографічній базі даних (ЛБД) загальновживаної лексики української мови “Словники України” дорівнює 32. Звідси й отримано ступінь двійки ( 8х32 = 256).

Класичний програмний МА виконується послідовно по буквам, починаючи з закінчення, знаходження основи словоформи та формування послідовності омонімів аналізованого слова. При цьому для кожного омоніма формується множина морфологічних характеристик.

Зазначимо, що у загальному випадку лише “ідеальна” апаратна реалізація дозволяє уникнути роздільного аналізу закінчення і основи словоформи.

Таким чином, якщо умовно розташувати на площині осі X реалізацію МА класичним програмним способом, то описана вище "ідеальна" апаратна реалізація буде розташована по осі *У*, а всі інші реалізації будуть розташовані між ними і є множиною Парето рішень реалізації АМП. Графічно це представлено на рис. 14.

Узагальнена схема АЗ реалізації алгоритму МА для деякого рішення $T_r, Q_r$ представлена на рис. 15. На ньому прийнято такі позначення:

$m$ – кількість слів у аналізованому тексті. Ця послідовність формується на етапі графематичного аналізу, записується в пам’ять слів тексту і є вихідними даними для МА;

$C_1^1, C_2^1, ..., C_n^1$ – максимальна кількість букв (символів) у словах аналізованих текстів;

$C_1^3, C_2^3, ..., C_{q1}^3, ..., C_{qs_1}^3, ...C_{qs_t}^3$ – перша літера слова та $qs$ груп поєднань символів (починаючи з другого), що формуються на основі статистичних характеристик та заданих обмежень на обладнання;

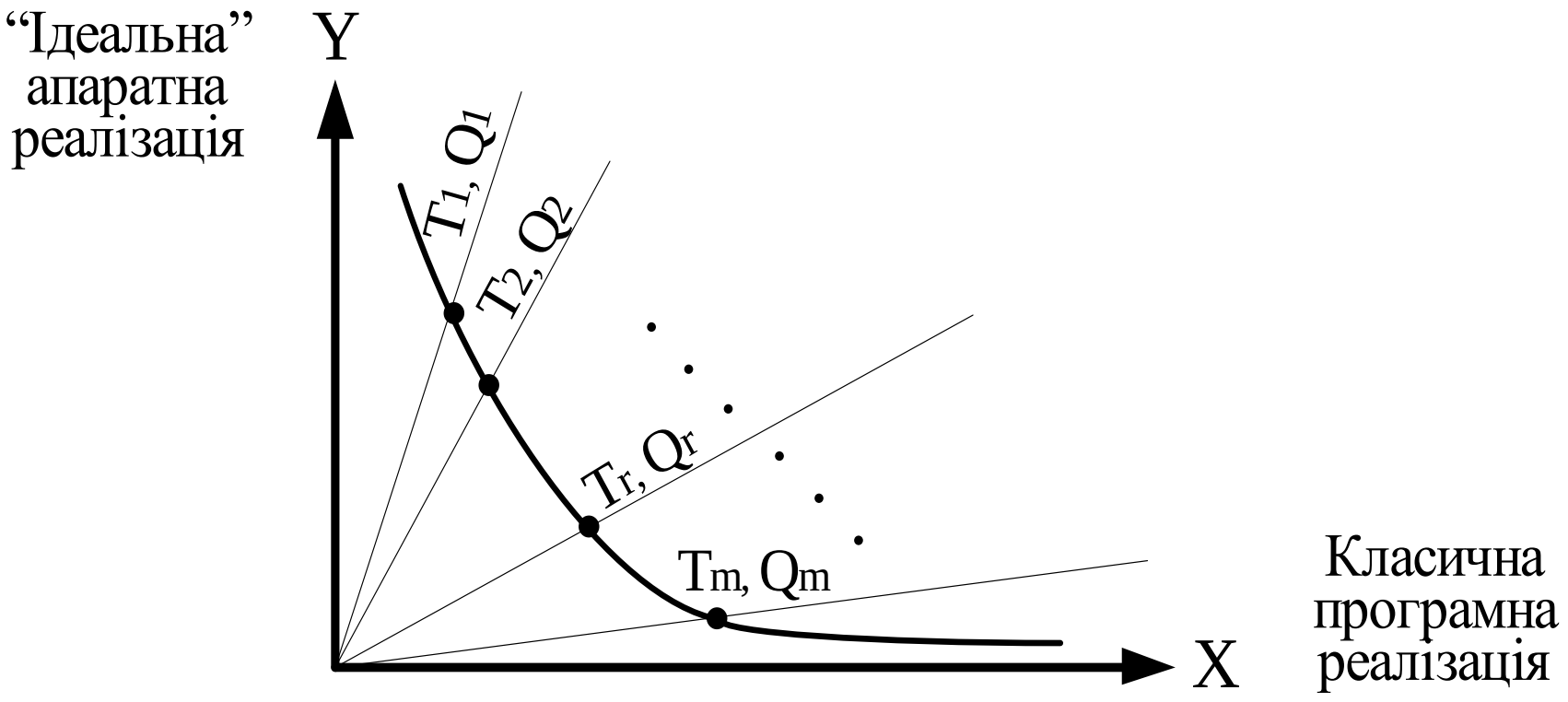

Рис. 14. Множина Парето рішень реалізації
морфологічного процесора

$A_r^{ст}$ – старші розряди адреси пам’яті слів;

$A_r^{мл}$ – молодші розряди адреса пам’яті слів.

Суть підходу до побудови схеми полягає в "усіканні" адресного простору, необхідного для "ідеальної" реалізації, до адресного простору пам’яті, представленої на стандартному устаткуванні. І цьому слугує четвертий рівень на рис. 15.

### 5.3.4 ***Структурна організація АМП для оброблення ЛКТ різного об’єму*** [10–11]

Структурна організація АЗ морфологічного аналізу природномовних текстів, що становлять деякий ЛКТ, і витрати обладнання сильно залежать від статистичних характеристик заданого корпусу текстів, зокрема, від кількості вживаних словоформ *К*, їх середньої довжини, кількості поєднань символів (починаючи з другого), що перекривають середню довжину, і ряду інших.

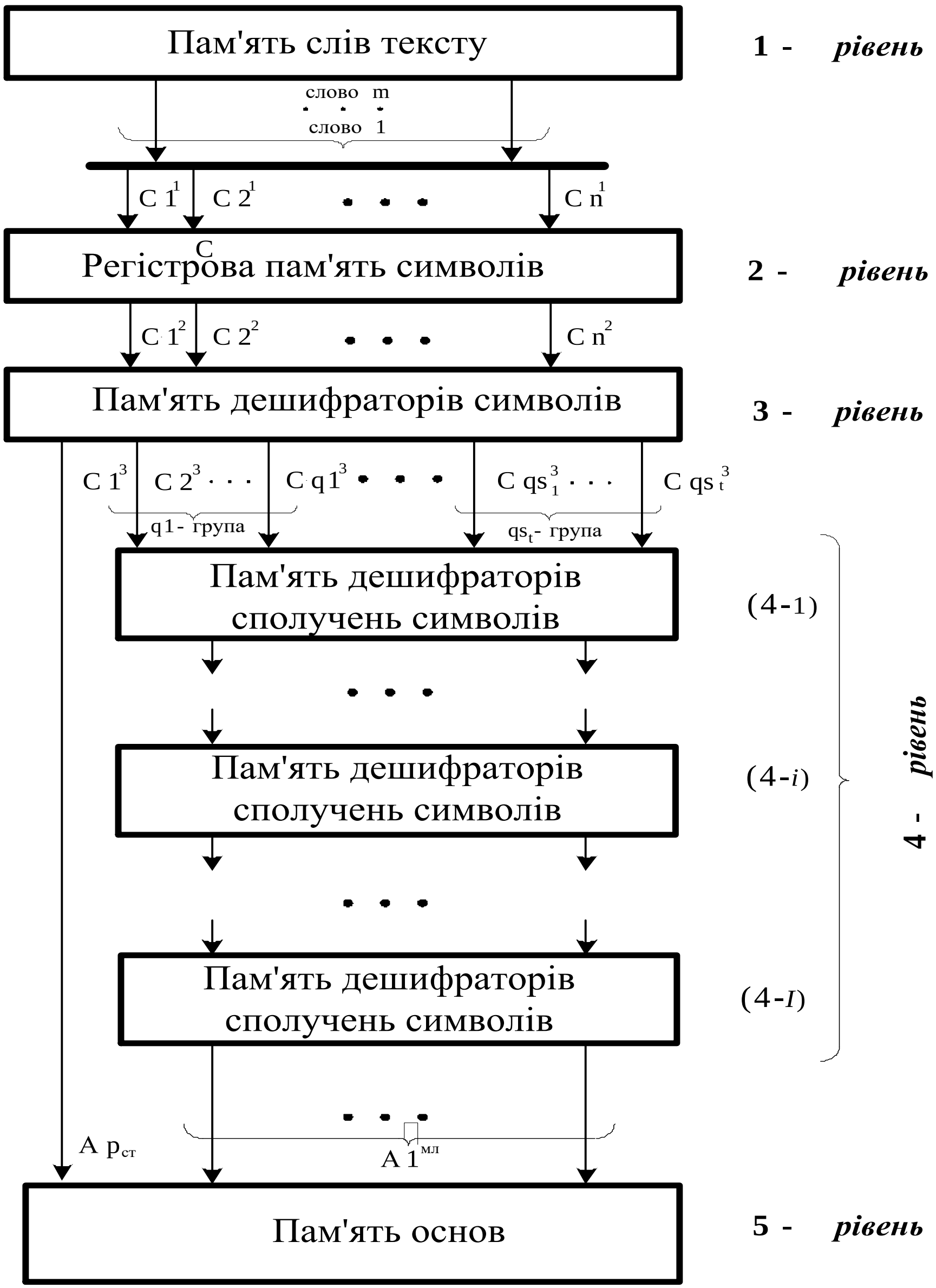

Рис. 15. Узагальнена схема апаратних засобів МА

Діаграма залежності кількості основ від їхньої довжини (кількості символів в основі) для загальновживаної лексики української мови наведена на рис. 16. Крива на діаграмі найближче апроксимується експоненційною функцією виду:

$$f(x) = 33600e^{-\frac{(x-9)^2}{16}}$$

з величиною достовірності апроксимації R2=0,982. Обчислення величини $R^2$ в Microsoft Excel показано нижче:

$$R^2 = 1 - \frac{SSE}{SST},$$

де $SSE = \sum\left(Y_j - \hat{Y}_j\right)^2$ і $SST = \left(\sum Y_j^2\right) - \frac{\left(\sum Y_j\right)^2}{n}$ .

Суть задачі побудови АМП зводиться до скорочення апаратного обладнання (розглядатимемо лише стандартне обладнання на платах з ПЛІС, тому що розробка спеціального обладнання є самостійною науково-технічною проблемою). Дослідження показали, що достатньо забезпечити незалежну адресацію поєднань символів, що перекривають середню довжину словоформ (або основ) заданого ЛКТ.

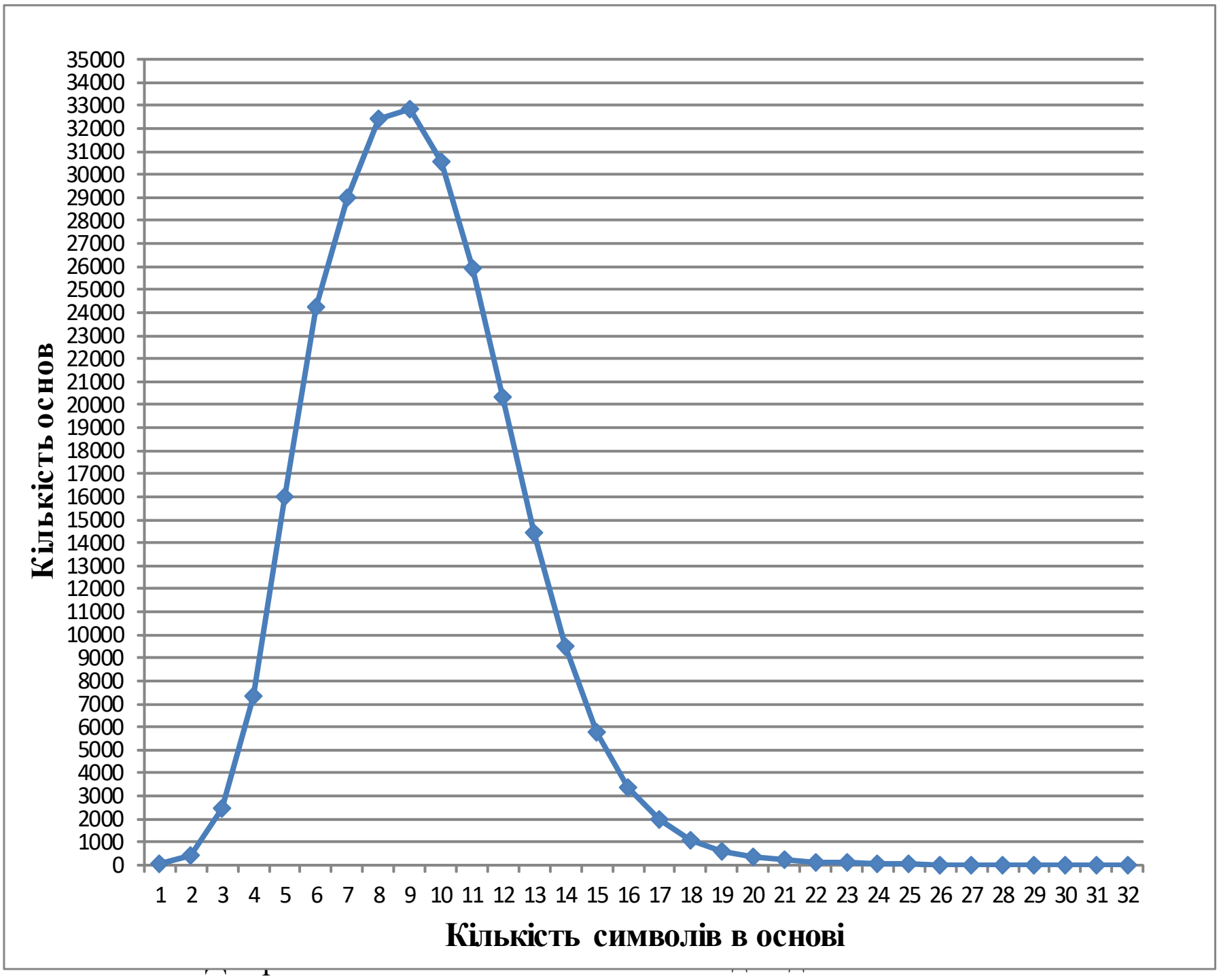

Рис. 16. Діаграма залежності кількості основ від їх довжини

Наприклад, для ЛКТ загальновживаної лексики української мови (всі основи представлені у лексикографічній базі даних “Словники України”), у якій середня довжина слова становить 9,27 символа, необхідно забезпечити перекриття до десятого символу та більше. Також потрібно врахувати необ-хідність адресації (до 4 розрядів) різних форм дієслова, що мають однакові послідовності символів та перекривають . Адресацію інших поєднань символів можна "збирати" за схемою АБО. Статистичні характеристики для довільного ЛКТ обчислюються в Microsoft Excel стандартними функціями після перетворення тексту в таблицю в Microsoft Word.

Нижче буде розглянуто розробку АМП для трьох варіантів ЛКТ:

– варіант А – ЛКТ, що включає загальновживану лексику української мови. Статистичні характеристики наведено у табл. 1;

– варіант В – ЛКТ з онтологічного інжинірингу. Статистичні характери-стики наведено у табл. 2;

– варіант С – ЛКТ з онтолого-керованих інформаційних систем загаль-ного призначення. Статистичні характеристики наведено у табл. 3.

Таблиця 1

Кількісні показники сполучень символів для основ ЛКТ, варіант А

Загальна кількість основ – 259209, середня довжина основи – 9,27

| **C2** | **C3** | **C4** | **C5** | **C6** | **C7** | **C8** | **C9** | **C10** | **C11** | **C32** |
|---|---|---|---|---|---|---|---|---|---|---|
| 35 | 38 | 37 | 36 | 6 | 6 | 35 | 35 | 35 | 34 | 1 |
| **C1-C9** | **C1-C10** | **C2-C3** | **C2-C4** | **C2-C5** | **C2-C6** | **C2-C7** | **C2-C8** | **C2-C9** | **C2-C10** | **C2-C32** |
| 206647 | 224417 | 789 | 26450 | 32208 | 76421 | 119831 | 152789 | 176251 | 193052 | 223140 |
| **C3-C32** | **C4-C5** | **C4-C32** | **C5-C6** | **C5-C7** | **C5-C8** | **C5-C9** | **C5-C10** | **C5-C11** | **C5-C16** | **C5-C32** |
| 179001 | 934 | 125805 | 908 | 7950 | 27987 | 48550 | 61356 | 69450 | 83290 | 83961 |
| **C6-C7** | **C6-C8** | **C6-C9** | **C7-C10** | **C7-C11** | **C7-C32** | **C8-C9** | **C8-C32** | **C9-C11** | **C10-C11** | **C8-C10** |
| 835 | 6796 | 21543 | 16290 | 23929 | 35226 | 759 | 22790 | 4074 | 649 | 4989 |
| **C11-C12** | **C11-C13** | **C11-C16** | **C11-C32** | **C13-C14** | **C14-C16** | **C14-C32** | **C15-C16** | **C17-C18** | **C17-C19** | **C17-C32** |
| 594 | 2733 | 6125 | 6639 | 501 | 1364 | 2156 | 388 | 293 | 581 | 703 |
| **C19-C20** | **C20-C22** | **C21-C22** | **C23-C24** | **C23-C25** | **C25-C26** | **C26-C28** | **C27-C28** | **C29-C30** | **C29-C31** | **C31-C32** |
| 198 | 206 | 110 | 66 | 72 | 28 | 17 | 14 | 6 | 6 | 2 |

Таблиця 2

Кількісні показники сполучень символів для основ ЛКТ, варіант В

Загальна кількість основ – 32055, середня довжина основи – 8,54

| **C2** | **C3** | **C4** | **C5** | **C6** | **C7** | **C8** | **C9** | **C10** | **C11** | **C31** |
|---|---|---|---|---|---|---|---|---|---|---|
| 38 | 40 | 40 | 36 | 38 | 36 | 35 | 38 | 39 | 38 | 1 |
| **C1-C9** | **C1-C10** | **C2-C3** | **C2-C4** | **C2-C5** | **C2-C6** | **C2-C7** | **C2-C8** | **C2-C9** | **C2-C10** | **C2-C31** |
| 25301 | 27666 | 548 | 3135 | 8095 | 12656 | 16655 | 20293 | 23278 | 25615 | 29221 |
| **C3-C32** | **C4-C5** | **C4-C31** | **C5-C6** | **C5-C7** | **C5-C8** | **C5-C9** | **C5-C10** | **C5-C11** | **C5-C16** | **C5-C31** |
| 25966 | 674 | 20909 | 624 | 3236 | 7016 | 9902 | 12078 | 13635 | 15542 | 15674 |
| **C6-C7** | **C6-C8** | **C6-C9** | **C7-C10** | **C7-C11** | **C7-C31** | **C8-C9** | **C8-C31** | **C9-C11** | **C10-C11** | |
| 568 | 2736 | 5469 | 4130 | 5385 | 7306 | 499 | 4820 | 1362 | 380 | |
| **C11-C12** | **C11-C13** | **C11-C16** | **C11-C32** | **C13-C14** | **C14-C16** | **C15-C16** | **C17-C18** | **C17-C19** | **C17-C31** | |
| 332 | 771 | 1233 | 1380 | 236 | 334 | 162 | 103 | 144 | 200 | |
| **C19-C20** | **C20-C22** | **C21-C22** | **C23-C24** | **C23-C25** | **C25-C26** | **C26-C28** | **C27-C28** | **C29-C30** | **C29-C31** | |
| 69 | 68 | 41 | 25 | 27 | 11 | 7 | 4 | 2 | 2 | |

Таблиця 3

Кількісні показники сполучень символів для основ ЛКТ, варіант С

Загальна кількість слів – 9406, середня довжина слова – 9,35

| | | | | | | | | | | |
|---|---|---|---|---|---|---|---|---|---|---|
| **C2** | **C3** | **C4** | **C5** | **C6** | **C7** | **C8** | **C9** | **C10** | **C11** | **C31** |
| 38 | 40 | 40 | 36 | 38 | 35 | 34 | 38 | 39 | 37 | 1 |
| **C1-C9** | **C1-C10** | **C2-C3** | **C2-C4** | **C2-C5** | **C2-C6** | **C2-C7** | **C2-C8** | **C2-C9** | **C2-C10** | **C2-C31** |
| 7203 | 7980 | 404 | 1461 | 2648 | 3653 | 4767 | 5930 | 6969 | 7748 | 9172 |
| **C3-C32** | **C4-C5** | **C4-C31** | **C5-C6** | **C5-C7** | **C5-C8** | **C5-C9** | **C5-C10** | **C5-C11** | **C5-C16** | **C5-C31** |
| 8707 | 479 | 7696 | 463 | 1699 | 3024 | 4079 | 4862 | 5443 | 6187 | 6304 |
| **C6-C7** | **C6-C8** | **C6-C9** | **C7-C10** | **C7-C11** | **C7-C31** | **C8-C9** | **C8-C31** | **C9-C11** | **C10-C11** | |
| 434 | 1468 | 2505 | 1905 | 2426 | 3301 | 356 | 2324 | 735 | 270 | |
| **C11-C12** | **C11-C13** | **C11-C16** | **C11-C32** | **C13-C14** | **C14-C16** | **C15-C16** | **C17-C18** | **C17-C19** | **C17-C31** | |
| 236 | 451 | 696 | 823 | 183 | 242 | 136 | 92 | 121 | 170 | |
| **C19-C20** | **C20-C22** | **C21-C22** | **C23-C24** | **C23-C25** | **C25-C26** | **C26-C28** | **C27-C28** | **C29-C30** | **C29-C31** | |
| 63 | 64 | 40 | 24 | 26 | 10 | 7 | 4 | 2 | 2 | |

Різні типи плат з ПЛІС, їх опис та технічні характеристики представлені на веб-сайтах www.hitechglobal.com/boards/allboards.htm та www.hilinx.com/products/boards_kits.htm. На рис. 17 представлена блок-діаграма плати HTG-V4PCIE, на якій проводилося моделювання АМП для варіанта С.

Моделювання АМП для варіанта С виконано в системі САПР ПЛІС Xilinx ISE 8.2i з використанням плати, на якій встановлені наступні апаратні засоби, доступні для користувача та необхідні, зокрема, для практичної реалізації АМП: 1) кристал ПЛІС Virtex-4, який містить 376 блоків надшвидкої RAM (СОЗУ) 18Kbx1, з можливістю організації від 16Kbx1 до 512x36 біт (www.hilinx.com/products/boards_kits/virtex6.htm); 2) зовнішня (по відношенню до кристала ПЛІС) пам’ять RAM – два незалежні блоки 64Mx16 біт на яких реалізовані пам’яті основ та закінчень.

*Структура варіанту А*

Як було зазначено вище, вихідними даними для проєктування АМП є задані статистичні характеристики ЛКТ та апаратне обладнання.

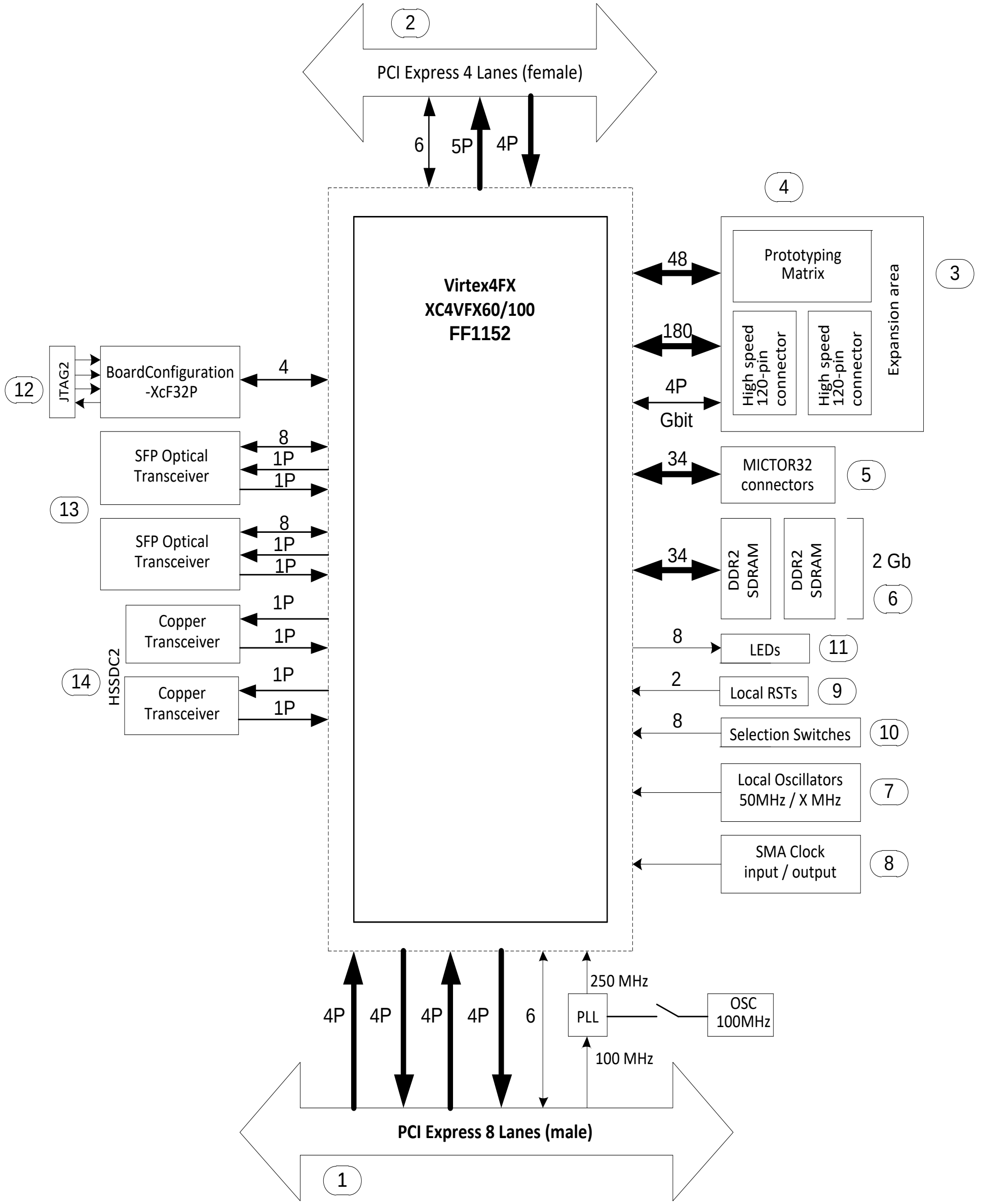

Рис. 17. Блок-діаграма плати HTG-V4PCIE

При морфологічному аналізі послідовності слів, що становлять деякий ПМТ, перший символ кожного слова (а ним може бути лише літера) аналізується окремо, тому що за його значенням визначається ряд граматичних показників, таких як: перше слово речення, абревіатура, співвідношення слова із заданою ПМ та низку інших. Для української мови першими літерами слів можуть бути 30 літер, отже, для їхньої адресації необхідно 5 розрядів (мала і велика літери вважаються однією літерою, а їхнє розрізнення визначається в окремій мікропрограмі). Аналіз першої літери слова ділить основну пам’ять слів на 32 сегменти, при цьому 30 сегментів відводяться під індекси відповідних літер, а 2 сегменти вільні, і їх обсяг достатній для зберігання всіх морфологічних характеристик (результатів обчислення) слів заданого ЛКТ. У "точці гіперпростору" деякого слова зберігається індекс-посилання на відповідну адресу в сегменті результатів. Тому розрядність даних пам’яті може бути порівняно невеликою – 16-20 розрядів.

Далі необхідно вибрати поєднання символів для 4-го рівня дешифраторів символів (рис. 15). Вони вибираються виходячи з апаратного обладнання, встановленого на заданій платі ПЛІС, зокрема, залежать від об’ємів внутрішнього СОЗУ і зовнішньої RAM. Зрозуміло, що чим більше символів увійдуть у поєднання, тим меншими будуть апаратурні витрати і вищий відсоток використання обладнання (пам’яті) плати. Четвертий рівень структурної схеми АМП, як правило, реалізується на внутрішніх СОЗУ ПЛІС, а їх об’єм порівняно невеликий: 18-36 Кбіт з доступним адресним простором 14–15 розрядів. На третьому рівні дешифрації за допомогою спеціальних алгоритмічних та технічних рішень 8-бітний код символу можна скоротити до 6 біт. Отже, для трьох символів, включених у поєднання, необхідна адресація 18 розрядів. Виконаємо розрахунок необхідного обладнання для поєднання символів С2–С4 у словах ЛКТ варіанта А відповідно до рис. 17 та таблицею 1. В якості плати ПЛІС обрана плата HTG – V6HXT – X16PCIE – 565 фірми Hitech Global (www.hitechglobal.com/boards/allboards.htm), на якій встановлена ПЛІС серії Virtex6, що має у своєму складі 912 СОЗУ розрядністю 32Кх1 біт пам’ять до 8Гіга 16-ти розрядних слів.

Кількість сполучень символів дорівнює 26450 (див. табл. 1, поєднання С2-С4), отже, розрядність даних RAM SS_1 дорівнює 15.

На один біт дешифрації потрібно 8 СОЗУ (відсутні 3 розряди до 18 біт, необхідні адресації поєднання з 3 символів), а їх загальна кількість для RAM SS_1 дорівнює 8х15=120. Це найбільша кількість СОЗУ для рівня 4-1 (рис. 18). Для поєднань символів С5–С7 та С8–С10 необхідно меншу кількість СОЗУ. Отже, для зазначених трьох поєднань символів (що забезпечують незалежну адресацію символів слів) необхідно близько третини із загальної кількості СОЗУ.

Архітектурно-структурна організація АЗ морфологічного аналізу для ЛКТ варіанта А представлена на рис. 18.

Як видно з таблиць 2 і 3 статистичних характеристик ЛКТ варіантів В і С, структурні схеми АМП для зазначених варіантів мало чим відрізняються один від одного. По суті, для варіанта С потрібно менше СОЗУ, що автоматично знижує відсоток використання обладнання, встановленого на платі. При цьому недоцільно використовувати плати ПЛІС більш ранніх версій, так як у них розрядність адрес СОЗУ менше, що спричинить необхідність збільшення кількості таких СОЗУ, збільшення тривалості синхросигналів процесора та й обсяг зовнішньої пам’яті RAM може бути критичним. Дослідження показали, що розробляти АМП із зовнішньою RAM, обсяг якої менший ніж (32–64)Мх16 біт, недоцільно: при цьому різко скорочується ефективність переведення операторів програмного рівня на мікропрограмний.

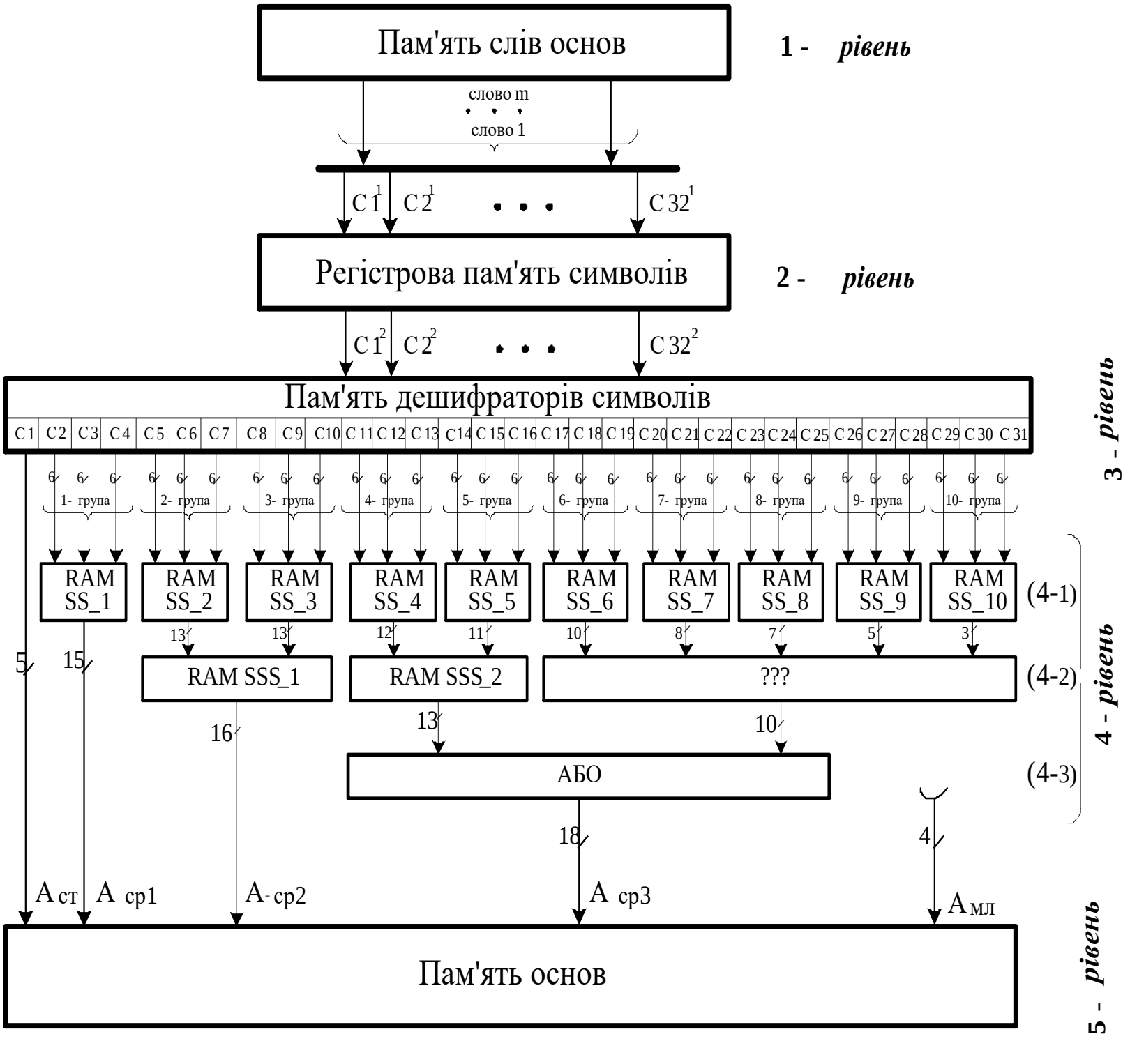

Рис. 18. Архітектурно-структурна організація АЗ МА для ЛКТ варіанта А

*Структури варіантів В і С*

Структурна схема АМП варіантів В і С представлена на рис. 19.

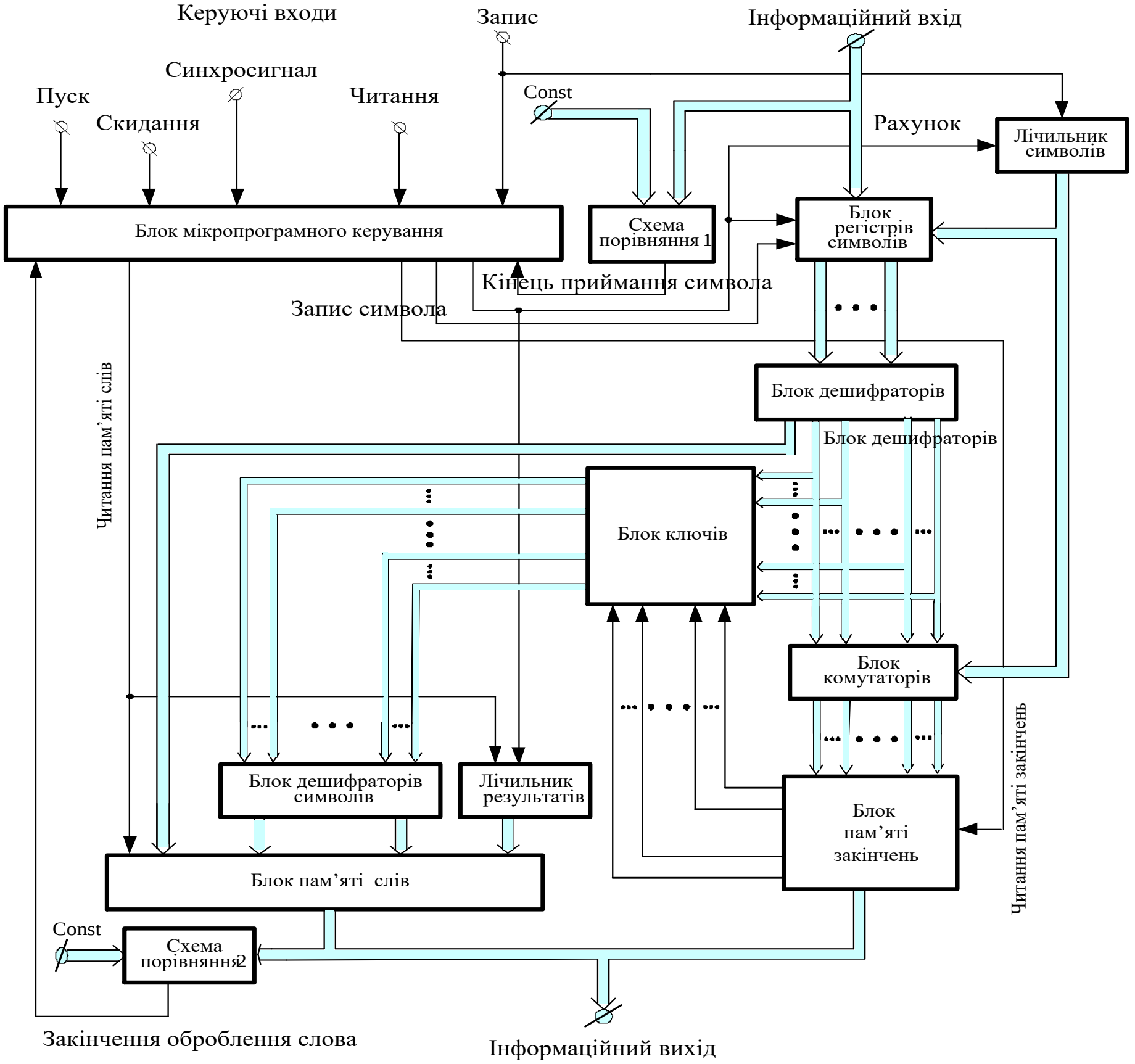

Рис. 19. Структурна схема АМП варіантів В і С

Робота АМП починається з приходу на вхід "Скидання" блоку мікропрограмного управління (БМУ) сигналу "Скидання = 1", який ініціює в ньому внутрішній керуючий сигнал "Скидання". Цей сигнал встановлює в "нульовий" стан блок регістрів символів, регістр мікрокоманд та лічильники символів та адрес результату. Потім АМП переходить в режим очікування сигналу "Пуск=1". З його приходом АМП чекає на перший символ вхідного слова ("Запис=1"). По його приходу на інформаційні входи блоку регістрів символів та першої схеми порівняння подається 8-бітний код першого символу (розглядається байтове кодування символів, наприклад Win 1251), та БМУ видає сигнал "Зпс=1". Номер символу запису формується лічильником символів, виходи якого керують дешифратором. Виходи останнього є керуючими сигналами запису у відповідний регістр символу.

8-бітний код першого символу з виходу РГС1 дешифрується в дешифраторі, з виходу якого 5-розрядний код формує старші адреси пам'яті слів. При цьому об'єм останньої розбивається на 32 сегменти.

Аналогічним чином в блок регістрів символів записуються всі символи вхідного слова. При цьому після приходу чергового символу у схемі порівняння виконується порівняння "код вхідного символу тотожний коду символу закінчення передачі символів вхідного слова" (це може бути, наприклад, код "09H", що означає "Пробіл"), який постійно присутній на другому інформаційному вході першої схеми порівняння. Після приходу коду символу закінчення передачі символів вхідного слова на виході першої схеми порівняння встановлюється сигнал "1", який надходить до відповідного керуючого входу БМУ.

Після цього БМУ переходить до інтерпретації алгоритму аналізу символів вхідного слова (рис. 20). Спочатку аналізуються символи, які, можливо, належать до закінчення вхідного слова. При цьому угруповання символів є важливим для перших інформаційних входів блоку ключів і не враховується для інформаційних входів блоку комутаторів.

Розглянемо алгоритм аналізу закінчення. При цьому в лічильнику символів буде записано код "01H", який надходить на вхід мультиплексорів, на виходи яких будуть передані коди "0, 0, ..., Cn", що передаються на адресні входи пам'яті закінчень.

У комірці пам'яті закінчень з адресою "0, 0, …, Cn" записано:

– якщо символ С*n* не є закінченням та словом без основи, то на других інформаційних виходах пам'яті закінчень буде код "NOP" (немає операції), а на перших – код "0, 0, …, 0", тобто на виходах блоку ключів всі символи "С2, …, С*n*" (символи всіх груп q) будуть заблоковані. Блокування символу означає, що у відповідних розрядах виходів блоку ключів, виходах блоку дешифраторів поєднань символів та середніх адрес пам'яті основ будуть коди "0, 0, …, 0";

– якщо символ С*n* є закінченням та словом без основи, то будуть вибрані адреси відповідних комірок пам'яті закінчень та пам'яті основ, у яких зберігаються результати для закінчення та слова Сn відповідно.

Якщо вхідне слово складається з двох літер (символів) С1 та С2 (вони ж при аналізі закінчень інтерпретуються як С*n*-1 та С*n*), то блоком ключів символ С2 не буде заблокований, і його дешифрований код через блок дешифраторів поєднань символів надійде на середні адреси пам'яті основ. На виходах блоку комутаторів будуть присутні коди "0, …, С*n*-1, С*n*", і в пам’яті закінчень буде обрано адресу результату аналізу можливого закінчення “С*n*-1, С*n*”. На відповідному виході других інформаційних виходів пам’яті закінчень буде зчитано код результату. При цьому керуючий вихід БМУ “Читання пам’яті закінчень” встановлюється в “1”.

Аналогічно виконується аналіз довільного ланцюжка символів “С1, …, С*n*-1, С*n*”.

Після аналізу закінчення та основи вхідного слова БМУ переходить до інтерпретації мікропрограми видачі результату аналізу. Спочатку на інформаційний вихід АМП передається результат аналізу основи вхідного слова. При цьому керуючий вихід БМУ "Читання пам’яті слів" встановлюється в "1", що дозволяє читання пам’яті основ і виконує лічення в лічильнику адрес результату, або вибирати послідовні (+1") комірки результату. Кількість комірок, в яких зберігається результат, є змінною величиною і залежать від конкретної основи. Кінцеві комірки кожного такого результату містять коди, наприклад, “0D0AH”, що означає кінець передачі результату проаналізованої основи. При цьому інформаційні виходи пам’яті слів підключені до першого інформаційного входу другої схеми порівняння, вихід якої, встановлений "1", надходить на відповідний керуючий вхід БМУ і сигналізує про закінчення передачі результату основи.

Потім на інформаційний вихід АМП передається код комірки, який містить результат аналізу закінчення (при цьому керуючий сигнал БМУ "Читання пам’яті закінчень = 1" активний).

При кожній передачі слова результату на інформаційний вихід АМП в блоці мікропрограмного управління аналізується керуючий вхід “Читання=1”, який сигналізує про закінчення передачі чергового слова результату.

При завершенні передачі на інформаційний вихід АМП кодів всіх комірок результату БМУ на своєму керуючому виході встановлює внутрішній сигнал “Скидання=1”, який встановлює в “нуль” відповідні регістри і лічильники, і алгоритм роботи АМП переходить в режим очікування прийому чергового слова для аналізу.

На описану вище структуру АМП отримано патент на винахід [16].

На рис. 21 наведено діаграму залежності часу роботи АМП (кількості тактів оброблення) від довжини слова, яке аналізується, (кількості символів у слові).

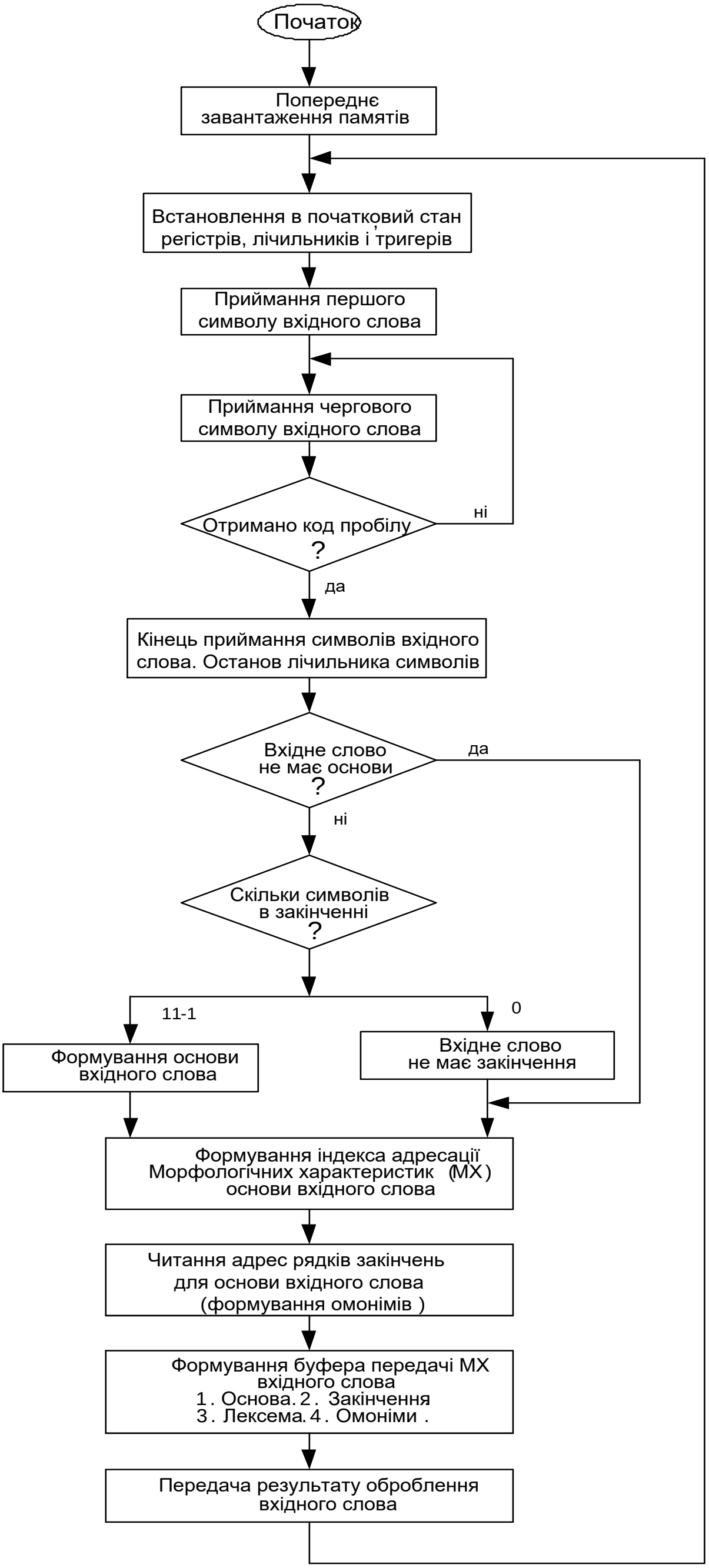

Рис. 20 Алгоритм роботи АМП

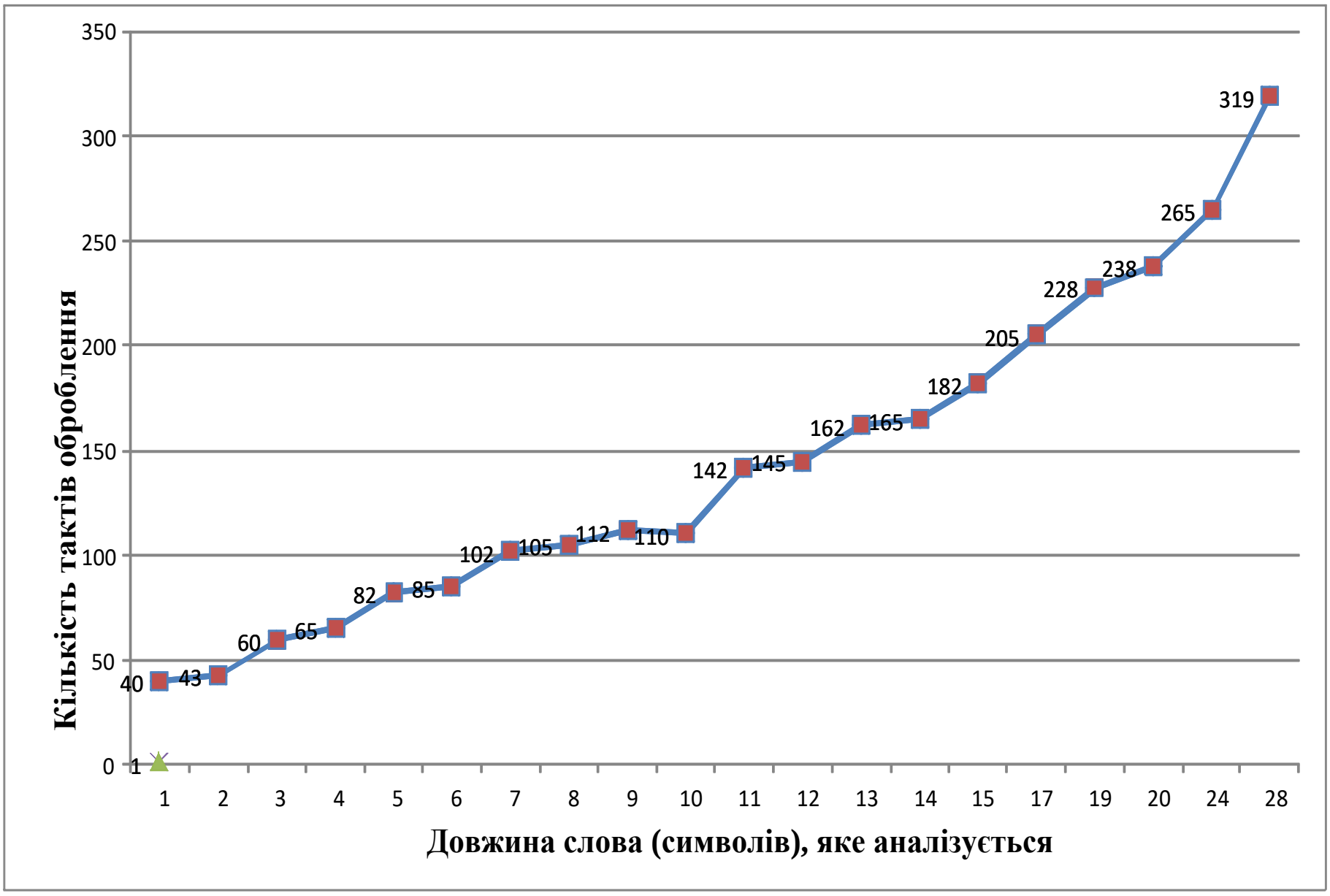

Рис. 21. Діаграма залежності часу роботи АМП від довжини слова

### 5.3.5 *Оцінки складності структурної реалізації АМП*

Вище було розглянуто проєктування АМП першого типу (з використанням ПЛІС-технології) для трьох варіантів лінгвістичних корпусів текстів, наведено продуктивність такого АМП порівняно із програмним способом реалізації морфологічного аналізу для загальновживаної лексики української мови.

У табл. 4 наведено показники моделювання АМП.

Таблиця 4

Порівняльний аналіз показників морфологічного оброблення для програмної й апаратної реалізацій АМП

| Вид реалізації | Середня довжина слова | Час оброблення (мкс) | Збільшення продуктивності для апаратної реалізації (разів) |
|---|---|---|---|
| Програмний | 9 | 937 | **252** |
| Апаратний | 9 | 3,72 | |

Видається доцільним розглянути узагальнену архітектуру АМП другого типу (спеціалізованої розробки) та порівняти витрати обладнання та продуктивність для першого та другого типів реалізації АМП. Така архітектура представлена на рис. 22. Для неї прийнято такі угоди.

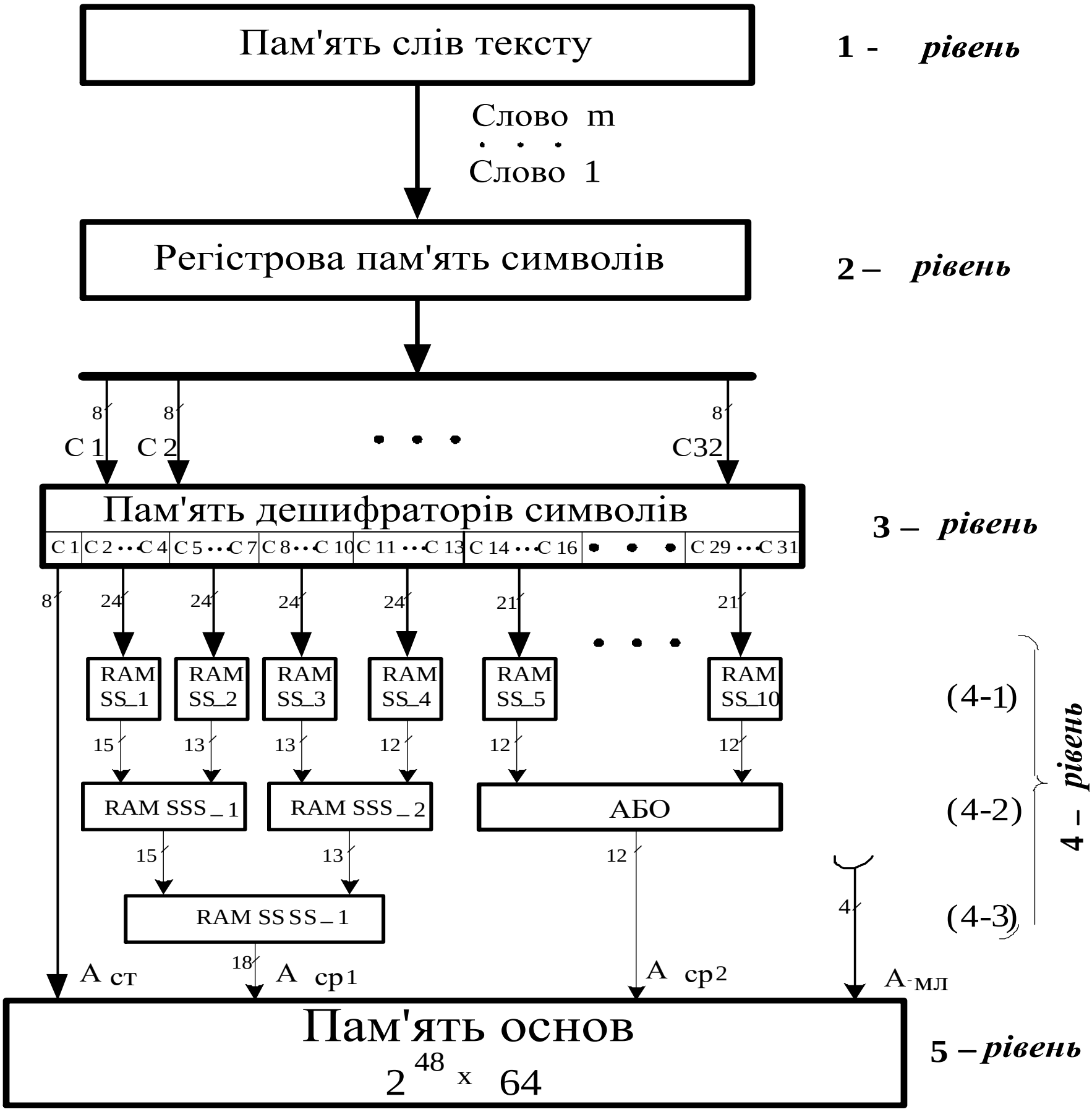

Рис. 22. Узагальнена архітектура АМП другого типу

Дослідження показали, що схеми управління пам’яттю АМП займають порівняно невелику частину обладнання від загального об’єму, і тому їх можна не враховувати.

Пам’ять слів тексту займає 214 комірок, що дорівнює середньому за обсягом науково-технічному тексту.

Реєстрова пам’ять символів та пам’ять дешифраторів символів розраховані на максимальну довжину слів загальновживаної лексики української мови (32 символи).

При дешифрації символів враховується 8-бітовий код, а не 6, як розглядалося в АМП першого типу, що дозволяє обробляти тексти українською, російською та англійською мовами, а також враховувати низку спеціальних символів.

Порівняння виконано для витрат обладнання на реалізацію АМП для ЛКТ варіанта А відповідно до архітектурно-структурної організації, представленої на рис. 18.

Як видно з діаграми (рис. 16), кількість слів завдовжки 14 символів і більше різко скорочується порівняно з кількістю слів меншої довжини. Тому можна прийняти низку обмежень, найбільш істотним з яких є "збирання по АБО" виходів пам’ятей поєднань символів SS_5 – SS_10.

Результати порівняльного аналізу витрат обладнання АМП першого та другого типів для морфологічного аналізу слів загальновживаної лексики української мови представлені у табл. 5.

З таблиці видно, що витрати пам’яті для АМП другого типу істотно нижчі порівняно з витратами пам’яті для АМП першого типу при одночасному підвищенні швидкодії на порядок та розширенні функціональних можливостей. Це пов’язано з проблемно-орієнтованою структурною організацією АМП і вибором для кожного рівня архітектури необхідних за об’ємом і розрядністю чипів пам’яті. При цьому, як зазначалося вище, складність розроблення АМП другого типу значно вище.

Таблиця 5

Порівняльний аналіз витрат обладнання (пам’яті) на реалізацію АМП з використанням ПЛІС і спеціалізованого розроблення

| Тип обладнання | Витрати пам’яті (біт) по рівням ієрархії $m$=1,6•$10^4$ | | | | | | Всього по рівням 1–4 | 5-ур-нь |
|---|---|---|---|---|---|---|---|---|
| | 1-ур-нь | 2-ур-нь | 3-ур-нь | 4-уровень | | | | |
| | | | | 4-1 | 4-2 | 4-3 | | |
| На базі ПЛІС | 4,2•$10^6$ | 2,6•$10^2$ | 4,7•$10^4$ | 2,5•$10^7$ | 1,2•$10^9$ | – | 1,23•$10^9$ | 2,9•1017 |
| Спеціалізоване розроблення | 4,2•$10^6$ | 2,6•$10^2$ | 5,4•104 | 1,1•$10^9$ | 4,5•$10^9$ | 4,8•$10^9$ | 10,41•$10^9$ | 2,8•$10^{14}$ |

| Продуктивність АМП у порівнянні з програмним способом реалізації морфологічного аналізу (разів) | |
|---|---|
| З використанням ПЛІС | **2,5•$10^2$** |
| Спеціалізоване обладнання | **2,6•$10^3$** |

**Висновки за розділом**

1. Виконано аналіз особливостей комп’ютерного оброблення лінгвістичного корпусу текстів надвеликих об’ємів, який показав, що для застосунків, які працюють у реальному режимі часу, програмної реалізації лінгвістичного (і особливо морфологічного) аналізу недостатньо, оскільки частина інформації може бути необробленою. Тому завдання побудови апаратних лінгвістичних процесорів є актуальним, і його вирішення дозволить: по-перше – скоротити терміни надання користувачеві оперативної інформації (без втрати частини інформації та зниження її актуальності) для прийняття рішень; по-друге – якісно підвищити рівень лінгвістичних досліджень за рахунок можливості врахування більшої кількості параметрів обробки.

2. Запропоновано метод багаторівневих проекцій розроблення структурної організації апаратного лінгвістичного процесора з урахуванням онто-логічного підходу, суть якого полягає у поетапному відображенні алгоритмів оброблення інформації у мережі операційних автоматів апаратного лінгвістичного процесора. Запропоновано модель апаратних засобів реалізації апаратного лінгвістичного процесора у вигляді асинхронних комбінаційних автоматів з пам’яттю, що дозволяє виконувати паралельне оброблення словоформ речення, що допускає можливість реконфігурації структури апаратного лінгвістичного процесора на фізичному рівні як інструменту налаштування (перебудови) на оброблення лінгвістичного корпусу текстів заданої предметної області або розв’язання задачі предметно-проблемної орієнтації апаратних засобів.

3. Розроблено функціональну схему паралельного апаратного лінгвістичного процесора, яка включає підсистеми графематичного, морфологічного, синтаксичного та поверхнево-семантичного аналізу. Для кожної з них розроблено граф-схеми алгоритмів та відповідні процедури.

4. Розроблено архітектурно-структурну організацію апаратних засобів, які реалізують етап морфологічного аналізу для трьох варіантів лінгвістичного корпусу текстів: загальновживаної лексики української мови, онтологічного інжинірингу та онтолого-керованих інформаційних систем загального призначення.

5. Виконаний порівняльний аналіз показників морфологічної обробки для програмної та апаратної реалізацій морфологічного аналізу показав, що підвищення продуктивності для апаратних засобів, виконаних за технологією програмовних логікових матриць, склало два порядки, а для апаратних засобів структурно-орієнтованої розробки – три порядки. Перевагами апаратного морфологічного процесора другого типу порівняно з апаратним морфологічним процесором першого типу є зниження на три порядки об’єму пам’яті, підвищення на порядок швидкодії та розширення функціональних можливостей.

**Перелік посилань до розділу 5**

1. Палагин А.В., Крывый С.Л., Петренко Н.Г. Онтологические методы и средства обработки предметных знаний. Луганск: изд-во ВНУ им. В. Даля, 2012. 323 с.
2. Рыков В.В. Управление знаниями. URL: http://rykkypc2.narod.ru/part2.doc. (дата звернення: 17.05.2011).
3. Гаазе-Рапопорт М.Г., Поспелов Д.А. Структура исследований в области искусственного интеллекта. М.: Радио и связь, 1992. С. 5–20.
4. Палагін О.В., Кривий С.Л., Петренко М.Г., Бібіков Д.С. Алгебро-логічний підхід до аналізу та обробки текстової інформації. Проблеми програмування. Спеціальний випуск. 7-а міжн. наук.-практ. конф. з програмування “УкрПРОГ′2010”. Україна, Київ, 25-27 травня, 2010 р. № 2, 3. С. 318–329. URL: http://dspace.nbuv.gov.ua/handle/123456789/14629. (дата звернення: 23.03.2022).
5. Лингвистический процессор для сложных информационных систем Апресян Ю.Д. и др. М.: Наука, 1992. 256 с.
6. Леонтьева Н.Н. К теории автоматического понимания естественных текстов. Часть 1. Моделирование системы "мягкого понимания" текста: информационно-лингвистическая модель. М., МГУ, 2000. 43 с.
7. Леонтьева Н.Н. К теории автоматического понимания естественных текстов. Часть 2. Семантические словари: состав, структура, методика создания. М.: МГУ, 2001. 41 с.
8. Леонтьева Н.Н. Автоматическое понимание текстов: системы, модели, ресурсы: учеб. пособие для студ. лингвист. вузов. М.: Академия, 2006. 303 с.
9. Палагин А.В., Опанасенко В.Н. Реконфигурируемые вычислительные системы: моногрфия. К.: Просвіта, 2006. 280 с.
10. Палагин А.В., Петренко Н.Г. Методологические основы разработки лингвистического процессора для обработки лингвистических корпусов текстов сверхбольших объемов. I. – УСиМ. – 2014. – № 2. – С. 44–57. URL: http://usim.org.ua/arch/2014/2/9.pdf. (дата звернення: 23.03.2023).
11. Палагин А.В., Петренко Н.Г. Методологические основы разработки лингвистического процессора для обработки лингвистических корпусов текстов сверхбольших объемов. II. – УСиМ. – 2014. – № 3. – С. 18–27. URL: http://usim.org.ua/arch/2014/3/4.pdf. (дата звернення: 23.03.2023).
12. Палагин А.В. К решению основной задачи эмуляции. УСиМ, 1980. №3. С. 24–28.
13. Kurgaev, A.F., Petrenko, N.G. Processor structure design. Cybern Syst

Anal 31, 618–625 (1995). DOI: https://doi.org/10.1007/BF02366417.

14. Петренко М.Г., Палагін О.В., Величко В.Ю., Кривий С.Л. Розробка методів та засобів онтолого-лінгвістичного аналізу природномовних об’єктів. Київ: Ін-т кібернетики НАН України, 2009. 38 с. (Препринт / НАН України, Ін-т кібернетики ім. В. М. Глушкова; 2009-2).
15. Палагін О.В., Світла С.Ю., Петренко М.Г., Величко В.Ю. Про один підхід до аналізу та розуміння природномовних об’єктів. *Комп’ютерні засоби, мережі та системи*, 2008. №7. С. 128–137.
16. Пристрій морфологічного аналізу природномовних текстів: пат. 104225 Україна, МПК (2012.01). G06F 15/00, G06F 15/16 (2006.01). заявл. 19.06.12; опубл. 10.01.2014, бюл. №1. 13 с.

## Загальні висновки

Монографія присвячена вирішенню сукупності актуальних науково-прикладних проблем, основними з яких є розроблення системології трансдисциплінарних наукових досліджень і побудови наукової картини світу, або що теж саме – глобальної системи трансдисциплінарних знань. Зокрема, розглянуто питання ефективної підтримки трансдисциплінарних наукових досліджень, які передбачають формування та системний аналіз сервіс-орієнтованої парадигми ноосферогенезу, що задається ланцюжком: ноосферогенез – трансдисциплінарність – інформатика – онтологічна концепція – наукова картина світу – перспективні інформаційні технології. Сутнісна функція, місце та послідовність концептів у цьому ланцюжку чітко визначені і, по суті, становлять методологічну основу сучасних наукових досліджень як основи розвитку цивілізації. Ноосферологія – це цілісна сукупність знань і гармонійна взаємодія у системі “Людина – Природа” під керівництвом свідомості людини та волі людини. Розвиток NBIC-кластера конвергенції відкриває широкі, поки що повністю не оцінені можливості глобального знання-орієнтованого Інтернету, а з ним і всієї сучасної цивілізації. Очевидно, цей розвиток йтиме шляхом створення спочатку прикладних розподілених систем у конкретних предметних областях (Інтернет-речей, Smart-системи в телемедицині, екологічному моніторингу, інформаційному супроводі товарів і послуг, енергетичних системах, комунальних службах тощо). Центральне місце в них займуть Grid-, Blockchain-технології та Cloud-computing, а також віртуальні організації, структури та сервіси. Показано, що сутність трансдисциплінарного підходу до дослідження комплексних науково-технічних проблем полягає в ефективному забезпеченні двоєдності концепцій поглиблення конкретних знань у предметній області, з одного боку, та розширення охоплення проблеми, виходячи з реальності єдності світу, та прагнення відтворити цілісну наукову картину світу – з іншого.

Онтологічний підхід до представлення й інтеграції наукових знань дозволяє створити ефективні засоби побудови систем і технологічну базу системології трансдисциплінарної взаємодії і онтологічного інжинірингу, а відповідний інструментарій дозволяє побудувати мовно-онтологічну картину світу (різновидність лексикографічної системи), розглядаючи її як складову наукової картини світу, яка є основою трансдисциплінарної концепції наукових досліджень. У цьому випадку мовно-онтологічна картина світу виконує функції категоріальної надбудови баз знань в конкретних предметних областях й інтегрованих сховищах знань. Доречно відмітити, що онтологічна парадигма почала і розвивалась практично одночасно з віртуальною. На сьогодення в повсякденну практику увійшли такі поняття, як віртуальний світ, віртуальна організація, віртуальна лабораторія, віртуальна система, віртуальна адресація тощо.

Запропоновано методологічні засади системно-онтологічного аналізу предметно-орієнтованої області знань, розроблена методика формалізованого проектування онтології предметної області та пов'язаного з нею проблемного простору, що включає онтологію об'єктів, онтологію процесів і онтологію задач. Методика враховує категоріальний рівень знань і використовує в якості джерела знань лінгвістичний корпус текстів.

Запропоновано архітектуру онтологічної системи оброблення баз даних наукових публікацій та алгоритми його функціонування на підготовчому й основному етапах. Наведено приклади запитів до бази даних наукових публікацій, які демонструють працездатність онтологічної системи, яка дає науковцю змогу значно пришвидшити отримання необхідної когнітивно-структурованої інформації із своїх джерел. Надалі слід розширити використання в розробці технологій, таких як когнітивні семантика та графіка, мультимедійне подання інформації, орієнтовані на ефективну підтримку процесів екстракції і/або генерації нових знань.

Розроблено метод побудови онтолого-керованих лінгвістичних процесорів який забезпечує налаштування на різні предметні області. Властивість онтологічного керування передбачає включення в алгоритм семантичного аналізу активну взаємодію з інформаційною структурою мовно-онтологічної картини світу, яка підвищує ступінь точності розпізнавання семантичних конструкцій речень тексту. Метод забезпечує підвищення на 2-3 порядки продуктивність при виконанні процедур лінгвістичного аналізу корпусів текстів великого об'єму, що особливо важливо в системах реального часу. Отриманий ефект заснований на обліку статистичних характеристик повної множини словоформ даної мови (при побудові архітектури і структури лінгвістичного процесора).

Розглянуто питання побудови знання-орієнтованих інформаційних систем з онтолого-керованою архітектурою та формальних комп'ютерних онтологій конкретних ПдО, що можуть бути ефективно використані при розробці системології трансдисциплінарних наукових досліджень, створенні концептуально-понятійного каркасу відповідних наукових теорій та інтегрованої системи знань.

Палагін Олександр Васильович
Петренко Микола Григорович
Кривий Сергій Лук'янович
Бойко Микола Олександрович
Малахов Кирило Сергійович

# ОНТОЛОГІЧНЕ ОБРОБЛЕННЯ ТРАНСДИСЦИПЛІНАРНИХ ПРЕДМЕТНИХ ЗНАНЬ

**Монографія**

# Scientific publication (issue)
# Monograph

Published in the author's edition

| | |
|---|---|
| Oleksandr Palagin | Academician of the National Academy of Sciences of Ukraine, DSc (Doctor of Sciences in Technical Sciences), PhD, Professor, Honored Inventor of Ukraine, Deputy Director for Research of the Glushkov Institute of Cybernetics of the National Academy of Sciences of Ukraine, Head of the Microprocessor Technology Lab.<br>https://orcid.org/0000-0003-3223-1391<br>palagin_a@ukr.net |
| Mykola Petrenko | DSc (Doctor of Sciences in Technical Sciences), Lead Researcher, Microprocessor Technology Lab, Glushkov Institute of Cybernetics of the National Academy of Sciences of Ukraine.<br>http://orcid.org/0000-0001-6440-0706<br>petrng@ukr.net |
| Sergii Kryvyi | DSc (Doctor of Sciences in Physics and Mathematics), Professor of Computer Science and Cybernetics department of Taras Shevchenko National University of Kyiv.<br>https://orcid.org/0000-0003-4231-0691<br>sl.krivoi@gmail.com |
| Mykola Boyko | MSc, Researcher, Microprocessor Technology Lab, Glushkov Institute of Cybernetics of the National Academy of Sciences of Ukraine.<br>https://orcid.org/0000-0003-1723-5765<br>xel@ukr.net |
| Kyrylo Malakhov | MSc, Researcher, Backend developer, DevOps engineer, Microprocessor Technology Lab, Glushkov Institute of Cybernetics of the National Academy of Sciences of Ukraine, Member of the expert subgroup on technical issues and architecture of telemedicine within the Interdepartmental Working Group for the development of the concept of implementation of telemedicine in Ukraine.<br>https://orcid.org/0000-0003-3223-9844<br>k.malakhov@incyb.kiev.ua; k.malakhov@outlook.com<br>https://linktr.ee/malakhovks |

## ONTOLOGY-DRIVEN PROCESSING OF TRANSDISCIPLINARY DOMAIN KNOWLEDGE